\documentclass[natbib,numbook,smallextended]{svjour3}

\smartqed  

\usepackage{amsmath}
\usepackage{amssymb}
\usepackage{amsfonts}
\usepackage{mdwlist}
\usepackage{xcolor}
\usepackage[colorlinks=true,linkcolor=blue,citecolor=blue,urlcolor=blue]{hyperref}

\usepackage{graphicx}

\allowdisplaybreaks

\newcommand{\be}{\begin{equation}}
\newcommand{\ee}{\end{equation}}
\newcommand{\dst}{\displaystyle}

\newcommand{\md}{\mathrm{d}}
\newcommand{\mD}{\mathrm{D}}
\newcommand{\grad}{\mathrm{grad}}

\newcommand{\xa}{\mathbf{x}_a}
\newcommand{\pa}{\mathbf{p}_a}

\newcommand{\bx}{{\bf x}}
\newcommand{\bp}{{\bf p}}
\newcommand{\bn}{{\bf n}}
\newcommand{\bm}{{\overline{m}}}

\newcommand{\vecn}{\mathbf{n}}
\newcommand{\vecp}{\mathbf{p}}

\newcommand{\vecr}{\mathbf{r}}
\newcommand{\vecx}{\mathbf{x}}
\newcommand{\vecy}{\mathbf{y}}
\newcommand{\vecz}{\mathbf{z}}

\newcommand{\pipi}{\mathbf{p}_1^2}
\newcommand{\pipip}{(\mathbf{p}_1^2)}
\newcommand{\piipii}{\mathbf{p}_2^2}
\newcommand{\piipiip}{(\mathbf{p}_2^2)}
\newcommand{\pipii}{(\mathbf{p}_1\cdot\mathbf{p}_2)}
\newcommand{\npi}{(\mathbf{n}_{12}\cdot\mathbf{p}_1)}
\newcommand{\npii}{(\mathbf{n}_{12}\cdot\mathbf{p}_2)}

\newcommand{\np}{(\mathbf{n}\cdot\mathbf{p})}

\newcommand{\npj}{({\bf n}_{12}\cdot{\bf p}_2)}
\newcommand{\pjpj}{{\bf p}_2^2}
\newcommand{\pjpjp}{({\bf p}_2^2)}
\newcommand{\pipj}{({\bf p}_1\cdot{\bf p}_2)}

\newcommand{\picSin}{((\mathbf{p}_1 \times \mathbf{S}_1) \cdot \mathbf{n}_{1 2})}
\newcommand{\piicSin}{((\mathbf{p}_2 \times \mathbf{S}_1) \cdot \mathbf{n}_{1 2})}
\newcommand{\piicSiin}{((\mathbf{p}_2 \times \mathbf{S}_2) \cdot \mathbf{n}_{1 2})}
\newcommand{\picSiin}{((\mathbf{p}_1 \times \mathbf{S}_2) \cdot \mathbf{n}_{1 2})}
\newcommand{\picSipii}{((\mathbf{p}_1 \times \mathbf{S}_1) \cdot \mathbf{p}_2)}
\newcommand{\SiSii}{(\mathbf{S}_1 \cdot \mathbf{S}_2)}

\newcommand{\Sipi}{(\mathbf{S}_1 \cdot \mathbf{p}_1)}
\newcommand{\Siipii}{(\mathbf{S}_2 \cdot \mathbf{p}_2)}
\newcommand{\Sipii}{(\mathbf{S}_1 \cdot \mathbf{p}_2)}
\newcommand{\Siipi}{(\mathbf{S}_2 \cdot \mathbf{p}_1)}
\newcommand{\Sin}{(\mathbf{S}_1 \cdot \mathbf{n}_{1 2})}
\newcommand{\Siin}{(\mathbf{S}_2 \cdot \mathbf{n}_{1 2})}

\newcommand{\SiSi}{\mathbf{S}_1^2}

\newcommand{\SipiNW}{({\mathbf{S}}_1 \cdot \mathbf{p}_1)}

\newcommand{\SipiiNW}{({\mathbf{S}}_1 \cdot \mathbf{p}_2)}

\newcommand{\SinNW}{({\mathbf{S}}_1 \cdot {\mathbf{n}}_{1 2})}

\newcommand{\pinNW}{(\mathbf{p}_1 \cdot {\mathbf{n}}_{1 2})}
\newcommand{\piinNW}{(\mathbf{p}_2 \cdot {\mathbf{n}}_{1 2})}

\newcommand{\SiSiNW}{{\mathbf{S}}_1^2}

\newcommand{\pitt}[2]{\pi^{#1#2}_{\rm TT}}
\newcommand{\piti}[2]{\widetilde{\pi}^{#1#2}}
\newcommand{\htt}[2]{h^{\rm TT}_{#1#2}}
\newcommand{\ghtt}[3]{h^{\rm TT}_{#1#2,#3}}

\newcommand{\relab}[1]{r_{#1}}
\newcommand{\rel}{\relab{12}}

\newcommand{\vmom}[1]{\mathbf{p}_{#1}}

\newcommand{\vnun}{\mathbf{n}_{12}}

\newcommand{\scpm}[2]{(#1\cdot#2)}

\newcommand{\vspin}[1]{\mathbf{S}_{#1}}

\newcommand{\naap}{n_{{a}a'}}
\newcommand{\mapr}{m_{a'}}
\newcommand{\raap}{r_{{a}a'}}
\newcommand{\sumbp}{\sum_{b'\ne a'}}
\newcommand{\suma}{\sum_a}
\newcommand{\sumb}{\sum_{b\ne a}}

\newcommand{\Pf}{{\mathrm{Pf}}}

\newcommand{\me}{\mathrm{e}}
\newcommand{\mi}{\mathrm{i}}

\newcommand{\rL}{\mathcal{L}}
\newcommand{\rG}{\mathcal{G}}

\newcommand{\rF}{\mathcal{F}}

\newcommand{\gE}{{\gamma_\textrm{E}}}
\newcommand{\sphys}{s_\textrm{phys}}

\newcommand{\heobeff}{H^{\text{EOB}}_{\text{eff}}}
\newcommand{\heobeffr}{\hat{H}^{\text{EOB}}_{\text{eff}}}

\journalname{Living Reviews in Relativity}

\begin{document}

\title{Hamiltonian formulation of general relativity and post-Newtonian dynamics of compact binaries%
  \thanks{This article is a revised version of \url{https://doi.org/10.1007/s41114-018-0016-5}}
  }
\titlerunning{Hamiltonian formulation of GR and PN dynamics of compact binaries}

\author{Gerhard Sch\"afer \and Piotr Jaranowski}

\institute{
G.\ Sch\"afer \at
Friedrich-Schiller-Universit\"at Jena, Jena, Germany\\
\email{gerhard.schaefer@uni-jena.de}
\and
P. Jaranowski \at
University of Bia{\l}ystok, Bia{\l}ystok, Poland\\
\email{p.jaranowski@uwb.edu.pl}
}

\date{Received: / Accepted:}

\maketitle

\begin{abstract}

Hamiltonian formalisms provide powerful tools for the computation of approximate analytic solutions of the Einstein field equations.
The post-Newtonian computations of the explicit analytic dynamics and motion of compact binaries
are discussed within the most often applied Arnowitt--Deser--Misner formalism.
The obtention of autonomous Hamiltonians is achieved by the transition to Routhians.
Order reduction of higher derivative Hamiltonians results in standard Hamiltonians.
Tetrad representation of general relativity is introduced for the tackling of compact binaries with spinning components.
Compact objects are modeled by use of Dirac delta functions and their derivatives.
Consistency is achieved through transition to $d$-dimensional space and application of dimensional regularization.
At the fourth post-Newtonian level, tail contributions to the binding energy show up for the first time.
The conservative dynamics of binary systems finds explicit presentation and discussion
through the fifth post-Newtonian order for spinless masses.
For masses with spin Hamiltonians are known through (next-to)$^3$-leading-order spin-orbit and spin-spin couplings
as well as through next-to-leading order cubic and quartic in spin interactions. Parts of those are given explicitly.
Tidal-interaction Hamiltonians are considered through (next-to)$^2$-leading post-Newtonian order.
The radiation reaction dynamics is presented explicitly
through the third-and-half post-Newtonian order for spinless objects,
and, for spinning bodies, to leading-order in the spin-orbit and spin1-spin2 couplings.
The most important historical issues get pointed out.


\end{abstract}

\setcounter{tocdepth}{3}
\tableofcontents

\section{Introduction}
\label{sec:intro}

Before entering the very subject of the article, namely the Hamiltonian treatment of the dynamics of compact binary systems
within general relativity (GR) theory, some historical insight will be supplied.
The reader may find additional history, e.g.,
in \cite{damour2-83,damour3-87}, \cite{futamase-07}, \cite{blanchet-14}, \cite{porto-16}, \cite{levi-20}.

\subsection{Early history (1916--1960)}

The problem of motion of many-body systems is an important issue in GR (see, e.g., \citealp{damour2-83,damour3-87}).
Earliest computations were performed by Droste, de~Sitter, and Lorentz in the years 1916--1917,
at the first post-Newtonian (1PN) order of approximation of the Einstein field equations,
i.e., at the order $n=1$, where $(1/c^2)^n$ corresponds to the $n$th post-Newtonian (PN) order with $n=0$ being the Newtonian level.
Already in the very first paper, where Droste calculated the 1PN gravitational field for a many-body system \citep{droste-16},
there occurred a flaw in the definition of the rest mass $m$ of a self-gravitating body of volume $V$
(we follow the Dutch version; the English version contains an additional misprint),
reading, in the rest frame of the body, indicated in the following by $\dot{=}$,
\be
\label{mdef1}
m\quad \stackrel{\text{\scriptsize{Droste 1916}}}{=}
\int_V \md^3 x\,\varrho~\dot{=} \int_V \md^3 x\,\varrho_*\left(1-\frac{3U}{c^2}\right),
\ee
where the ``Newtonian'' mass density $\varrho_*=\sqrt{-g}\varrho u^0/c$
[$g=\det(g_{\mu\nu})$, $u^0$ is the time component of the four-velocity field $u^{\mu}$, $u^{\mu}u_{\mu}=-c^2$]
fulfills the metric-free continuity equation
\be
\partial_t \varrho_* + {\rm div}(\varrho_*{\bf v}) = 0,
\ee
where ${\bf v} = (v^i)$ is the Newtonian velocity field (with $v^i = cu^i/u^0$).
The Newtonian potential $U$ is defined by
\be
\Delta U = -4\pi G \varrho_*,
\ee
with the usual boundary condition for $U$ at infinity: $\lim_{|\mathbf{r}|\to\infty}U(\mathbf{r},t)=0$.
Let us stress again that the definition \eqref{mdef1} is \emph{not} correct.
The correct expression for the rest mass contrarily reads, at the 1PN level,
\be
m ~\dot{=} \int_V \md^3x\, \varrho_*\left(1 + \frac{1}{c^2}\left(\Pi -\frac{U}{2}\right)\right),
\ee
with specific internal energy $\Pi$. For pressureless (dust-like) matter
(for a dust-like body $\Pi=0$, but then the potential term $U$ has to disappear too,
because of the internal pressure-gravity balance: a pressureless body cannot show up internal gravity),
the correct 1PN expression is given by
\be
m = \int_V \md^3x\, \varrho_* ~\dot{=} \int_V \md^3x \sqrt{\det(g_{ij})}\,\varrho = \int_V \md V \varrho,
\ee
where $\md V\equiv\sqrt{\det(g_{ij})}\,\md^3x$.

The error in question slept into second of two sequential papers by \cite{sitter1-16,sitter2-16,sitter-16-e}
when calculating the 1PN equations of motion for a many-body system.
Luckily, that error had no influence on the de Sitter precession of the Moon orbit around the Earth in the gravitational field of the Sun.
The error became identified (at least for dusty matter) by \cite{eddington-38}.
On the other side, \cite{levicivita1-37} used the correct rest mass formula for dusty bodies.
Einstein criticized the calculations by Levi-Civita because he was missing pressure for stabilizing the bodies.
Hereupon, Levi-Civita argued with the ``effacing principle'', inaugurated by Brillouin,
that the internal structure should have no influence on the external motion.
The 1PN gravitational field was obtained correctly by Levi-Civita but errors occurred in the equations of motion
including self-acceleration and wrong periastron advance (\citealp{levicivita2-37}, \citealp{damour-88}).
Full clarification was achieved by \cite{eddington-38}, letting aside the unstable interior of their dusty balls.
Interestingly, in a 1917 paper by Lorentz and Droste (in Dutch),
the correct 1PN Lagrangian of a self-gravitating many-body system of fluid balls was obtained but never properly recognized.
Only in 1937, for the edition of the collected works by Lorentz, it became translated into English \citep{lorentz-17}.
A full-fledged calculation made by \citet*{einstein-38}---posed in the spirit of Hermann Weyl
by making use of surface integrals around field singularities---convincingly achieved the 1PN equations of motion,
nowadays called Einstein--Infeld--Hoffmann (EIH) equations of motion.
In the publication seamless following \citet*{einstein-38},
\cite{robertson-38} derived the 1PN periastron advance based on the EIH equations of motion.
Some further refining work by Einstein and Infeld appeared in the 1940s.
\cite{fichtenholz-50} computed the Lagrangian and Hamiltonian out of the EIH equations.
A consistent fluid ball derivation of the EIH equations has been achieved by \cite{fock-39},
\cite{petrova-49} (delayed by World War~II), and \cite{papapetrou-51} (see also \citealp{fock-59}).

In the 1950s, Infeld and Pleba\'nski rederived the EIH equations of motion with the aid of Dirac $\delta$-functions as field sources
by postulating the properties of Infeld's ``good'' $\delta$-function
(\citealp{infeld-54,infeld-57}, \citealp{infeld-60};
see Sect.~\ref{subs:RHreg} of our review for more details).
Also in the 1950s, the Dirac $\delta$-function became applied to the post-Newtonian problem of motion of spinning bodies by \cite{tulczyjew-59},
based on the seminal work by \cite{mathisson-37,mathisson-10},
with the formulation of a general relativistic gravitational skeleton structure of extended bodies.
Equations of motion for spinning test particles had been obtained before by \cite{papapetrou-51a} and \cite{corinaldesi-51}.
Further in the 1950s, another approach to the equations-of-motion problem, called fast-motion or 
post-Minkowskian (PM) approximation, which is particularly useful for the treatment of high-speed scattering problems,
was developed and elaborated by \cite{bertotti-56} and \cite{kerr-59a,kerr-59b,kerr-59c}, at the 1PM level.
First results at the 2PM level were obtained by \cite{bertotti-60}.

\subsection{History on Hamiltonian results}

Hamiltonian frameworks are powerful tools in theoretical physics
because of their capacity of full-fledged structural exploration
and efficient application of mathematical theories
(see, e.g., \citealp{holm-85}, \citealp{alexander-87}, \citealp{vinti-98}, \citealp{boccaletti-04,boccaletti-02}).
Most importantly, Hamiltonians generate the time evolution of all quantities in a physical theory.
For closed systems, the total Hamiltonian is conserved in time.
Together with the other conserved quantities, total linear momentum and total angular momentum,
which are given by very simple universal expressions,
and the boost vector, which is connected with the Hamiltonian density (which defines ``centre-of-energy vector'') and the total linear momentum,
the total Hamiltonian is one of the generators of the globally operating Poincar\'e or inhomogeneous Lorentz group.
A natural ingredient of a Hamiltonian formalism is the (3+1)-splitting of spacetime in space and time.
Consequently Hamiltonian formalisms allow transparent treatments of both initial value problems and Newtonian
limits. Finally, for solving equations of motion, particularly in approximation schemes, Hamiltonian frameworks naturally fit into the
powerful Lie-transform technique based on action-angle variables
(\citealp{hori-66}, \citealp{kinoshita-78}, \citealp{vinti-98}, \citealp{boccaletti-04,boccaletti-02}, \citealp{tessmer-13}).
Lie series are also very useful when treating canonical transformations with usual canonical variables
(see, e.g., \citealp{blumlein-20b,blumlein-20c,blumlein-21}).

Additionally we refer to an important offspring of the Hamiltonian framework,
the effective-one-body (EOB) approach, which will find its presentation in an upcoming \textit{Living Reviews} article by Thibault Damour.
References in the present article referring to EOB are particularly
\cite{buonanno-99,buonanno-00}, \cite{damour2-00}, \cite{damour3-01}, \cite{damour2-08}, \cite{damour-15}, \cite{damour2-16}.

The focus of the present article is on the Hamiltonian formalism of GR as developed by Arnowitt, Deser, and Misner (ADM)
\citep{arnowitt-59,arnowitt1-60,arnowitt2-60}, with its Routhian modification \citep{jaranowski-98,jaranowski-98-e}
(where the matter is treated in Hamiltonian form and the field in the Lagrangian one)
and classical-spin generalization (\citealp{steinhoff2-09}, \citealp{steinhoff-11}),
and with application to the problem of motion of binary systems with compact components including proper rotation (spin)
and rotational deformation (quadratic in the spin variables);
for other approaches to the problem of motion in GR, see the reviews by \cite{futamase-07}, \cite{blanchet-14}, \cite{porto-16}.
The review article by \citet*{arnowitt-62} gives a thorough account of the ADM formalism
(see also \citealp{regge-74} for the discussion about asymptotics).
In this formalism, the final Hamiltonian, nowadays called ADM Hamiltonian,
is given in form of a volume integral of the divergence of a vector over three-dimensional spacelike hypersurface,
which can also naturally be represented as surface integral at flat spatial infinity $i^0$.

It is also interesting to give insight into other Hamiltonian formulations of GR,
because those are closely related to the ADM approach but differently posed.
Slightly ahead of ADM, \cite{dirac-58,dirac-59} had developed a Hamiltonian formalism for GR,
and slightly afterwards, \cite{schwinger1-63,schwinger2-63}.
Schwinger's approach starts from tetrad representation of GR
and ends up with a different set of canonical variables and, related herewith, different coordinate conditions.
Dirac has developed his approach with some loose ends toward the final Hamiltonian
(see Sect.~\ref{sec:2-1} below and also, e.g., \citealp{deser-04}),
but the coordinate conditions introduced by him---nowadays called Dirac gauge---are often used, mainly in numerical relativity.
A subtle problem in all Hamiltonian formulations of GR is the correct treatment of surface terms at spacelike infinity
which appear in the asymptotically flat spacetimes.
In 1967, this problem has been clearly addressed by \cite{dewitt-67}
and later, in 1974, full clarification has been achieved by \cite{regge-74}.
For a short comparison of the three canonical formalisms in question,
the Dirac, ADM, and Schwinger ones, see \cite{schaefer-14}.

The first authors who had given the Hamiltonian as two-dimensional surface integral at $i^0$ on three-dimensional spacelike hypersurfaces were ADM.
Of course, the representation of the total energy as surface integral was known before,
particularly through the Landau--Lifshitz gravitational stress-energy-pseudotensor approach.
Schwinger followed the spirit of ADM. He was fully aware of the correctness of his specific calculations modulo surface terms only which finally 
became fixed by asymptotic Lorentz invariance considerations.
He presented the Hamiltonian (as well as the other generators of the Lorentz group) as two-dimensional surface integrals. 
Only one application of the Schwinger approach by somebody else than Schwinger himself is known to the authors
(apart from \citealp{faddeev-82} who presented Einstein's theory of gravitation in the Schwinger canonical variables).
It is the paper by Kibble in 1963 in which the Dirac spin-1/2 field found a canonical treatment within GR \citep{kibble-63}.
This paper played a crucial role in the implementation of classical spin into the ADM framework by \cite{steinhoff2-09} and \cite{steinhoff-11}
(details can be found in Sect.~\ref{sec:spinADM} of the present article).

The ADM formalism is the most often used Hamiltonian framework in the analytical treatment of the problem of motion of gravitating compact objects.
The main reason for this is surely the very well adapted coordinate conditions for explicit calculations introduced by \citet*{arnowitt3-60}
(generalized isotropic coordinates; nowadays, for short, often called ADMTT coordinates,
albeit the other coordinates introduced by \citealp{arnowitt-62}, are ADMTT too),
though also in Schwinger's approach similar efficient coordinate conditions could have been introduced \citep{schaefer-14}.
Already \cite{kimura-61} started application of the ADM formalism to gravitating point masses at the 1PN level.
In 1974, that research activity culminated in a 2PN Hamiltonian for binary point masses obtained by
\cite{ohta1-74,ohta2-74}, based on earlier work by \cite{hiida-72}.
However, one coefficient of their Hamiltonian was not correctly calculated and the Hamiltonian as such was not clearly identified,
i.e., it was not clear to which coordinate system it referred to.
In 1985, full clarification has been achieved in a paper by \cite{damour1-85} relying on the observation by \cite{schaefer-84}
that the perturbative use of the equations of motion on the action level implies
that coordinate transformations have been applied; also see \cite{barker-84,barker-86}.
In addition, \cite{damour1-85} showed how to correctly compute the delicate
integral ($U^\textrm{TT}$) which had been incorrectly evaluated by \cite{hiida-72}, \cite{ohta1-74,ohta2-74},
and made contact with the first fully correct calculation of the 2PN dynamics of binary systems (in harmonic coordinates)
by \cite{damour1-81}, \cite{damour-82} in 1981--1982.
The 2PN periastron advance for binary systems has been obtained for the first time by \cite{damour-87};
generalized by adding to it the effect of the leading-order spin-orbit coupling, in 1988 \citep{damour-88}.

In \cite{schaefer2-83}, the leading-order 2.5PN radiation reaction force for $n$-body systems was derived by using the ADM formalism.
The same force expression had already been obtained earlier by \cite{schaefer-82} within coordinate conditions
closely related to the ADM ones---actually identical with the ADM conditions through 1PN and at 2.5PN order---and then again by \cite{schaefer1-83},
as quoted in \cite{poisson-14}, based on a different approach but in coordinates identical to the ADM ones at 2.5PN order.
The 2PN Hamiltonian shown by \cite{schaefer-82} and taken from \cite{ohta2-74}, apart from the erroneous coefficient mentioned above,
is the ADM one as discussed above (the factor 7 in the static part therein has to be replaced by 5),
and in the definition of the reaction force in the centre-of-mass system,
a misprinted factor 2 is missing, i.e.\ $2{\bf F}={\bf F}_1-{\bf F}_2$.
The detailed calculations were presented in \cite{schaefer-85};
and in \cite{schaefer-86}, a further ADM-based derivation by use of a PM approximation scheme has been performed.
At 2PN level, the genuine 3-body potential was derived by \cite{schaefer-87}.
However, in the reduction of a 4-body potential derived by \cite{ohta-73,ohta1-74,ohta2-74} to three bodies
made by \cite{schaefer-87} some combinatorical shortcomings slept in,
which were identified and corrected by \cite{lousto-08}, and later by \cite{galaviz-11} in different form.
The $n$-body 3.5PN non-autonomous radiation reaction Hamiltonian\footnote{In such a particle Hamiltonian,
the field degrees of freedom are treated as independent from the particle variables,
rendering the particle Hamiltonian an explicit function of time.}
was obtained by the authors in \cite{jaranowski-97},
confirming energy balance results in \cite{blanchet-89},
and the equations of motion out of it were derived by \cite{koenigsdoerffer-03}.

Additionally within the ADM formalism, for the first time in 2001,
the conservative 3PN dynamics for compact binaries has been fully obtained by Damour and the authors,
by also for the first time making extensive use of the dimensional regularization technique\footnote{
Dimensional regularization was originally introduced by \cite{bollini-72a,bollini-72b} and \cite{thooft-72}.}
\citep{damour1-01} (for an earlier mentioning of application of dimensional regularization
to classical point particles, see \citealp{damour-80,damour2-83};
and for an earlier $n$-body static result,
i.e.\ a result valid for vanishing particle momenta and vanishing reduced canonical variables of the gravitational field,
not based on dimensional regularization, see \citealp{kimura-72}).
Only by performing all calculations in a $d$-dimensional space
the regularization has worked out fully consistently in the limit $d\to3$
(later on, a $d$-dimensional Riesz kernel calculation has been performed too, \citealp{damour3-08}).
In purely 3-dimensional space computations two coefficients, denoted by $\omega_{\rm kinetic}$ and $\omega_{\rm static}$,
could not be determined by analytical three-dimensional regularization.
The coefficient $\omega_{\rm kinetic}$ was shown to be fixable by insisting on global Lorentz invariance
and became thus calculable with the aid of the Poincar\'e algebra
(with value 41/24) \citep{damour3-00,damour3-00-e}.\footnote{
L.\ Blanchet (private communication) and P.\ Bizo\'n and A.\ Staruszkiewicz (private communication)
suggested to the authors of \cite{damour3-00} that the coefficient $\omega_\text{kinetic}$
should be fixable by insisting on global Lorentz invariance.
It found explicit verification by \cite{jaranowski-00}.
L.\ Blanchet had obtained the analytical value of $\omega_\text{kinetic}$
and communicated the three-digit approximate value 1.71 of $\omega_\text{kinetic}$ before completion of \cite{damour3-00}.
Derivation of $\omega_\text{kinetic}$ in harmonic coordinates by \cite{blanchet2-00,blanchet1-01}
crucially relies on the extended Hadamard regularization method, see Sect.\ \ref{subs:EHreg} below.}
The first evaluation of the value of $\omega_{\rm static}$ (namely $\omega_{\rm static}=0$)
was obtained by \cite{jaranowski1-99,jaranowski-00} by assuming a matching with the Brill--Lindquist
initial-value configuration of two black holes. The correctness of this value
(and thereby the usefulness of considering that the Brill--Lindquist initial-value data represent a relevant configuration of two black holes)
was later confirmed by dimensional regularization \citep{damour1-01}.
Explicit analytical solutions for the motion of compact binaries through 2PN order
were derived by \cite{damour-88} and \cite{schaefer2-93,schaefer2-93-e},
and through 3PN order by \cite{memmesheimer-05},
extending the seminal 1PN post-Keplerian parametrization proposed by \cite{damour-85}.

Quite recently, the 4PN binary dynamics has been successfully derived,
using dimensional regularization and sophisticated far-zone matching
(\citealp{jaranowski-12,jaranowski-13}, \citealp{damour-14}, \citealp{jaranowski-15}).
Let us remark in this respect that the linear in $G$ (Newtonian gravitational constant)
part can be deduced to all PN orders from the 1PM Hamiltonian derived by \cite{ledvinka-08}.
For the first time, the contributions to 4PN Hamiltonian were obtained by the authors in \cite{jaranowski-12} through $G^2$ order,
including additionally all log-terms at 4PN going up to the order $G^5$.
Also the related energy along circular orbits was obtained as function of orbital frequency.
The application of the Poincar\'e algebra by \cite{jaranowski-12}
clearly needed the noncentre-of-mass Hamiltonian, though only the centre-of-mass one was published.
By \cite{jaranowski-13}, all terms became calculated with the exception of terms in the Hamiltonian
linear in the symmetric mass ratio $\nu\equiv m_1m_2/(m_1+m_2)^2$
(where $m_1$ and $m_2$ denote the masses of binary system components)
and of the orders $G^3$, $G^4$, and $G^5$.
Those terms are just adding up to the log-terms mentioned above.
However, taking a numerical self-force solution for circular orbits
in the Schwarzschild metric into account,
already the innermost (or last) stable circular orbit
could be determined numerically through 4PN order by \cite{jaranowski-13}.

The computations by \cite{jaranowski-12,jaranowski-13,jaranowski-15}
are all based on a straightforward use of the PN expansion,
and are thereby a priori only valid in the near zone.
The formal extension of the 4PN-level near-zone computation to the full space
implies the appearance of infrared (IR) divergences (linked to the formal limit $r\to\infty$).
The regularization of these IR divergences is unambiguous,
except for a single 4PN-level ambiguity coefficient, denoted by $C$ in \cite{damour-14},
linked to the arbitrariness in the IR regulator scale $s$
entering within a logarithm (see Eq.\ (3.7) in \citealp{damour-14}).
The value of $C$ ($C=-1681/1536$) was, however, uniquely determined in \cite{damour-14}
by combining several other previous results:
1) the understanding that the IR effect responsible for this logarithmic ambiguity
was in precise agreement with a nonlocal 4PN tail effect
discovered long ago \cite{blanchet-88}---and recovered within the ADM formalism by \cite{damour-16};
2) the ``first law of binary black-hole mechanics'' by \cite{letiec-12}
allowing one to link the energy-angular-momentum function $E(j,\nu)$ to the redshift along circular orbits;
and, most importantly from the conceptual point of view,
3) a computation, at first order in the symmetric mass ratio $\nu$, of the redshift by \cite{bini-13},
obtained by using an analytical representation of the (linear in $\nu$) metric perturbation
in terms of series of hypergeometric functions \citep{mano-96}.
The crucial point is that the latter analytical representation
incorporated a precise matching between the near-zone metric and the far-zone one,
thereby providing  the ``beyond-PN'' information needed for the analytical determination of the value of $C$.
Previous results obtained by \cite{letiec-12} and \cite{barausse-12},
based on {\it numerical} self-force computations \citep{blanchet-10},
had given an approximate numerical knowledge of a PN expansion coefficient equivalent to the knowledge of $C$.
Applications of 4PN Hamiltonian dynamics for bound and unbound orbits
were performed by \cite{damour-15}, \cite{bini-17}.

For spinning bodies, counting spin as 0.5PN effect,
the 1.5PN spin-orbit and 2PN spin-spin Hamiltonians were derived by \cite{barker-75,barker-79},
where the given quadrupole-moment-dependent part can be regarded as representing spin-squared terms for extended bodies
(notice the presence of the tensor product of two unit vectors pointing each to the spin direction
in the quadrupole-moment-dependent Hamiltonians).
For an observationally important application of the spin-orbit dynamics, see \cite{damour-88}.
In 2008, the 2.5PN spin-orbit Hamiltonian was successfully calculated by \citet{damour1-08},
and the 3PN spin1-spin2 and spin1-spin1 binary black-hole Hamiltonians by \cite{steinhoff1-08,steinhoff2-08,steinhoff3-08}.
The 3PN spin1-spin1 Hamiltonian for binary neutron stars was obtained by \cite{hergt-10}.
The 3.5PN spin-orbit and 4PN spin1-spin2 Hamiltonians were obtained by \cite{hartung1-11,hartung2-11}
(also see \citealp{hartung-13} and \citealp{levi-14}).
The 4PN spin1-spin1 Hamiltonian was presented in \cite{levi3-16}.
Based on the Dirac approach, the Hamiltonian of a spinning test-particle in the Kerr metric has been obtained by \cite{barausse-09,barausse-09-erratum}.
The canonical Hamiltonian for an extended test body in curved spacetime, to quadratic order in spin, was derived by \cite{vines-16}.
Finally, the radiation-reaction Hamiltonians from the leading-order spin-orbit and spin1-spin2 couplings
have been derived by \cite{steinhoff3-10} and \cite{wang-11}.

\subsection{More recent history on non-Hamiltonian results}

At the 2PN level of the equations of motion, the Polish school founded by Infeld succeeded in getting many expressions
whereby the most advanced result was obtained by \cite{ryten-61} in her MSc thesis from 1961
using as model for the source of the gravitational field Infeld's ``good $\delta$-function''.
Using the same source model as applied by Fock and Petrova,
\cite{kopeikin-85} and \cite{grishchuk-86} derived the 2PN and 2.5PN equations of motion for compact binaries.
However, already in 1982, Damour and Deruelle had obtained the 2PN and 2.5PN equations of motion for compact binaries,
using analytic regularization techniques \citep{damour-82,damour1-83,damour2-83}
(for another such derivation see \citealp{blanchet-98},
who additionally got the metric coefficients at the 2.5PN accuracy).
Also \cite{ohta-88} should be mentioned for a Fokker action derivation of the 2PN dynamics.
Regarding the coordinate conditions used in the papers quoted in the present subsection, treating spinless particles,
all are based on the harmonic gauge with the exceptions of the ones with a Hamiltonian background and those by Ryte\'n or Ohta and Kimura.

The two-point-mass equations of motion at 3PN order in harmonic coordinates were obtained complete
with the exception of one parameter called $\lambda$ (equivalent to $\omega_\text{static}$, see above)
by \cite{blanchet1-00,blanchet2-00} (see also \citealp{deandrade-01} and \citealp{blanchet-03}).
The derivation used the modified version of the Hadamard regularization called the extended Hadamard regularization
(\citealp{blanchet1-01,blanchet2-01}, see Sect.\ \ref{subs:EHreg} of our review for more details).
This regularization was not able to resolve the problem of the ambiguity parameter $\lambda$,
but gives a final result physically equivalent to that of dimensional regularization, except for the unknown value of this parameter.
Using the technique of Einstein, Infeld, and Hoffmann (EIH),
\cite{itoh-03} and \cite{itoh-04} succeeded in deriving the 3PN equations of motion for compact binaries,
and \cite{blanchet-04} derived the same 3PN equations of motion based on dimensional regularization.

The 3.5PN equations of motion were derived within several independent approaches:
by \cite{pati-02} using the method of direct integration of the relaxed Einstein equations (DIRE) developed by \cite{pati-00},
by \cite{nissanke-05} applying Hadamard self-field regularization,
by \cite{itoh-09} using the EIH technique,
and by \cite{galley-12} within the effective field theory (EFT) approach.
Radiation recoil effects, starting at 3.5PN order, have been discussed by
\cite{bekenstein-73}, \cite{fitchett-83}, \cite{junker-92}, \cite{kidder-95}, \cite{blanchet-05}.

\cite{bernard-16} calculated the 4PN Fokker action for binary point-mass systems
and found a nonlocal-in-time Lagrangian inequivalent to the Hamiltonian obtained by \cite{damour-14}.
On the one hand, the local part of the result of \cite{bernard-16}
differed from the local part of the Hamiltonian of \cite{damour-14} only in a few terms.
On the other hand, though the nonlocal-in-time part of the action in \cite{bernard-16} was the same as the one in \cite{damour-14,damour-15}, \cite{bernard-16}
advocated to treat it (notably for deriving the conserved energy, and deriving
its link with the orbital frequency) in a way which was inequivalent to
the one in \cite{damour-14,damour-15}. It was then shown by \cite{damour-16} that:
(i) the treatment of the nonlocal-in-time part in \cite{bernard-16} was not correct,
and that (ii) the difference in local-in-time terms was composed of a combination of gauge terms and of a new ambiguity structure
which could be fixed either by matching to \cite{damour-14,damour-15}
or by using the results of self-force calculations in the Schwarzschild metric.
In their recent articles \citep{bernard-17a,bernard-17b}
Blanchet and collaborators have recognized that the criticisms of \cite{damour-16} were founded,
and, after correcting their previous claims and using results on periastron precession first derived by \cite{damour-15,damour-16},
have obtained full equivalence with the earlier derived ADM results.
Let us emphasize that \cite{marchand-17} (also see \citealp{bernard-17b}) have presented
the first self-contained calculation of the full 4PN dynamics (not making any use of self-force results),
which confirms again the correctness of the 4PN dynamics first obtained by \cite{damour-14}.
That calculation is based on asymptotic expansion of the radiative gravitational field in $d$ dimensions
with matching equations to be regularized first analytically and then dimensionally.
An application of the 4PN dynamics for bound orbits was performed by \cite{bernard-17a}.

The application of EFT approach to PN calculations, devised by \cite{goldberger1-06,goldberger2-06},
has also resulted in PN equations of motion for spinless particles up to the 3PN order
(\citealp{gilmore-08}, \citealp{kol-09}, \citealp{foffa-11}).
At the 4PN level, \cite{foffa-13} calculated a quadratic in $G$ higher-order Lagrangian,
the published version of which was found in agreement with \cite{jaranowski-12}.
The quintic in $G$ part of the 4PN Lagrangian was derived within the EFT approach by \cite{foffa-17}
(with its 2016 arXiv version corrected by \citealp{damour-17}).
\cite{galley-16} got the 4PN nonlocal-in-time tail part.
Then \cite{porto-17a} and \cite{porto-17b} performed a deeper analysis of IR divergences in PN expansions.
Recently, \cite{foffa1-19} and \cite{foffa2-19} succeeded for the first time
with a purely $d$-dimensional derivation of the 4PN dynamics,
without use of any additional regularizations.
This again shows the power of dimensional regularization in PN calculations,
which have been established for the first time at 3PN order by \cite{damour1-01}.

The 1.5PN spin-orbit dynamics was derived in Lagrangian form by \cite{tulczyjew-59} and \cite{damour-82}.
The 2PN spin-spin equations of motion were derived by \cite{death1-75,death2-75},
and \cite{thorne-85}, respectively, for rotating black holes.
The 2.5PN spin-orbit dynamics was successfully tackled by \cite{tagoshi-01},
and by \cite{faye-06}, using harmonic coordinates approach.
Within the EFT approach, \cite{porto-10} and \cite{levi1-10} succeeded in determining the same coupling (also see \citealp{perrodin-11}).
The 3PN spin1-spin2 dynamics was successfully tackled by \cite{porto1-08,porto1-10}
(based on \citealp{porto1-06}, \citealp{porto2-06}) and by \cite{levi2-10},
and the 3PN spin1-spin1 one, again by \cite{porto2-08},
but given in 2010 only in fully correct form \citep{porto2-10}.
For the 3PN spin1-spin1 dynamics, also see \cite{bohe-15}.
The most advanced results for spinning binaries can be found
in \cite{levi-12}, \cite{marsat-13}, \cite{bohe-13}, \cite{marsat-15}, \cite{levi1-16,levi2-16,levi3-16},
reaching 3.5PN and 4PN levels (also see \citealp{steinhoff-17}).
Finally, the radiation-reaction dynamics of the leading-order spin-orbit and spin1-spin2 couplings
have been obtained by \cite{wang-07} and \cite{zeng-07}, based on the DIRE method \citep{will-05}
(see also \citealp{maia1-17,maia2-17}, where the EFT method became applied).
For a review of spin effects in the radiation field, see \cite{blanchet-14}.

\subsection{Most recent history since 2019}

The year 2019 can be regarded as the beginning of the epoch of the calculation of conservative PN approximations beyond 4PN.
These calculations have been dominated by the EFT approach in the treatment of the gravitational field,
working with Lagrangians and action functionals based on harmonic coordinates.
Only at the end of the field calculations, having at hand effective Lagrangians and actions for the matter sources,
the transition to effective Hamiltonians for the particles takes place.
Hereof the effective EOB Hamiltonians can be constructed which are extremely useful objects
for applications and comparisons of different approaches.
Bound binary systems were the first to be addressed at 5PN with calculations of
static potential contributions by \cite{foffa-19} and \cite{blumlein-20b}.
\cite{blumlein-20a} checked their approach by calculating the complete 4PN Hamiltonian for the binary dynamics.

For the calculation of binary dynamics at 5PN and beyond
a new strategy was devised by \cite{bini-19},
later coined ``tutti frutti'' (TF) approach \citep{bini-21}.
This strategy combines various analytical approximation methods: PN (post-Newtonian), PM (post-Minkowskian),
MPM (multipolar post-Minkowskian), EFT (effevtive field theory), SF (gravitational self force),
EOB (effective one body), and Delaunay averaging.
Binary Hamiltonians at 5PN order have been derived by \cite{bini-20}
and by \cite{blumlein-21,blumlein-22}.
Up to three rational numbers, the results do agree.
Details are given in Sect.\ \ref{subsubsec:results5pn}.
The TF approch has become leading through the 6PN order presenting
almost complete (with 4 coefficients still unknown) 6PN effective EOB Hamiltonian
\citep{bini-20b,bini-20c,bini-21}; also see \cite{blumlein-20c,blumlein-21b}.
In Sect.\ \ref{subsubsec:results6pn}, PN-knowledge through 6PN order can be found.

Based on the PM approach, scattering calculations became more and more important
in the determination of the binary Hamiltonian. Here, a new powerful approach entered,
based on advanced calculations of scattering amplitudes using
generalized unitarity, double-copy construction, eikonal resummation, and advanced multiloop integration methods,
in the beginning resulting straight with an ordinary centre-of-mass 2PM binary Hamiltonian
in isotropic gauge (isotropic coordinates for the canonical momentum) \citep{cheung-18},
followed by the first computation of the 3PM two-body Hamiltonian in \cite{bern-19,bern-19b};
also see \cite{kalin-20}, using standard EFT techniques.
Quite recently, the 4PM binary Hamiltonian became available, see \cite{bern-21,bern-22};
also see \cite{dlapa-22a,dlapa-22b}.
Evidently, the $n$PM-order level controls all terms in the corresponding PN approximation through $(n-1)$PN order.
Binary scattering is usually treated in the action language, so Hamiltonians are close by.
The problem is to make sure that the PN parts of the straightforwardly obtained PM Hamiltonians
are a priori applicable to bound binary systems because of different boundary conditions,
see, e.g., \cite{kalin-20}.

Recently, the NNNLO quadratic-in-spin \citep{mandal-22b,kim-22b},
the NLO cubic-in-spin \citep{levi-21a,levi-22},
as well as the quartic-in-spin NLO \citep{levi-21b,levi-23} Hamiltonians were derived;
also the spin-orbit gravitational couplings got obtained
through the NNNLO level \citep{antonelli-20,levi-21c,mandal-22a,kim-22a}, all based on EFT methods.
The complete Hamiltonian for spinning binary systems at 1PM order,
exact to all orders in momentum and spin expansions, was derived in \cite{chung-20}
(also see \citealp{lee-23} for comparison of \citealp{chung-20} with other results).
At the 2PM order, binary dynamics through the fifth power of spin was considered in \cite{bern-23}.

Regarding tidal interactions, Hamiltonians through NNLO post-Newtonian \citep{henry-20a,henry-20b}
and NLO post-Minkowskian \citep{cheung-20,kalin-20b} order corrections are available, again based on EFT
(see also \citealp{bern-21b}).
The Wilson coefficients for rotational deformations, our $C_{Q_a}$,
are called $C^{(0)}_{{\rm E}{\rm S}^2}$ by \cite{mandal-22b} and for tidal ones,
$C^{(2)}_{{\rm E}^2}$, $C^{(0)}_{{\rm E}^2{\rm S}^2}$.
The rotational coefficient starts at the 2PN level
[i.e.\ at $\mathcal{O}(c^{-2}c^{-1}c^{-1})=\mathcal{O}(c^{-4})$, where spins are counted of order $\mathcal{O}(c^{-1})$],
whereas tidal coefficients enter from NNNLO on
[i.e.\ at $\mathcal{O}\big((c^{-2})^3c^{-2}c^{-1}c^{-1}\big)=\mathcal{O}(c^{-10})$, what corresponds to the 5PN level].
Relativistic theory of tidal Love numbers was presented in \cite{binnington-09};
in a post-Newtonian setting, including Hamiltonian constructions,
the leading-order relativistic theory of tides has been developed by \cite{vines-13}.
Effective one-body description of tidal effects was given in \cite{damour-10};
dynamical tides in general relativity were treated in \cite{steinhoff-16}.
More details on tidal interactions can be found in Sect.\ \ref{sec:tidal}.

\subsection{Notation and conventions}

In this article, Latin indices from the mid alphabet are running from 1 to 3 (or $d$ for an arbitrary number of space dimensions),
Greek indices are running from 0 to 3 (or $d$ for arbitrary space dimensions), whereby $x^0=ct$.
We denote by $\mathbf{x}=(x^i)$ ($i\in\{1,\ldots,d\}$) a point in the $d$-dimensional Euclidean space $\mathbb{R}^d$
endowed with a standard Euclidean metric defining a scalar product (denoted by a dot).
For any spatial $d$-dimensional vector $\mathbf{w}=(w^i)$
we define $|{\bf w}|\equiv\sqrt{{\bf w}\cdot{\bf w}}\equiv\sqrt{\delta_{ij}w^iw^j}$,
so $|\cdot|$ stands here for the Euclidean length of a vector,
$\delta_{ij}=\delta^i_j$ denotes Kronecker delta.
The partial differentiation with respect to $x^\mu$ is denoted by $\partial_\mu$ or by a comma, i.e., $\partial_\mu\phi\equiv\phi_{,\mu}$,
and the partial derivative with respect to time coordinates $t$ is denoted by $\partial_t$ or by an overdot, $\partial_t\phi\equiv\dot{\phi}$.
The covariant differentiation is generally denoted by $\nabla$,
but we may also write $\nabla_\alpha(\cdot)\equiv(\cdot)_{||\alpha}$ for spacetime
or $\nabla_i(\cdot)\equiv(\cdot)_{;i}$ for space variables, respectively.
The signature of the $(d+1)$-dimensional metric $g_{\mu\nu}$ is $+(d-1)$.
The Einstein summation convention is adopted.
The speed of light is denoted by $c$ and $G$ is the Newtonian gravitational constant.

We use the notion of a \emph{tensor density}. The components of a tensor density of weight $w$,
$k$ times contravariant and $l$ times covariant,
transform, when one changes one coordinate system to another, by the law
[see, e.g., p.\ 501 in \cite{misner-73} or, for more general case,
Sects.\ 3.7--3.9 and 4.5 in \cite{plebanski-06},
where however definition of the density weight differs by sign from the convention used by us;
note the primed notation is on the indices, not on the main symbol]
\be
\label{tdensity1}
\mathcal{T}^{\alpha'_1\ldots\alpha'_k}_{\beta'_1\ldots\beta'_l}
= \left(\frac{\partial x'}{\partial x}\right)^{-w} {x^{\alpha'_1}}_{,\alpha_1}\ldots{x^{\alpha'_k}}_{,\alpha_k}
{x^{\beta_1}}_{,\beta'_1}\ldots{x^{\beta_l}}_{,\beta'_l} \mathcal{T}^{\alpha_1\ldots\alpha_k}_{\beta_1\ldots\beta_l},
\ee
where $({\partial x'}/{\partial x})$ is the Jacobian of the transformation $x\to x'(x)$.
For example, determinant of the metric $g\equiv\det(g_{\mu\nu})$ is a scalar density of weight $+2$.
The covariant derivative of the tensor density of weight $w$, $k$ times contravariant and $l$ times covariant,
is computed according to the rule
\begin{align}
\label{tdensity2}
\nabla_\gamma \mathcal{T}^{\alpha_1\ldots\alpha_k}_{\beta_1\ldots\beta_l}
&= \partial_\gamma \mathcal{T}^{\alpha_1\ldots\alpha_k}_{\beta_1\ldots\beta_l}
- w \Gamma^\rho_{\rho\gamma} \mathcal{T}^{\alpha_1\ldots\alpha_k}_{\beta_1\ldots\beta_l}
\nonumber\\&\quad
+ \sum_{i=1}^k \Gamma^{\alpha_i}_{\rho_i\gamma} \mathcal{T}^{\alpha_1\ldots\rho_i\ldots\alpha_k}_{\beta_1\ldots\beta_l}
- \sum_{j=1}^l \Gamma^{\rho_j}_{\beta_j\gamma} \mathcal{T}^{\alpha_1\ldots\alpha_k}_{\beta_1\ldots\rho_j\ldots\beta_l}.
\end{align}
For the often used case when $\mathcal{T}^{\alpha_1\ldots\alpha_k}_{\beta_1\ldots\beta_l}=|g|^{w/2}T^{\alpha_1\ldots\alpha_k}_{\beta_1\ldots\beta_l}$
(where $T^{\alpha_1\ldots\alpha_k}_{\beta_1\ldots\beta_l}$ is a tensor $k$ times contravariant and $l$ times covariant),
Eq.~\eqref{tdensity2} implies that the covariant derivative of $\mathcal{T}^{\alpha_1\ldots\alpha_k}_{\beta_1\ldots\beta_l}$
can be computed by means of the rule,
\be
\label{tdensity3}
\nabla_{\gamma}\mathcal{T}^{\alpha_1\ldots\alpha_k}_{\beta_1\ldots\beta_l}
= T^{\alpha_1\ldots\alpha_k}_{\beta_1\ldots\beta_l} \nabla_{\gamma}|g|^{w/2}
+ |g|^{w/2} \nabla_{\gamma} T^{\alpha_1\ldots\alpha_k}_{\beta_1\ldots\beta_l}
= |g|^{w/2} \nabla_{\gamma} T^{\alpha_1\ldots\alpha_k}_{\beta_1\ldots\beta_l},
\ee
because
\be
\label{tdensity4}
\nabla_{\gamma}|g|^{w/2} = \partial_{\gamma}|g|^{w/2} - w \Gamma^{\rho}_{\rho\gamma}|g|^{w/2} = 0.
\ee

Letters $a,b$ ($a,b=1,2$) are particle labels,
so $\xa=(x_a^i)\in\mathbb{R}^d$ denotes the position of the $a$th point mass.
We also define ${\bf r}_a\equiv\mathbf{x}-\xa$, $r_a\equiv|{\bf r}_a|$, $\vecn_a\equiv{\bf r}_a/r_a$;
and for $a\ne b$, ${\bf r}_{ab}\equiv\xa-\mathbf{x}_b$, $r_{ab}\equiv|{\bf r}_{ab}|$, $\vecn_{ab}\equiv{\bf r}_{ab}/r_{ab}$.
The linear momentum vector of the $a$th particle is denoted by $\pa=(p_{ai})$,
and $m_a$ denotes its mass parameter.
We abbreviate Dirac delta distribution $\delta(\mathbf{x}-\xa)$ by $\delta_a$ (both in $d$ and in 3 dimensions);
it fulfills the condition $\int\md^dx\,\delta_a=1$.

Thinking in terms of dimensions of space, $d$ has to be an integer,
but whenever integrals within dimensional regularization get performed, we allow $d$ to become an arbitrary complex number
[like in the analytic continuation of factorial $n!=\Gamma(n+1)$ to $\Gamma(z)$].
A thorough introduction to dimensional regularization can be found in Chapter 4 of \cite{collins-84}.

\section{Hamiltonian formalisms of GR}
\label{sec:Hformulations}

The presented Hamiltonian formalisms do all rely on a $(3+1)$ splitting of spacetime metric $g_{\mu\nu}$ in the following form:
\be
\label{3+1}
\md s^2 = g_{\mu\nu}\md x^{\mu}\md x^{\nu}
= -(Nc\,\md t)^2 + \gamma_{ij}(\md x^i +  N^ic\,\md t)(\md x^j + N^jc\,\md t),
\ee
where
\be
\gamma_{ij} \equiv g_{ij}, \quad
N \equiv (-g^{00})^{-1/2}, \quad
N^i = \gamma^{ij}N_j \quad\text{with}\quad N_i \equiv g_{0i},
\ee
here $\gamma^{ij}$ is the inverse metric of $\gamma_{ij}$ ($\gamma_{ik}\gamma^{kj}=\delta_i^j$), $\gamma\equiv{\rm det}(\gamma_{ij})$;
lowering and raising of spatial indices is with $\gamma_{ij}$.
The splitting \eqref{3+1},
and the associated explicit 3+1 decomposition of Einstein's equations,
was first introduced by \cite{bruhat-56}.
The notations $N$ and $N^i$ are due to \citet*{arnowitt-62} and their names,
respectively ``lapse'' and ``shift'' functions, are due to \cite{wheeler-64}.
Let us note the useful relation between the determinants
$g\equiv\det(g_{\mu\nu})$ and $\gamma$:
\be
g = -N^2 \gamma.
\ee

We restrict ourselves to consider only \textit{asymptotically flat spacetimes}
and we employ \textit{quasi-Cartesian coordinate systems} $(t,x^i)$
which are characterized by the following asymptotic spacelike behaviour
(i.e., in the limit $r\to\infty$ with $r\equiv\sqrt{x^ix^i}$ and $t$ = const)
of the metric coefficients:
\begin{align}
N &= 1+O(1/r), \quad N^i=O(1/r), \quad \gamma_{ij}=\delta_{ij}+O(1/r),
\\[1ex]
N_{,i} &= O(1/r^2), \quad N^i_{,j}=O(1/r^2), \quad \gamma_{ij,k}=O(1/r^2).
\end{align}

\cite{dewitt-67} and later, in a more refined way, \cite{regge-74} explicitly showed
that the Hamiltonian which generates all Einsteinian field equations can be put into the form,
\begin{align}
\label{ADMh}
H[\gamma_{ij},\pi^{ij},N,N^i;q^A,\pi_A] &= \int \md^3x\, (N{\mathcal{H}} - c N^i {\mathcal{H}}_i)
\nonumber\\&\quad
+ \frac{c^4}{16\pi G}\oint_{i^0} \md S_i\,\partial_j (\gamma_{ij} - \delta_{ij} \gamma_{kk}),
\end{align}
wherein $N$ and $N^i$ operate as Lagrangian multipliers
and where $\mathcal{H}$ and ${\mathcal{H}}_i$ are Hamiltonian and momentum densities, respectively;
$i^0$ denotes spacelike flat infinity.
They depend on matter canonical variables $q^A,\pi_A$
(through matter Hamiltonian density ${\mathcal{H}}_{\textrm{m}}$ and matter momentum density ${\mathcal{H}}_{\textrm{m}i}$)
and read
\begin{eqnarray}
\label{Hden1}
{\mathcal{H}} &\equiv& \frac{c^4}{16\pi G}\left[-\gamma^{1/2} R
+ \frac{1}{\gamma^{1/2}}\left(\gamma_{ik}\gamma_{jl}\pi^{ij}\pi^{kl}-\frac{1}{2}\pi^2\right)\right]
+ {\mathcal{H}}_{\textrm{m}},
\\
\label{Mden1}
{\mathcal{H}}_i &\equiv& \frac{c^3}{8\pi G}\gamma_{ij}\nabla_k\pi^{jk} + {\mathcal{H}}_{\textrm{m}i},
\end{eqnarray}
where $R$ is the intrinsic curvature scalar of the spacelike hypersurfaces of constant-in-time slices $t=x^0/c$ = const;
the ADM canonical field momentum is given by the density $\dst\frac{c^3}{16\pi G}\pi^{ij}$, where
\be
\label{pibyK}
\pi_{ij} \equiv -\gamma^{1/2} (K_{ij}-K\gamma_{ij}),
\ee
with $K\equiv\gamma^{ij}K_{ij}$,
where  $K_{ij}=-N \Gamma^0_{ij}$ is the extrinsic curvature of  $t$ = const slices,
$\Gamma^0_{ij}$ denote Christoffel symbols; $\pi\equiv\gamma_{ij}\pi^{ij}$;
$\nabla_k$ denotes the three-dimensional covariant derivative (with respect to $\gamma_{ij}$).
The given densities are densities of weight one with respect to three-dimensional coordinate transformations.
Let us note the useful formula for the density of the three-dimensional scalar curvature of the surface $t$ = const:
\begin{align}
\label{3R}
\sqrt{\gamma} R &= \frac{1}{4} \sqrt{\gamma} \Big(\big(\gamma^{ij}\gamma^{lm}-\gamma^{il}\gamma^{jm}\big)\gamma^{kn}
+ 2\big(\gamma^{il}\gamma^{km}-\gamma^{ik}\gamma^{lm}\big)\gamma^{jn}\Big)\gamma_{ij,k}\gamma_{lm,n}
\nonumber\\&\quad
+ \partial_i\big(\gamma^{-1/2}\partial_j (\gamma \gamma^{ij})\big).
\end{align}
The matter densities ${\mathcal{H}}_{\textrm{m}}$ and ${\mathcal{H}}_{\textrm{m}i}$
are computed from components of the matter energy-momentum tensor $T^{\mu\nu}$ by means of formulae
\begin{align}
\label{mHden1}
{\mathcal{H}}_{\textrm{m}} &= \sqrt{\gamma}\,T^{\mu\nu}n_\mu n_\nu = \sqrt{\gamma}\, N^2 T^{00},
\\[1ex]
\label{mMden1}
{\mathcal{H}}_{\textrm{m}i} &= -\sqrt{\gamma}\,T^{\mu}_i n_\mu = \sqrt{\gamma}\, N T^0_i,
\end{align}
where $n_\mu=(-N,0,0,0)$ is the timelike unit covector orthogonal to the spacelike hypersurfaces $t$ = const.
Opposite to what the right-hand sides of Eqs.\ \eqref{mHden1}--\eqref{mMden1} seem to suggest,
the matter densities must be independent on lapse $N$ and shift $N^i$
and expressible in terms of the dynamical matter and field variables
$q^A$, $\pi_A$, $\gamma_{ij}$ only ($\pi^{ij}$ does not show up for matter
which is minimally coupled to the gravitational field).
The variation of \eqref{ADMh} with respect to $N$ and $N^i$ yields the \textit{constraint equations}
\be
\label{CEqs}
{\mathcal{H}} = 0 \quad \text{and} \quad {\mathcal{H}}_i = 0.
\ee

The most often applied Hamiltonian formalism employs the following coordinate choice
made by ADM (which we call ADMTT gauge),
\be
\label{ADMTTa}
\pi^{ii} = 0, \qquad 3\partial_j\gamma_{ij} - \partial_i\gamma_{jj} = 0
\quad \text{or} \quad \gamma_{ij} = \psi \delta_{ij} + h_{ij}^{\rm TT},
\ee
where the TT piece $h_{ij}^{\rm TT}$ is transverse and traceless, i.e., it satisfies $\partial_jh_{ij}^{\rm TT}=0$ and $h_{ii}^{\rm TT}=0$.
The TT piece of any field function can be computed by means of the TT projection operator defined as follows
\be
\label{defTT3}
\delta^{{\rm TT}kl}_{ij} \equiv \frac{1}{2}(P_{il}P_{jk} + P_{ik}P_{jl} - P_{kl}P_{ij}),
\quad
P_{ij} \equiv \delta_{ij} - \partial_i\partial_j\Delta^{-1},
\ee
where $\Delta^{-1}$ denotes the inverse of the flat space Laplacian,
which is taken without homogeneous solutions for source terms
decaying fast enough at infinity (in 3-dimensional or, if not, then in generalized $d$-dimensional space).
The nonlocality of the TT-operator $\delta^{{\rm TT}kl}_{ij}$ is just the gravitational analogue of
the well-known nonlocality of the Coulomb gauge in the electrodynamics.

Taking into account its gauge condition as given in Eq.~\eqref{ADMTTa},
the field momentum $\dst\frac{c^3}{16\pi G}\pi^{ij}$ can be split into its longitudinal and TT parts, respectively,
\be
\pi^{ij} = \tilde{\pi}^{ij} + \pi^{ij}_{\rm TT},
\ee
where the TT part $\pi^{ij}_{\rm TT}$ fulfills the conditions $\partial_j\pi^{ij}_{\rm TT}=0$ and $\pi^{ii}_{\rm TT}=0$
and where the longitudinal part $\tilde{\pi}^{ij}$ can be expressed in terms of a vectorial function $V^i$,
\be
\tilde{\pi}^{ij} = \partial_i V^j +\partial_j V^i - \frac{2}{3}\delta^{ij}\partial_k V^k.
\ee
It is also convenient to parametrize the field function $\psi$ from Eq.~\eqref{ADMTTa} in the following way
\be
\label{psibyphi}
\psi = \left(1 + \frac{1}{8}\phi\right)^4.
\ee

The independent field variables are $\pi^{ij}_{\rm TT}$ and $h_{ij}^{\rm TT}$.
Already \cite{kimura-61} used just this presentation for applications.
The Poisson bracket for the independent degrees of freedom reads
\begin{align}
\{F(\vecx),G(\vecy)\} &\equiv \frac{16\pi G}{c^3}
\int\md^3z\, \Bigg\{ \frac{\delta F(\vecx)}{\delta h^{\rm TT}_{ij}(\vecz)}
\bigg(\delta^{{\rm TT}kl}_{ij}(\vecz)\frac{\delta G(\vecy)}{\delta \pi^{kl}_{\rm TT}(\vecz)}\bigg)
\nonumber\\[1ex]&\quad
- \frac{\delta G(\vecy)}{\delta h^{\rm TT}_{ij}(\vecz)}
\bigg(\delta^{{\rm TT}kl}_{ij}(\vecz)\frac{\delta F(\vecx)}{\delta \pi^{kl}_{\rm TT}(\vecz)}\bigg) \Bigg\},
\end{align}
where $\delta F(\vecx)/(\delta f(\vecz))$ denotes the functional (or Fr\'echet) derivative.
ADM gave the Hamiltonian in fully reduced form, i.e., after having applied (four) constraint equations \eqref{CEqs}
and (four) coordinate conditions \eqref{ADMTTa}. It reads
\begin{align}
\label{ADMHred}
H_\textrm{red}[h_{ij}^{\rm TT},\pi^{ij}_{\rm TT};q^A,\pi_A]
&= \frac{c^4}{16\pi G}\oint_{i^0} \md S_i\, \partial_j (\gamma_{ij} - \delta_{ij} \gamma_{kk})
\nonumber\\[1ex]
&= \frac{c^4}{16\pi G}\int\md^3x\, \partial_i\partial_j (\gamma_{ij} - \delta_{ij} \gamma_{kk}).
\end{align}
The reduced Hamiltonian generates the field equations of the two remaining metric coefficients
(eight metric coefficients are determined by the four constraint equations
and four coordinate conditions combined with four otherwise degenerate field equations for the lapse and shift functions).
By making use of \eqref{psibyphi} the reduced Hamiltonian \eqref{ADMHred} can be written as
\be
\label{ADMHred2}
H_\textrm{red}[h_{ij}^{\rm TT},\pi^{ij}_{\rm TT};q^A,\pi_A]
= -\frac{c^4}{16\pi G} \int\md^3x\, \Delta\phi[h_{ij}^{\rm TT},\pi^{ij}_{\rm TT};q^A,\pi_A].
\ee

\subsection{Hamiltonian formalisms of Dirac and Schwinger}
\label{sec:2-1}

Dirac had chosen the following coordinate system, called ``maximal slicing'' because of the field momentum condition,
\be
\pi \equiv \gamma_{ij}\pi^{ij} = 0, \quad
\partial_j(\gamma^{1/3}\gamma^{ij}) = 0.
\ee
The reason for calling the condition $\pi=2K\gamma^{1/2}=0$ ``maximal slicing'' is because
the congruence of the timelike unit vectors $n^{\mu}$ normal to the $t$ = const hypersurfaces (slices)---as such irrotational---is free of expansion
(notice that $\nabla_\mu n^{\mu}=-K$).
Hereof it immediately follows that a finite volume in any slice gets unchanged by a small timelike deformation of the slice
which vanishes on the boundary of the volume, i.e.\ an extremum principle holds (see, e.g., \citealp{york-79}).
The corresponding independent field variables are
(no implementation of the three differential conditions!)
\be
\label{fvarDirac}
\tilde{\pi}^{ij} = \Big(\pi^{ij}-\frac{1}{3}\gamma^{ij}\pi\Big)\gamma^{1/3},
\quad
\tilde{g}_{ij} = \gamma^{-1/3}\gamma_{ij},
\ee
with the algebraic properties $\gamma_{ij}\tilde{\pi}^{ij}=0$ and $\det(\tilde{g}_{ij})=1$.
To leading order linear in the metric functions, the Dirac gauge coincides with the ADM gauge.
The reduction of the Dirac form of dynamics to the independent tilded degrees of freedom
has been performed by \cite{regge-74},
including a fully satisfactory derivation of the Hamiltonian introduced by Dirac.
The Poisson bracket for the Dirac variables reads
\begin{align}
\{F(\mathbf{x}),G(\mathbf{y})\} &\equiv \frac{16\pi G}{c^3}\int\md^3z\,\Bigg\{
\tilde{\delta}^{kl}_{ij}(\mathbf{z}) \left(
\frac{\delta F(\mathbf{x})}{\delta\tilde{g}_{ij}(\mathbf{z})}\frac{\delta G(\mathbf{y})}{\delta\tilde{\pi}^{kl}(\mathbf{z})}
- \frac{\delta G(\mathbf{y})}{\delta\tilde{g}_{ij}(\mathbf{z})}\frac{\delta F(\mathbf{x})}{\delta\tilde{\pi}^{kl}(\mathbf{z})}\right)
\nonumber\\&\quad
+ \frac{1}{3}\Big(\tilde{\pi}^{ij}(\mathbf{z})\tilde{g}^{kl}(\mathbf{z}) - \tilde{\pi}^{kl}(\mathbf{z})\tilde{g}^{ij}(\mathbf{z})\Big)
\frac{\delta F(\mathbf{x})}{\delta \tilde{\pi}^{ij}(\mathbf{z})}\frac{\delta G(\mathbf{y})}{\delta \tilde{\pi}^{kl}(\mathbf{z})} \Bigg\},
\end{align}
with
\be
\tilde{\delta}^{kl}_{ij} \equiv \frac{1}{2}(\delta^k_i\delta^l_j + \delta^l_i\delta^k_j) - \frac{1}{3}\tilde{g}_{ij}\tilde{g}^{kl},
\quad \tilde{g}^{ij} = \gamma^{1/3}\gamma^{ij},
\quad \tilde{g}_{ij}\tilde{g}^{jk} = \delta^k_i.
\ee

The Hamiltonian proposed by Dirac results from the expression
\be
\label{HDirac}
H_{\rm D}[\tilde{g}_{ij},\tilde{\pi}^{ij},q^A,\pi_A]
= -\int\md^3x~c\,N^i \mathcal{H}_i
- \frac{c^4}{16\pi G} \int\md^3x~\partial_i\big(\gamma^{-1/2}\partial_j (\gamma \gamma^{ij})\big),
\ee
which itself results from Eq.~\eqref{ADMh}
under imposing the Hamiltonian constraint ${\mathcal{H}}=0$ [see Eq.~\eqref{CEqs}] as identity,
replacing in \eqref{ADMh} the surface term with another but equivalent surface term,
and implementing the Dirac variables from Eq.~\eqref{fvarDirac},
which are the independent variables under the maximal slicing condition.
The further reduction, the one with implementing the coordinate conditions on the hypersurfaces,
goes via the Dirac brackets as follows.

The fixation of the coordinates in the hypersurface through $\partial_j\tilde{g}^{ij}=0$
results in Dirac brackets in phase space of the form \citep{dirac-59}
\begin{align}
\{F&(\mathbf{x}),G(\mathbf{y})\}_{\rm D} \equiv \{F(\mathbf{x}),G(\mathbf{y})\}
+ \int\md^3z\int\md^3z'\,C_i^j(\mathbf{z},\mathbf{z}')
\nonumber\\&
\times \Big(\big\{F(\mathbf{x}),\partial_k\tilde{g}^{ik}(\mathbf{z})\big\}\big\{{\cal{H}}_j(\mathbf{z}'),G(\mathbf{y})\big\}
- \big\{F(\mathbf{x}),{\cal{H}}_j(\mathbf{z}')\big\}\big\{\partial_k\tilde{g}^{ik}(\mathbf{z}),G(\mathbf{y})\big\}\Big),
\end{align}
where the matrix $C_m^l(\mathbf{z}'',\mathbf{z}')$ is defined by
\be
\int\md^3z'\, C_m^l(\mathbf{z}'',\mathbf{z}')
\{{\cal{H}}_l(\mathbf{z}'),\partial_k\tilde{g}^{nk}(\mathbf{z})\}
= \delta_m^n\delta(\mathbf{z}-\mathbf{z}'').
\ee
It obeys the differential equation
\be
\tilde{g}^{ij}(\mathbf{x})\partial_i\partial_jC_m^n(\mathbf{x}',\mathbf{x})
+ \frac{1}{3}\tilde{g}^{nk}(\mathbf{x})\partial_k\partial_lC_m^l(\mathbf{x}',\mathbf{x})
= \delta_m^n\delta(\mathbf{x}-\mathbf{x}').
\ee

When using Dirac brackets the momentum constraint reads [see Eq.~\eqref{CEqs}]
\be
{\cal{H}}_i = \frac{c^3}{8\pi G}\big(\tilde{\pi}^{jk}\partial_i\tilde{g}_{jk}
- 2\partial_k(\tilde{\pi}^{jk}\tilde{g}_{ji})\big) + {\cal{H}}_{{\rm m} i} = 0,
\ee
and the corresponding coordinate conditions $\partial_j\tilde{g}^{ij} = 0$
can be treated as strong equations, because for an arbitrary functional $F$
\be
\{F,{\cal{H}}_i\}_{\rm D} = 0, \quad
\{F,\partial_j\tilde{g}^{ij}\}_{\rm D} = 0.
\ee
Thus, applying Dirac brackets,
\be
\label{HDirac2}
H_{\rm D}[\tilde{g}_{ij},\tilde{\pi}^{ij},q^A,\pi_A] = -\frac{c^4}{16\pi G}
\int\md^3x\, \partial_i\big(\gamma^{-1/2}\partial_j(\gamma^{2/3} \tilde{g}^{ij})\big)
\ee
holds.

For the determination of the surface term in Eq.~\eqref{HDirac2}
only the determinant $\gamma$ of the metric must be expressed by
independent field variables \eqref{fvarDirac}.
This can be done through the differential equation
\be
-\frac{c^4}{4\pi G} \tilde{g}^{ij}\partial_i\partial_j\kappa = \frac{c^4}{16\pi G}
\Big( \frac{1}{\kappa^3}\tilde{g}_{ij}\tilde{g}_{kl} \tilde{\pi}^{ik}\tilde{\pi}^{jl} + B \Big)
+ {\cal{H}}_{\rm m},
\qquad \kappa^6 = \gamma,
\ee
resulting from the Hamiltonian constraint, first equation in Eq.~\eqref{CEqs}, with
\be
B = \frac{1}{4}\kappa(\partial_i\tilde{g}_{jk})(\partial_l\tilde{g}_{mn})
\tilde{g}^{jm}(\tilde{g}^{kn}\tilde{g}^{il}- 2\tilde{g}^{in}\tilde{g}^{lk})
- \frac{2}{\kappa}(\partial_i\kappa)(\partial_j\kappa)\tilde{g}^{ij}.
\ee

Schwinger proposed still another set of canonical field variables $(q^{ij},\Pi_{ij})$,
for which the Hamiltonian and momentum densities have the form
\begin{align}
\label{HdenSchwinger}
\mathcal{H} &\equiv  \frac{c^4}{16\pi G}\gamma^{-1/2} \Big(-\frac{1}{4}q^{mn}\partial_m q^{kl}\partial_n q^{kl}
- \frac{1}{2}q_{ln}\partial_m q^{kl}\partial_k q^{mn}
\nonumber\\[1ex]&\quad
- \frac{1}{2}q^{kl}\partial_k \mbox{ln}(q^{1/2})\partial_l \mbox{ln}(q^{1/2})
+ \partial_i\partial_j q^{ij} + q^{ik}q^{jl}\Pi_{ij}\Pi_{kl} - (q^{ij}\Pi_{ij})^2 \Big) + {\mathcal{H}}_{\textrm{m}},
\\[2ex]
{\mathcal{H}}_i &\equiv \frac{c^3}{16\pi G}\Big(-\Pi_{lm}\partial_iq^{lm}
+ \partial_i(2\Pi_{lm}q^{lm}) - \partial_l(2\Pi_{im}q^{lm})\Big) + {\mathcal{H}}_{\textrm{m}i},
\end{align}
where $\Pi_{ij}\equiv-\gamma^{-1}(\pi_{ij} - \frac{1}{2}\pi \gamma_{ij})$, $q^{ij}\equiv\gamma \gamma^{ij}$, $q\equiv\gamma^2$;
Schwinger's canonical field momentum $\dst\frac{c^3}{16\pi G}\Pi_{ij}$ is just $\dst\frac{c^3}{16\pi G}\gamma^{-1/2} K_{ij}$.
The Poisson bracket for the Schwinger variables does have the same structure as the one for the ADM variables.
The Schwinger's reduced Hamiltonian has the form
\be
\label{HSchwinger}
H_{\rm S} = -\frac{c^4}{16\pi G}\oint_{i^0} \md S_i\, \partial_j q^{ij}
= -\frac{c^4}{16\pi G}\int\md^3x\, \partial_i\partial_j q^{ij}.
\ee
If Schwinger had chosen coordinate conditions
corresponding to those introduced above in Eqs.\ \eqref{ADMTTa}
(ADM also introduced another set of coordinate conditions to which Schwinger adjusted), namely
\be
\Pi_{ii} = 0, \quad q^{ij} = \varphi \delta_{ij} + f^{ij}_{\rm TT},
\ee
a similar simple technical formalism convenient for practical calculations
would have resulted with the independent field variables $\Pi_{ij}^{\rm TT}$ and $f^{ij}_{\rm TT}$.
To our best knowledge, only the paper by \cite{kibble-63} delivers an application of Schwinger's formalism, apart from Schwinger himself,
namely a Hamiltonian formulation of the Dirac spinor field in gravity.
Much later, \cite{nelson-78} completed the same task within the tetrad-generalized Dirac formalism \citep{dirac-62}.

Notice that the Dirac Hamiltonian \eqref{HDirac2} shows first derivatives of the metric coefficients only,
plugging in the Hamiltonian constraint.
The same holds with the Hamiltonian proposed by Schwinger, see Eq.~ \eqref{HSchwinger}
and the Eq.~\eqref{HdenSchwinger} on-shell, i.e.\ after application of the Hamiltonian constraint.
The Hamiltonians \eqref{ADMHred}, \eqref{HDirac2}, and \eqref{HSchwinger} are identical as global objects
because their integrands differ by total divergences which do vanish after integration.

\subsection{Derivation of the ADM Hamiltonian}
\label{subsec:ADMHderivation}

The ADM Hamiltonian was derived via the generator of field and spacetime-coordinates variations.
Let the generator of general field variations be defined as
(it corresponds to the generator $G\equiv p_i\,\delta x^i$ of the point-particle dynamics in classical mechanics
with the particle's canonical momentum $p_i$ and position $x^i$)
\be
G_{\rm field} \equiv \frac{c^3}{16\pi G} \int \md^3x\, \pi^{ij} \delta \gamma_{ij}.
\ee
Let the coefficients of three space-metric $\gamma_{ij}$ be fixed by the relations \eqref{ADMTTa},
then the only free variations left are
\be
G_{\rm field} = \frac{c^3}{16\pi G} \int \md^3x\, \pi^{ij}_{\rm TT} \delta h_{ij}^{\rm TT}
+  \frac{c^3}{16\pi G} \int \md^3x\, \pi^{jj} \delta\psi
\ee
or, modulo a total variation,
\be
G_{\rm field} = \frac{c^3}{16\pi G} \int \md^3x\, \pi^{ij}_{\rm TT} \delta h_{ij}^{\rm TT}
- \frac{c^3}{16\pi G} \int \md^3x\, \psi \delta \pi^{jj}. 
\ee
It is consistent with the Einstein field equations in space-asymptotically flat space-time with quasi-Cartesian coordinates to put
[the mathematically precise meaning of this equation is detailed in the Appendix B of \cite{arnowitt1-60}]
\be
ct = - \frac{1}{2}\Delta^{-1} \pi^{jj},
\ee
which results in, dropping total space derivatives,
\be
G_{\rm field} = \frac{c^3}{16\pi G} \int \md^3x\, \pi^{ij}_{\rm TT} \delta h_{ij}^{\rm TT}
+  \frac{c^4}{8\pi G} \int \md^3x\, \Delta \psi\, \delta t.
\ee
Hereof the Hamiltonian easily follows in the form
\be
H = -\frac{c^4}{8\pi G} \int\md^3x\,\Delta\psi,
\ee
which can also be written, using the form of the three-metric from Eq.~\eqref{ADMTTa},
\be
H = \frac{c^4}{16\pi G} \int \md^3x\, \partial_i\partial_j (\gamma_{ij} - \delta_{ij}\gamma_{kk}).
\ee
This expression is valid also in case of other coordinate conditions \citep{arnowitt-62}.
For the derivation of the generator of space translations,
the reader is referred to \cite{arnowitt-62} or, equivalently, to \cite{schwinger1-63}.

\section{The ADM formalism for point-mass systems}
\label{sec:ADMformalism}

\subsection{Reduced Hamiltonian for point-mass systems}
\label{subsec:PMredH}

In this section we consider the ADM canonical formalism applied to a system of self-gravitating nonrotating point masses (particles).
The energy-momentum tensor of such system reads
\be
\label{PMemt1}
T^{\alpha\beta}(x^\gamma) = \sum_a m_a c \int_{-\infty}^\infty
\frac{u_a^\alpha u_a^\beta}{\sqrt{-g}} \delta^{(4)}\big(x^\mu-x_a^\mu(\tau_a)\big)\md\tau_a,
\ee
where $m_a$ is the mass parameter of $a$th point mass ($a=1,2,\ldots$ labels the point masses),
$u_a^\alpha\equiv\md x_a^\alpha/\md\tau_a$ (with $c\,\md\tau_a=\sqrt{-g_{\mu\nu}\md x_a^\mu \md x_a^\nu}$) is the four-velocity
along the worldline $x^\mu=x_a^\mu(\tau_a)$ of the $a$th particle.
After performing the integration in \eqref{PMemt1} one gets
\be
\label{PMemt2}
T^{\alpha\beta}(\mathbf{x},t)
= \sum_a m_a c \frac{u_a^\alpha u_a^\beta}{u_a^0 \sqrt{-g}} \delta^{(3)}\big(\mathbf{x}-\mathbf{x}_a(t)\big),
\ee
where $\mathbf{x}_a=(x_a^i)$ is the position three-vector of the $a$th particle.
The linear four-momentum of the $a$th particle equals $p_a^\alpha\equiv m_a u_a^\alpha$,
and the three-momentum canonically conjugate to the position $\mathbf{x}_a$ comes out to be $\mathbf{p}_a=(p_{ai})$,
where $p_{ai}=m_a u_{ai}$.

The action functional describing particles-plus-field system reads
\be
\label{AF1}
S = \int\md t \left(\frac{c^3}{16\pi G} \int\md^3x\, \pi^{ij} \partial_t \gamma_{ij}
+ \sum_a p_{ai} \dot{x}_a^i  - H_0\right),
\ee
where $\dot{x}_a^i\equiv\md x_a^i/\md t$.
The asymptotic value 1 of the lapse function
enters as prefactor of the surface integral in the Hamiltonian $H_0$,
which takes the form
\be
H_0 = \int \md^3x\, (N{\mathcal{H}} - cN^i {\mathcal{H}}_i)
+ \frac{c^4}{16\pi G} \oint_{i^0}\md S_i\,\partial_j (\gamma_{ij} - \delta_{ij} \gamma_{kk}),
\ee
where the so-called super-Hamiltonian density $\mathcal{H}$ and super-momentum density $ {\mathcal{H}}_i$
can be computed by means of Eqs.\ \eqref{Hden1}--\eqref{Mden1}, \eqref{mHden1}--\eqref{mMden1}, and \eqref{PMemt2}.
They read [here we use the abbreviation $\delta_a$ for $\delta^{(3)}(\mathbf{x}-\mathbf{x}_a)$]
\begin{align}
\label{Hdensity}
\mathcal{H} &= \frac{c^4}{16\pi G} \left[\frac{1}{\gamma^{1/2}}\left(\pi^i_j\pi^j_i-\frac{1}{2}\pi^2\right) - \gamma^{1/2} R\right]
+ \sum_a c \left(m_a^2c^2 + \gamma_a^{ij}p_{ai}p_{aj}\right)^{1/2} \delta_a,
\\
\label{Pdensity}
\mathcal{H}_i &= \frac{c^3}{8\pi G} \nabla_j \pi^j_i + \sum_a p_{ai}\delta_a,
\end{align}
where $\gamma_a^{ij}\equiv\gamma_{\textrm{reg}}^{ij}(\xa)$ is the finite part of the inverse metric
evaluated at the particle position, which can be perturbatively and, using dimensional regularization, unambiguously defined
(see Sects.~\ref{subs:RHreg}, \ref{subs:DR} below and Appendix A~4 of \citealp{jaranowski-15}).

The evolutionary part of the field equations is obtained
by varying the action functional \eqref{AF1} with respect to the field variables $\gamma_{ij}$ and $\pi^{ij}$.
The resulting equations read
\begin{align}
\label{evFE1}
\gamma_{ij,0} &= 2N \gamma^{-1/2} \left(\pi_{ij} - \frac{1}{2} \pi \gamma_{ij} \right) + \nabla_iN_j + \nabla_jN_i,
\\[1ex]
\label{evFE2}
\pi^{ij}_{~,0} &= -N \gamma^{1/2} \left(R^{ij} - \frac{1}{2}\gamma^{ij}R\right)
+ \frac{1}{2}N\gamma^{-1/2}\gamma^{ij}\left(\pi^{mn}\pi_{mn} - \frac{1}{2}\pi^2\right)
\nonumber\\[1ex]&\quad
- 2 N \gamma^{-1/2}\left(\pi^{im}\pi^j_m - \frac{1}{2}\pi\pi^{ij} \right)
+ \nabla_m(\pi^{ij}N^m)
- (\nabla_m N^i)\pi^{mj}
\nonumber\\[1ex]&\quad
- (\nabla_mN^j)\pi^{mi}
+ \frac{1}{2} \sum_a N_a \gamma^{ik}_a p_{ak} \gamma^{jl}_a p_{al} \left(\gamma^{mn}_ap_{am}p_{an} + m^2_ac^2\right)^{-1/2}\delta_a.
\end{align}
The constraint part of the field equations results from varying the action \eqref{AF1}
with respect to $N$ and $N^i$. It has the form
\be
\label{CEqs2}
{\mathcal{H}} = 0, \qquad {\mathcal{H}}_i = 0.
\ee
The variation of the action \eqref{AF1} with respect to $\mathbf{x}_a$ and $\mathbf{p}_a$
leads to equations of motion for the particles,
\begin{align}
\dot{p}_{ai} &= -\frac{\partial}{\partial x_a^i} \int\md^3x\, (N{\mathcal{H}} - cN^k {\mathcal{H}}_k)
\nonumber\\[1ex]
&= c p_{aj} \frac{\partial N^j_a}{\partial x_a^i}
- c \left(m_a^2c^2 + \gamma^{kl}_a p_{ak}p_{al} \right)^{1/2} \frac{\partial N_a}{\partial x_a^i}
\nonumber\\[1ex]&\quad
- \frac{c N_a}{2\left(m_a^2c^2 + \gamma^{mn}_a p_{am}p_{an} \right)^{1/2}}\frac{\partial \gamma^{kl}_a}{\partial x_a^i}p_{ak}p_{al},
\\[2ex]
\dot{x}_a^i &= \frac{\partial}{\partial p_{ai}} \int\md^3x\, \left(N{\mathcal{H}} - cN^k {\mathcal{H}}_k\right)
\nonumber\\[1ex]
&= \frac{c N_a \gamma_a^{ij}p_{aj}}{\left(m_a^2c^2 + \gamma^{kl}_a p_{ak}p_{al}\right)^{1/2}} - cN^i_a.
\end{align}
Notice the involvement of lapse and shift functions in the equations of motion.
Both the lapse and shift functions, four functions in total,
get determined by the application of the four coordinate conditions \eqref{ADMTTa}
to the field equations \eqref{evFE1} and \eqref{evFE2}.

The reduced action, which is fully sufficient for the derivation of the dynamics of the particles and the gravitational field,
reads (only the asymptotic value 1 of the shift function survives)
\be
S = \int\md t \left[\frac{c^3}{16\pi G} \int\md^3x\, \pi_{\rm TT}^{ij} \partial_t h^{\rm TT}_{ij}
+ \sum_a p_{ai} \dot{x}_a^i - H_{\textrm{red}} \right],
\ee
where both the constraint equations \eqref{CEqs2} and the coordinate conditions \eqref{ADMTTa} are taken to hold.
The reduced Hamilton functional $H_{\textrm{red}}$ is given by
\be
H_{\textrm{red}}[\xa,\pa,h^{\rm TT}_{ij},\pi_{\rm TT}^{ij}]
= -\frac{c^4}{16\pi G} \int \md^3x\,\Delta\phi[\xa,\pa,h^{\rm TT}_{ij},\pi_{\rm TT}^{ij}].
\ee
The remaining field equations read
\be
\label{FEqs}
\frac{c^3}{16\pi G}\partial_t\pi_{\textrm{TT}}^{ij}
= -\delta^{\textrm{TT} {ij}}_{kl} \frac{\delta H_{\textrm{red}}}{\delta h_{kl}^{\textrm{TT}}},
\quad
\frac{c^3}{16\pi G}\partial_t h_{ij}^{\textrm{TT}}
= \delta^{\textrm{TT} {kl}}_{ij} \frac{\delta H_{\textrm{red}}}{\delta \pi_{\textrm{TT}}^{kl}},
\ee
and the equations of motion for the point masses take the form
\be
\dot{p}_{ai} = -\frac{\partial H_{\textrm{red}}}{\partial x_a^i},
\quad\quad \dot{x}_a^i = \frac{\partial H_{\textrm{red}}}{\partial p_{ai}}.
\ee
Evidently, there is no involvement of lapse and shift functions in the equations of motion
and in the field equations for the independent degrees of freedom (\citealp{arnowitt2-60}, \citealp{kimura-61}).

\subsection{Routh functional}
\label{subsec:Routhian}

The Routh functional (or Routhian) of the system is defined by 
\be
\label{ADMRouthian}
R\left[\xa,\pa,h^{\rm TT}_{ij},\partial_t h^{\rm TT}_{ij}\right]
\equiv H_{\textrm{red}} - \frac{c^3}{16\pi G} \int \md^3x\,\pi^{ij}_{\rm TT}\,\partial_t h^{\rm TT}_{ij}.
\ee
This functional is a Hamiltonian for the point-mass degrees of freedom,
and a Lagrangian for the independent gravitational field degrees of freedom.
Within the post-Newtonian framework it was first introduced by \cite{jaranowski-98,jaranowski-98-e}.
The evolution equation for the gravitational field degrees of freedom reads
\be
\label{RFE}
\frac{\dst \delta}{\delta h^{\rm TT}_{ij}(\bx,t)}\int R(t')\,\md t' = 0.
\ee
The Hamilton equations of motion for the two point masses take the form
\be
\dot{p}_{ai} = - \frac{\partial R}{\partial x^i_a},
\quad
\dot{x}^i_a = \frac{\partial R}{\partial p_{ai}}.
\ee

For the following treatment of the conservative part of the dynamics only,
we will make now a short model calculation revealing the structure and logic behind the treatment.
Let's take a Routhian of the form $R(q,p;\xi,\dot\xi)$. Then the action reads
\be
S[q,p;\xi] = \int \big(p\dot{q} - R(q,p;\xi,\dot\xi)\big)\md t.
\ee
Its variation through the independent variables gives
\begin{align}
\delta S &= \int \bigg[\frac{\md}{\md t}(p\delta q) + \left(\dot{q}-\frac{\partial R}{\partial p}\right)\delta p
+ \left(-\dot{p}-\frac{\partial R}{\partial q}\right)\delta q
\nonumber\\[1ex]&\quad\quad
- \left(\frac{\partial R}{\partial\xi}-\frac{\md}{\md t}\frac{\partial R}{\partial\dot\xi}\right)\delta\xi
- \frac{\md}{\md t}\left(\frac{\partial R}{\partial \dot\xi}\delta\xi\right)\bigg]\md t.
\end{align}
Going on-shell with the $\xi$-dynamics yields
\be
\delta S = \int \left[\frac{\md}{\md t}(p\delta q) + \left(\dot{q}-\frac{\partial R}{\partial p}\right)\delta p
+ \left(-\dot{p} -\frac{\partial R}{\partial q}\right)\delta q\right]\md t
- \left(\frac{\partial R}{\partial \dot\xi}\delta\xi\right)^{+\infty}_{-\infty}.
\ee
The vanishing of the last term means---thinking in terms of $h^{\rm TT}_{ij}$ and $\dot{h}^{\rm TT}_{ij}$,
i.e.\ considering the term $(\int\md^3x\,\pi^{ij}_{\rm TT}\,\delta h^{\rm TT}_{ij})^{+\infty}_{-\infty}$
on the solution space of the field equations (``on-field-shell'')---that as much incoming as outgoing radiation has to be present,
or time-symmetric boundary conditions have to be applied. Thus in the Fokker-type procedure no dissipation shows up.
Assuming a leading-order-type prolongation (allowing additions of only first time derivatives of $q$ and $p$)
of the form $R=R(q,p,\dot{q},\dot{p})$,
the autonomous dynamics can be deduced from the variation
\be
\delta S = \int \left[\frac{\md}{\md t}(p\delta q) + \left(\dot{q}-\frac{\delta R}{\delta p}\right)\delta p
+ \left(-\dot{p}-\frac{\delta R}{\delta q}\right)\delta q \right]\md t,
\ee
where the Euler--Lagrange derivative $\delta A/\delta z\equiv\partial A /\partial z-\md(\partial A/\partial\dot{z})/\md t$ has been introduced.

Having explained that, the {\it conservative} part of the binary dynamics is given by the higher-order Hamiltonian
equal to the on-field-shell Routhian,
\begin{align}
\label{matterHho}
H_{\textrm{con}}[&\xa,\pa,\dot{\bx}_a,\dot{\bp}_a,\ldots]
\nonumber\\[1ex]
&\equiv R\big[\xa,\pa,h^{\rm TT}_{ij}(\xa,\pa,\dot{\bx}_a,\dot{\bp}_a,\ldots),\dot{h}^{\rm TT}_{ij}(\xa,\pa,\dot{\bx}_a,\dot{\bp}_a,\ldots)\big],
\end{align}
where the field variables $h^{\rm TT}_{ij}$, $\dot{h}^{\rm TT}_{ij}$ were ``integrated out'',
i.e., replaced by their solutions as functionals of particle variables.
The conservative equations of motion defined by the higher-order Hamiltonian \eqref{matterHho} read
\be
\dot{p}_{ai}(t) = -\frac{\delta}{\delta x^i_a(t)} \int H_{\textrm{con}}(t')\,\md t',
\quad
\dot{x}^i_a(t) = \frac{\delta}{\delta p_{ai}(t)} \int H_{\textrm{con}}(t')\,\md t',
\ee
where the functional derivative is given by
\be
\frac{\delta}{\delta z(t)} \int H_{\textrm{con}}(t')\,\md t'
= \frac{\partial H_{\textrm{con}}}{\partial z(t)}
- \frac{\md}{\md t}\frac{\partial H_{\textrm{con}}}{\partial\dot{z}(t)} + \cdots,
\ee
with $z=x^i_a$ or $z=p_{ai}$.
\cite{schaefer-84} and \cite{damour-91} show that time derivatives of $\xa$ and $\pa$ in the higher-order Hamiltonian \eqref{matterHho}
can be eliminated by the use of lower-order equations of motion, leading to an ordinary Hamiltonian,
\be
\label{matterHho1}
H_{\textrm{con}}^{\textrm{ord}}[\xa,\pa] = H_{\textrm{con}}[\xa,\pa,\dot{\bx}_a(\xa,\pa),\dot{\bp}_a(\xa,\pa),\ldots].
\ee
Notice the important point that the two Hamiltonians $H_{\textrm{con}}$ and $H_{\textrm{con}}^{\textrm{ord}}$
do not belong to the same coordinate system.
Therefore, the Hamiltonians $H_{\textrm{con}}$ and $H_{\textrm{con}}^{\textrm{ord}}$ and their variables
should have, say, primed and unprimed notations which usually however does not happen in the literature
due to a slight abuse of notation.

A formal PN expansion of the Routh functional in powers of $1/c^2$ is feasible to all PN orders. With the aid of the definition 
$\dst h^{\rm TT}_{ij}\equiv\frac{16\pi G}{c^4}\hat{h}^{\rm TT}_{ij}$, we may write
\be
R\left[\xa,\pa, h^{\rm TT}_{ij}, \partial_th^{\rm TT}_{ij}\right] - \sum_a m_a c^2 
= \sum_{n=0}^{\infty} \frac{1}{c^{2n}} R_n\big[\xa,\pa, \hat{h}^{\rm TT}_{ij}, \partial_t{\hat{h}}^{\rm TT}_{ij}\big].
\ee
Hereof, the field equation for $h^{\rm TT}_{ij}$ results in a PN-series form,
\be
\label{FEqhTT}
\left(\Delta-\frac{1}{c^2}\partial_t^2\right)\hat{h}^{\rm TT}_{ij} = \sum_{n=0}^{\infty} \frac{1}{c^{2n}}
D^{\rm TT}_{(n)ij}\big[\bx,\xa,\pa, \hat{h}^{\rm TT}_{kl}, \partial_t{\hat{h}}^{\rm TT}_{kl}\big].
\ee
This equation must now be solved step by step using either retarded integrals
for getting the whole dynamics or time-symmetric ones for only the conservative dynamics defined by $H_{\textrm{con}}$,
which themselves have to be expanded in powers of $1/c$.
In higher orders, however, non-analytic in $1/c$ log-terms do show up (see, e.g., \citealp{damour-14,damour-16}).

To calculate the reduced Hamiltonian of Eq.~\eqref{ADMHred2} for a many-particle system
one has to perturbatively solve for $\phi$ and $\tilde{\pi}^ {ij}$ the constraint equations
$\mathcal{H}=0$ and $\mathcal{H}_i=0$ with the densities $\mathcal{H}$, $\mathcal{H}_i$
defined in Eqs.\ \eqref{Hdensity}--\eqref{Pdensity}.
Then the transition to the Routhian of Eq.~\eqref{ADMRouthian} is straightforward
using the second equation in \eqref{FEqs}.
The expansion of the Hamiltonian constraint equation up to $c^{-10}$ leads to the following equation
[in this equation and in the next one we use units $c=1$, $G=1/(16\pi)$]:\footnote{
Equations \eqref{Hcon3} and \eqref{Pcon3} are taken from \cite{jaranowski-98,jaranowski-98-e}
and they are enough to calculate 3PN-accurate two-point-mass Hamiltonian.
In \cite{jaranowski-15} one can find higher-order PN expansion of constraint equations,
performed in $d$ dimensions, necessary to compute 4PN Hamiltonian.}
\begin{align}
\label{Hcon3}
-\Delta\phi &= \sum_a \bigg[
1-\frac{1}{8}\phi+\frac{1}{64}\phi^2-\frac{1}{512}\phi^3+\frac{1}{4096}\phi^4
\nonumber\\&\quad
+ \left(\frac{1}{2}-\frac{5}{16}\phi+\frac{15}{128}\phi^2-\frac{35}{1024}\phi^3\right)
\frac{{\bf p}_a^2}{m_a^2}
\nonumber\\&\quad
+ \left(-\frac{1}{8}+\frac{9}{64}\phi-\frac{45}{512}\phi^2\right)
\frac{({\bf p}_a^2)^2}{m_a^4}
+\left(\frac{1}{16}-\frac{13}{128}\phi\right)\frac{({\bf p}_a^2)^3}{m_a^6}
-\frac{5}{128}\frac{({\bf p}_a^2)^4}{m_a^8}
\nonumber\\&\quad
+\left(-\frac{1}{2}+\frac{9}{16}\phi+\frac{1}{4}\frac{{\bf p}_a^2}{m_a^2}\right)
\frac{p_{ai}p_{aj}}{m_a^2}{\htt ij}
-\frac{1}{16}\left({\htt ij}\right)^2
\bigg] m_a\delta_a
\nonumber\\&\quad
+\left(1+\frac{1}{8}\phi\right)\left({\piti ij}\right)^2
+\left(2+\frac{1}{4}\phi\right){\piti ij}{\pitt ij}
+\left({\pitt ij}\right)^2
\nonumber\\&\quad
+\left[\left(-\frac{1}{2}+\frac{1}{4}\phi-\frac{5}{64}\phi^2\right)\phi_{,{ij}}
+\left(\frac{3}{16}-\frac{15}{128}\phi\right)\phi_{,i}\phi_{,j}
+2{\piti ik}{\piti jk}\right]{\htt ij}
\nonumber\\&\quad
+\left(\frac{1}{4}-\frac{7}{32}\phi\right)\left({\ghtt ijk}\right)^2
+\left(\frac{1}{2}+\frac{1}{16}\phi\right){\ghtt ijk}{\ghtt ikj}
\nonumber\\&\quad
+ \Delta\left[\left(-\frac{1}{2}+\frac{7}{16}\phi\right)\left({\htt ij}\right)^2\right]
-\left[\frac{1}{2}\phi{\htt ij}{\ghtt ikj}
+\frac{1}{4}\phi_{,k}\left({\htt ij}\right)^2\right]_{,k}
\nonumber\\&\quad
+ \mathcal{O}(c^{-12}).
\end{align}
The expansion of the momentum constraint equation up to $c^{-7}$ reads
\begin{align}
\label{Pcon3}
{\piti ij}_{,j} &= \left(-\frac{1}{2}+\frac{1}{4}\phi-\frac{5}{64}\phi^2\right) \sum_a p_{ai}\delta_a
+ \left(-\frac{1}{2}+\frac{1}{16}\phi\right)\phi_{,j}{\piti ij}
\nonumber\\&\quad
- \frac{1}{2}\phi_{,j}{\pitt ij} - {\piti jk}_{,k}{\htt ij}
+ {\piti jk}\left(\frac{1}{2}{\ghtt jki}-{\ghtt ijk}\right)
+ \mathcal{O}(c^{-8}).
\end{align}
In the Eqs.\ \eqref{Hcon3} and \eqref{Pcon3} dynamical field variables $h^{\rm TT}_{ij}$ and $\pi^{ij}_{\rm TT}$
are counted as being of the orders $1/c^4$ and $1/c^5$, respectively [cf.\ Eq.~\eqref{FEqhTT}].

\subsection{Poincar\'e invariance}
\label{subsec:Poincare}

In asymptotically flat spacetimes the Poincar\'e group is a global symmetry group.
Its generators $P^{\mu}$ and $J^{\mu\nu}$
are realized as functions $P^\mu(\xa,\pa)$ and $J^{\mu\nu}(\xa,\pa)$ on the many-body phase-space.
They are conserved on shell and fulfill the Poincar\'e algebra relations
for the Poisson bracket product (see, e.g., \citealp{regge-74}),
\begin{align}
\{ P^{\mu}, P^{\nu} \} &= 0,
\\[1ex]
\{ P^{\mu}, J^{\rho\sigma} \} &= -\eta^{\mu\rho} P^{\sigma} + \eta^{\mu\sigma} P^{\rho},
\\[1ex]
\{ J^{\mu\nu}, J^{\rho\sigma} \} &= -\eta^{\nu\rho} J^{\mu\sigma} + \eta^{\mu\rho} J^{\nu\sigma}
+ \eta^{\sigma\mu} J^{\rho\nu} - \eta^{\sigma\nu} J^{\rho\mu},
\end{align}
where the Poisson brackets are defined in an usual way,
\be
\{A,B\} \equiv \sum_a \left( \frac{\partial A}{\partial x_a^i}\frac{\partial B}{\partial p_{ai}}
- \frac{\partial A}{\partial p_{ai}}\frac{\partial B}{\partial x_a^i} \right).
\ee

The meaning of the components of $P^{\mu}$ and $J^{\mu\nu}$ is as follows:
the time component $P^0$ (i.e., the total energy)
is realized as the Hamiltonian $H\equiv c P^0$,
$P^i=P_i$ is linear momentum,
$J^i\equiv\frac{1}{2}\varepsilon^{ikl}J_{kl}$
[with $\varepsilon^{ijk}\equiv\varepsilon_{ijk}\equiv\frac{1}{2}(i-j)(j-k)(k-i)$,
$J_{kl}=J^{kl}$, and $J_{ij}=\varepsilon_{ijk}J^k$] is angular momentum,
and Lorentz boost vector is $K^i\equiv J^{i0}/c$.
The boost vector represents the constant of motion associated with the centre-of-mass theorem
and can further be decomposed as $K^i=G^i-t\,P^i$ (with $G_i=G^i$).
In terms of three-dimensional quantities
the Poincar\'e algebra relations read (see, e.g., \citealp{damour3-00,damour3-00-e})
\begin{align}
\label{PoiRel1}
\{ P_i, H \} &= 0,
\quad \{ J_i, H \} = 0,
\\[1ex]
\{ J_i, P_j \} &= \varepsilon_{ijk} \, P_k,
\quad \{ J_i, J_j \} = \varepsilon_{ijk} \, J_k,
\\[1ex]
\{ J_i, G_j \} &= \varepsilon_{ijk} \, G_k,
\\[1ex]
\label{PA}
\{ G_i, H \} &= P_i,
\\[1ex]
\{ G_i, P_j \} &= \frac{1}{c^2}\,H\,\delta_{ij},
\\[1ex]
\label{PoiRel6}
\{ G_i, G_j \} &= -\frac{1}{c^2}\,\varepsilon_{ijk}\,J_k.
\end{align}
The Hamiltonian $H$ and the centre-of-mass vector $G^i$
have the integral representations
\begin{align}
\label{defHint}
H &= -\frac{c^4}{16 \pi G} \int\md^3x\,\Delta\phi = -\frac{c^4}{16 \pi G} \oint_{i^0} r^2 \md\Omega\,\vecn\cdot\nabla\phi,
\\[1ex]
\label{defGint}
G^i &= -\frac{c^2}{16 \pi G} \int\md^3x\, x^i \Delta \phi
= -\frac{c^2}{16 \pi G} \oint_{i^0} r^2\md\Omega\, n^j (x^i \partial_j - \delta_{ij})\phi,
\end{align}
where $\vecn\,r^2\md\Omega$ ($\vecn$ is the outward radial unit vector)
is the two-dimensional surface-area element at $i^0$.
The two quantities $H$ and $G^i$ are the most involved ones of those entering the Poincar\'e algebra.

The Poincar\'e algebra has been extensively used in the calculations of PN Hamiltonians
for spinning binaries \citep{hergt1-08,hergt2-08}.
Hereby the most useful equation was \eqref{PA},
which tells that the total linear momentum has to be a total time derivative.
This equation was also used by \cite{damour3-00,damour3-00-e} to fix the so called ``kinetic ambiguity''
in the 3PN ADM two-point-mass Hamiltonian without using dimensional regularization.
In harmonic coordinates, the kinetic ambiguity got fixed by a Lorentzian version of the Hadamard regularization
based on the Fock--de Donder approach \citep{blanchet2-01}.

The explicit form of the generators $P^\mu(\xa,\pa)$ and $J^{\mu\nu}(\xa,\pa)$
(i.e., $\mathbf{P}$, $\mathbf{J}$, $\mathbf{G}$, and $H$) for two-point-mass systems is given in Appendix~\ref{app:ncom}
with 4PN accuracy.

The global Lorentz invariance results in the following useful expressions (see, e.g., \citealp{rothe-10}, \citealp{georg-15}).
Let us define the quantity $\mathcal{M}$ through the relation
\be
\label{ME}
\mathcal{M}c^2 \equiv \sqrt{H^{2}-{\bf{P}}^{2}c^{2}}
\quad \mbox{or} \quad
H = \sqrt{\mathcal{M}^{2}c^{4}+{\bf{P}}^{2}c^{2}},
\ee
and let us introduce the canonical centre of the system vector ${\bf X}$ (with components $X^i = X_i$),
\be
{\bf X} \equiv \frac{{\bf G}c^2}{H}
+ \frac{1}{\mathcal{M}\left(H+\mathcal{M}c^{2}\right)}\left({\bf J}
- \left(\frac{{\bf G}c^2}{H}\times{\bf P}\right)\right)\times{\bf P}.
\ee
Then the following commutation relations are fulfiled:
\begin{align}
\left\{X_i,\, P_j \right\} &= \delta_{ij}, \quad \left\{X_i,\, X_j \right\}= 0, \quad \left\{P_i,\, P_j \right\} = 0,
\\[1ex]
\left\{\mathcal{M},\, P_i \right\} &= 0, \quad \left\{\mathcal{M},\, X_i \right\} = 0,
\\
\left\{\mathcal{M},\, H \right\} &= 0, \quad  \left\{P_i,\, H \right\} = 0, \quad  \frac{H}{c^2}\left\{X_i,\, H \right\} = P_i.
\end{align}
The commutation relations clearly show the complete decoupling of
the internal dynamics from the external one by making use of the canonical variables.
The equations \eqref{ME} additionally indicate that $\mathcal{M}^2$ is simpler (or, more primitive)
than $\mathcal{M}$, cf.\ \cite{georg-15}.
A centre-of-energy vector can be defined by $X_E^i = X_{Ei} = c^2 G^i / H = c^2 G_i / H$.
This vector, however, is not a canonical position vector, see, e.g., \cite{hanson-74}.

In view of our later treatment of particles with spin,
let us decompose the total angular momentum $J^{\mu\nu}$ of a single object
into orbital angular momentum $L^{\mu\nu}$ and spin $S^{\mu\nu}$,
both of them being anti-symmetric tensors,
\be
J^{\mu\nu} = L^{\mu\nu} + S^{\mu\nu}.
\ee
The orbital angular momentum tensor is given by
\be
L^{\mu\nu} = Z^{\mu}P^{\nu} - Z^{\nu}P^{\mu},
\ee
where $Z^{\mu}$ denotes 4-dimensional position vector (with $Z^0 = ct$).
The splitting in space and time results in
\be
J^{ij} = Z^iP^j - Z^jP^i + S^{ij},
\quad
J^{i0} = Z^iH/c - P^ict + S^{i0}.
\ee

Remarkably, relativity tells us that any object with mass $\mathcal{M}$, spin length $S$, and positive energy density
must have extension orthogonal to its spin vector of radius of at least $S/(\mathcal{M}c)$ (see, e.g., \citealp{misner-73}).
Clearly then, the position vector of such an object is not given a priori but must be defined.
As the total angular momentum should not depend on the fixation of the position vector,
the notion of spin must depend on the fixation of the position vector and vice versa.
Thus, imposing a \emph{spin supplementary condition} (SSC) fixes the position vector.
We enumerate here the most often used SSCs
(see, e.g., \citealp{fleming-65}, \citealp{hanson-74}, \citealp{barker-79}).
\begin{enumerate}
\item[(i)] Covariant SSC (also called Tulczyjew-Dixon SSC):
\be
P_{\nu}S^{\mu\nu} = 0.
\ee
The variables corresponding to this SSC are denoted in Sect.~7
by $Z^i = z^i$, $S^{ij}$, and $P^i = p^i$.
\item[(ii)] Canonical SSC (also called Newton-Wigner SSC):
\be
(P_{\nu} + \mathcal{M}c\,n_{\nu})S^{\mu\nu} = 0,
\quad \mathcal{M}c = \sqrt{-P_{\mu}P^{\mu}},
\ee
where $n_{\mu} = (-1,0,0,0)$, $n_{\mu}n^{\mu} = -1$.
The variables corresponding to this SSC are denoted in Sect.~7
by $\hat{z}^i$, $\hat{S}^{ij}$, and $P^i$.
\item[(iii)] Centre-of-energy SSC (also called Corinaldesi-Papapetrou SSC):
\be
n_{\nu}S^{\mu\nu} = 0.
\ee
Here the boost vector takes the form of a spinless object, $K^i = Z^iH/c^2 - P^it = G^i - P^it$.
\end{enumerate}

\subsection{Poynting theorem of GR}
\label{subsec:Poynting}

Let us start with the following local identity,
having structure of a Poynting theorem for GR in local form,
\be
\label{Poynting1}
-\dot{h}^{\textrm{TT}}_{ij} \Box h^{\textrm{TT}}_{ij}
= -\partial_k\left(\dot{h}^{\textrm{TT}}_{ij}h^{\textrm{TT}}_{ij,k}\right)
+ \frac{1}{2}\partial_t \left[(\dot{h}^{\textrm{TT}}_{ij}/c)^2 + (h^{\textrm{TT}}_{ij,k})^2\right],
\ee
where $\Box\equiv-\partial_t^2/c^2+\Delta$ denotes the d'Alembertian.
Integrating this equation over whole space gives, assuming past stationarity,
\be
\label{Poynting2}
-\int_{V_\infty}\md^3x\, \dot{h}^{\textrm{TT}}_{ij} \Box h^{\textrm{TT}}_{ij}
= \frac{1}{2} \int_{V_\infty}\md^3x\,\partial_t \left[(\dot{h}^{\textrm{TT}}_{ij}/c)^2 + (h^{\textrm{TT}}_{ij,k})^2\right],
\ee
where $V_\infty$ is just another expression for ${\mathbb R}^3$.
Notice that the far or wave zone\footnote{For precise definition of wave zone see, e.g.,
section III in \cite{thorne-80} or sections 1.3 and 3.2.3 in \cite{thorne-83}.
We do not separate here the local wave zone from the distant wave zone as, e.g., in \cite{thorne-80};
for specific tail terms, see our subsection \ref{subsec:radfield}.}
is understood as area of the $t$ = const slice where
gravitational waves are decoupled from their source and do freely propagate outwards,
what means that the relation
$h^{\textrm{TT}}_{ij,k}=-(n^k/c)\dot{h}^{\textrm{TT}}_{ij}+\mathcal{O}(r^{-2})$
($r$ being the radial coordinate) is fulfilled in the far zone
at distances $r\gg\lambda/(2\pi)$ from the source,
where $\lambda$ is characteristic wavelength of gravitational waves emitted by the source.
We always use $t$ = const slices, where our Hamiltonians are defined on,
and explore physical processes by going from one slice to another one located in the close-by future.
This suffices to discriminate radiation from non-radiation for any given approximation.
Spacelike infinity ($i^0$) is enough for posing reliable boundary conditions,
timelike infinity is not needed, neither for the future nor for the past
(past stationarity simply replaces past infinity).
Integration of Eq.~\eqref{Poynting1} over the volume $V_{\rm fz}$
enclosed by its outer boundary located in the far zone (fz)
with $\md s_k = n^k r^2 \md\Omega$ surface-area element of the two-surface of integration
with $\md\Omega$ as the solid-angle element, yields
\be
\label{Poynting3}
-\int_{V_{\rm fz}}\md^3x\, \dot{h}^{\textrm{TT}}_{ij} \Box h^{\textrm{TT}}_{ij}
= -\oint_{\rm fz} \md s_k\,\dot{h}^{\textrm{TT}}_{ij}h^{\textrm{TT}}_{ij,k}
+ \frac{1}{2} \int_{V_{\rm fz}}\md^3x\,
\partial_t \left[(\dot{h}^{\textrm{TT}}_{ij}/c)^2 + (h^{\textrm{TT}}_{ij,k})^2\right].
\ee
To make sure that the surface integral (say over a sphere of radius $r_{\text{fz}}$)
in the above equation is not zero,
we have to assume that $r_{\text{fz}}$ is located in the far zone,
where real wave propagation happens,
i.e.\ behind the wave front of the out-propagating wave.
Of course, as the system is stationary in the remote past,
the wave front has still infinite distance to $i^0$.

Combining Eqs.\ \eqref{Poynting2} and \eqref{Poynting3} together, one gets
\begin{align}
\label{Poynting5}
-\int_{(V_\infty- V_{\rm fz})}\md^3x\, &\dot{h}^{\textrm{TT}}_{ij} \Box h^{\textrm{TT}}_{ij}
= \oint_{\rm fz}\md s_k\,\dot{h}^{\textrm{TT}}_{ij}h^{\textrm{TT}}_{ij,k}
\nonumber\\[1ex]&\quad
+ \frac{1}{2} \int_{(V_\infty-V_{\rm fz})}\md^3x\,
\partial_t \left[(\dot{h}^{\textrm{TT}}_{ij}/c)^2 + (h^{\textrm{TT}}_{ij,k})^2\right].
\end{align}
The volume $(V_\infty–V_{\rm fz})$ is meant for $t$ = const and thus reaches $i^0$;
it embraces the radial coordinates $r_{\rm bfz}\lesssim r \lesssim+\infty$,
where $r_{\rm bfz}$ denotes the beginning of the far zone.
In the following we drop the left side of this equation as negligibly small
[of the relative order $\lambda/(2\pi r_{\text{fz}})$,
where $r_{\text{fz}}$ is located in the far zone].
Indeed, we can assume that the source term for $\Box h^{\textrm{TT}}_{ij}$,
which follows from the Routhian field equation \eqref{RFE},
decays at least as $1/r^3$ for $r\to\infty$ (for isolated systems,
all source terms for $\Box h^{\textrm{TT}}_{ij}$ decay at least as $1/r^4$ if not TT-projected;
the TT-projection may raise the decay to $1/r^3$, e.g.\ TT-projection of Dirac delta function).
Additionally, $\dot{h}^{\textrm{TT}}_{ij}$ decays as $1/r$,
so the integrand on the left side decays in total as $1/r^4$.
This results in
\be
\label{Poynting6}
\frac{c^3}{32\pi G}\oint_{\rm fz} \md\Omega\, r^2 (\dot{h}_{ij}^{\rm TT})^2
= \frac{c^2}{32\pi G}\frac{\md}{\md t} \int_{(V_\infty - V_{\rm fz})}\md^3x\, (\dot{h}_{ij}^{\rm TT})^2,
\ee
with meaning that the energy flux through a surface in the far zone
equals the growth of gravitational energy beyond that surface.

\subsection{Near-zone energy loss and far-zone energy flux}
\label{subsec:nfzones}

The change in time of the matter Routhian reads, assuming ${\mathcal R}$ to be \emph{local} in the gravitational field,
\be
\label{nfz1}
\frac{\md R}{\md t} = \frac{\partial R}{\partial t}
= \int\md^3x\, \frac{\partial {\mathcal R}}{\partial h^{\textrm{TT}}_{ij}} \dot{h}^{\textrm{TT}}_{ij}
+ \int\md^3x\, \frac{\partial {\mathcal R}}{\partial h^{\textrm{TT}}_{ij,k}} \partial_k \dot{h}^{\textrm{TT}}_{ij}
+ \int\md^3x\, \frac{\partial {\mathcal R}}{\partial\dot{h}^{\textrm{TT}}_{ij}} \ddot{h}^{\textrm{TT}}_{ij},
\ee
where
\be
\label{nfz2}
R(\xa,\pa,t) \equiv \int\md^3x\,
\mathcal{R}(\xa,\pa,h^{\textrm{TT}}_{ij}(t),h^{\textrm{TT}}_{ij,k}(t),\dot{h}^{\textrm{TT}}_{ij}(t)).
\ee
The equation for $\md R/\md t$ is valid provided the equations of motion
\be
\label{nfz3}
{\dot p}_{ai} = -\frac{\partial R}{\partial x_a^i},
\quad
{\dot x}_a^i = \frac{\partial R}{\partial p_{ai}}
\ee
hold. Furthermore, we have
\begin{align}
\label{nfz4}
&\int\md^3x\, \frac{\partial {\mathcal R}}{\partial h^{\textrm{TT}}_{ij,k}} \partial_k \dot{h}^{\textrm{TT}}_{ij}
+ \int\md^3x\, \frac{\partial {\mathcal R}}{\partial\dot{h}^{\textrm{TT}}_{ij}} \ddot{h}^{\textrm{TT}}_{ij}
= \int\md^3x\, \partial_k\left(\frac{\partial {\mathcal R}}{\partial h^{\textrm{TT}}_{ij,k}} \dot{h}^{\textrm{TT}}_{ij}\right)
\nonumber\\[1ex]
&+ \frac{\md}{\md t} \int\md^3x\, \frac{\partial {\mathcal R}}{\partial\dot{h}^{\textrm{TT}}_{ij}} \dot{h}^{\textrm{TT}}_{ij}
- \int\md^3x\, \partial_k\left(\frac{\partial {\mathcal R}}{\partial h^{\textrm{TT}}_{ij,k}}\right) \dot{h}^{\textrm{TT}}_{ij}
- \int\md^3x\, \frac{\md}{\md t}\left(\frac{\partial {\mathcal R}}{\partial\dot{h}^{\textrm{TT}}_{ij}}\right) \dot{h}^{\textrm{TT}}_{ij}.
\end{align}
The canonical field momentum is given by 
\be
\label{nfz5}
\frac{c^3}{16\pi G} \pi^{ij}_{\textrm{TT}}
= -\delta^{\textrm{TT}ij}_{kl}\frac{\partial {\mathcal R}}{\partial \dot{h}^{\textrm{TT}}_{kl}}.
\ee
Performing the Legendre transformation
\be
\label{nfzones}
H = R + \frac{c^3}{16\pi G} \int\md^3x\,\pi^{ij}_{\textrm{TT}}\dot{h}^{\textrm{TT}}_{ij},
\quad \mbox{or} \quad
R = H - \frac{c^3}{16\pi G} \int\md^3x\,\pi^{ij}_{\textrm{TT}}\dot{h}^{\textrm{TT}}_{ij},
\ee
the energy loss equation takes the form
[using Eq.~\eqref{nfz1} together with \eqref{nfz4} and \eqref{nfz5}]
\begin{align}
\label{nfzones1a}
\frac{\md H}{\md t} &= \int\md^3x\, \partial_k\left(\frac{\partial {\mathcal R}}{\partial h^{\textrm{TT}}_{ij,k}} \dot{h}^{\textrm{TT}}_{ij}\right)
+ \int\md^3x\, \frac{\partial {\mathcal R}}{\partial h^{\textrm{TT}}_{ij}} \dot{h}^{\textrm{TT}}_{ij}
\nonumber\\[1ex]&\quad
- \int\md^3x\, \partial_k\left(\frac{\partial {\mathcal R}}{\partial h^{\textrm{TT}}_{ij,k}}\right) \dot{h}^{\textrm{TT}}_{ij}
- \int\md^3x\, \frac{\md}{\md t}\left(\frac{\partial {\mathcal R}}{\partial\dot{h}^{\textrm{TT}}_{ij}}\right) \dot{h}^{\textrm{TT}}_{ij}.
\end{align}
Application of the field equations
\be
\frac{\partial {\mathcal R}}{\partial h^{\textrm{TT}}_{ij}}
- \partial_k\left(\frac{\partial {\mathcal R}}{\partial h^{\textrm{TT}}_{ij,k}}\right)
- \frac{\md}{\md t}\left(\frac{\partial\mathcal{R}}{\partial\dot{h}^{\textrm{TT}}_{ij}}\right) = 0
\ee
yields, assuming past stationarity [meaning that at any finite time $t$
no radiation can have reached spacelike infinity,
so the first (surface) term in the right-hand side of Eq.~\eqref{nfzones1a} vanishes],
\be
\frac{\md H}{\md t} = 0.
\ee
The Eq.~\eqref{Poynting6} shows that the Eq.~\eqref{nfzones} infers,
employing the leading-order quadratic field structure of ${\mathcal R}$
\big[$\mathcal{R}=-(1/4)(c^2/(16\pi G))(\dot{h}^{\textrm{TT}}_{ij})^2+\cdots$; see Eq.~\eqref{1PMRfield}\big],
\be
\label{nfzones2}
\frac{\md}{\md t}\left(R - \int_{V_\textrm{fz}}\md^3x\,\frac{\partial {\mathcal R}}{\partial \dot{h}^{\textrm{TT}}_{ij}}\dot{h}^{\textrm{TT}}_{ij}\right)
= -\cal{L},
\ee
where
\be
\label{luminosity}
{\mathcal L} = -\frac{c^4}{32\pi G} \oint_{\rm fz} \md s_k h^{\textrm{TT}}_{ij,k} \dot{h}^{\textrm{TT}}_{ij}
= \frac{c^3}{32\pi G} \oint_{\rm fz} \md\Omega\, r^2 (\dot{h}^{\textrm{TT}}_{ij})^2
\ee
is the well known total energy flux (or luminosity) of gravitational waves.
The Eq.~\eqref{nfzones2} can be put into the energy form, again employing the leading-order quadratic field structure of ${\mathcal R}$,
\be
\label{nfzones3}
\frac{\md}{\md t}\left(H - \frac{c^2}{32\pi G}\int_{(V_\infty-V_\textrm{fz})}\md^3x\,(\dot{h}^{\textrm{TT}}_{ij})^2\right) = -\cal{L}.
\ee
Note that the integral over $V_\infty-V_\textrm{fz}$ changes with time for radiating sources
because more and more radiation is entering the volume $V_\infty-V_\textrm{fz}$,
whereas the integral over $V_\textrm{fz}$ changes on secular damping-time scales only
because for stationary time-sections the volume $V_\textrm{fz}$ is filled with constant amount of radiation energy.

Taking into account the Eqs.~\eqref{Hcon3} and \eqref{defHint}
we find that the second term in the parenthesis of the left side of Eq.~\eqref{nfzones3}
exactly subtracts the corresponding terms from pure $(h^{\rm TT}_{ij,k})^2$
and $(\pi_{\rm TT}^{ij})^2$ expressions therein.
This improves, by one order in radial distance, the large distance decay of the integrand of the integral of the whole left side of Eq.~\eqref{nfzones3},
which runs over the whole hypersurface $t$ = const.
We may now perform near- and far-zone PN expansions of the left and right sides of the Eq.~\eqref{nfzones3}, respectively.
Though the both series are differently defined---on the left side,
expansion in powers of $1/c$ around fixed time $t$ of an energy expression which is time differentiated;
on the right side, expansion in powers of $1/c$ around fixed retarded time $t-r/c$---the expansions cannot contradict each other
as long as they are not related term by term.
For the latter relation we must keep in mind that PN expansions are instantaneous expansions
so that the two times, $t$ and $t-r/c$, are  not allowed to be located too far apart from each other.
This means that we have to read off the radiation right when it enters far zone.
Time-averaging of the expressions on the both sides of Eq.~\eqref{nfzones3} over several wave periods
[see text below Eq.~\eqref{balance2}] makes the difference between the two times negligible
as it should be if one is interested in a one-to-one correspondence between the terms on the both sides.
The Newtonian and 1PN wave generation processes were explicitly shown to fit into this scheme by \cite{koenigsdoerffer-03}.

\subsection{Radiation field}
\label{subsec:radfield}

In the far zone, the multipole expansion of the transverse-traceless (TT) part of the gravitational field,
obtained by algebraic projection with
\begin{align}
P_{ijkl}(\mathbf{n}) &\equiv \frac{1}{2}\Big(P_{ik}(\mathbf{n})P_{jl}(\mathbf{n})
+ P_{il}(\mathbf{n})P_{jk}(\mathbf{n}) - P_{ij}(\mathbf{n})P_{kl}(\mathbf{n}\Big),
\\[1ex]
P_{ij}(\mathbf{n}) &\equiv \delta_{ij} - n_i n_j,
\end{align}
where $\vecn\equiv \vecx/r$ ($r\equiv|\vecx|$) is the unit vector in the direction from the source to the far away observer,
reads (see, e.g., \citealp{thorne-80}, \citealp{blanchet-14})
\begin{align}
h_{ij}^{\rm TT\,fz}(\mathbf{x},t) &= \frac{G}{c^4} \frac{P_{ijkm}(\mathbf{n})}{r}
\sum_{l=2}^{\infty} \left\{\left(\frac{1}{c^2}\right)^{\frac{l-2}{2}} \frac{4}{l!}\,
\mbox{M}^{(l)}_{kmi_3 \ldots i_l}\left(t-\frac{r_*}{{c}}\right)~N_{i_3 \ldots i_l} \right.
\nonumber\\[1ex]&\quad
+ \left. \left(\frac{1}{{c^2}}\right)^{\frac{l-1}{2}} \frac{8l}{(l+1)!}\,
\varepsilon_{pq(k} \mbox{S}^{(l)}_{m)pi_3 \ldots i_l}\left(t-\frac{r_*}{{c}}\right)~n_q~N_{i_3 \dots i_l}\right\},
\end{align}
where $N_{i_3 \dots i_l}\equiv n^{i_3}\ldots n^{i_l}$
and where $\mbox{M}^{(l)}_{i_1i_2i_3\ldots i_l}$ and $\mbox{S}^{(l)}_{i_1i_2i_3\ldots i_l}$ denote the $l$th time derivatives
of the symmetric and tracefree (STF) radiative mass-type and current-type multipole moments, respectively.
The term with the leading mass-quadrupole tensor takes the form (see, e.g., \citealp{schaefer-90})
\begin{align}
\mbox{M}_{ij}^{(2)}&\left(t-\frac{r_*}{c}\right) = \widehat{\mbox{M}}_{ij}^{(2)}\left(t - \frac{r_*}{c}\right)
\nonumber\\[1ex]&\quad
+ \frac{2Gm}{c^3} \int_0^{\infty} \md v \left[\ln\left(\frac{v}{2b}\right) + \kappa\right]
\widehat{\mbox{M}}^{(4)}_{ij}\left(t-\frac{r_*}{c}-v\right)
+ {\mathcal O}\left(\frac{1}{c^4}\right),
\end{align}
with
\be
r_* = r + \frac{2Gm}{c^2}\mbox{ln}\left(\frac{r}{cb}\right) + {\mathcal O}\left(\frac{1}{c^3}\right)
\ee
showing the leading-order tail term of the quadrupole radiation
(the gauge dependent relative phase constant $\kappa$ between direct and tail term was not explored by \citealp{schaefer-90};
for more details see, e.g., \citealp{blanchet-93} and \citealp{blanchet-14}).
Notice the modification of the standard PN expansion through tail terms.
This expression nicely shows that also multipole expansions in the far zone do induce PN expansions.
The mass-quadrupole tensor $\widehat{\mbox{M}}_{ij}$ is just the standard Newtonian one.
Higher-order tail terms up to ``tails-of-tails-of-tails'' can be found in \cite{marchand-16}.
Leading-order tail terms result from the backscattering of the leading-order outgoing radiation,
the ``tails-of-tails'' from their second backscattering, and so on.

Through 1.5PN order, the luminosity expression \eqref{luminosity} takes the form
\be
{\mathcal{L}}(t) = \frac{G}{5c^5}\left\{\mbox{M}^{(3)}_{ij}\mbox{M}^{(3)}_{ij}
+ \frac{1}{c^2}\left[\frac{5}{189}\mbox{M}^{(4)}_{ijk}\mbox{M}^{(4)}_{ijk}
+ \frac{16}{9}\mbox{S}^{(3)}_{ij}\mbox{S}^{(3)}_{ij}\right] \right\}.
\ee
On reasons of energy balance in asymptotically flat space,
for any coordinates or variables representation of the Einstein theory,
the time-averaged energy loss has to fulfill a relation of the form
\be
\label{balance2}
-\left\langle\frac{\dst\md{\mathcal{E}}\left(t-{r_*}/{c}\right)}{\md t}\right\rangle
= \left\langle{\mathcal{L}}(t)\right\rangle,
\ee
where the time averaging procedure takes into account typical periods of the system
(i.e.\ it is averaging over several periods of the lowest frequency mode,
usually called ``averaging over several wavelengths''; see, e.g., \citealp{thorne-80}).
Generalizing our considerations after Eq.~\eqref{nfzones3}
we may take the observation time $t$ much larger than the time,
say $t_{\rm bfz}$, the radiation enters the far or wave zone,
even larger than the damping time of the radiating system,
by just freely transporting the radiation power along the null cone with tacitly assuming
$\langle{\cal L}(t,r)\rangle=\langle{\cal L}(t_{\rm bfz},r_{\rm bfz})\rangle$,
where $t-t_{\rm bfz}=(r-r_{\rm bfz})/c>0$.
Coming back to Eq.~\eqref{nfzones3}, time averaging on the left side of Eq.~\eqref{nfzones3}
eliminates total time derivatives of higher PN order, so-called Schott terms,
and transforms them into much higher PN orders.
The both sides of the equation \eqref{balance2} are gauge (or, coordinate) invariant.
We stress that the Eq.~\eqref{balance2} is valid for bound systems.
In case of scattering processes, a coordinate invariant quantity is the emitted total energy.

\sloppy The energy flux to $n$PN order in the far zone
implies energy loss to \mbox{$(n+2.5)$PN} order in the near zone.
The leading-order 2.5PN energy loss is usually called ``Newtonian''
because only the Newtonian source dynamics contributes;
corresponding notions are applied to the higher order PN fluxes.
Hereof it follows that energy-loss calculations are quite efficient via energy-flux calculations \citep{blanchet-14}.
In general, only after averaging over orbital periods the both expressions do coincide.
In the case of circular orbits, however, this averaging procedure is not needed.

\section{Applied regularization techniques}

The most efficient source model for analytical computations
of many-body dynamics in general relativity are point masses (or particles)
represented through Dirac delta functions. If internal degrees of freedom are
come into play, derivatives of the delta functions must be incorporated into the source.
Clearly, point-particle sources in field theories introduce field singularities, which must be regularized in computations.
Two aspects are important: (i) the differentiation of singular functions
(i.e.\ functions which are not infinitely differentiable),
and (ii) the integration of singular functions, either to new (usually also singular) functions or to the final Routhian/Hamiltonian.
The item (ii) relates to the integration of the field equations
and the item (i) to the differentiation of their (approximate) solutions.
On consistency reasons, differentiation and integration must commute.

The most efficient strategy developed for computation of higher-order PN point-particle Hamiltonians
relies on performing a 3-dimensional full computation in the beginning
(using Riesz-implemented Hadamard regularization defined later in this section)
and then correcting it by a $d$-dimensional one around the singular points,
as well the local ones (UV divergences) as the one at infinity (IR divergences).
A $d$-dimensional full computation is not needed.
At higher than the 2PN level 3-dimensional computations with analytical Hadamard and Riesz regularizations
show up ambiguities which require a more powerful treatment. The latter is dimensional regularization.
The first time this strategy was successfully applied in the context of general relativity
was in the 3PN dynamics of binary point particles \citep{damour1-01};
IR divergences did not appear therein, those enter from the 4PN level on only, the same 
as the nonlocal-in-time tail terms to which they are connected.
At 4PN order, using different regularization methods for the treatment of IR divergences \citep{jaranowski-15},
an ambiguity parameter was left which, however, got fixed by matching to self-force
calculations in the Schwarzschild metric (\citealp{letiec-12}, \citealp{bini-13}, \citealp{damour-14}).

The regularization techniques needed to perform PN calculations up to (and including) 4PN order,
are described in detail in Appendix~A of \cite{jaranowski-15}.

\subsection{Distributional differentiation of homogeneous functions}

Besides appearance of UV divergences, another consequence of employing Dirac-delta sources
is necessity to differentiate homogeneous functions using an enhanced (or distributional) derivative,
which comes from standard distribution theory (see, e.g., Sect.~3.3 in Chapter~III of \citealp{gelfand-64}).

Let $f$ be a real-valued function defined in a neighbourhood of the origin of $\mathbb{R}^3$.
$f$ is said to be a \emph{positively homogeneous function of degree $\lambda$},
if for any number $a>0$
\be
f(a\,{\bf x}) = a^\lambda\,f({\bf x}).
\ee
Let $k:=-\lambda-2$. If $\lambda$ is an integer and if $\lambda\le-2$
(i.e., $k$ is a nonnegative integer), then the partial derivative of $f$ with
respect to the coordinate $x^i$ has to be calculated by means of the formula
\be
\label{DisDiff}
\partial_i f({\bf x}) = \partial_{\underline i}f({\bf x})
+ \frac{(-1)^k}{k!} \frac{\partial^k\delta({\bf x})}{\partial x^{i_1}\cdots\partial x^{i_k}}
\times \oint_\Sigma \md\sigma_i\,f({\bf x}')\,x'^{i_1}\cdots x'^{i_k},
\ee
where $\partial_i f$ on the lhs denotes the derivative of $f$ considered as a distribution,
while $\partial_{\underline i}f$ on the rhs denotes the derivative of $f$ considered as a function
(which is computed using the standard rules of differentiation),
$\Sigma$ is any smooth close surface
surrounding the origin and $\md\sigma_i$ is the surface element on $\Sigma$.

The distributional derivative does not obey the Leibniz rule. It 
can easily be seen by considering the distributional partial derivative of the 
product $1/r_a$ and $1/r_a^2$. Let us suppose that the Leibniz rule is 
applicable here:
\be
\partial_i{\frac{1}{r_a^3}}
= \partial_i{\left(\frac{1}{r_a}\frac{1}{r_a^2}\right)}
= \frac{1}{r_a^2}\, \partial_i{\frac{1}{r_a}}
+ \frac{1}{r_a}\, \partial_i{\frac{1}{r_a^2}}.
\ee
The right-hand side of this equation can be computed using standard differential calculus (no terms 
with Dirac deltas), whereas computing the left-hand side one obtains some term 
proportional to $\partial_i\delta_a$. The distributional differentiation is necessary when one differentiates
homogeneous functions under the integral sign.
For more details, see Appendix A~5 in \cite{jaranowski-15}.

\subsection{Riesz-implemented Hadamard regularization}
\label{subs:RHreg}

The usage of Dirac $\delta$-functions to model point-mass sources of gravitational field leads
to occurence of UV divergences, i.e., the divergences near the particle locations $\xa$,
as $r_a\equiv|\mathbf{x}-\xa|\to0$. To deal with them, \cite{infeld-54,infeld-57}, \cite{infeld-60} introduced ``good'' $\delta$-functions,
which, besides having the properties of ordinary Dirac $\delta$-functions, also satisfy the condition
\be
\label{aReg1}
\frac{1}{|{\bf x} - {\bf x}_0|^k}\delta({\bf x}-{\bf x}_0) = 0,
\quad k = 1,\ldots,p,
\ee
for some positive integer $p$
(in practical calculations one takes $p$ large enough to take all singularities appearing in the calculation into account).
They also assumed that the ``tweedling of products'' property is always satisfied
\be
\label{aReg2}
\int \md^3x\, f_1({\bf x})f_2({\bf x})\delta({\bf x} - {\bf x}_0)
= f_{1 \rm reg}({\bf x}_0) f_{2 \rm reg}({\bf x}_0),
\ee
where ``reg'' means regularized value of the function at its singular point
(i.e., ${\bf x}_0$ in the equation above) evaluated by means of the rule \eqref{aReg1}.

A natural generalization of the rule \eqref{aReg1}
is the concept of ``partie finie'' value of function at its singular point, defined as
\be
\label{aReg3}
f_{\rm reg}({\bf x}_0) \equiv \frac{1}{4\pi} \int\md\Omega \, a_0(\vecn),
\ee
with (here $M$ is some non-negative integer)
\be
\label{aReg4}
f({\bf x} = {\bf x}_0 + \epsilon \vecn) = \sum_{m=-M}^{\infty} a_m(\vecn)\epsilon^m,
\quad
\vecn \equiv \frac{{\bf x}-{\bf x}_0}{|{\bf x}-{\bf x}_0|}.
\ee
Defining, for a function $f$ singular at $\mathbf{x}={\bf x}_0$,
\be
\label{aReg5}
\int\md^3x f({\bf x})\delta({\bf x}-{\bf x}_0) \equiv f_{\rm reg}({\bf x}_0),
\ee
the ``tweedling of products'' property \eqref{aReg2} can be written as
\be
\label{aReg6}
(f_1 f_2)_{\rm reg}({\bf x}_0) = f_{1 \rm reg}({\bf x}_0) f_{2 \rm reg}({\bf x}_0).
\ee
The above property is generally wrong for arbitrary singular functions $f_1$ and $f_2$.
In the PN calculations problems with fulfilling this property begin at the 3PN order.
This is one of the reasons why one should use dimensional regularization.

The Riesz-implemented Hadamard (RH) regularization was developed
in the context of deriving PN equations of motion of binary systems
by \cite{jaranowski-97,jaranowski-98,jaranowski-98-e}
to deal with locally divergent integrals computed in three dimensions.
The method is based on the Hadamard ``partie finie'' and the Riesz analytic continuation procedures.

The RH regularization relies on multiplying the full integrand, say $i({\bf x})$,
of the divergent integral by a regularization factor,
\be
\label{regfactor}
i({\bf x}) \longrightarrow i({\bf x})\Big(\frac{r_1}{s_1}\Big)^{\epsilon_1} \Big(\frac{r_2}{s_2}\Big)^{\epsilon_2},
\ee
and, after integration, studying the double limit $\epsilon_1\to0$, $\epsilon_2\to0$
within analytic continuation in the complex $\epsilon_1$ and $\epsilon_2$ planes
(here $s_1$ and $s_2$ are arbitrary three-dimensional UV regularization scales).
Let us thus consider such integral performed over the whole space ${\mathbb R}^3$
and let us assume that it develops \emph{only local poles} (so it is convergent at spatial infinity).
The value of the integral, after performing the RH regularization in three dimensions,
has the structure (this is the most general structure in the calculation
of conservative Hamiltonians up to and including 4PN order)
\begin{align}
\label{IRHuv}
I^{\text{RH}}&(3;\epsilon_1,\epsilon_2) \equiv
\int_{{\mathbb R}^3} i({\mathbf x})
\Big(\frac{r_1}{s_1}\Big)^{\epsilon_1} \Big(\frac{r_2}{s_2}\Big)^{\epsilon_2}\,\md^3x
\nonumber\\[1ex]
&= A + c_1 \Big(\frac{1}{\epsilon_1} + \ln\frac{r_{12}}{s_1} \Big)
+ c_2 \Big(\frac{1}{\epsilon_2} + \ln\frac{r_{12}}{s_2} \Big)
+ \mathcal{O}(\epsilon_1,\epsilon_2).
\end{align}
\sloppy{Let us mention that in the PN calculations regularized integrands
$i({\mathbf x})(r_1/s_1)^{\epsilon_1}(r_2/s_2)^{\epsilon_2}$
depend on $\mathbf{x}$ only through $\mathbf{x}-\mathbf{x}_1$ and $\mathbf{x}-\mathbf{x}_2$,
so they are translationally invariant. This explains why the regularization result \eqref{IRHuv}
depends on $\mathbf{x}_1$ and $\mathbf{x}_2$ only through $\mathbf{x}_1-\mathbf{x}_2$.}

In the case of an integral over ${\mathbb R}^3$ developing poles \emph{only at spatial infinity}
(so it is locally integrable)
it would be enough to use a regularization factor of the form $(r/r_0)^\epsilon$
(where $r_0$ is an IR regularization scale), but it is more convenient to use the factor
\be
\Big(\frac{r_1}{r_0}\Big)^{a\epsilon} \Big(\frac{r_2}{r_0}\Big)^{b\epsilon}
\ee
and, after integration, study the limit $\epsilon\to0$.
Let us denote the integrand again by $i(\mathbf{x})$.
The integral, after performing the RH regularization in three dimensions,
has the structure
\be
\label{inf1}
I^{\text{RH}}(3;a,b,\epsilon)
\equiv \int_{{\mathbb R}^3} i({\mathbf x}) \Big(\frac{r_1}{r_0}\Big)^{a\epsilon} \Big(\frac{r_2}{r_0}\Big)^{b\epsilon}\,\md^3x
= A - c_\infty \bigg(\frac{1}{(a+b)\epsilon} + \ln\frac{r_{12}}{r_0} \bigg) + \mathcal{O}(\epsilon).
\ee

Many integrals appearing in PN calculations were computed using a famous formula derived in \cite{riesz-49}
in $d$ dimensions. It reads
\be
\int\md^dx\, r^{\alpha}_1r^{\beta}_2 = \pi^{d/2} \frac{\Gamma(\frac{\alpha +d}{2}) \Gamma(\frac{\beta +d}{2}) \Gamma(-\frac{\alpha+\beta+d}{2}) }
{\Gamma(-\frac{\alpha}{2})\Gamma(-\frac{\beta}{2})\Gamma(\frac{\alpha +\beta+2d}{2})}r^{\alpha + \beta + d}_{12}.
\ee
To compute the 4PN-accurate two-point-mass Hamiltonian one needs to employ a generalization of the three-dimensional
version of this formula for integrands of the form $r^{\alpha}_1r^{\beta}_2(r_1+r_2+r_{12})^\gamma$.
Such formula was derived by \cite{jaranowski-98,jaranowski-98-e} and also there an efficient way of implementing both formulae
to regularize divergent integrals was proposed (it employs prolate spheroidal coordinates in three dimensions).
See Appendix A~1 of \cite{jaranowski-15} for details
and Appendix~A of \cite{hartung-13} for generalization of this procedure to $d$ space dimensions.

\subsection{Extended Hadamard regularization}
\label{subs:EHreg}

A specific variant of 3-dimensional Hadamard regularization called the \emph{extended Hadamard regularization} (EHR)
was devised by \cite{blanchet1-00,blanchet2-01}.
It was used by \cite{blanchet2-00,blanchet1-01} at the 3PN-level computations
of two-point-mass equations of motion in harmonic coordinates.

The basic idea of EHR is to associate to any function $F\in\mathcal{F}$,
where the set $\mathcal{F}$ comprises functions which are smooth on $\mathbb{R}^3$
except for the two points (around which they admit a power-like singular expansion),
a partie-finie \emph{pseudo-function} $\mathrm{Pf}F$,
which is a linear form acting on functions from $\mathcal{F}$:
\be
\label{EHreg1}
\langle\mathrm{Pf}F,G\rangle := \mathrm{Pf}_{s_1,s_2}\int\md^3x\,FG, \quad \text{for any $G\in\mathcal{F}$},
\ee
where $\mathrm{Pf}_{s_1,s_2}$ on the right-hand side means partie finie of the divergent integral
[see Eq.\ (3.1) in \cite{blanchet1-00} and the text around for the definition];
it depends on two---one per each singularity---arbitrary regularization scales $s_1$ and $s_2$.
The Dirac $\delta$-functions $\delta_a$ are represented by the pseudo-functions $\mathrm{Pf}\delta_a$
defined by
\be
\label{EHreg2}
\langle\mathrm{Pf}\delta_a,G\rangle := G_{\rm reg}({\bf x}_a), \quad \text{for any $G\in\mathcal{F}$},
\ee
where the regularized value $G_{\rm reg}({\bf x}_a)$ of function at its singular point
is defined in Eqs.\ \eqref{aReg3}--\eqref{aReg4} above.
The product $F\delta_a$ is represented by another pseudo-function $\mathrm{Pf}(F\delta_a)$ such that
\be
\label{EHreg3a}
\langle\mathrm{Pf}(F\delta_a),G\rangle := (FG)_{\rm reg}({\bf x}_a), \quad \text{for any $G\in\mathcal{F}$}.
\ee
As a consequence, in general
\be
\label{EHreg3b}
\mathrm{Pf}(F\delta_a) \ne F_{\textrm{reg}}(\mathbf{x}_a)\mathrm{Pf}\delta_a.
\ee

Another ingredient of the EHR relies on the specific treatment of partial derivatives of singular functions.
To ensure the possibility of integration by parts,
one requires that $\langle\partial_i(\mathrm{Pf}F),G\rangle=-\langle\partial_i(\mathrm{Pf}G),F\rangle$
for any functions $F,G\in\mathcal{F}$.
This requirement leads to the following definition of the partial derivative of the pseudo-function:
\be
\label{EHreg4}
\partial_i(\mathrm{Pf}F) = \mathrm{Pf}(\partial_i F) + \mathrm{D}_i[F],
\ee
where $\mathrm{Pf}(\partial_i F)$ denotes the ordinary derivative of $F$ viewed as a pseudo-function,
and $\mathrm{D}_i[F]$ is the purely distributional part with support concentrated on $\mathbf{x}_1$ or $\mathbf{x}_2$
[see Sects.\ VII--IX of \cite{blanchet1-00} for more details].
The derivative $\mathrm{D}_i[F]$ is an extended distributional derivative
which differs in general from the usual Schwartz derivative introduced in Eq.~\eqref{DisDiff} above.
Let us quote the results
\be
\label{EHreg5}
\mathrm{D}_i\Big[\frac{1}{r_1}\Big] = 2\pi \mathrm{Pf}(r_1 n_1^i \delta_1),
\quad
\mathrm{D}_{ij}\Big[\frac{1}{r_1}\Big] = -\frac{4\pi}{3} \mathrm{Pf}\Big(\delta^{ij}+\frac{15}{2}\hat{n}_1^{ij}\Big)\delta_1,
\ee
where $\hat{n}_1^{ij}\equiv n_1^in_1^j-\frac{1}{3}\delta^{ij}$.
The Schwartz derivative \eqref{DisDiff} of $\partial_i(1/r_1)$ contains no distributional part,
whereas distributional part of $\partial_i\partial_j(1/r_1)$ equals $-(4\pi/3)\delta^{ij}\delta_1$.

There is no known generalization of the EHR definitions \eqref{EHreg3a} and \eqref{EHreg4} to generic $d$-dimensional case.
Moreover, these definitions disagree with the dimensional-regularization rules.
\begin{enumerate}
\item[(i)] In generic $d$ dimensions one can always use
\be
F^{(d)}(\bx)\delta^{(d)}(\bx-\bx_a) = F^{(d)}_{\textrm{reg}}(\bx_a)\delta^{(d)}(\bx-\bx_a),
\ee
where $F^{(d)}$ is the $d$-dimensional version of $F$.
This leads to the following dimensional-regularization rule [see Sect.\ III~A in \cite{blanchet-04}]:
\be
\big[F(\bx)\delta^{(3)}(\bx-\bx_a)\big]_{\text{reg}}
:= \big(\lim_{d\to3}F^{(d)}_{\textrm{reg}}(\bx_a)\big)\delta^{(3)}(\bx-\bx_a).
\ee
The property \eqref{EHreg3b} disagrees with this.
\item[(ii)] The extended differentiation \eqref{EHreg4},
when applied to smooth functions of compact support,
coincides with Schwartz differentiation \eqref{DisDiff}.
However, in the 3PN-level computations performed by \cite{blanchet2-00,blanchet1-01}
it operated with other singular functions
and gave the results different from the results obtained by applying Schwartz differentiation.
The definition \eqref{DisDiff} of Schwartz differentiation is valid in $d$ dimensions (see Sect.\ \ref{subsubs:DisDiff-d} above),
which supports the use of this definition also in the limit of three dimensions.
\end{enumerate}

The computation using the EHR constitutes an approach very different from dimensional regularization,
following a different route which could not be combined with the latter.
This can be clearly seen in the paper by \cite{blanchet-04}
on dimensional-regularization completion of the 3PN equations of motion in harmonic coordinates
[see the paragraph containing Eq.\ (1.8) and Sect.\ III~D there].
Before applying dimensional regularization the authors of \cite{blanchet-04}
had to subtract from the 3-dimensional results of \cite{blanchet2-00,blanchet1-01}
all contributions, which were direct consequences of the use of EHR.
However, \cite{blanchet2-00,blanchet1-01} have shown that at the 3PN level
the difference between the final results of EHR and dimensional regularization computations
of two-point-mass equations of motion can be described in terms of one constant ambiguity parameter (they called $\lambda$).

\cite{yang-13} have recently developed the theory of ``thick distributions''
in higher dimensions $n$ (where $n$ is an integer larger than 1).
This theory is connected with the extended Hadamard regularization, but is not equivalent to the latter.

\subsection{Dimensional regularization}
\label{subs:DR}

It was first shown by \cite{damour1-01}, that the unambiguous treatment of UV divergences in the current context
requires usage of dimensional regularization (see, e.g., \citealp{collins-84}).
It was used both in the Hamiltonian approach and in the one using the Einstein field equations in harmonic coordinates
(\citealp{damour1-01}, \citealp{blanchet-04}, \citealp{jaranowski-13}, \citealp{damour-14}, \citealp{jaranowski-15},
\citealp{bernard-16}, \citealp{bernard-17b}, \citealp{marchand-17}, \citealp{foffa1-19},\citealp{foffa2-19}).
The dimensional regularization preserves the law of ``tweedling of products'' \eqref{aReg6}
and gives all involved integrals, particularly the inverse Laplacians, a unique definition.

\subsubsection{$D$-dimensional ADM formalism}

Dimensional regularization (DR) needs the representation of the Einstein field equation for arbitray space dimensions, say $d$
for the dimension of space and $D=d+1$ for the spacetime dimension. In the following, $G_D = G_{\rm N} \ell_0^{d-3}$ will denote 
the gravitational constant in $D$-dimensional spacetime and $G_{\rm N}$ the standard Newtonian one,
$\ell_0$ is the DR scale relating both constants.

The unconstraint Hamiltonian takes the form
\be
H = \int\md^dx\,(N {\cal{H}} - c N^i {\cal{H}}_i)
+ \frac{c^4}{16\pi G_D}\oint_{i^0}\md^{d-1}S_i\,\partial_j(\gamma_{ij}-\delta_{ij}\gamma_{kk}),
\ee
where $\md^{d-1}S_i$ denotes the $(d-1)$-dimensional surface element.
The Hamiltonian and the momentum constraint equations
written for many-point-particle systems are given by
\begin{align}
\label{dHcon}
\sqrt{\gamma}\,R &= \frac{1}{\sqrt{\gamma}} \left(\gamma_{ik}\gamma_{j\ell}\pi^{ij}\pi^{k\ell} - \frac{1}{d-1}(\gamma_{ij}\pi^{ij})^2 \right)
\nonumber\\[1ex]&\quad
+ \frac{16\pi G_D}{c^3}\sum_a (m_a^2c^4 + \gamma_a^{ij}p_{ai}p_{aj})^{\frac{1}{2}}\delta_a,
\\[1ex]
\label{dMcon}
-\nabla_j\pi^{ij} &= \frac{8\pi G_D}{c^3}\sum_a \gamma_a^{ij}p_{aj}\delta_a.
\end{align}
The gauge (or coordiante) ADMTT conditions read
\be
\gamma_{ij} = \left(1 + \frac{d-2}{4(d-1)}\phi\right)^{4/(d-2)} \delta_{ij} + h^{\rm TT}_{ij},
\quad \pi^{ii}=0,
\ee
where 
\be
h^{\rm TT}_{ii} = 0, \quad \partial_j h^{\rm TT}_{ij} = 0.
\ee
The field momentum $\pi^{ij}$ splits into its longitudinal and TT parts, 
respectively,
\be
\pi^{ij} = \tilde{\pi}^{ij} + \pi_{\rm TT}^{ij}\,,
\ee
where the longitudinal part $\tilde{\pi}^{ij}$ can be expressed in terms of a vectorial function $V^i$,
\be
\tilde{\pi}^{ij} = \partial_i V^j +\partial_j V^i - \frac{2}{d}\delta^{ij}\partial_k V^k,
\ee
and where the TT part satisfies the conditions,
\be
\pi_{\rm TT}^{ii} = 0, \quad \partial_j \pi_{\rm TT}^{ij} = 0.
\ee

The reduced Hamiltonian of the particles-plus-field system takes the form
\be
H_\text{red}\big[\xa,\pa,h^{\rm TT}_{ij},\pi_{\rm TT}^{ij}\big]
= -\frac{c^4}{16\pi G_D}\int \md^dx\,\Delta\phi\big[\xa,\pa,h^{\rm TT}_{ij},\pi_{\rm TT}^{ij}\big].
\ee
The equations of motion for the particles read 
\be
\dot{\mathbf{x}}_a = \frac{\partial H_\text{red}}{\partial{{\mathbf{p}}_a}},
\quad
\dot{\mathbf{p}}_a = -\frac{\partial H_\text{red}}{\partial{{\mathbf{x}}_a}},
\ee
and the field equations for the independent degrees of freedom are given by
\be
\frac{\partial}{\partial t}h^{\rm TT}_{ij} = \frac{16\pi G_D}{c^3}\,\delta^{\textrm{TT}kl}_{ij}\frac{\delta H_\text{red}}{\delta \pi_{\rm TT}^{kl}},
\qquad
\frac{\partial}{\partial t}\pi_{\rm TT}^{ij} = -\frac{16\pi G_D}{c^3}\,\delta^{\textrm{TT}ij}_{kl}\frac{\delta H_\text{red}}{\delta h^{\rm TT}_{kl}},
\ee
where the $d$-dimensional TT-projection operator is defined by
\begin{align}
\label{defTT}
\delta^{\textrm{TT}ij}_{kl} &\equiv \frac{1}{2}(\delta_{ik}\delta_{jl}+\delta_{il}\delta_{jk})
- \frac{1}{d-1}\delta_{ij}\delta_{kl}
\nonumber\\[1ex]&\quad
- \frac{1}{2}(\delta_{ik}\partial_{jl}+\delta_{jl}\partial_{ik}+\delta_{il}\partial_{jk}+\delta_{jk}\partial_{il})\Delta^{-1}
\nonumber\\[1ex]&\quad
+ \frac{1}{d-1}(\delta_{ij}\partial_{kl}+\delta_{kl}\partial_{ij})\Delta^{-1}
+ \frac{d-2}{d-1}\partial_{ijkl}\Delta^{-2}.
\end{align}

Finally, the Routh functional is defined as
\be
R\big[\xa,\pa,h^{\rm TT}_{ij},\dot{h}^{\rm TT}_{ij}\big]
\equiv H_\text{red}\big[\xa,\pa, h^{\rm TT}_{ij}, \pi_{\rm TT}^{ij}\big]
-\frac{c^3}{16\pi G_D}\int\md^dx\,\pi_{\rm TT}^{ij}\dot{h}^{\rm TT}_{ij},
\ee
and the fully reduced matter Hamiltonian for the conservative dynamics reads
\be
H[\xa,\pa] \equiv R\big[\xa,\pa,h^{\rm TT}_{ij}(\xa,\pa),\dot{h}^{\rm TT}_{ij}(\xa,\pa)\big].
\ee

\subsubsection{Local and asymptotic dimensional regularization}

The technique developed by \cite{damour1-01} to control local (or UV) divergences
boils down to the computation of the difference
\be
\lim_{d\to3} H^\text{loc}(d) - H^\text{RH loc}(3),
\ee
where $H^\text{RH loc}(3)$ is the ``local part'' of the Hamiltonian
obtained by means of the three-dimensional RH regularization
[it is the sum of all integrals of the type $I^{\text{RH}}(3;\epsilon_1,\epsilon_2)$
introduced in Eq.~\eqref{IRHuv}],
$H^\text{loc}(d)$ is its $d$-dimensional counterpart.

\cite{damour1-01} showed that to find the DR correction to the integral $I^{\text{RH}}(3;\epsilon_1,\epsilon_2)$ of Eq.~\eqref{IRHuv}
related with the local pole at, say, $\mathbf{x}=\mathbf{x}_1$,
it is enough to consider only this part of the integrand $i({\bf x})$ which develops
logarithmic singularities in three dimensions, i.e., which locally behaves like $1/r_1^3$,
\be
\label{UVexp3}
i(\mathbf{x}) = \cdots + \tilde{c}_1(\mathbf{n}_1)\,r_1^{-3} + \cdots,
\quad \text{when}\ \mathbf{x}\to\mathbf{x}_1.
\ee
Then the pole part of the integral \eqref{IRHuv}
related with the singularity at $\mathbf{x}=\mathbf{x}_1$
can be recovered by RH regularization of the integral of $\tilde{c}_1(\mathbf{n}_1)\,r_1^{-3}$
over the ball $\mathbb{B}(\mathbf{x}_1,{\ell_1})$ of radius $\ell_1$ surrounding the particle $\mathbf{x}_1$.
The RH regularized value of this integral reads
\be
I_1^{\text{RH}}(3;\epsilon_1)
\equiv \int_{\mathbb{B}(\mathbf{x}_1,{\ell_1})} \tilde{c}_1(\mathbf{n}_1) \, r_1^{-3}
\Big(\frac{r_1}{s_1}\Big)^{\epsilon_1} \, \md^3 {\mathbf r}_1
= c_1 \int_0^{\ell_1} r_1^{-1} \Big(\frac{r_1}{s_1}\Big)^{\epsilon_1}\,\md r_1,
\ee
where $c_1/(4\pi)$ is the angle-averaged value of the coefficient $\tilde{c}_1(\mathbf{n}_1)$.
The expansion of the integral $I_1^{\text{RH}}(3;\epsilon_1)$ around $\epsilon_1=0$ equals
\be
I_1^{\text{RH}}(3;\epsilon_1) = c_1\Big(\frac{1}{\epsilon_1}
+ \ln\frac{\ell_1}{s_1}\Big) + \mathcal{O}(\epsilon_1).
\ee

The idea of the technique developed by \cite{damour1-01}
relies on replacing the RH-regularized value of the three-dimensional integral $I_1^{\text{RH}}(3;\epsilon_1)$
by the value of its $d$-dimensional version $I_1(d)$.
One thus considers the $d$-dimensional counterpart of the expansion \eqref{UVexp3}.
It reads
\be
i(\mathbf{x}) = \cdots + \ell_0^{k(d-3)}\tilde{\mathfrak{c}}_1(d;{\mathbf n}_1)\,r_1^{6-3d} + \cdots,
\quad \text{when}\ \mathbf{x}\to\mathbf{x}_1.
\ee
Let us note that the specific exponent $6-3d$ of $r_1$ visible here
follows from the $r_1\to0$ behaviour of the (perturbative) solutions of the $d$-dimensional constraint equations \eqref{dHcon}--\eqref{dMcon}.
The number $k$ in the exponent of $\ell_0^{k(d-3)}$ is related with the momentum-order of the considered term
[e.g., at the 4PN level the term with $k$ is of the order of $\mathcal{O}(p^{10-2k})$, for $k=1,\ldots,5$;
such term is proportional to $G_D^k$]. The integral $I_1(d)$ is defined as
\be
\label{I1def}
I_1(d) \equiv \ell_0^{k(d-3)} \int_{\mathbb{B}(\mathbf{x}_1,{\ell_1})}
\tilde{\mathfrak{c}}_1(d;{\mathbf n}_1) \, r_1^{6-3d}\, \md^d {\mathbf r}_1
= \mathfrak{c}_1(d) \int_0^{\ell_1} r_1^{5-2d}\,\md r_1,
\ee
where $\mathfrak{c}_1(d)/\big(\Omega_{d-1}\ell_0^{k(d-3)}\big)$
($\Omega_{d-1}$ stands for the area of the unit sphere in $\mathbb{R}^d$)
is the angle-averaged value of the coefficient $\tilde{\mathfrak{c}}_1(d;{\mathbf n}_1)$,
\be
\label{c1def}
\mathfrak{c}_1(d) \equiv \ell_0^{k(d-3)} \oint_{\mathbb{S}^{d-1}(\mathbf{0},1)}
\tilde{\mathfrak{c}}_1(d;{\mathbf n}_1)\,\md\Omega_{d-1}.
\ee
One checks that always there is a smooth connection between $\mathfrak{c}_1(d)$ and its three-dimensional counterpart $c_1$,
\be
\label{limitc1IR}
\lim_{d\to3} \mathfrak{c}_1(d) = \mathfrak{c}_1(3) = c_1.
\ee
The radial integral in Eq.~\eqref{I1def} is convergent if the real part $\Re(d)$ of $d$ fulfills the condition $\Re(d)<3$.
Making use of the expansion $\mathfrak{c}_1(d) = \mathfrak{c}_1(3+\varepsilon) = c_1 + \mathfrak{c}'_1(3)\varepsilon + \mathcal{O}(\varepsilon^2)$,
where $\varepsilon\equiv d-3$, the expansion of the integral $I_1(d)$ around $\varepsilon=0$ reads
\be
\label{UVcorr11}
I_1(d) = -\frac{\ell_1^{-2\varepsilon}}{2\varepsilon}\mathfrak{c}_1(3+\varepsilon)
= -\frac{c_1}{2\varepsilon} -\frac{1}{2} \mathfrak{c}'_1(3)
+ c_1 \ln\ell_1 + \mathcal{O}(\varepsilon).
\ee
Let us note that the coefficient $\mathfrak{c}'_1(3)$ usually depends on $\ln r_{12}$
and it has the structure
\be
\mathfrak{c}'_1(3) = \mathfrak{c}'_{11}(3) + \mathfrak{c}'_{12}(3) \ln\frac{r_{12}}{\ell_0} + 2c_1 \ln\ell_0,
\ee
where $\mathfrak{c}'_{12}(3)=(2 - k)c_1$ [what can be inferred knowing the dependence
of $\mathfrak{c}_1(d)$ on $\ell_0$ given in Eq.~\eqref{c1def}].
Therefore the DR correction also changes the terms $\propto\ln{r_{12}}$.

The DR correction to the RH-regularized value of the integral
$I^{\text{RH}}(3;\epsilon_1,\epsilon_2)$ relies on replacing this integral by
\be
I^{\text{RH}}(3;\epsilon_1,\epsilon_2) + \Delta I_1 + \Delta I_2,
\ee
where
\be
\Delta I_a \equiv I_a(d) - I_a^{\text{RH}}(3;\epsilon_a),
\quad a=1,2.
\ee
Then one computes the double limit
\begin{align}
\label{IRHcorr}
\lim_{\substack{\epsilon_1\to 0 \\ \epsilon_2\to 0}}
\Big(I^{\text{RH}}&(3;\epsilon_1,\epsilon_2) + \Delta I_1 + \Delta I_2\Big)
\nonumber\\
&= A - \frac{1}{2} \big(\mathfrak{c}'_{11}(3) + \mathfrak{c}'_{21}(3)\big)
- \frac{1}{2}\big(\mathfrak{c}'_{12}(3) + \mathfrak{c}'_{22}(3)\big) \ln\frac{r_{12}}{\ell_0}
\nonumber\\[1ex]&\quad
+ \big(c_1 + c_2\big)\left(- \frac{1}{2\varepsilon} + \ln\frac{r_{12}}{\ell_0}\right)+ \mathcal{O}(\varepsilon).
\end{align}
Note that all poles $\propto1/\epsilon_1,1/\epsilon_2$
and all terms depending on radii $\ell_1$, $\ell_2$ or scales $s_1$, $s_2$ cancel each other.
The result \eqref{IRHcorr} is as if all computations were fully done in $d$ dimensions.

In the DR correcting UV divergences in the 3PN two-point-mass Hamiltonian performed by \cite{damour1-01},
after collecting all terms of the type \eqref{IRHcorr} together, all poles $\propto 1/(d-3)$ cancel each other.
This is not the case for the UV divergences of the 4PN two-point-mass Hamiltonian derived by \cite{jaranowski-15}.
As explained in Sect.~VIII~D of \cite{jaranowski-15}, after collecting all terms of the type \eqref{IRHcorr},
one has to add to the Hamiltonian a \emph{unique} total time derivative
to eliminate all poles $\propto 1/(d-3)$ (together with $\ell_0$-dependent logarithms).

The above described technique of the DR correcting of UV divergences can easily be transcribed to control IR divergences.
This is done by  the replacement of the integrals
\be
\int_{\mathbb{B}(\mathbf{x}_a,{\ell_a})}\md^dx\,i({\bf x})
\ee
by the integral
\be
\int_{\mathbb{R}^d\setminus \mathbb{B}(\mathbf{0},R)}\md^dx\,i({\bf x}),
\ee
where $\mathbb{B}(\mathbf{0},R)$ means a large ball of radius $R$ (with the centre at the origin $\mathbf{0}$ of the coordinate system),
and by studying expansion of the integrand $i({\bf x})$ for $r\to\infty$.
This technique was not used to regularize IR divergences
in the computation of the 4PN two-point-mass Hamiltonian by \cite{damour-14} and \cite{jaranowski-15}.
This was so because this technique applied only to the instantaneous part of the 4PN Hamiltonian
is not enough to get rid of the IR poles in the limit $d\to3$.
For resolving IR poles it was necessary to observe
that the IR poles have to cancel with the UV poles from the tail part of the Hamiltonian
(what can be achieved e.g.\ after implementing the so-called zero-bin subtraction in the EFT framework, see \citealp{porto-17a}).

Another two different approaches were employed by \cite{damour-14} and \cite{jaranowski-15}
to regularize IR divergences in the instantaneous part of the 4PN Hamiltonian
(see Appendix A~3 in \citealp{jaranowski-15}):
(i) modifying the behavior of the function $h^\mathrm{TT}_{(6)ij}$ at infinity,\footnote{
This approach is described in Appendix A~3~a of \cite{jaranowski-15},
where Eqs.\ (A40)--(A42) are misprinted: $(r/s)^B\ddot{h}^\text{TT}_{(4)ij}$
should be replaced by $\big[(r/s)^B\ddot{h}^\text{TT}_{(4)ij}\big]^\text{TT}$.
The Eq.\ (3.6) in \cite{damour-14} is the correct version of Eq.\ (A40) in \cite{jaranowski-15}.}
(ii) implementing a $d$-dimensional version of Riesz-Hadamard regularization.
Both approaches were developed in $d$ dimensions, but the final results of using any of them in the limit $d\to3$
turned out to be identical with the results of computations performed in $d=3$ dimensions.
Moreover, the results of the two approaches were different in the limit $d\to3$,
what indicated the ambiguity of IR regularization,
discussed in detail by \cite{jaranowski-15} and fixed by \cite{damour-14}.
This IR ambiguity can be expressed in terms of only one unknown parameter,
because the results of two regularization approaches, albeit different,
have exactly the same structure with only different numerical prefactors.
This prefactor can be treated as the ambiguity parameter.
The full 4PN Hamiltonian was thus computed up to a single ambiguity parameter
and it was used to calculate, in a gauge invariant form,
the energy of two-body system along circular orbits as a function of frequency.
The ambiguity parameter was fixed by comparison of part of this formula
[linear in the symmetric mass ratio $\nu$, see Eq.~\eqref{massDef} below for the definition]
with the analogous 4PN-accurate formula for the particle in the Schwarzschild metric which included self-force corrections.

Analogous ambiguity was discovered in 4PN-acccurate calculations
of two-body equations of motion done by \cite{bernard-16} in harmonic coordinates,
where also analytic regularization\footnote{The analytic regularization in \cite{bernard-16}
is a finite part procedure based on analytic continuation in $B$ of a regulator $(r/r_0)^B$.}
of the IR divergences of the instantaneous part of the dynamics was performed.
However, the computations made by \cite{bernard-16} faced also a second ambiguity (\citealp{damour-16}, \citealp{bernard-17a}),
which must come from their different (harmonic instead of ADMTT) gauge condition
and the potentiality of analytic regularization not to preserve gauge (in contrast to dimensional regularization).
The first method of analytic regularization applied by \cite{damour-14} and \cite{jaranowski-15}
is manifest ADMTT gauge preserving.
Finally, \cite{marchand-17} and \cite{bernard-17b} successfully applied in harmonic-coordinates approach
$d$-dimensional regularization all-over.
However, it is worth to emphasize that in intermediate steps their derivation makes crucial use of an auxiliary regulator parameter $\eta$,
entering as a factor $r^\eta$ multiplying the formal expansions of the source.
The confidence in the procedure stems from the fact that the occurring poles in $\eta$ do cancel each other in $d$ dimensions.
On the other side, the obtained crucial rational number in the tail action,
41/60 or 41/30 depending on representation, was already derived within pure $d$-dimensional calculations
by \cite{foffa2-13} and \cite{galley-16} based on the EFT formalism.
Yet only quite recently, a complete pure dimensional-regularization calculation has been achieved by \cite{foffa1-19,foffa2-19},
where use has been made of the zero-bin subtraction method for interrelated UV and IR poles,
as discussed in view of the 4PN approximation by \cite{porto-17b} and \cite{porto-17a}.

\subsubsection{Distributional differentiation in $d$ dimensions}
\label{subsubs:DisDiff-d}

One can show that the formula \eqref{DisDiff} for distributional differentiation of homogeneous functions
is also valid (without any change) in the $d$-dimensional case. It leads, e.g., to equality
\be
\label{DisDiffD1}
\partial_i \partial_j r^{2-d} = (d-2)\frac{d\,n^i n^j-\delta_{ij}}{r^d} - \frac{4\pi^{d/2}}{d\,\Gamma(d/2 -1)}\delta_{ij}\delta.
\ee
To overcome the necessity of using distributional differentiations
it is possible to replace Dirac $\delta$-function
by the class of analytic functions introduced in \cite{riesz-49},
\be
\delta_{\epsilon}(\bx) \equiv \frac{\Gamma((d-\epsilon)/2)}{\pi^{d/2}2^{\epsilon}\Gamma(\epsilon/2)}r^{\epsilon - d},
\ee
resulting in the Dirac $\delta$-function in the limit
\be
\delta = \lim_{\epsilon \rightarrow 0}\delta_{\epsilon}.
\ee
On this class of functions, the inverse Laplacian operates as
\be
\Delta^{-1} \delta_{\epsilon} = -\delta_{\epsilon + 2},
\ee
and instead of \eqref{DisDiffD1} one gets
\be
\partial_i \partial_j r^{\epsilon + 2 - d} = (d-2-\epsilon)\frac{(d-\epsilon)n^i n^j-\delta_{ij}}{r^{d-\epsilon}}.
\ee
There is no need to use distributional differentiation here, so no $\delta$-functions are involved.

Though the replacements in the stress-energy tensor density of $\delta_a$ through $\delta_{\epsilon_a}$ (with $a=1,2$)
do destroy the divergence freeness of the stress-energy tensor and thus the integrability conditions of the Einstein theory,
the relaxed Einstein field equations (the ones which result after imposing coordinate conditions)
do not force the stress-energy tensor to be divergence free and can thus be solved without problems.
The solutions one gets do not fulfill the complete Einstein field equations but in the final limits $\epsilon_a\to0$
the general coordinate covariance of the theory is manifestly recovered.
This property, however, only holds if these limits are taken \emph{before} the limit $d=3$ is performed \citep{damour3-08}.

\section{Point-mass representations of spinless black holes}
\label{sec:BH}

This section is devoted to an insight of how black holes, the most compact objects in GR, can be represented by point masses. 
On the other side, the developments in the present section show that point masses, interpreted as fictitious point masses
(analogously to image charges in the electrostatics), allow to represent black holes.
Later on, in the section on approximate Hamiltonians for spinning binaries, neutron stars will also be considered,
taking into account their different rotational deformation.
Tidal deformations are considered in Sect.~\ref{sec:tidal}.

The simplest black hole is a Schwarzschildian one which is isolated and non-rotating.
Its metric is a static solution of the \emph{vacuum} Einstein field 
equations. In isotropic coordinates, the Schwarzschild metric reads (see, e.g., \citealp{misner-73})
\be
\md s^2 = -\left(\frac{\dst 1-\frac{GM}{2rc^2}}{\dst 1+\frac{GM}{2rc^2}}\right)^2 c^2 \md t^2
+ \left(1+\frac{GM}{2rc^2}\right)^4 \md{\bf x}^2,
\ee
where $M$ is the gravitating mass of the black hole and $(x^1,x^2,x^3)$ are Cartesian coordinates in $\mathbb{R}^3$
with $r^2 = (x^1)^2 + (x^2)^2 + (x^3)^2$ and $\md{\bf x}^2 = (\md x^1)^2 + (\md x^2)^2 + (\md x^3)^2$.
The origin of the coordinate system $r=0$ is not located where the Schwarzschild singularity $R=0$,
with $R$ the radial Schwarzschild coordinate, is located,
rather it is located on the other side of the Einstein--Rosen bridge, at infinity, where space is flat.
The point $r=0$ does not belong to the three-dimensional spacelike curved manifold,
so we do have an open manifold excluding the point $r=0$,
a so-called ``puncture'' manifold (see, e.g., \citealp{brandt-97}, \citealp{cook-05}).
However, as we shall see below, the Schwarzschild metric can be contructed
with the aid of a Dirac $\delta$ function with support at $r=0$,
located in a conformally related flat space of dimension smaller than three.
Distributional sources with support at the Schwarzschild singularity
are summarized and treated by \cite{pantoja-02}, \cite{heinzle-02}.

A two black hole initial value solution of the vacuum Einstein field equations
is the time-symmetric Brill--Lindquist one (\citealp{brill-63}, \citealp{lindquist-63}),
\be
\label{BLmetric}
\md s^2 = -\left(\frac{\dst 1-\frac{\beta_1G}{2r_1c^2}-\frac{\beta_2G}{2r_2c^2}}{\dst 1+\frac{\alpha_1G}{2r_1c^2}+\frac{\alpha_2G}{2r_2c^2}}\right)^2 c^2\md t^2
+ \left(1+\frac{\alpha_1G}{2r_1c^2}+\frac{\alpha_2G}{2r_2c^2}\right)^4\md{\bf x}^2,
\ee
where ${\bf r}_a\equiv{\bf x}-{\bf x}_a$ and $r_a\equiv|\vecr_a|$ ($a=1,2$),
the coefficients $\alpha_a$ and $\beta_a$ can be found in \cite{jaranowski-02}
(notice that $h^\mathrm{TT}_{ij}=0$, $\pi^{ij}=0$, and, initially, $\partial_t r_a=0$).
Its total energy results from the ADM surface integral
[this is the reduced ADM Hamiltonian from Eq.~\eqref{ADMHred} written for the metric \eqref{BLmetric}]
\be
E_{\textrm{ADM}} = -\frac{c^4}{2\pi G} \oint_{i_0}\md S_i\, \partial_i\Psi = (\alpha_1 + \alpha_2)c^2,
\ee
where $\md S_i=n^ir^2\md\Omega$ is a two-dimensional surface-area element
(with unit radial vector $n^i\equiv{x^i/r}$ and solid angle element $\md\Omega$) and
\be
\Psi \equiv 1 + \frac{\alpha_1G}{2r_1c^2} + \frac{\alpha_2G}{2r_2c^2}.
\ee
Introducing the inversion map $\mathbf{x}\rightarrow\mathbf{x}'$ defined by \cite{brill-63}
\be
{\bf r}'_1 \equiv  {\bf r}_1 \frac{\alpha_1^2G^2}{4c^4r_1^2}
\quad\Longrightarrow\quad
{\bf r}_1 =  {\bf r}'_1 \frac{\alpha_1^2G^2}{4c^4r_1'^{2}},
\ee
where ${\bf r}'_1\equiv{\bf x}'-{\bf x}_1$, $r'_1\equiv|{\bf x}'-{\bf x}_1|$,
the three-metric $\md l^2 = \Psi^4 \md{\bf x}^2$ transforms into
\be
\md l^2 = \Psi'^4 \md{\bf x}'^2,
\quad\textrm{with}\quad \Psi' \equiv 1+\frac{\alpha_1G}{2r'_1c^2}+\frac{\alpha_1\alpha_2G^2}{4r_2 r'_1 c^4},
\ee
where ${\bf r}_2 =  {\bf r}'_1 \alpha_1^2G^2/(4c^4r_1'^{2}) + {\bf r}_{12}$
with ${\bf r}_{12}\equiv{\bf x}_{1} - {\bf x}_{2}$.
From the new metric function $\Psi'$ the proper mass of the throat 1 results in,
\be
m_1 \equiv -\frac{c^2}{2\pi G}\oint_{i_0^1} \md S'_i\, \partial'_i \Psi'
= \alpha_1 + \frac{\alpha_1\alpha_2G}{2r_{12}c^2},
\ee
where $i_0^1$ denotes the black hole's 1 own spacelike infinity.
Hereof the ADM energy comes out in the form,
\be
E_{\text{ADM}} = (m_1 + m_2)c^2 - G \frac{\alpha_1\alpha_2}{r_{12}},
\ee
where
\be
\alpha_a = \frac{m_a-m_b}{2} + \frac{c^2r_{ab}}{G}\left(\sqrt{1+\frac{m_a+m_b}{c^2r_{ab}/G}
+ \left(\frac{m_a-m_b}{2c^2r_{ab}/G}\right)^2} - 1 \right).
\ee
This construction, as performed by \cite{brill-63},
is a purely geometrical (or vacuum) one without touching singularities.
Recall that this energy belongs to an initial value solution of the Einstein constraint equations
with vanishing of both $h^{\rm TT}_{ij}$ and particle together with field momenta.
In this initial conditions spurious gravitational waves are included.

In the following we will show how the vacuum Brill--Lindquist solution
can be obtained with Dirac $\delta$-function source terms located at $r_1=0$ and $r_2=0$
in a conformally related three-dimensional flat space.
To do this we will formulate the problem in $d$ space dimensions
and make analytical continuation in $d$ of the results down to $d=3$.
The insertion of the stress-energy density for point masses
into the Hamiltonian constraint equation yields,
for $p_{ai}=0$, $h^{\rm TT}_{ij}=0$, and $\pi^{ij}=0$,
\be
\label{DDBL1}
-\Psi \Delta\phi = \frac{16\pi G}{c^2} \sum_a m_a \delta_a,
\ee
where $\Psi$ and $\phi$ parametrize the space metric,
\be
\label{DDBL2}
\gamma_{ij} = \Psi^{4/(d-2)}\delta_{ij}, \quad \Psi\equiv 1 + \frac{d-2}{4(d-1)}\phi.
\ee
If the lapse function $N$ is represented by
\be
\label{DDBL3}
N \equiv \frac{\chi}{\Psi},
\ee
an equation for $\chi$ results of the form
(using the initial-data conditions $p_{ai}=0$, $h^{\rm TT}_{ij}=0$, $\pi^{ij}=0$),
\be
\label{DDBL4}
\Psi^2 \Delta \chi =\frac{4\pi G}{c^2} \frac{d-2}{d-1}\chi \sum_a m_a\delta_a.
\ee

With the aid of the relation
\be
\label{DDBL5}
\Delta\frac{1}{r_a^{d-2}} = -\frac{4\pi^{d/2}}{\Gamma(d/2-1)}\delta_a
\ee
it is easy to show that for $1<d<2$ the equations for $\Psi$ and $\chi$ do have well-defined solutions.
To obtain these solutions we employ the ansatz
\be
\label{DDBL6}
\phi = \frac{4G}{c^2}\frac{\Gamma(d/2-1)}{\pi^{d/2-1}}\left(\frac{\alpha_1}{r^{d-2}_1}+\frac{\alpha_2}{r^{d-2}_2}\right),
\ee
where $\alpha_1$ and $\alpha_2$ are some constants.
After plugging the ansatz \eqref{DDBL6} into Eq.~\eqref{DDBL1}
we compare the coefficients of the Dirac $\delta$-functions on both sides of the equation.
For point mass 1 we get
\be
\bigg(1 + \frac{G(d-2)\Gamma(d/2-1)}{c^2(d-1)\pi^{d/2-1}}
\Big(\frac{\alpha_1}{r^{d-2}_1}+\frac{\alpha_2}{r^{d-2}_2}\Big)\bigg) \alpha_1 \delta_1 = m_1 \delta_1.
\ee
After taking $1<d<2$, one can perform the limit $r_1 \rightarrow 0$ for the coefficient of $\delta_1$
in the left-hand-side of the above equation,
\be
\bigg(1 + \frac{G(d-2)\Gamma(d/2-1)}{c^2(d-1)\pi^{d/2-1}} \frac{\alpha_2}{r^{d-2}_{12}}\bigg) \alpha_1 \delta_1 = m_1 \delta_1.
\ee
Going over to $d=3$ by arguing that the solution is analytic in $d$ results in the relation
\be
\alpha_a = \frac{m_a}{\dst 1+ \frac{G}{2c^2}\frac{\alpha_b}{r_{ab}}},
\ee
where $b \ne a$ and $a,b = 1,2$. The ADM energy is again given by, in the limit $d=3$, 
\be
E_{\textrm{ADM}} = (\alpha_1 + \alpha_2) c^2.
\ee

Here we recognize the important aspect that although the metric may describe close binary black holes with strongly deformed apparent horizons,
the both black holes can still be generated by point masses in conformally related flat space.
This is the justification for our particle model to be taken as model for orbiting black holes.
Obviously black holes generated by point masses are orbiting black holes without spin, i.e., Schwarzschild-type black holes.
The representation of a Schwarzschild-type black hole in binary--black-hole systems
with one Dirac $\delta$-function seems not to be the only possibility.
As shown by \cite{jaranowski2-00}, binary--black-hole configurations defined through isometry-conditions at the apparent horizons \citep{misner-63}
need infinitely many Dirac $\delta$-functions per each one of the black holes.
Whether or not those black holes are more physical is not known.
It has been found by \cite{jaranowski1-99} that the expressions for ADM energy
of the two kinds of binary black holes do agree through 2PN order,
and that at the 3PN level the energy of the Brill--Lindquist binary black holes
is additively higher by $G^4m_1^2m_2^2(m_1+m_2)/(8c^6r^4_{12})$,
i.e.\ the Misner configuration seems stronger bound\footnote{
This could be an issue for the effacing principle as discussed in subsection 1.1.}.
The same paper has shown that the spatial metrics of both binary--black-hole configurations coincide through 3PN order,
and that at least through 5PN order they can be made to coincide by shifts of black-hole position variables.

\section{Post-Newtonian Hamilton dynamics of nonspinning compact binaries}
\label{sec:PNbinaries}

In this section we collect explicit results on Hamilton dynamics of binaries made of compact and nonspinning bodies.
Up to the 4PN order the Hamiltonian of binary point-mass systems is explicitly known
and it can be written as the sum
\begin{align}
\label{4PNHt}
H[\xa,\pa,t] &= \sum_a m_a c^2 + H_{\textrm{N}}(\xa,\pa)
+ \frac{1}{c^2} H_{\textrm{1PN}}(\xa,\pa)
+ \frac{1}{c^4} H_{\textrm{2PN}}(\xa,\pa)
\nonumber\\[1ex]&\quad
+ \frac{1}{c^5} H_{\textrm{2.5N}}(\xa,\pa,t)
+ \frac{1}{c^6} H_{\textrm{3PN}}(\xa,\pa)
+ \frac{1}{c^7} H_{\textrm{3.5PN}}(\xa,\pa,t)
\nonumber\\[1ex]&\quad
+ \frac{1}{c^8} H_{\textrm{4PN}}[\xa,\pa]
+ \mathcal{O}(c^{-9}).
\end{align}
This Hamiltonian is the PN-expanded reduced ADM Hamiltonian of point-masses plus field system;
the nontrivial procedure of reduction is described in Sects.\ \ref{subsec:PMredH} and \ref{subsec:Routhian} of this review.
The non-autonomous dissipative Hamiltonians $H_{\textrm{2.5PN}}(\xa,\pa,t)$ and $H_{\textrm{3.5PN}}(\xa,\pa,t)$
are written as explicitly depending on time because they depend on the gravitational field variables
(see Sect.~\ref{subsec:PNdissipation} for more details).
The dependence of the 4PN Hamiltonian $H_{\textrm{4PN}}$ on $\xa$ and $\pa$ is both pointwise and functional
(and this is why we have used square brackets for arguments of $H_{\textrm{4PN}}$).

We will display here the conservative Hamiltonians $H_{\textrm{N}}$ to $H_{\textrm{4PN}}$ in the centre-of-mass reference frame,
relegating their generic, noncentre-of-mass forms, to Appendix~\ref{app:ncom}.
In the ADM formalism the centre-of-mass reference frame is  defined by the simple requirement
\be
\mathbf{p}_1 + \mathbf{p}_2 = \mathbf{0}.
\ee
Here we should point out that at the 3.5PN order for the first time recoil arises, hence the conservation of linear momentum is violated
[see, e.g., \citealp{fitchett-83} (derivation based on wave solutions of linearized field equations)
and \citealp{junker-92} (derivation based on wave solutions of non-linear field equations)].
This however has no influence on the energy through 6.5PN order, if $\mathbf{P}\equiv\mathbf{p}_1+\mathbf{p}_2=\mathbf{0}$ holds initially,
because up to 3PN order the Eq.~\eqref{ME} is valid
and the change of the Hamiltonian $H$ caused by nonconservation of $\mathbf{P}$ equals
$(\md H/\md t)|_{\mathcal{M}=\text{const}}
=\big((c^2/H)\mathbf{P}\big)_{\rm 3PN}\cdot(\md\mathbf{P}/\md t)_{\rm 3.5PN}=0$
[where $\mathcal{M}$ is defined in Eq.~\eqref{ME}] through 6.5PN order.

Let us define
\be
\label{massDef}
M \equiv m_1 + m_2,
\quad
\mu \equiv \frac{m_1m_2}{M},
\quad
\nu \equiv \frac{\mu}{M},
\ee
where the symmetric mass ratio $0 \le \nu \le 1/4$, with $\nu = 0$ being the test-body case and $\nu = 1/4$ for equal-mass binaries.
It is convenient to introduce reduced (or rescaled) variables $\vecr$ and $\vecp$ (together with the rescaled time variable $\hat{t}$),
\be
\label{defRedVar}
\vecr \equiv \frac{{\bf x}_1 - {\bf x}_2}{GM},
\quad
\vecn \equiv \frac{\bf r}{|{\bf r}|},
\quad
{\bf p} \equiv \frac{{\bf p}_1}{\mu} = -\frac{{\bf p}_2}{\mu},
\quad
p_r \equiv \vecn\cdot{\bf p},
\quad
\hat{t} \equiv \frac{t}{GM},
\ee
as well as the reduced Hamiltonian
[note that $H=\mathcal{M}c^2$, see Eq.~\eqref{ME}]
\be
\label{redHdef}
\hat{H} \equiv \frac{H-Mc^2}{\mu}.
\ee

\subsection{Conservative Hamiltonians through 4PN order}
\label{subsec:Hcon}

The conservative reduced 4PN-accurate two-point-mass Hamiltonian in the centre-of-mass frame reads
\begin{align}
\hat{H}[\vecr,\vecp] &= \hat{H}_{\textrm{N}}(\vecr,\vecp)
+ \frac{1}{c^2} \hat{H}_{\textrm{1PN}}(\vecr,\vecp)
+ \frac{1}{c^4} \hat{H}_{\textrm{2PN}}(\vecr,\vecp)
\nonumber\\[1ex] &\quad
+ \frac{1}{c^6} \hat{H}_{\textrm{3PN}}(\vecr,\vecp)
+ \frac{1}{c^8} \hat{H}_{\textrm{4PN}}[\vecr,\vecp].
\end{align}
The Hamiltonians $\hat{H}_{\textrm{N}}$ through $\hat{H}_{\textrm{3PN}}$ are local in time.
They explicitly read
\be
\hat{H}_{\textrm{N}}(\vecr,\vecp) = \frac{p^2}{2} - \frac{1}{r},
\ee
\be
\hat{H}_{\textrm{1PN}}(\vecr,\vecp)
= \frac{1}{8} (3{\nu} -1) p^4
- \frac{1}{2}\left[(3+{\nu}) p^2 + {\nu} p^2_r\right]\frac{1}{r}
+ \frac{1}{2r^2},
\ee
\begin{align}
\hat{H}_{\textrm{2PN}}(\vecr,\vecp) &= \frac{1}{16}(1-5{\nu}+5\nu^2)p^6
\nonumber\\[1ex] &\quad
+ \frac{1}{8}\big[(5-20{\nu} - 3\nu^2)p^4 - 2\nu^2p^2_rp^2 - 3\nu^2 p^4_r\big]\frac{1}{r}
\nonumber\\[1ex] &\quad
+ \frac{1}{2}[(5+8{\nu} )p^2 + 3{\nu} p_r^2]\frac{1}{r^2}
- \frac{1}{4}(1+3{\nu})\frac{1}{r^3},
\end{align}
\begin{align}
\hat{H}_{\textrm{3PN}}(\vecr,\vecp) &= \frac{1}{128}(-5+35{\nu} - 70\nu^2 + 35 \nu^3) p^8
\nonumber\\[1ex] &\quad
+ \frac{1}{16}\Big[(-7+42{\nu} - 53\nu^2 - 5\nu^3)p^6 +(2-3\nu)\nu^2p_r^2p^4
\nonumber\\[1ex] &\quad
+ 3(1-\nu)\nu^2p_r^4p^2 - 5\nu^3p_r^6\Big]\frac{1}{r}
+ \Big[\frac{1}{16}(-27 + 136{\nu} + 109\nu^2)p^4
\nonumber\\[1ex] &\quad
+ \frac{1}{16}(17+30\nu){\nu} p_r^2p^2
+ \frac{1}{12}(5+43\nu){\nu} p_r^4\Big]\frac{1}{r^2}
\nonumber\\[1ex] &\quad
+ \Bigg[\left(-\frac{25}{8} + \left(\frac{1}{64} \pi^2 - \frac{335}{48}\right){\nu} - \frac{23}{8} \nu^2\right)p^2
\nonumber\\[1ex] &\quad
+ \left(-\frac{85}{16} - \frac{3}{64}\pi^2 - \frac{7}{4} \nu\right){\nu} p^2_r\Bigg] \frac{1}{r^3}
\nonumber\\[1ex] &\quad
+ \left[\frac{1}{8} + \left(\frac{109}{12} - \frac{21}{32} \pi^2\right) {\nu}\right] \frac{1}{r^4}.
\end{align}

The total 4PN Hamiltonian $\hat{H}_{\textrm{4PN}}[\vecr,\vecp]$
is the sum of the local-in-time piece $\hat{H}_{\textrm{4PN}}^{\textrm{local}}(\vecr,\vecp)$
and the piece $\hat{H}_{\textrm{4PN}}^{\textrm{nonlocal}}[\vecr,\vecp]$ which is nonlocal in time:
\be
\label{H4PNsum1}
\hat{H}_{\textrm{4PN}}[\vecr,\vecp] = \hat{H}_{\textrm{4PN}}^{\textrm{local}}(\vecr,\vecp)
+ \hat{H}_{\textrm{4PN}}^{\textrm{nonlocal}}[\vecr,\vecp].
\ee
The local-in-time 4PN Hamiltonian $\hat{H}_{\textrm{4PN}}^{\textrm{local}}(\vecr,\vecp)$ reads
\begin{align}
\label{H4PNlocal1}
\hat{H}_{\textrm{4PN}}^{\textrm{local}}&(\vecr,\vecp)
= \left(
\frac{7}{256}
-\frac{63}{256}\nu
+\frac{189}{256}\nu^2
-\frac{105}{128}\nu^3
+\frac{63}{256}\nu^4
\right)p^{10}
\nonumber\\ &
+ \Bigg\{
\frac{45}{128} p^8
-\frac{45}{16} p^8\nu
+\left(
\frac{423}{64} p^8
-\frac{3}{32} p_r^2 p^6
-\frac{9}{64} p_r^4 p^4
\right)\nu^2
\nonumber\\ &
+ \left(
-\frac{1013}{256} p^8
+\frac{23}{64} p_r^2 p^6
+\frac{69}{128} p_r^4 p^4
-\frac{5}{64} p_r^6 p^2
+\frac{35}{256} p_r^8
\right)\nu^3
\nonumber\\&
+ \left(
-\frac{35}{128} p^8
-\frac{5}{32} p_r^2 p^6
-\frac{9}{64} p_r^4 p^4
-\frac{5}{32} p_r^6 p^2
-\frac{35}{128} p_r^8
\right)\nu^4
\Bigg\}\frac{1}{r}
\nonumber\\&
+ \Bigg\{
\frac{13}{8} p^6
+ \left(
-\frac{791}{64}p^6
+\frac{49}{16} p_r^2 p^4
-\frac{889}{192} p_r^4 p^2
+\frac{369}{160} p_r^6
\right)\nu
\nonumber\\&
+ \left(
\frac{4857}{256} p^6
-\frac{545}{64} p_r^2 p^4
+\frac{9475}{768} p_r^4 p^2
-\frac{1151}{128} p_r^6
\right)\nu^2
\nonumber\\&
+ \left(
\frac{2335}{256} p^6
+\frac{1135}{256} p_r^2 p^4
-\frac{1649}{768} p_r^4 p^2
+\frac{10353}{1280} p_r^6
\right)\nu^3
\Bigg\}\frac{1}{r^2}
\nonumber\\[1ex] &\
+ \Bigg\{ \frac{105}{32} p^4
+ \left[ \left(\frac{2749}{8192}\pi^2-\frac{589189}{19200}\right) p^4
+ \left(\frac{63347}{1600} - \frac{1059}{1024}\pi^2\right) p_r^2 p^2 
\right.  \nonumber \\[1ex] & \left. 
+ \left(\frac{375}{8192}\pi^2-\frac{23533}{1280}\right) p_r^4 \right]\nu
+ \bigg[ \left(\frac{18491}{16384}\pi^2 - \frac{1189789}{28800}\right) p^4
\nonumber\\[1ex]&
- \left(\frac{127}{3} + \frac{4035}{2048}\pi^2\right) p_r^2 p^2
+ \left(\frac{57563}{1920} - \frac{38655}{16384}\pi^2 \right) p_r^4
\bigg]\nu^2
\nonumber\\[1ex]&
+ \bigg(
-\frac{553}{128} p^4
-\frac{225}{64} p_r^2 p^2
-\frac{381}{128} p_r^4
\bigg)\nu^3
\Bigg\}\frac{1}{r^3}
\nonumber\\[1ex]&
+ \Bigg\{
\frac{105}{32}p^2
+ \left[  \left(\frac{185761}{19200} - \frac{21837}{8192}\pi^2\right) p^2
+ \left(\frac{3401779}{57600} - \frac{28691}{24576}\pi^2\right) p_r^2 \right]\nu
\nonumber\\[1ex]&
+ \left[ \left(\frac{672811}{19200} - \frac{158177}{49152}\pi^2\right) p^2
+ \left(-\frac{21827}{3840} + \frac{110099}{49152}\pi^2\right) p_r^2 \right]\nu^2
\Bigg\}\frac{1}{r^4}
\nonumber\\&
+ \Bigg\{
-\frac{1}{16}
+ \left({-\frac{169199}{2400} + \frac{6237}{1024}\pi^2}\right) \, \nu
+ \left(-\frac{1256}{45} + \frac{7403}{3072}\pi^2\right)\,\nu^2
\Bigg\}\frac{1}{r^5}.
\end{align}

The time-symmetric but nonlocal-in-time Hamiltonian $\hat{H}_{\textrm{4PN}}^{\textrm{nonlocal}}[\vecr,\vecp]$
is related with the leading-order tail effects \citep{damour-14}. It equals
\be
\label{H4PNnonloc1}
\hat{H}_{\textrm{4PN}}^{\textrm{nonlocal}}[\vecr,\vecp]
= -\frac{1}{5}\frac{G^2}{\nu c^8} \dddot{I}_{\!ij}(t)
\times \Pf_{2r_{12}/c} \int_{-\infty}^{+\infty} \frac{\md \tau}{\vert \tau \vert} \dddot{I}_{\!ij}(t+\tau),
\ee
where $\Pf_T$ is a Hadamard partie finie with time scale $T\equiv2r_{12}/c$ and where
$\dddot{I}_{\!ij}$ denotes a third time derivative of the Newtonian quadrupole moment $I_{ij}$ of the binary system,
\be
I_{ij} \equiv \sum_a m_a \left(x_a^i x_a^j - \frac{1}{3}\delta^{ij}\xa^2\right).
\ee
The Hadamard partie finie operation is defined as \citep{damour-14}
\be
\Pf_T \int_0^{+\infty} \frac{\md v}{v}g(v) \equiv \int_0^T \frac{\md v}{v}[g(v)-g(0)]
+ \int_T^{+\infty} \frac{\md v}{v}g(v).
\ee
Let us also note that in reduced variables the quadrupole moment $I_{ij}$
and its third time derivative $\dddot{I}_{\!ij}$ read
\be
\label{QMdddot}
I_{ij} = (GM)^2 \mu \left(r^ir^j-\frac{1}{3}{\bf r}^2\delta^{ij}\right),
\quad
\dddot{I}_{\!ij} = -\frac{\nu}{Gr^2}
\left( 4 n^{\langle i} p_{j\rangle} - 3\np n^{\langle i}n^{j\rangle} \right),
\ee
where $\langle\cdots\rangle$ denotes a symmetric tracefree projection
and where in $\dddot{I}_{\!ij}$ the time derivatives $\dot{\vecr}$, $\ddot{\vecr}$, and $\dddot{\vecr}$
were eliminated by means of Newtonian equations of motion.

From the reduced conservative Hamiltonians displayed above,
where a factor of $1/\nu$ is factorized out
[through the definition \eqref{redHdef} of the reduced Hamiltonian],
the standard test-body dynamics is very easily obtained, simply by putting $\nu=0$.
The conservative Hamiltonians $\hat{H}_{\textrm{N}}$ through $\hat{H}_{\textrm{4PN}}$ serve as basis of the EOB approach,
where with the aid of a canonical transformation the two-body dynamics is put into test-body form
of an effective particle moving in deformed Schwarzschild metric,
with $\nu$ being the deformation parameter (\citealp{buonanno-99,buonanno-00}, \citealp{damour2-00,damour-15}).
These Hamiltonians, both directly and through the EOB approach,
constitute an important element in the construction of templates
needed to detect gravitational waves emitted by coalescing compact binaries.
Let us stress again that the complete 4PN Hamiltonian has been obtained only in 2014 \citep{damour-14},
based on earlier calculations (\citealp{blanchet-88}, \citealp{bini-13}, \citealp{jaranowski-13})
and a work published later \citep{jaranowski-15}.

\subsection{Nonlocal-in-time tail Hamiltonian at 4PN order}

The nonlocal-in-time tail Hamiltonian at the 4PN level
(derived and applied by \citealp{damour-14} and \citealp{damour-15}, respectively)
is the most subtle part of the 4PN Hamiltonian. It certainly deserves some discussion.
Let us remark that though the tail Hamiltonian derived in 2016 by \cite{bernard-16}
was identical with the one given in \cite{damour-14},
the derivation there of the equations of motion and the conserved energy
was incorrectly done, as detailed by \cite{damour-16},
which was later confirmed by \cite{bernard-17a}.

The 4PN-level tail-related contribution to the action reads
\be
S^\textrm{tail}_\textrm{4PN} = -\int {H}^{\rm tail}_\textrm{4PN}(t)\, \md t,
\ee
where the 4PN tail Hamiltonian equals
\be
{H}^\textrm{tail}_\textrm{4PN}(t) = -\frac{G^2M}{5c^8} \dddot{I}_{\!ij}(t)\,
{\rm Pf}_{2r(t)/c}\int_{-\infty}^{\infty}\frac{\md v}{|v|}\dddot{I}_{\!ij}(t+v).
\ee
Because formally
\be
\dddot{I}_{\!ij}(t+v) = \exp\left(v\frac{\md}{\md t}\right)\dddot{I}_{\!ij}(t),
\ee
the tail Hamiltonian can also be written as
\begin{align}
{H}^\textrm{tail}_\textrm{4PN}(t) &= - \frac{G^2M}{5c^8} \dddot{I}_{\!ij}(t)\,
{\rm Pf}_{2r(t)/c}\int_0^{\infty}\frac{\md v}{v}\left[\dddot{I}_{\!ij}(t+v)+\dddot{I}_{\!ij}(t-v)\right]
\nonumber\\[1ex]
&= - \frac{2G^2M}{5c^8} \dddot{I}_{\!ij}(t)\,
{\rm Pf}_{2r(t)/c}\int_0^{\infty}\frac{\md v}{v} \mbox{cosh}\left(v\frac{\md}{\md t}\right)\dddot{I}_{\!ij}(t).
\end{align}
Another writing of the tail Hamiltonian is 
\be
H^\textrm{tail}_\textrm{4PN}(t) = -\frac{2G^2M}{5c^8} \dddot{I}_{\!ij}(t)\,
{\rm Pf}_{2r(t)/c}\int_0^{\infty}\frac{\md v}{v} \cosh\left(vX(H_0)\right)\dddot{I}_{\!ij}(t)
\ee
with 
\be
X(H_0) \equiv \sum_i \left(\frac{\partial H_0}{\partial p_i(t)}\frac{\partial}{\partial x^i(t)}
- \frac{\partial H_0}{\partial x^i(t)}\frac{\partial}{\partial p_i(t)}\right),
\quad
H_0 = \frac{({\bf p}(t))^2}{2\mu} - \frac{GM\mu}{r(t)}.
\ee
This presentation shows that $H^\textrm{tail}_\textrm{4PN}$
can be constructed from positions and momenta at time $t$.

For circular orbits, $\dddot{I}_{\!ij}(t)$ is an eigenfunction of $\mbox{cosh}\left(v\frac{\md}{\md t}\right)$, reading
\be
\label{COcosh}
\cosh\left(v\frac{\md}{\md t}\right)\dddot{I}_{\!ij}(t)
= \cos\left(2v\Omega(t)\right)\dddot{I}_{\!ij}(t),
\ee
where $\Omega$ is the angular frequency along circular orbit ($p_r=0$),
\be
\label{COomegat}
\Omega(t) \equiv \dot{\varphi} = \frac{\partial H_0(p_{\varphi},r)}{\partial p_{\varphi}}
= \frac{p_{\varphi}(t)}{\mu r^2(t)},
\quad
H_0(p_{\varphi},r) = \frac{p_\varphi^2}{2\mu r^2} - \frac{GM\mu}{r}.
\ee
Notice the representation of $\Omega(t)$ as function of the still independent
(dynamical equation $\dot{p}_r=-\partial H_0/\partial r$ has not yet been used)
canonical variables $p_\varphi(t)$ and $r(t)$
(in \citealp{damour-14,damour-16}, a more concise representation for circular orbits has been applied,
based on the orbital angular momentum as only variable).
The somewhat complicated structure of Eq.~\eqref{COcosh} can be made plausible
by writing $\dst v\frac{\md}{\md t}=v\,\Omega(p_\varphi,r)\frac{\md}{\md\varphi}$, see Eq.~\eqref{COomegat},
and parametrizing the Eq.~\eqref{QMdddot} for circular orbits ($p_r=0$) with orbital angle $\varphi$.
The 4PN tail Hamiltonian for circular orbits can thus be written as
\begin{align}
\label{htailcirc}
H^\textrm{tail circ}_\textrm{4PN}(t) &= -\frac{2G^2M}{5c^8} \left(\dddot{I}_{\!ij}(t)\right)^2\,
{\rm Pf}_{2r(t)/c}\int_0^{\infty}\frac{\md v}{v}\cos\left(\frac{2p_{\phi}(t)}{\mu r^2(t)}v\right)
\nonumber \\[1ex]
&= \frac{2G^2M}{5c^8} \left(\dddot{I}_{\!ij}(t)\right)^2\,
\left[\ln\left(\frac{4p_{\phi}(t)}{\mu c r(t)}\right) + \gamma_\mathrm{E} \right],
\end{align}
where $\gamma_\mathrm{E}=0.577\ldots$ denotes Euler's constant.
This representation has been quoted and used by \cite{bernard-16}, see Eq.~(5.32) therein,
for a straightforward comparison of their tail results
with the tail results presented by \cite{damour-14}.

\subsection{Dynamical invariants of two-body conservative dynamics}
\label{subsec:dyninv}

The observables of two-body systems that can be measured from infinity by, say, gravitational-wave observations,
are describable in terms of dynamical invariants, i.e., functions which do not depend on the choice of phase-space coordinates.
Dynamical invariants are easily obtained within a Hamiltonian framework of integrable systems.

We start from the reduced conservative Hamiltonian $\hat{H}(\vecr,\vecp)$ in the centre-of-mass frame
(we are thus considering here a local-in-time Hamiltonian;
for the local reduction of a nonlocal-in-time 4PN-level Hamiltonian see Sect.\ \ref{subsubsec:dyninv4} below)
and we employ reduced variables $(\vecr,\vecp)$. The invariance of $\hat{H}(\vecr,\vecp)$ under time translations
and spatial rotations leads to the conserved quantities
\be
\label{EJreduced}
E \equiv \hat{H}(\vecr,\vecp), \quad
\mathbf{j} \equiv \frac{\mathbf{J}}{\mu GM} = \vecr\times\vecp,
\ee
where $E$ is the total energy and $\mathbf{J}$ is the total orbital angular momentum of the binary system in the centre-of-mass frame.
We further restrict considerations to the plane of the relative trajectory endowed with polar coordinates $(r,\phi)$
and we use Hamilton-Jacobi approach to obtain the motion.
To do this we separate the variables $\hat{t}\equiv t/(GM)$ and $\phi$
in the reduced planar action $\hat{S}\equiv S/(G\mu M)$, which takes the form
\be
\hat{S} = -E \hat{t} + j \phi + \int\sqrt{R(r,E,j)}\,\md r.
\ee
Here $j\equiv|\mathbf{j}|$ and the effective radial potential $R(r,E,j)$ is obtained by solving
the equation $E=\hat{H}(\vecr,\vecp)$ with respect to $p_r\equiv\vecn\cdot\vecp$, after making use of the relation
\be
\vecp^2 = (\vecn\cdot\vecp)^2 +  (\vecn\times\vecp)^2 = p_r^2 + \frac{j^2}{r^2}.
\ee

The Hamilton-Jacobi theory shows that the observables of the two-body dynamics
can be deduced from the (reduced) radial action integral
\be
\label{ir}
i_r(E,j) \equiv \frac{2}{2\pi} \int_{r_\textrm{min}}^{r_\textrm{max}} \sqrt{R(r,E,j)}\, \md r,
\ee
where the integration is defined from minimal to maximal radial distance.
The dimensionless parameter $k\equiv\Delta\Phi/(2\pi)$ (with $\Delta\Phi\equiv\Phi-2\pi$)
measuring the fractional periastron advance per orbit and the periastron-to-periastron period $P$
are obtained by differentiating the radial action integral:
\begin{align}
k &= -\frac{\partial i_r(E,j)}{\partial j} - 1,
\\[1ex]
P &= 2\pi GM\,\frac{\partial i_r(E,j)}{\partial E}.
\end{align}

It is useful to express the Hamiltonian as a function of the Delaunay (reduced) action variables
(see, e.g., \citealp{goldstein-81}) defined by
\be
n \equiv i_r + j = \frac{\cal N}{\mu GM},
\quad
j = \frac{J}{\mu GM},
\quad
m \equiv j_z = \frac{J_z}{\mu GM}.
\ee
The angle variables conjugate to $n$, $j$, and $m$ are, respectively:
the mean anomaly, the argument of the periastron, and the longitude of the ascending node.
In the quantum language, ${\cal N}/\hbar$ is the principal quantum number,
$J/\hbar$ the total angular-momentum quantum number, and $J_z/\hbar$ the magnetic quantum number.
They are adiabatic invariants of the dynamics and they are,
according to the Bohr-Sommerfeld rules of the old quantum theory,
(approximately) quantized in integers.
Knowing the Delaunay Hamiltonian $\hat{H}(n,j,m)$ one computes
the angular frequencies of the (generic) rosette motion of the binary system
by differentiating $\hat{H}$ with respect to the action variables. Namely,
\begin{align}
\omega_{\text{radial}} &= \frac{2\pi}{P}
= \frac{1}{GM} \frac{\partial\hat{H}(n,j,m)}{\partial n},
\\[1ex]
\omega_{\text{periastron}} &= \frac{\Delta\Phi}{P} = \frac{2\pi k}{P}
= \frac{1}{GM} \frac{\partial\hat{H}(n,j,m)}{\partial j}.
\end{align}
Here, $\omega_{\text{radial}}$ is the angular frequency of the radial motion,
i.e., the angular frequency of the return to the periastron,
while $\omega_{\text{periastron}}$ is the average angular frequency
with which the major axis advances in space.

\subsubsection{3PN-accurate results}
\label{subsubsec:dyninv3}

The dynamical invariants of two-body dynamics were computed by \cite{damour-88} at the 2PN level
and then generalized to the 3PN level of accuracy by \cite{damour1-00}.
We are displaying here 3PN-accurate formulae.
The periastron advance parameter $k$ reads\footnote{Let us note a misprint in Eq.\ (4.16) of \cite{damour1-00}:
the prefactor ``3'' in the term proportional to $i_3(\nu)$ should be removed.}
\begin{align}
k &= \frac{3}{c^2j^2} \Bigg\{ 1
+ \frac{1}{c^2} \left[ \frac{5}{4}(7-2\nu)\frac{1}{j^2} + \frac{1}{2}(5-2\nu)\,E \right]
\nonumber\\[1ex]&\quad
+ \frac{1}{c^4} \Bigg[ \frac{5}{2} \Bigg(\frac{77}{2}
+ \left(\frac{41}{64}\pi^2-\frac{125}{3}\right)\nu
+ \frac{7}{4}\nu^2\Bigg)\frac{1}{j^4}
\nonumber\\[1ex]&\quad
+ \Bigg(\frac{105}{2} + \left(\frac{41}{64}\pi^2-\frac{218}{3}\right)\nu + \frac{45}{6}\nu^2\Bigg)\frac{E}{j^2}
\nonumber\\[1ex]&\quad
+ \frac{1}{4}(5-5\nu+4\nu^2)\,E^2 \Bigg] + \mathcal{O}(c^{-6}) \Bigg\}.
\end{align}
The 3PN-accurate formula for the orbital period reads
\begin{align}
P &= \frac{2\pi GM}{(-2E)^{3/2}} \Bigg\{1 - \frac{1}{c^2} \frac{1}{4}(15-\nu) E
\nonumber\\[1ex]&\quad
+ \frac{1}{c^4} \left[\frac{3}{2}(5-2\nu)\frac{(-2E)^{3/2}}{j} -\frac{3}{32}(35+30\nu+3\nu^2)\,E^2 \right]
\nonumber\\[1ex]&\quad
+ \frac{1}{c^6} \Bigg[ \Bigg(\frac{105}{2} + \left(\frac{41}{64}\pi^2-\frac{218}{3}\right)\nu + \frac{45}{6}\nu^2\Bigg)\frac{(-2E)^{3/2}}{j^3}
\nonumber\\[1ex]&\quad
- \frac{3}{4}(5-5\nu+4\nu^2) \frac{(-2E)^{5/2}}{j}
\nonumber\\[1ex]&\quad
+ \frac{5}{128}(21-105\nu+15\nu^2+5\nu^3)\,E^3 \Bigg] + \mathcal{O}(c^{-8}) \Bigg\}.
\end{align}
These expressions have direct applications to binary pulsars \citep{damour-88}.
Explicit analytic orbit solutions of the conservative dynamics through 3PN order are given by \cite{memmesheimer-05}.
The 4PN periastron advance was first derived by \cite{damour-15,damour-16},
with confirmation provided in a later rederivation \citep{bernard-17a}; also see \cite{letiec-17}.

All conservative two-body Hamiltonians respect rotational symmetry,
therefore the Delaunay variable $m$ does not enter these Hamiltonians.
The 3PN-accurate Delaunay Hamiltonian reads \citep{damour1-00}
\begin{align}
\widehat{H}&(n,j,m) = -\frac{1}{2n^2} \bigg\{ 1 +
\frac{1}{c^2} \bigg( \frac{6}{j n}-\frac{1}{4}(15-\nu)\frac{1}{n^2} \bigg)
\nonumber\\[1ex]&
+ \frac{1}{c^4} \bigg( \frac{5}{2}(7-2\nu)\frac{1}{j^3 n}
+ \frac{27}{j^2 n^2}
- \frac{3}{2}(35-4\nu)\frac{1}{j n^3}
+ \frac{1}{8}(145-15\nu+\nu^2)\frac{1}{n^4} \bigg)
\nonumber\\[1ex]&
+ \frac{1}{c^6} \bigg[
\bigg(\frac{231}{2}+\Big(\frac{123}{64}\pi^2-125\Big)\nu+\frac{21}{4}\nu^2\bigg)\frac{1}{j^5 n}
+ \frac{45}{2}(7-2\nu)\frac{1}{j^4 n^2}
\nonumber\\[1ex]&
+\bigg(-\frac{303}{4}+\Big(\frac{1427}{12}-\frac{41}{64}\pi^2\Big)\nu-10\nu^2\bigg)\frac{1}{j^3 n^3}
- \frac{45}{2}(20-3\nu)\frac{1}{j^2 n^4}
\nonumber\\[1ex]&
+ \frac{3}{2}(275-50\nu+4\nu^2)\frac{1}{j n^5}
- \frac{1}{64}(6363-805\nu+90\nu^2-5\nu^3)\frac{1}{n^6}
\bigg]
\nonumber\\[1ex]&
+ \mathcal{O}(c^{-8}) \bigg\}.
\end{align}

Additional insight into the 3PN dynamics can be found in a paper by \cite{letiec-15},
where the first law of mechanics for binary systems of point masses \citep{letiec-12}
was generalized to generic eccentric orbits.

\subsubsection{Results at 4PN order}
\label{subsubsec:dyninv4}

The reduced 4PN Hamiltonian $\hat{H}_{\textrm{4PN}}[\vecr,\vecp]$ can be decomposed in two parts
in a way slightly different from the splitting shown in Eq.~\eqref{H4PNsum1}. Namely,
\be
\hat{H}_{\textrm{4PN}}[\vecr,\vecp] = \hat{H}^{\textrm{I}}_{\textrm{4PN}}(\vecr,\vecp;s)
+ \hat{H}^{\textrm{II}}_{\textrm{4PN}}[\vecr,\vecp;s],
\ee
where the first part is local in time while the second part is nonlocal in time;
$s\equiv{\sphys}/(GM)$ is a reduced scale with dimension of 1/velocity$^2$,
where $\sphys$ is a scale with dimension of a length.
The Hamiltonian $\hat{H}^{\textrm{I}}_{\textrm{4PN}}$
is a function of phase-space variables $(\vecr,\vecp)$ of the form 
\be
\label{eq2.3bis}
\hat{H}^{\textrm{I}}_{\textrm{4PN}}(\vecr,\vecp;s)
= \hat{H}^{\textrm{loc}}_{\textrm{4PN}}(\vecr,\vecp) + F(\vecr,\vecp)\ln\frac{r}{s},
\quad
F(\vecr,\vecp) \equiv \frac{2}{5}\frac{G^2}{\nu}(\dddot{I}_{\!ij})^2,
\ee
where the Hamiltonian $\hat{H}^{\textrm{loc}}_{\textrm{4PN}}$ is given in Eq.~\eqref{H4PNlocal1} above.
The Hamiltonian $\hat{H}^{\textrm{II}}_{\textrm{4PN}}$
is a functional of phase-space trajectories $(\vecr(t),\vecp(t))$,
\be
\label{H4PNnonloc2}
\hat{H}^{\textrm{II}}_{\textrm{4PN}}[\vecr,\vecp;s] = -\frac{1}{5}\frac{G^2}{\nu} \dddot{I}_{\!ij}(t)
\times \Pf_{2\sphys/c} \int_{-\infty}^{+\infty} \frac{\md \tau}{\vert \tau \vert} \dddot{I}_{\!ij}(t+\tau).
\ee
The nonlocal Hamiltonian $\hat{H}^{\textrm{II}}_{\textrm{4PN}}[\vecr,\vecp;s]$
differs from what is displayed in Eq.~\eqref{H4PNnonloc1} as the nonlocal part of the 4PN Hamiltonian.
There the nonlocal piece of $\hat{H}_{\textrm{4PN}}$ is defined
by taking as regularization scale in the partie finie operation entering Eq.~\eqref{H4PNnonloc1}
the time $2r_{12}/c$ instead of $2s_{\rm phys}/c$ appearing in \eqref{H4PNnonloc2}.
Thus the arbitrary scale $s_{\rm phys}$ enters both parts
$\hat{H}^{\textrm{I}}_{\textrm{4PN}}$ and $\hat{H}^{\textrm{II}}_{\textrm{4PN}}$
of $\hat{H}_{\textrm{4PN}}$, though it cancels out in the total Hamiltonian.
\cite{damour-15} has shown that modulo some nonlocal-in-time shift of the phase-space coordinates,
one can reduce a nonlocal dynamics defined by the Hamiltonian
$\hat{H}[\vecr,\vecp;s]\equiv\hat{H}_{\textrm{N}}(\vecr,\vecp)+\hat{H}^{\textrm{II}}_{\textrm{4PN}}[\vecr,\vecp;s]$
to an ordinary (i.e., local in time) one. We will sketch here this reduction procedure,
which employs the Delaunay form of the Newtonian equations of motion.
In the circular motion case things are much simpler
and we can directly perform the integral in the nonlocal Hamiltonian, Eq.~\eqref{htailcirc}.

It is enough to consider the planar case.
In that case the action-angle variables are $(\rL,\ell;\rG,g)$, using the standard notation of \cite{brouwer-61}
(with $\rL\equiv n$ and $\rG\equiv j$).
The variable $\rL$ is conjugate to the ``mean anomaly'' $\ell$, while $\rG$ is conjugate to the argument of the periastron $g=\omega$.
The variables $\rL$ and $\rG$ are related to the usual Keplerian variables $a$ (semimajor axis) and $e$ (eccentricity) via
\be
\label{4Del1}
\rL \equiv \sqrt{a},
\quad
\rG \equiv \sqrt{a(1-e^2)}.
\ee
By inverting \eqref{4Del1} one can express $a$ and $e$ as functions of $\rL$ and $\rG$:
\be
\label{4Del2}
a = \rL^2,
\quad
e = \sqrt{1 - \left(\frac{\rG}{\rL}\right)^2}.
\ee
We use here rescaled variables: in particular, $a$ denotes the rescaled semimajor axis $a\equiv a_{\rm phys}/(GM)$.
We also use the rescaled time variable $\hat{t}\equiv t_{\rm phys}/(GM)$
appropriate for the rescaled Newtonian Hamiltonian
\be
\label{4Del3}
\hat{H}_{\textrm{N}}(\rL) = \frac12 \, \vecp^2 - \frac1r = -\frac1{2 \rL^2}.
\ee
The explicit expressions of the Cartesian coordinates $(x,y)$ of a Newtonian motion in terms of action-angle variables are given by
\begin{align}
x (\rL , \ell ; \rG , g) &= \cos g \, x_0 - \sin g \, y_0,
\quad
y (\rL , \ell ; \rG , g) = \sin g \, x_0 + \cos g \, y_0,
\\[1ex]
x_0 &= a (\cos u - e),
\quad
y_0 = a \sqrt{1-e^2} \sin u,
\end{align}
where the ``eccentric anomaly'' $u$ is the function of $\ell$ and $e$ defined by solving Kepler's equation
\be
\label{Keq}
u - e \sin u = \ell .
\ee
The solution of Kepler's equation can be written in terms of Bessel functions:
\be
\label{u(l)}
u = \ell + \sum_{n=1}^\infty \frac{2}{n} J_n(ne) \sin (n \, \ell).
\ee
Note also the following Bessel-Fourier expansions of $\cos u$ and $\sin u$ [which directly enter $(x_0 , y_0)$ and thereby $(x,y)$]
\begin{align}
\cos u &= -\frac e2 + \sum_{n=1}^{\infty} \frac1n [J_{n-1} (ne) - J_{n+1} (ne)] \cos n \, \ell ,
\\[1ex]
\sin u &= \sum_{n=1}^{\infty} \frac1n [J_{n-1} (ne) + J_{n+1} (ne)] \sin n \, \ell .
\end{align}
For completeness, we also recall the expressions involving the ``true anomaly'' $f$ (polar angle from the periastron) and the radius vector $r$:
\begin{align}
r &= a (1-e \cos u) = \frac{a (1-e^2)}{1+e \cos f},
\\[1ex]
\frac{x_0}r &= \cos f = \frac{\cos u -e}{1-e \cos u},
\quad
\frac{y_0}r = \sin f = \frac{\sqrt{1-e^2} \sin u}{1-e \cos u}.
\end{align}
The above expressions allow one to evaluate the expansions of $x$, $y$,
and therefrom the components of the quadrupole tensor $I_{ij}$, as power series in $e$ and Fourier series in $\ell$.

Let us then consider the expression
\be
\label{calF}
\mathcal{F}(t,\tau) \equiv \dddot{I}_{\!ij}(t)\dddot{I}_{\!ij}(t+\tau),
\ee
which enters the nonlocal-in-time piece \eqref{H4PNnonloc2} of the Hamiltonian.
In order to evaluate the order-reduced value of $\mathcal{F}(t,\tau)$ one needs to use the equations of motion,
both for computing the third time derivatives of $I_{ij}$,
and for expressing the phase-space variables at time $t+\tau$ in terms of the phase-space variables at time $t$.
One employs the zeroth-order equations of motion following from the Newtonian Hamiltonian \eqref{4Del3},
\begin{align}
\frac{{\rm d}\ell}{{\rm d} \hat t} &= \frac{\partial\hat{H}_{\textrm{N}}}{\partial \rL} = \frac1{\rL^3} \equiv \Omega (\rL),
\quad
\frac{{\rm d} g}{{\rm d} \hat t} = \frac{\partial\hat{H}_{\textrm{N}}}{\partial \rG} = 0,
\\[1ex]
\frac{{\rm d}\rL}{{\rm d} \hat t} &= - \frac{\partial\hat{H}_{\textrm{N}}}{\partial \ell} = 0,
\quad
\frac{{\rm d} \rG}{{\rm d}\hat t} = - \frac{\partial\hat{H}_{\textrm{N}}}{\partial g} = 0,
\end{align}
where $\Omega (\rL) \equiv \rL^{-3}$ is the ($\hat{t}$-time) rescaled  Newtonian (anomalistic) orbital frequency $\Omega = G M \Omega_{\rm phys}$
(it satisfies the rescaled third Kepler's law: $\Omega = a^{-3/2}$).
The fact that $g$, $\rL$, and $\rG$ are constant and that $\ell$ varies linearly with time,
makes it easy to compute $\dddot I_{\!ij} (t+\tau)$ in terms of the values of $(\ell,g,\rL,\rG)$ at time $t$.
It suffices to use (denoting by a prime the values at time $t' \equiv t+\tau$)
\be
\ell' \equiv \ell (t+\tau) = \ell(t) + \Omega (\rL) \hat\tau,
\ee
where $\hat\tau \equiv \tau/(GM)$, together with $g' = g$, $\rL' = \rL$, and $\rG' = \rG$.
The order-reduced value of $\mathcal{F}(t,\tau)$ is given by
(using $\md/\md\hat t = \Omega\,\md/\md\ell$)
\be
\label{calF2}
\mathcal{F}(\ell,\hat\tau) = \bigg(\frac{\Omega (\rL)}{GM}\bigg)^6 \,
\frac{\md^3 I_{ij}}{\md\ell^3}(\ell) \frac{\md^3 I_{ij}}{\md\ell^3}(\ell+\Omega (\rL) \hat\tau).
\ee
Inserting the expansion of $I_{ij} (\ell)$ in powers of $e$ and in trigonometric functions of $\ell$ and $g$,
yields $\rF$ in the form of a series of monomials of the type
\be
\label{expF}
\rF(\ell,\hat\tau) = \sum_{n_1 , n_2 , \pm n_3} C_{n_1 n_2 n_3}^{\pm} \, e^{n_1} \cos (n_2 \, \ell \pm n_3 \, \Omega \, \hat\tau),
\ee
where $n_1$, $n_2$, $n_3$ are natural integers.
(Because of rotational invariance, and of the result $g' = g$, there is no dependence of $\rF$ on $g$.)

All the terms in the expansion \eqref{expF} containing a nonzero value of $n_2$ will,
after integrating over $\hat\tau$ with the measure $\md\hat\tau/\vert\hat\tau\vert$ as indicated in Eq.~\eqref{H4PNnonloc2},
generate a corresponding contribution to $\hat{H}^{\textrm{II}}_{\textrm{4PN}}$ which varies with $\ell$ proportionally to $\cos (n_2\,\ell)$.
One employs now the standard Delaunay technique:
any term of the type $A(\rL)\cos(n\ell)$ in a first-order perturbation
$\varepsilon H_1(\rL,\ell)\equiv\hat{H}^{\textrm{II}}_{\textrm{4PN}}(\rL,\ell)$
of the leading-order Hamiltonian $H_0(\rL)\equiv H_{\textrm{N}}(\rL)$
can be eliminated by a canonical transformation with generating function of the type
$\varepsilon\mathfrak{g}(\rL,\ell)\equiv\varepsilon B(\rL) \sin (n\ell)$. Indeed,
\be
\delta_\mathfrak{g} H_1 = \{ H_0 (\rL) , \mathfrak{g} \} = - \frac{\partial H_0 (\rL)}{\partial \rL} \, \frac{\partial\mathfrak{g}}{\partial \ell}
= - n \, \Omega (\rL) \, B(\rL) \cos (n\ell),
\ee
so that the choice $B = A/(n \, \Omega)$ eliminates the term $A \cos (n\ell)$ in $H_1$.
This shows that all the periodically varying terms (with $n_2 \ne 0$) in the expansion \eqref{expF} of $\rF$
can be eliminated by a canonical transformation.
Consequently one can simplify the nonlocal part $\hat{H}^{\textrm{II}}_{\textrm{4PN}}$ of the 4PN Hamiltonian
by replacing it by its $\ell$-averaged value,
\be
\label{avh2a}
\hat{\bar{H}}^{\textrm{II}}_{\textrm{4PN}}(\rL,\rG;s)
\equiv \frac{1}{2\pi}\int_0^{2\pi}\md\ell\,\hat{H}^{\textrm{II}}_{\textrm{4PN}}[\vecr,\vecp;s]
= -\frac15 \, \frac{G^2}{\nu c^8} \, {\rm Pf}_{2s/c}
\int_{-\infty}^{+\infty} \frac{{\rm d} \hat\tau}{\vert \hat\tau \vert} \, \bar\rF ,
\ee
where $\bar\rF$ denotes the $\ell$-average of $\rF (\ell , \hat\tau)$
[which is simply obtained by dropping all the terms with $n_2 \ne 0$ in the expansion (\ref{expF})].
This procedure yields an averaged Hamiltonian $\hat{\bar{H}}^{\textrm{II}}_{\textrm{4PN}}$ which depends only on $\rL$, $\rG$ (and $s$)
and which is given as an expansion in powers of $e$
(because of the averaging this expansion contains only even powers of $e$).
\cite{damour-15} derived the $\ell$-averaged Hamiltonian as a power series of the form\footnote{
Here $\me=2.718\ldots$ should be distinguished from the eccentricity $e$.}
\be
\label{avh2b}
\hat{\bar{H}}^{\textrm{II}}_{\textrm{4PN}}(\rL,\rG;s)
= \frac{4}{5}\frac{\nu}{c^8\rL^{10}} \sum_{p=1}^\infty p^6 |\hat{I}^p_{ij}(e)|^2 \ln\left(2p\frac{\me^\gE s}{c\rL^3}\right),
\ee
where $\hat{I}^p_{ij}(e)$ are coefficients in the Bessel-Fourier expansion of the dimensionless reduced quadrupole moment
$\hat{I}_{ij}\equiv I_{ij}/[(GM)^2\mu a^2]$,
\be
\hat{I}_{ij}(\ell,e) = \sum_{p=-\infty}^{+\infty} \hat{I}_{ij}^p(e) \me^{\mi p \ell}.
\ee
Equation \eqref{avh2b} is the basic expression for the transition
of the tail-related part of the 4PN dynamics to the EOB approach \citep{damour-15}.

For another approach to the occurrence and treatment of the $(\ell,\ell')$-structure in nonlocal-in-time Hamiltonians
the reader is referred to \cite{damour-16} (therein, $\ell$ is called $\lambda$).
Generalized quasi-Keplerian parametrization for eccentric orbits at 4PN order was studied in \cite{cho-22}
(ignoring certain oscillatory terms arising due to 4PN tail effects).

\subsubsection{Results at 5PN order}
\label{subsubsec:results5pn}

To compactify the expressions for higher-order PN Hamiltonians it is most convenient
to go over to the canonically equivalent Hamiltonians of the EOB formalism
\citep{buonanno-99,buonanno-00} (let us remind that the EOB approach is not in the scope of this review).
Within this formalism the $n$PN-accurate Hamiltonian $H_{\le\text{$n$PN}}(x,p)$ of the two-body system, in the centre of mass frame,
is replaced by the \emph{real} (i.e.\ giving the evolution equations with respect to the real ADM time coordinate $t_{\text{ADM}}$
and the real two-body energy) and \emph{improved} (i.e.\ representing a nonperturbative resummed estimate of the PN Hamiltonian)
Hamiltonian $H^\text{improved}_\text{real}(x'(x,p),p'(x,p))$ \citep{buonanno-00}.
The Hamiltonian $H^\text{improved}_\text{real}$ is related to the \emph{effective} EOB Hamiltonian $\heobeff$
through the equation \citep{damour2-00}
\be
\label{EOBeq1}
\frac{\heobeff}{\mu c^2} = \frac{(H^\text{improved}_\text{real})^2 -m_1^2c^4 - m_2^2c^4}{2m_1m_2c^4},
\ee
resulting in the useful representation of $H^\text{improved}_\text{real}$ in terms of $\heobeff$,
\be
\label{EOBeq2}
H^\text{improved}_\text{real} = Mc^2\sqrt{1 + 2\nu \left(\heobeffr - 1\right)},
\ee
where $\heobeffr:=\heobeff/(\mu c^2)$ denotes the \emph{reduced} effective EOB Hamiltonian.
In turn, the EOB effective Hamiltonian is defined as $\heobeff:=-c\,p'_0$,
where $p'_0$ is the solution of a general mass-shell condition of the form
\begin{align}
\label{EOBeq3}
g^{\mu\nu}_{\rm eff}(x')p'_{\mu}p'_{\nu} + Q(x',p'_r) = -\mu^2c^2,
\end{align}
where the scalar $Q$ denotes contributions which are at least quartic in momenta;
one can reduce the dependence of $Q$ on momenta to a dependence on the sole radial momentum $p'_r$.
The spherically symmetric effective metric $g_{\mu\nu}^{\rm eff}$
is a $\nu$-dependent deformation of Schwarzschild metric,
\begin{align}
\label{EOBeq4}
g_{\mu\nu}^{\rm eff}\md x'^\mu \md x'^\nu &= -A(r';\nu) c^2 \md t'^2
+ \big(A(r';\nu)\bar{D}(r';\nu)\big)^{-1}\md r'^2
\nonumber\\&\qquad
+ r'^2 (\md\theta'^2 + \sin^2\theta'\,\md\phi'^2).
\end{align}
Solving Eq.~\eqref{EOBeq3} [with the metric \eqref{EOBeq4}] with respect to $p'_0$
gives the reduced effective EOB Hamiltonian of the form
\be
\label{EOBeq5}
\heobeffr(x',p';\nu) = \sqrt{A(u;\nu)\Big(1 + \hat{p}'^2
+ \Big(A(u;\nu)\bar{D}(u;\nu)-1\Big)\hat{p}'^2_r + \hat{Q}(u,\hat{p}'_r;\nu)\Big)},
\ee
where $\hat{Q}=Q/(\mu c^2)$, $u:=GM/(r'c^2)$, $\hat{p}'_r:=p'_r/(\mu c)$,
$\hat{p}':=p'/(\mu c)$ with $p':=\sqrt{p'^2_r+p'^2_\theta/r'^2+p'^2_\phi/(r'^2\sin^2\theta')}$.

The 5PN-accurate PN expansions of the potentials $A$, $\bar{D}$, and $\hat{Q}$ read
[let us note that $u=\mathcal{O}(c^{-2})$ and $p'_r=\mathcal{O}(c^{-1})$]
\begin{subequations}
\label{EOBpot}
\begin{align}
A(u;\nu) &= 1 + \sum_{k=1}^4 a_k(\nu) u^k
+ \sum_{k=5}^6 \big( a_k^\text{c}(\nu)+a_k^\text{ln}(\nu)\ln u \big) u^k,
\\[2ex]
\bar{D}(u;\nu) &= 1 + \sum_{k=2}^3 \bar{d}_k(\nu) u^k
+ \sum_{k=4}^5 \big( \bar{d}_k^\text{c}(\nu)+{\bar{d}}_k^\text{ln}(\nu)\ln u \big) u^k,
\\[2ex]
Q(u,p'_r;\nu) &= \Big(q_{42}(\nu)u^2 + q_{43}(\nu)u^3 + \big(q_{44}^{\text{c}}(\nu)+q_{44}^{\text{ln}}(\nu) \ln u\big)u^4\Big) p'^4_r
\nonumber\\[1ex]&\quad
+ \Big(q_{62}(\nu)u^2 + \big(q_{63}^{\text{c}}(\nu)+q_{63}^{\text{ln}}(\nu) \ln u\big)u^3\Big) p'^6_r
\nonumber\\[1ex]&\quad
+ \Big(q_{81}(\nu)u + q_{82}(\nu)u^2 \Big) p'^8_r.
\end{align}
\end{subequations}

Up to the 3PN level, the coefficients read as follows \citep{buonanno-99,damour2-00}:
\begin{subequations}
\begin{align}
&\text{At 0PN:}\quad a_1(\nu) = -2,
\\[2ex]
&\text{at 1PN:}\quad a_2(\nu) = 0,
\\[2ex]
&\text{at 2PN:}\quad a_3(\nu) = 2\nu, \quad \bar{d}_2(\nu) = 6\nu,
\\[2ex]
&\text{at 3PN:}\quad a_4(\nu) = \left(\frac{94}{3}-\frac{41}{32}\pi^2\right)\nu,
\quad \bar{d}_3(\nu) = 52\nu - 6\nu^2,
\nonumber\\[1ex]&\phantom{\text{at 3PN:}\quad}
q_{42}(\nu) = 8\nu - 6\nu^2.
\end{align}
\end{subequations}
At the 4PN level, the coefficients read \citep{damour-15,bini-20}
\begin{subequations}
\begin{align}
a_5^\text{c}(\nu) &= \left( \frac{2275}{512}\pi^2 - \frac{4237}{60} + \frac{128}{5}\gE + \frac{256}{5}\ln2 \right) \nu
+ \left( \frac{41}{32}\pi^2 - \frac{221}{6} \right)\nu^2,
\\[1ex]
a_5^\text{ln}(\nu) &= \frac{64}{5}\nu,
\\[1ex]
\bar{d}_4^\text{c}(\nu) &= \left( -\frac{533}{45} - \frac{23761}{1536}\pi^2 + \frac{1184}{15}\gE - \frac{6496}{15}\ln2 + \frac{2916}{5}\ln3 \right)\nu
\nonumber\\&\quad
+ \left(\frac{123}{16}\pi^2 - 260\right)\nu^2,
\\[1ex]
\bar{d}_4^\text{ln}(\nu) &= \frac{592}{15}\nu,
\\[1ex]
q_{43}(\nu) &= \left( -\frac{5308}{15} + \frac{496256}{45}\ln2 - \frac{33048}{5}\ln3 \right) \nu
- 83 \nu^2 + 10 \nu^3,
\\[1ex]
q_{62}(\nu) &= \left(-\frac{827}{3} - \frac{2358912}{25}\ln2 + \frac{1399437}{50}\ln3
+ \frac{390625}{18}\ln5 \right)\nu
\nonumber\\&\quad
- \frac{27}{5}\nu^2 + 6\nu^3,
\\[1ex]
q_{81}(\nu) &= \left(-\frac{35772}{175} + \frac{21668992}{45}\ln2 + \frac{6591861}{350}\ln3
- \frac{27734375}{126}\ln5 \right)\nu.
\end{align}
\end{subequations}

At the 5PN level, solution with unique numerical prefactors is not available.
The TF approach yields all 5PN-order coefficients of the EOB potentials \eqref{EOBpot}
except for numerical prefactors of two terms proportional to $\nu^2$
entering the coefficients $a_6^\text{c}(\nu)$ and $\bar{d}_5^\text{c}(\nu)$.
Also, \cite{blumlein-22,blumlein-22e} disagree with obtained by \cite{bini-19,bini-20}
local contribution to a term proportional to $\nu^2$ in the coefficient $q_{44}^\text{c}(\nu)$.
The coefficients of the 5PN-order EOB potentials read \citep{bini-19,bini-20}
\begin{subequations}
\begin{align}
a_6^\text{c}(\nu) &= \left(-\frac{1066621}{1575} + \frac{246367}{3072}\pi^2
- \frac{14008}{105}\gE - \frac{31736}{105}\ln2 + \frac{243}{7}\ln3 \right){\nu}
\nonumber\\&\quad
+ a_{62}\nu^2 + 4\nu^3,
\\[1ex]
a_6^\text{ln}(\nu) &= -\frac{7004}{105}\nu - \frac{144}{5}\nu^2,
\\[2ex]
\bar{d}_5^\text{c}(\nu) &= \left(\frac{294464}{175} - \frac{63707}{512}\pi^2 - \frac{2840}{7}\gE + \frac{120648}{35}\ln2 - \frac{19683}{7}\ln3\right){\nu}
\nonumber\\&\quad
+ \bar{d}_{52}\nu^2 + \left(\frac{1069}{3} - \frac{205}{16}\pi^2\right) {\nu^3},
\\[1ex]
\bar{d}_5^\text{ln}(\nu) &= -\frac{1420}{7}\nu - \frac{3392}{15}\nu^2,
\\[2ex]
q_{44}^\text{c}(\nu) &= \bigg( \frac{1295219}{350} - \frac{93031}{1536}\pi^2 + \frac{10856}{105}\gE
- \frac{40979464}{315}\ln2 + \frac{14203593}{280}\ln3
\nonumber\\&\quad
+ \frac{9765625}{504}\ln5 \bigg){\nu}
+ q_{442}\nu^2
+ \left(640 - \frac{615}{32}\pi^2\right)\nu^3,
\\[1ex]
q_{44}^\text{ln}(\nu) &= \frac{5428}{105}\nu - \frac{592}{5}\nu^2,
\\[2ex]
q_{63}^\text{c}(\nu) &= \bigg( \frac{2613083}{1050} + \frac{6875745536}{4725}\ln2 - \frac{23132628}{175}\ln3
- \frac{101687500}{189}\ln5 \bigg)\nu
\nonumber\\&\quad
+ \bigg( \frac{159089}{75} - \frac{4998308864}{1575}\ln2 - \frac{45409167}{350}\ln3
+ \frac{26171875}{18}\ln5 \bigg){\nu^2}
\nonumber\\&\quad
+ 116\nu^3 – 14\nu^4,
\\[2ex]
q_{63}^\text{ln}(\nu) &= 0,
\\[2ex]
q_{82}(\nu) &= \bigg( \frac{5790381}{2450} - \frac{16175693888}{1575}\ln2
- \frac{393786545409}{156800}\ln3
\nonumber\\&\quad
+ \frac{875090984375}{169344}\ln5 + \frac{13841287201}{17280}\ln7 \bigg)\nu
\nonumber\\&\quad
+ \bigg( \frac{870976}{525} + \frac{703189497728}{33075}\ln2
+ \frac{332067403089}{39200}\ln3
\nonumber\\&\quad
- \frac{468490234375}{42336}\ln5
- \frac{13\,841\,287\,201}{4\,320}\ln7 \bigg)\nu^2
+ \frac{24}{7}{\nu^3} - 6\nu^4.
\end{align}
\end{subequations}
The nonlocal part of the potential $q_{82}$ was computed in Appendix~G of \cite{bini-20c}.

The non computed in \cite{bini-19,bini-20} prefactors $a_{62}$ and $\bar{d}_{52}$
enter the local-in-time parts of the EOB potentials,
\be
a_{62} = a_{62}^\text{nloc} + a_{62}^\text{loc},
\quad
\bar{d}_{52} = \bar{d}_{52}^\text{nloc} + \bar{d}_{52}^\text{loc},
\ee
where the prefactors $a_{62}^\text{nloc}$ and $\bar{d}_{52}^\text{nloc}$
related with the nonlocal-in-time parts are well confirmed
and equal [see Table IV in \cite{bini-20}]
\begin{subequations}
\begin{align}
a_{62}^\text{nloc} &= \frac{64}{5} - \frac{288}{5}\gE
+ \frac{928}{35}\ln2 - \frac{972}{7}\ln3,
\\
\bar{d}_{52}^\text{nloc} &= \frac{67\,736}{105} -\frac{6\,784}{15}\gE
- \frac{326\,656}{21}\ln2 + \frac{58\,320}{7}\ln3.
\end{align}
\end{subequations}
The EFT approach by \cite{blumlein-21,blumlein-22} gives,
\begin{subequations}
\begin{align}
a_{62}^\text{loc} &= a_{62(\text{rat})}^\text{loc} + a_{62(\pi^2)}^\text{loc},
\quad
a_{62(\text{rat})}^\text{loc} = -\frac{584881}{525},
\quad
a_{62(\pi^2)}^\text{loc} = \frac{25911}{256}\pi^2,
\\[2ex]
\bar{d}_{52}^\text{loc} &= \bar{d}_{52(\text{rat})}^\text{loc} + \bar{d}_{52(\pi^2)}^\text{loc},
\quad
\bar{d}_{52(\text{rat})}^\text{loc} = -\frac{10442728}{1575},
\quad
\bar{d}_{52(\pi^2)}^\text{loc} = \frac{306545}{512}\pi^2.
\end{align}
\end{subequations}
The coefficients $a_{62(\pi^2)}^\text{loc}$ and $\bar{d}_{52(\pi^2)}^\text{loc}$ are confirmed by TF.

The computed in \cite{bini-19,bini-20} prefactor $q_{442}$
is the sum of the local-in-time and the nonlocal-in-time parts,
\be
q_{442} = q_{442}^\text{nloc} + q_{442}^\text{loc},
\ee
where the nonlocal-in-time part $q_{442}^\text{nloc}$ reads
[see Table IV in \cite{bini-20}]
\be
q_{442}^\text{nloc} = \frac{74\,436}{35} -\frac{1\,184}{5}\gE
+ \frac{33\,693\,536}{105}\ln2 - \frac{6\,396\,489}{70}\ln3 - \frac{9\,765\,625}{126}\ln5.
\ee
The local-in-time part $q_{442}^\text{loc}$ equals
\be
q_{442}^\text{loc} = q_{442(\text{rat})}^\text{loc} + q_{442(\pi^2)}^\text{loc},
\quad
q_{442(\pi^2)}^\text{loc} = \frac{31\,633}{512}\pi^2,
\ee
where the transcendental part $q_{442(\pi^2)}^\text{loc}$
is confimed by both \cite{bini-19,bini-20} and \cite{blumlein-22,blumlein-22e}.
However, the rational part $q_{442(\text{rat})}^\text{loc}$ has incompatible values
according to \cite{bini-19,bini-20} (TF) and \cite{blumlein-22,blumlein-22e} (BMMS),
\be
q_{442(\text{rat})}^\text{loc TF} = -\frac{9\,367}{15},
\quad
q_{442(\text{rat})}^\text{loc BMMS} = -\frac{1\,252\,924}{1\,575}.
\ee
Agreement between the TF and BMMS results could be achieved
by a possibly missing conservative quadratic radiation-reaction (anti-symmetric)$^2$ term mentioned in \cite{bini-21},
which could lead to the following change of the TF Hamiltonian \citep{blumlein-22,blumlein-22e},
\be
\delta H^{\rm (reac)^2}_{\rm rad} = a\,\nu^2 p'^4_r u^4, \quad a\in\mathbb{R}.
\ee
The agreement would be achieved for $a=-168/5$ \citep{blumlein-22,blumlein-22e}.

The genuine (i.e., not the 1PN corrections coming from 4PN level)
local and nonlocal tail Hamiltonians at the 5PN order are
(\citealp{foffa-20,bini-21,almeida-21,blumlein-21,blumlein-22})
\begin{subequations}
\begin{align}
\label{5PNnloctail1}
H^{\rm tail, nloc}_{5\rm PN}
&= -\frac{G\mathcal{M}}{c^3} {\rm Pf}_{2r_{12}/c}
\int_{-\infty}^{\infty}\frac{\md\tau}{|\tau|}{\cal{F}}_{\rm 1PN}^{\rm split}(t,t + \tau),
\\[1ex]
\label{5PNnloctail2}
H^{\rm tail, loc}_{\rm 5PN}
&= -\frac{G\mathcal{M}}{c^3} \left(
R_{\rm oct,e}{\cal{F}}^{{\rm split}, MQ^2}_{\rm 1PN}(t,t)
+ R_{\rm quad,m} {\cal{F}}^{{\rm split}, MJ^2}_{\rm 1PN}(t,t) \right).
\end{align}
\end{subequations}
Here, $\mathcal{M}$ denotes the total ADM conserved mass-energy of the binary system
[$\mathcal{M}=M+\mathcal{O}(c^{-2})$]
and the indices $MQ^2$ and $MJ^2$ are denoting the mass-type (or electric-type) octupole-moment ($Q_{ijk}$)
and the spin-type (or magnetic-type) quadrupole-moment ($J_{ij}$) contributions, respectively, and
\begin{subequations}
\begin{align}
{\cal{F}}_{\rm 1PN}^{\rm split}(t,t') &= \frac{G}{c^5}\frac{1}{c^2} \left(
\frac{1}{189}Q^{(4)}_{ijk}(t)Q^{(4)}_{ijk}(t')
+ \frac{16}{45}J^{(3)}_{ij}(t)J^{(3)}_{ij}(t') \right),
\\[2ex]
R^{\rm TF}_{\rm oct,e} &= R^{\rm FS}_{\rm oct,e} = R^{\rm BMMS}_{\rm oct,e} = \frac{82}{35},
\\[1ex]
R^{\rm TF}_{\rm quad,m} &= R^{\rm AFS}_{\rm quad,m} = R^{\rm BMMS}_{\rm quad,m} = \frac{49}{20}.
\end{align}
\end{subequations}
We have used here the notation $f^{(n)}(t)\equiv\md f(t)/\md t^n$
to denote the $n$-th derivative with respect to time $t$.

The magnetic-type quadrupole moment $J_{ij}=J_{ji}$ comes in
via the most subtle form $\frac{1}{2}R_{0iab}\epsilon_{abj}J_{ij}$, valid in 3 dimensions only.
Its $d$-dimensional generalization needs the avatar $J_{i|ab}$,
antisymmetric with respect to $i$ and $b$, $J_{i|ab}=-J_{b|ai}$,
that satisfies the cyclic identity $J_{i|ab}+J_{a|bi}+J_{b|ia}=0$.
It reads \citep{henry-21,bini-21}
\begin{align}
J_{i|ab} = \nu(m_2-m_1) \bigg(&\Big(x^ix^a - \frac{{\bf x} \cdot {\bf x}}{d-1}\delta^{ia}\Big){\rm v}_b
- \Big(x^ax^b-\frac{{\bf x} \cdot {\bf x}}{d-1}\delta^{ab}\Big)v_i
\nonumber\\[1ex]&
- \frac{{\bf x} \cdot {\bf v}}{d-1}(x^i\delta^{ab} - x^b\delta^{ia})\bigg).
\end{align}
Then $\epsilon_{abj}J_{ij} \equiv J_{b|ia}$,
\be
J^{(3)}_{ij} J^{(3)}_{ij} \rightarrow \frac{1}{2}J^{(3)}_{i|ab} J^{(3)}_{i|ab}.
\ee

The following relations have been derived within TF \citep{bini-21},
using $R_{\rm oct,e}$ and $R_{\rm quad,m}$,
\begin{subequations}
\begin{align}
a_{62}^\text{loc} &= \frac{25\,911}{256}\pi^2 + R_{a_6}(C_{QQL},C_{QQQ_1},C_{QQQ_2}),
\\[1ex]
\bar{d}_{52}^\text{loc} &= \frac{306\,545}{512}\pi^2 + R_{d_5}(C_{QQL},C_{QQQ_1},C_{QQQ_2}),
\end{align}
\end{subequations}
where $R_{a_6}$ and $R_{d_5}$ are given rational-valued functions
of the three numerical constants $C_{QQA}$ ($A = L,Q_1,Q_2$)
which are defined by specific terms in the effective action for the radiation-type graviton exchange:
\begin{subequations}
\begin{align}
S_{QQL} &= C_{QQL}G^2 \int\md t\, Q^{(4)}_{il}Q^{(3)}_{jl}\epsilon_{ijk}L_k,
\\[1ex]
S_{QQQ_1} &= C_{QQQ_1}G^2 \int\md t\, Q^{(4)}_{il}Q^{(4)}_{jl}Q_{ij},
\\[1ex]
S_{QQQ_2} &= C_{QQQ_2}G^2 \int\md t\, Q^{(3)}_{il}Q^{(3)}_{jl}Q^{(2)}_{ij},
\end{align}
\end{subequations}
with values all having been calculated by
\cite{foffa-20,foffa-20e}, \cite{blumlein-22,blumlein-22e}, and \cite{almeida-23}
using in-out and in-in (or, closed-time) formalisms, respectively,\footnote{
The numerical value $-1/30$ of the coefficient $C^{\rm AFS}_{QQL}$ computed in \cite{almeida-23}
corrects the value $-8/15$ obtained by means of an incomplete computation by \cite{foffa-20,foffa-20e}.
The value $-1/30$ was also recently confirmed by \cite{henry-23} by means of the Fokker Lagrangian method
and dimensional regularization.}
\begin{subequations}
\begin{align}
C^{\rm AFS}_{QQL} &= -\frac{1}{30} = \frac{1}{16} C^{\rm BMMS}_{QQL},
\\[1ex]
C^{\rm mem FS}_{QQQ_1} &= -\frac{1}{15} = \frac{4}{3}C^{\rm mem BMMS}_{QQQ_1},
\\[1ex]
C^{\rm cont BMMS}_{QQQ_1} &= \frac{1}{8},
\\[1ex]
C^{\rm mem FS}_{QQQ_2} &= -\frac{4}{105}
= \frac{4}{3}C^{\rm mem BMMS}_{QQQ_2}.
\end{align}
\end{subequations}
The abbreviations ``mem'' and ``cont'' denote so-called memory and contact terms, respectively.

In terms of doubled in-in position variables,
$x^i_{a,1}$ (moving forward in time) and $x^i_{a,2}$ (moving backward in time),
with then $x^i_{a,-} = (x^i_{a,1} - x^i_{a,2})/\sqrt{2}$
and $x^i_{a,+} = (x^i_{a,1} + x^i_{a,2})/\sqrt{2}$ or, alternatively,
$x^i_{a,-} = x^i_{a,1} - x^i_{a,2}$ and $x^i_{a,+} = (x^i_{a,1} + x^i_{a,2})/2$,
the action functionals obtained in respectively \cite{blumlein-22,blumlein-22e} and \cite{almeida-22} coincide.
The classical limit reads $x^i_{a,1} = x^i_{a,2} = x^i_a$.
In the extractions of classical dynamics information, however,
\cite{blumlein-22,blumlein-22e} and \cite{almeida-22} did obtain different results.

By TF \citep{bini-21}, the following constraint equation is derived
from the condition on scattering-angles $\chi^{\rm cons,EFT}_4 - \chi^{\rm cons,TF}_4 = 0$ of conservative dynamics,
where $\chi^{\rm cons,TF}_4$ is based on a general rule on mass-polynomiality \citep{damour-20}
that terms proportional to $\nu^2$ are not present,
\be
0 = \frac{2973}{350} - \frac{69}{2}C_{QQL} + \frac{253}{18} C_{QQQ_1} + \frac{85}{9} C_{QQQ_2},
\ee
where the pure rational number is obtained for a specific value of $q_{44}$.
That condition gets fulfilled by neither the values from \cite{foffa-20,foffa-20e}
nor those from \cite{blumlein-22,blumlein-22e}.
Also \cite{almeida-22} does not stay in agreement.

To sum up: on the local-in-time level, the 5PN EOB numerical coefficients
$a_{62(\text{rat})}^\text{loc}$, $\bar{d}_{52(\text{rat})}^\text{loc}$, and $q_{442(\text{rat})}^\text{loc}$
are still controversial.

\subsubsection{Results at 5.5PN order}
\label{subsubsec:results5.5pn}

Half-integer-power PN contributions to \emph{conservative} two-body dynamics
start at the 5.5PN order \citep{shah-14,blanchet-14b}.
The complete 5.5PN conservative Hamiltonian comes from the second-order tail
(i.e., tail-of-tail or tail$^2$) effects and it reads \citep{damour-15,bini-20}
\be
\label{5.5PN}
H_{5.5 \rm PN}^{{\rm tail}^2,\rm nloc} = -\frac{107}{210}\frac{G^2\mathcal{M}^2}{c^6}
\int_{-\infty}^{\infty}\frac{\md\tau}{\tau}[{\cal{G}}^{\rm split}(t,t + \tau)
- {\cal{G}}^{\rm split}(t,t - \tau)],
\ee
where
\be
{\cal{G}}^{\rm split}(t,t') := \frac{G}{5c^5} Q^{(3)}_{ij}(t) Q^{(4)}_{ij}(t').
\ee
The contribution of the 5.5PN Hamiltonian $H_{5.5 \rm PN}^{{\rm tail}^2,\rm nloc}$
to an effective EOB dynamics was computed in \citep{damour-15,bini-20}.

\subsubsection{Results at 6PN order}
\label{subsubsec:results6pn}

The TF approach succeeded with 6PN level to some remarkable extent \citep{bini-20b,bini-20c}
and the EFT approach to some part \citep{blumlein-20c,blumlein-21b}.
Only four numerical coefficients of the EOB representation of the 6PN dynamics are unknown
[two of them are prefactors of terms proportional to $\nu^2$ and $\nu^3$ in the potential $A(u;\nu)$,
the remaining two are prefactors of terms proportional to $\nu^2$
entering the $\bar{D}(u;\nu)$ and $Q(u,p'_r;\nu)$ potentials].
Each of these coefficients is predicted to be the sum of a rational number and a transcendental number.

The nonlocal-in-time 6PN Hamiltonian is known explicitly and reads
\begin{align}
H^{\rm tail, nloc}_{6\rm PN} &= -\frac{G\mathcal{M}}{c^3}{\rm Pf}_{2r_{12}/c}
\int_{-\infty}^{\infty}\frac{\md\tau}{|\tau|}{\cal{F}}_{\rm 2PN}^{\rm split}(t,t + \tau),
\\[1ex]
{\cal{F}}_{\rm 2PN}^{\rm split}(t,t') &= \frac{G}{c^5}\frac{1}{c^4} \left(
\frac{1}{9072} Q^{(5)}_{ijkl}(t) Q^{(5)}_{ijkl}(t')
+ \frac{1}{84} J^{(4)}_{ijk}(t) J^{(4)}_{ijk}(t') \right),
\end{align}
$Q_{ijkl}$ and $J_{ijk}$ denoting mass-type hexadecapole and magnetic-type octupole moments.
The $R$-coefficients, cf.\ \eqref{5PNnloctail2}, of the corresponding local-in-time part
are known, even through all PN orders, see \cite{almeida-21}.
Not known are many other local-in-time expressions.
All these expressions contribute to the four coefficients listed at the beginning of this subsection.

\subsection{The innermost stable circular orbit}
\label{subsec:isco}

The innermost stable circular orbit (ISCO) of a test-body orbiting in the Schwarzschild metric is located at $R=6MG/c^2$, in Schwarzschild coordinates. 
Within a Hamiltonian formalism the calculation of the ISCO for systems made of bodies of comparable masses is rather straightforward.
It is relevant to start with the discussion of dynamics of a two-body system along circular orbits.

The centre-of-mass conservative Hamiltonian $\hat{H}(\vecr,\vecp)$ can be reduced to circular orbits
by setting $p_r = \vecn\cdot\vecp = 0$ and $\vecp^2 = j^2/r^2$, then $\hat{H}=\hat{H}(r,j)$.
Moreover, $\partial\hat{H}(r,j)/\partial r = 0$ along circular orbits, what gives the link between $r$ and $j$, $r=r(j)$.
Finally the energy $\hat{E}^\textrm{circ}$ along circular orbits can be expressed
as a function of $j$ only, $\hat{E}^\textrm{circ}(j)\equiv\hat{H}(r(j),j)$.
The link between the (reduced) centre-of-mass energy $\hat{E}^\textrm{circ}$ and the (reduced) angular momentum $j$
is explicitly known up to the 4PN order. It reads (\citealp{bini-13}, \citealp{damour-14})
\begin{align}
\label{COEj}
\hat{E}^\textrm{circ}&(j;\nu) = -\frac{1}{2j^2} \Bigg\{ 1 + \bigg(\frac{9}{4}+\frac{1}{4}\nu\bigg)\frac1{j^2} 
+ \bigg(\frac{81}{8} - \frac{7}{8}\nu + \frac{1}{8}\nu^2\bigg)\frac{1}{j^4}
\nonumber\\[1ex]&
+ \bigg[ \frac{3861}{64} + \bigg(\frac{41\pi^2}{32}-\frac{8833}{192}\bigg) \nu - \frac{5}{32}\nu^2 + \frac{5}{64}\nu^3 \bigg] \frac{1}{j^6}
\nonumber\\[1ex]&
+ \bigg[ \frac{53703}{128} + \bigg(\frac{6581\pi^2}{512}-\frac{989911}{1920}-\frac{64}{5}\bigg(2\gamma_\mathrm{E}+\ln\frac{16}{j^2}\bigg)\bigg)\nu
\nonumber\\[1ex]&
+ \bigg(\frac{8875}{384}-\frac{41\pi^2}{64}\bigg)\nu^2-\frac{3}{64}\nu^3+\frac{7}{128}\nu^4\bigg] \frac{1}{j^8}
+ \mathcal{O}(j^{-10}) \Bigg\}.
\end{align}
An important observational quantity is the angular frequency of circular orbits, $\omega_\textrm{circ}$.
It can be computed as
\be
\label{COomega}
\omega_\textrm{circ} = \frac{1}{GM} \frac{\md \hat{E}^\textrm{circ}}{\md j}.
\ee
It is convenient to introduce the coordinate-invariant dimensionless variable
(which can also serve as small PN expansion parameter)
\be
\label{COx}
x \equiv \left(\frac{GM\omega_\textrm{circ}}{c^3}\right)^{2/3}.
\ee
Making use of Eqs.\ \eqref{COomega} and \eqref{COx}
it is not difficult to translate the link of Eq.~\eqref{COEj}
into the dependence of the energy $\hat{E}^\textrm{circ}$ on the parameter $x$.
The 4PN-accurate formula reads (\citealp{bini-13}, \citealp{damour-14})
\begin{align}
\label{COEx}
\hat{E}^\textrm{circ}&(x;\nu) = -\frac{x}{2} \Bigg\{
1 - \bigg(\frac{3}{4} + \frac{\nu}{12} \bigg) x
+ \bigg(-\frac{27}{8} + \frac{19\nu}{8} - \frac{\nu^2}{24}\bigg) x^2
\nonumber\\[1ex]&
+ \bigg[ -\frac{675}{64} + \left(\frac{34445}{576}-\frac{205\pi^2}{96}\right)\nu -\frac{155\nu^2}{96} - \frac{35\nu^3}{5184} \bigg] x^3
\nonumber\\[1ex]&
+ \bigg[ -\frac{3969}{128} + \bigg(\frac{9037 \pi ^2}{1536}-\frac{123671}{5760}+\frac{448}{15}\big(2\gamma_\mathrm{E}+\ln(16 x)\big)\bigg)\nu
\nonumber\\[1ex]&
+ \left(\frac{3157\pi^2}{576} -\frac{498449}{3456}\right)\nu^2
+ \frac{301\nu^3}{1728} + \frac{77\nu^4}{31104} \bigg] x^4
+ \mathcal{O}(x^5) \Bigg\}.
\end{align}

In the test-mass limit $\nu\to0$ (describing motion of a test particle
on a circular orbit in the Schwarzschild spacetime)
the link $\hat{E}^\textrm{circ}(x;\nu)$ is exactly known,
\be
\label{COtm1}
\hat{E}^\textrm{circ}(x;0) = \frac{1-2x}{\sqrt{1-3x}} - 1.
\ee
The location $x_{\textrm{ISCO}}=1/6$ of the ISCO in the test-mass limit
corresponds to the minimum of the function $\hat{E}^\textrm{circ}(x;0)$,
i.e.
\be
\label{COtm2}
\frac{\md \hat{E}^\textrm{circ}(x;0)}{\md x}\bigg|_{x=x_{\textrm{ISCO}}} = 0.
\ee
Therefore the most straightforward way of locating the ISCO for $\nu>0$
relies on looking for the minimum of the function $\hat{E}^\textrm{circ}(x;\nu)$,
i.e., for a given value of $\nu$, the location of the ISCO is obtained by (usually numerically)
solving the equation $\md\hat{E}^\textrm{circ}(x;\nu)/(\md x)=0$ \citep{blanchet2-02}.
Equivalently the location of the ISCO can be defined as a solution of
the set of simultaneous equations $\partial\hat{H}(r,j)/\partial r = 0$ and $\partial^2\hat{H}(r,j)/\partial r^2 = 0$.
Both approaches are equivalent only for the \textit{exact} Hamiltonian $\hat{H}(r,j)$,
see however Sect.\ IV~A~2 in \cite{buonanno-03,buonanno-03-erratum} for subtleties related to equivalence of both approaches
when using post-Newtonian-accurate Hamiltonians.
With the aid of the latter method \cite{schafer-93} computed the $n$PN-accurate ISCO of the test mass in the Schwarzschild metric
through 9PN order in three different coordinate systems, obtaining three different results.
Clearly, the application of the first method only results in a $n$PN-accurate ISCO
described by parameters which are coordinate invariant.

Let us consider the 4PN-accurate expansion of the exact test-mass-limit formula \eqref{COtm1},
\be
\label{COtm3}
\hat{E}^\textrm{circ}(x;0) = -\frac{x}{2} \bigg(
1 - \frac{3}{4} x - \frac{27}{8} x^2 -\frac{675}{64} x^3 -\frac{3969}{128}x^4 +\mathcal{O}(x^5) \bigg).
\ee
Let us compute the succesive PN estimations of the exact ISCO frequency parameter $x_{\textrm{ISCO}}=1/6\cong0.166667$ in the test-mass limit,
by solving the equations $\md\hat{E}_\textrm{$n$PN}^\textrm{circ}(x;0)/(\md x)=0$ for $n=1,\ldots,4$,
where the function $\hat{E}_\textrm{$n$PN}^\textrm{circ}(x;0)$ is defined
as the $\mathcal{O}(x^{n+1})$-accurate truncation of the right-hand-side of Eq.~\eqref{COtm3}.
They read: 0.666667 (1PN), 0.248807 (2PN), 0.195941 (3PN), 0.179467 (4PN).
One sees that the 4PN prediction for the ISCO frequency parameter is still $\sim$8\% larger than the exact result.
This suggests that the straightforward Taylor approximants of the energy function $\hat{E}^\textrm{circ}(x;\nu)$
do not converge fast enough to determine satisfactorily the frequency parameter of the ISCO also in $\nu>0$ case,
at least for sufficiently small values of $\nu$. The extrapolation of this statement for larger $\nu$
is supported by the values of the ISCO locations in the equal-mass case ($\nu=1/4$),
obtained by solving the equations $\md\hat{E}_\textrm{nPN}^\textrm{circ}(x;1/4)/(\md x)=0$ for $n=1,\ldots,4$,
where the function $\hat{E}_\textrm{$n$PN}^\textrm{circ}(x;\nu)$ is now defined
as the $\mathcal{O}(x^{n+1})$-accurate truncation of the right-hand-side of Eq.~\eqref{COEx}.
For the approximations from 1PN up to 4PN the ISCO locations read
(\citealp{damour2-00}, \citealp{blanchet2-02}, \citealp{jaranowski-13}):
0.648649 (1PN), 0.265832 (2PN), 0.254954 (3PN), and 0.236599 (4PN)\footnote{The 4PN value of the ISCO frequency parameter
given here, 0.236599, is slightly different from the value 0.236597 published in \cite{jaranowski-13}.
The reason is that in \cite{jaranowski-13} the only then known approximate value 153.8803 of the linear-in-$\nu$ coefficient
in the 4PN-order term in Eq.~\eqref{COEx} was used, whereas the numerical exact value of this coefficient reads $153.8837968\cdots$.
From the same reason the 4PN ISCO frequency parameter determined by the $j$-method described below in this section, is equal 0.242967,
whereas the value published in \cite{jaranowski-13} reads 0.247515.}.

To overcome the problem of the slow convergence of PN expansions
several new methods of determination of the ISCO for comparable-mass binaries
were devised by \cite{damour2-00}.
They use different ``resummation'' techniques and are based on the consideration of gauge-invariant functions.
One of the methods, called the ``$j$-method'' by \cite{damour2-00},
employs the invariant function linking the angular momentum
and the angular frequency along circular orbits and uses Pad\'e approximants.
The ISCO is defined in this method as the minimum, for the fixed value of $\nu$,
of the function $j^2(x;\nu)$, where $j$ is the reduced angular momentum [introduced in Eq.~\eqref{EJreduced}].
The function $j^2(x;\nu)$ is known in the test-mass limit,
\be
j^2(x;0) = \frac{1}{x(1-3x)},
\ee
and its minimum coincides with the exact ``location" $x_{\textrm{ISCO}}=1/6$ of the test-mass ISCO.
The form of this function suggests to use Pad\'e approximants instead of direct Taylor expansions.
It also suggests to require that all used approximants have a pole for some $x_\textrm{pole}$,
which is related with the test-mass ``light-ring'' orbit occurring for $x_\textrm{lr}=1/3$
in the sense that $x_\text{pole}(\nu)\to1/3$ when $\nu\to0$.
The 4PN-accurate function $j^2(x;\nu)$ has the symbolic structure
$(1/x)(1+x+\ldots+x^4+x^4\ln x)$.
In the $j$-method the Taylor expansion at the 1PN level
with symbolic form $1+x$ is replaced by Pad\'e approximant of type (0,1),
at the 2PN level $1+x+x^2$ is replaced by (1,1) approximant,
at the 3PN level $1+x+x^2+x^3$ is replaced by (2,1) approximant,
and finally at the 4PN level $1+x+x^2+x^3+x^4$ is replaced by (3,1) Pad\'e approximant
[the explicit form of the (0,1), (1,1), and (2,1) approximants
can be found in Eqs.\ (4.16) of \citealp{damour2-00}].
At all PN levels the test-mass result is recovered exactly
and \cite{jaranowski-13} showed that the ISCO locations
resulting from 3PN-accurate and 4PN-accurate calculations almost coincide
for all values of $\nu$, $0\le\nu\le\frac{1}{4}$.
The ISCO locations in the equal-mass case $\nu=1/4$
for the approximations from 1PN up to 4PN are as follows \citep{jaranowski-13}:
0.162162 (1PN), 0.185351 (2PN), 0.244276 (3PN), 0.242967 (4PN).

\subsection{Dissipative Hamiltonians}
\label{subsec:PNdissipation}

To discuss dissipative Hamiltonians it is convenient to use the toy model from Sect.\ \ref{subsec:Routhian}
with the Routhian $R(q,p;\xi,\dot\xi)$ and its corresponding Hamiltonian $H(q,p;\xi,\pi)=R+\pi\dot\xi$.
The Hamilton equations of motion for the $(q,p)$ variables read
\be
\dot{p} = -\frac{\partial H}{\partial q} = -\frac{\partial R}{\partial q},
\quad \dot{q} = \frac{\partial H}{\partial p} = \frac{\partial R}{\partial p},
\ee
and the Euler-Lagrange equation for the $\xi$ variable is
\be
\frac{\partial R}{\partial \xi} - \frac{\md}{\md t}\frac{\partial R}{\partial \dot\xi} = 0.
\ee
Alternatively, the Hamilton equations of motion for the $(\xi,\pi)$ variables can be used.
Solutions of the Euler-Lagrange equation are functions $\xi=\xi(q,p)$.
Under those solutions, the Hamilton equations of motion for the $(q,p)$ variables become
\be
\dot{p} = -\frac{\partial R}{\partial q}\bigg|_{\xi=\xi(q,p)},
\quad \dot{q} = \frac{\partial R}{\partial p}\bigg|_{\xi=\xi(q,p)}.
\ee
These autonomous equations in the $(q,p)$ variables contain the full conservative and dissipative content of the $(q,p)$ dynamics.
The time-symmetric part of $R$ yields the conservative equations of motion
and the time-antisymmetric part the dissipative ones.
The conservative equations of motion agree with the Fokker-type ones showing the same boundary conditions for the $(\xi,\dot\xi)$ variables.
When going from the $(\xi,\dot\xi)$ variables to the field variables $h^{\rm TT}$ and $\dot{h}^{\rm TT}$,
those time-symmetric boundary conditions mean as much incoming as outgoing radiation.

To describe astrophysical systems one should use the physical boundary conditions of no incoming radiation and past stationarity.
Clearly, radiative dissipation happens now and the time-symmetric part of the whole dynamics makes the conservative part.
In linear theories the conservative part just results from the symmetric Green function $G_\text{s}$,
whereas the dissipative one comes from the antisymmetric Green function $G_\text{a}$,
which is a homogeneous solution of the wave equation.
They both together combine to the retarded Green function $G_{\rm ret}=G_{\rm s}+G_{\rm a}$,
with $G_{\rm s} = (1/2)(G_{\rm ret} + G_{\rm adv})$ and $G_{\rm a} = (1/2)(G_{\rm ret} - G_{\rm adv})$,
where $G_{\rm adv}$ denotes the advanced Green function.
In non-linear theories time-symmetric effects can also result from homogeneous solutions,
e.g., the tail contributions.

For a binary system, the leading-order direct and tail radiation reaction enter the Routhian in the form
\be
R^{\text{rr}}(\xa,\pa,t) = -\frac{1}{2}\,h^{\text{TT\,rr}}_{ij}(t)\,
\left(\frac{p_{1i}p_{1j}}{m_1} + \frac{p_{2i}p_{2j}}{m_2} - \frac{Gm_1m_2}{r_{12}} n_{12}^i n_{12}^j\right),
\ee
where $h^{\text{TT\,rr}}_{ij}(t)$ decomposes into a direct radiation-reaction term and a tail one \citep{damour-16},
\be
h^{\text{TT\,rr}}_{ij}(t) = -\frac{4G}{5c^5} \left({I}^{(3)}_{\!ij}(t)
+ \frac{4GM}{c^3} \int^\infty_0\md\tau\,{\rm ln}\left(\frac{c\tau}{2s_{\rm phys}}\right){I}^{(5)}_{\!ij}(t-\tau)\right).
\ee
The last term on the right side results in a Routhian, which reproduces the corresponding tail effects
in \cite{blanchet2-93} and \cite{galley-16}.

The conservative (time-symmetric) part in $h^{\text{TT\,rr}}_{ij}$ reads
\be
h^{\text{TT\,rr\,con}}_{ij}(t) = -\frac{8G^2M}{5c^8} {\rm Pf}_{2s_{\rm phys}/c}
\int^\infty_{-\infty} \frac{\md t'}{|t-t'|}\,{I}^{(4)}_{\!ij}(t'),
\ee
and the dissipative (time-antisymmetric) one equals
\be
h^{\text{TT\,rr\,dis}}_{ij}(t) = -\frac{4G}{5c^5} {I}^{(3)}_{\!ij}(t)
- \frac{8G^2M}{5c^8} {\rm Pf}_{2s_{\rm phys}/c} \int^\infty_{-\infty}\frac{\md t'}{t-t'}\,{I}^{(4)}_{\!ij}(t'),
\ee
where use has been made of the relations
\begin{align}
{\rm Pf}_{\tau_0} \int^\infty_{-\infty} \frac{\md t' f(t')}{|t-t'|} &=
\int^{\infty}_0 \md\tau ~ {\rm ln}\left(\frac{\tau}{\tau_0}\right)[f^{(1)}(t-\tau) - f^{(1)}(t+\tau)],
\\[1ex]
{\rm Pf}_{\tau_0} \int^\infty_{-\infty} \frac{\md t' f(t')}{t-t'} &=
\int^{\infty}_0 \md\tau ~{\rm ln}\left(\frac{\tau}{\tau_0}\right)[f^{(1)}(t-\tau) + f^{(1)}(t+\tau)].
\end{align}

The leading-order 2.5PN dissipative binary orbital dynamics is described by the non-autonomous Hamiltonian \citep{schaefer-95},
\be
\label{H2.5PN}
H_{\textrm{2.5PN}}(\xa,\pa,t) = \frac{2G}{5c^5}\,\dddot{I}_{\!ij}\big(x'^k_a(t)\big)\,
\left(\frac{p_{1i}p_{1j}}{m_1} + \frac{p_{2i}p_{2j}}{m_2} - \frac{Gm_1m_2}{r_{12}} n_{12}^i n_{12}^j\right),
\ee
where $I_{ij}$ is the Newtonian mass-quadrupole tensor,
\be
I_{ij}\big(x'^k_a(t)\big) \equiv \sum_a m_a \big(x'^i_a(t)x'^j_a(t) - \frac{1}{3}{\bf x}'^2_a(t)\delta_{ij}\big).
\ee
Only after the Hamilton equations of motion have been obtained
the primed position and momentum variables coming from $\dddot{I}_{\!ij}$ are allowed to be identified
with the unprimed position and momentum variables, also see \cite{galley-13}.
Generally, the treatment of dissipation with Hamiltonians or Lagrangians necessarily needs doubling of variables \citep{bateman-31}.
In quantum mechanics, that treatment was introduced by \cite{schwinger-61} and \cite{keldysh-65}.
In the EFT approach as well a doubling of variables is needed
if one wants to treat dissipative systems in a full-fledged manner at the action level
(see, e.g., \citealp{galley-12} and \citealp{galley-16}).
However, one should keep in mind that in quantum mechanics damping can also be treated without doubling of variables
by making use of the fact that the Feynman Green function $G_{\rm F}$,
the analogue of the retarded Green function of classical physics, decomposes into real and imaginary parts,
$G_{\rm F} = G_{\rm s} + (i/2)G^{(1)}$, where both $G_{\rm s}$ from above and $G^{(1)}$, Hadamard's elementary function,
are symmetric Green functions, $G^{(1)}$ solving homogeneous wave equation as $G_a$ does.
The imaginary part in e.g.\ the Eq.\ (8.7.57) in the book by \cite{brown-92}
yields nothing but the dipole radiation loss formula and this without any doubling of variables
(also see Sect.\ 9-4 in \citealp{feynman-65}).
Note, however, that the statement concerning the Feynman propagator
applies only to the calculation of the energy flux, not to that of the gravitational-wave amplitude.

Applications of the 2.5PN Hamiltonian can be found in, e.g.,
\cite{kokkotas-95}, \cite{ruffert-96}, \cite{buonanno-99}, \cite{gopakumar-08},
where in \cite{gopakumar-08} a transformation to the Burke-Thorne gauge (coordinate conditions) is performed.
More information on the 2.5PN dissipation can be found in \cite{damour2-87}.
The 3.5PN Hamiltonian for many point-mass systems is known too,
it is displayed in Appendix \ref{app:dissH} (\citealp{jaranowski-97}, \citealp{koenigsdoerffer-03}).
Recently the 4.5PN radiation-reaction acceleration for nonspinning binary
was computed using the EFT approach \citep{leibovich-23}.
Regarding gravitational spin interaction, see the next section,
also for this case radiation reaction Hamiltonians have been derived
through leading order spin-orbit and spin-spin couplings (\citealp{steinhoff3-10}, \citealp{wang-11}).
Recent related developments within the EFT formalism are found in \cite{maia1-17,maia2-17}.

Let us mention that the already cited article \cite{galley-16} contains two interesting results
improving upon and correcting an earlier article by \cite{foffa2-13}:
on the one hand it confirms the conservative part of the tail action,
particularly the additional rational constant 41/30 which corresponds to the famous 5/6 in the Lamb shift (see, e.g., \citealp{brown-00}),
and on the other side it correctly delivers the dissipative part of the tail interaction.
It is worth noting that in the both articles the involved calculations were performed in harmonic coordinates.

\section{Generalized ADM formalism for spinning objects}
\label{sec:spinADM}

In this section we review the relatively recent generalization of ADM formalism
describing dynamics of systems made of spinning point masses or, more precisely, pole-dipole particles.
We start from reviewing the generalization which is of fully reduced form
(i.e., without unresolved constraints, spin supplementary and coordinate conditions)
and which is valid to linear order in spin variables
(our presentation of linear-in-spins dynamics closely follows that of \cite{steinhoff2-09}).

\subsection{Dynamics linear in spins}
\label{subsec:spinlinearADM}

In this section Latin indices from the middle of the alphabet $i$, $j$, $k$, $\ldots$ are running through $\{1,2,3\}$.
We utilize three different reference frames here, denoted by different indices.
Greek indices refer to the \emph{coordinate frame} $(x^\mu)$ and have the values $\mu=0,1,2,3$.
Lower case Latin indices from the beginning of the alphabet refer to the \emph{local Lorentz frame}
with its associated tetrad fields $\big(e_a^\mu(x^\nu)\big)$
($e_a^\mu$ denotes thus the $\mu$ coordinate-frame component of the tetrad vector of label $a$),
while upper case ones denote the so-called \emph{body-fixed Lorentz frame}
with its associated ``tetrad'' $\big(\Lambda^{~a}_A(z^\mu)\big)$,
where $(z^\mu)$ denotes coordinate-frame components of the body's position
(so $\Lambda^{~a}_A$ is the $a$ local-Lorentz-frame component of the tetrad vector of label $A$).
The values of these Lorentz indices are marked by round and square brackets
as $a=(0),(i)$ and $A=[0],[i]$, respectively, e.g., $A=[0],[1],[2],[3]$.
The basics of the tetrad formalism in GR can be found in, e.g., Sect.\ 12.5 of \cite{weinberg-72}.

In GR, the coupling of a spinning object to a gravitational field,
in terms of a Lagrangian density, reads
\be
\label{spin1}
\mathcal{L}_M = \int \md\tau \left[ \left( p_{\mu}
- \frac{1}{2} S_{ab} \,\omega_{\mu}^{~ab}\right) \frac{\md z^{\mu}}{\md\tau}
+ \frac{1}{2} S_{ab} \frac{\delta \theta^{ab}}{\md\tau} \right] \delta^{(4)}(x^{\nu}-z^{\nu}(\tau)).
\ee
The linear momentum variable is $p_{\mu}$
and the spin tensor is denoted by $S_{ab}$.
The object's affine time variable is $\tau$
and $\delta^{(4)}(x^{\nu}-z^{\nu}(\tau))$ is the 4-dimensional Dirac delta function
(from now on we will abbreviate it to $\delta^{(4)}$).
The angle variables are represented by some Lorentz matrix satisfying
$\Lambda^{Aa} \Lambda^{Bb} \eta_{AB} = \eta^{ab}$
or $\Lambda_{Aa} \Lambda_{Bb} \eta^{ab} = \eta_{AB}$, where $\eta_{AB} = \mbox{diag}(-1,1,1,1) = \eta^{ab}$,
which must be respected upon infinitesimal Lorentz transformations (see \citealp{hanson-74}),
so $\delta\theta^{ab}\equiv\Lambda_{C}^{~a}\md\Lambda^{Cb}=-\delta\theta^{ba}$.
The Ricci rotation coefficients $\omega_{\mu}^{~ab}$ are given by
$\omega_{\mu\alpha\beta}=e_{a\alpha}e_{b\beta}\omega_{\mu}^{~ a b}=-\Gamma_{\beta\alpha\mu}^{(4)} + e_{\alpha,\mu}^c e_{c\beta}$,
with $\Gamma_{\beta\alpha\mu}^{(4)} = \frac{1}{2} (g_{\beta\alpha,\mu} + g_{\beta\mu,\alpha} - g_{\alpha\mu,\beta})$
as the 4-dimensional Christoffel symbols of the first kind with $g_{\mu\nu} = e_{a\mu} e_{b\nu} \eta^{ab}$
the 4-dimensional metric. As in \cite{hanson-74}, the matrix $\Lambda^{Ca}$ can be subjected
to right (or left) Lorentz transformations, which correspond to
transformations of the local Lorentz reference frame (or the body-fixed frame, respectively).
In the action \eqref{spin1} only a minimal coupling between spin variables and gravitational field is employed;
for more general (than minimal) couplings, the reader is referred to \cite{bailey-75}.

The matter constraints are given by, also in terms of a Lagrangian density,
\be
\mathcal{L}_C = \int\md\tau \left[ \lambda_1^a p^b S_{ab} + \lambda_{2[i]} \Lambda^{[i] a} p_a
- \frac{\lambda_3}{2} (p^2 + m^2c^2) \right] \delta^{(4)},
\ee
where $m$ is the constant mass of the object, $p^2\equiv p_{\mu}p^{\mu}$,
and $\lambda_1^a$, $\lambda_{2[i]}$, $\lambda_3$ are the Lagrange multipliers.
The constraint
\be
\label{covSSC}
p^b S_{ab} = 0
\ee
is called the \emph{spin supplementary condition} (SSC),
it states that in the rest frame the spin tensor contains the 3-dimensional spin $S_{(i)(j)}$ only
(i.e., the mass-dipole part $S_{(0)(i)}$ vanishes).\footnote{For more details about SSCs, see Sect.~\ref{subsec:Poincare} of our review.}
The conjugate constraint $\Lambda^{[i] a} p_a = 0$ ensures
that $\Lambda^{C a}$ is a pure 3-dimensional rotation matrix in the rest frame
(no Lorentz boosts), see \cite{hanson-74}.
Finally, the gravitational part is given by the usual Einstein-Hilbert Lagrangian density
\be
\mathcal{L}_G = \frac{c^4}{16\pi G} \sqrt{-g} R^{(4)},
\ee
where $g$ is the determinant of the 4-dimensional metric and $R^{(4)}$
is the 4-dimensional Ricci scalar. Using a second-order form of the
gravitational action, i.e., not varying the connection
independently, ensures that the torsion tensor vanishes, see, e.g.,
\cite{nelson-78}. The complete Lagrangian density is the sum
\be
\mathcal{L} = \mathcal{L}_G + \mathcal{L}_M + \mathcal{L}_C.
\ee
We assume space-asymptotic flatness as a boundary condition of the spacetime.
The total action is given in a second-order form,
where the Ricci rotation coefficients are not independent field degrees of freedom
and where no torsion of spacetime shows up. It reads
\be
W[e_{a\mu},z^{\mu},p_{\mu},\Lambda^{Ca},S_{ab},\lambda_1^a,\lambda_{2[i]},\lambda_3]
= \int\md t\,\md^3x\, \mathcal{L},
\ee
and must be varied with respect to the tetrad field $e_{a\mu}$,
the Lagrange multipliers $\lambda_1^a$, $\lambda_{2[i]}$, $\lambda_3$,
position $z^{\mu}$ and linear momentum $p_{\mu}$ of the object,
as well as with respect to angle-type variables $\Lambda^{Ca}$
and spin tensor $S_{ab}$ associated with the object.

Variation of the action $\delta W=0$ leads to the equations of motion for the matter variables
(here $\md$ and $\mD$ denote ordinary and covariant total derivatives, respectively\footnote{
Covariant derivative of an object with Greek index means application of the Christoffel symbols
(apart from $z^\mu$ which are four scalars),
in case of a small Latin index the application of the Ricci rotation coeffients,
and in case of a capital Latin index just the ordinary derivative of a scalar.})
\begin{align}
\label{EOM1} 
\frac{\mD S_{ab}}{\mD\tau} &= 0,
\quad
\frac{\mD \Lambda^{Ca}}{\mD\tau} = 0,
\quad
u^{\mu} \equiv \frac{\md z^\mu}{\md\tau} = \lambda_3 p^{\mu},
\\[2ex]
\label{EOM2}
\frac{\mD p_{\mu}}{\mD \tau} &= -\frac{1}{2} R_{\mu\rho ab}^{(4)} u^{\rho} S^{ab},
\end{align}
as well as to the usual Einstein equations with the stress-energy tensor
(cf.\ \citealp{tulczyjew-57} and Sect.~12.5 in \citealp{weinberg-72}\footnote{Especially Eq.\ (12.5.35) there.})
\begin{align}
\label{SET}
T^{\mu\nu} &= \frac{e^{\mu}_a}{\sqrt{-g}} \frac{\delta ( \mathcal{L}_M + \mathcal{L}_C )}{\delta e_{a \nu}}
\nonumber\\[1ex]
&= \int \md \tau \left[
\lambda_3 p^{\mu} p^{\nu} \frac{\delta^{(4)}}{\sqrt{-g}}
+ \bigg( u^{(\mu} S^{\nu)\alpha} \frac{\delta^{(4)}}{\sqrt{-g}} \bigg)_{||\alpha}
\right], 
\end{align}
where $R_{\mu\rho ab}^{(4)}$ is the 4-dimensional Riemann tensor in mixed indices,
$_{||\alpha}$ denotes the 4-dimensional covariant derivative.
Here it was already used that preservation of the constraints
in time requires $\lambda_1^a$ to be proportional to $p^a$
and $\lambda_{2[i]}$ to be zero, so that $\lambda_1^a$ and $\lambda_{2[i]}$
drop out of the matter equations of motion and the stress-energy tensor.
The Lagrange multiplier $\lambda_3 = \lambda_3(\tau)$
represents the reparametrization invariance of the action (notice $\lambda_3=\sqrt{-u^2}/m$).
Further, an antisymmetric part of the stress-energy tensor vanishes,
\be
\frac{1}{2} \int\md\tau \left( S^{\mu\nu} u^{\rho} \frac{\delta^{(4)}}{\sqrt{-g}} \right)_{||\rho}
= \frac{1}{2} \int\md\tau \frac{\mD S^{\mu\nu}}{\mD \tau} \frac{\delta^{(4)}}{\sqrt{-g}} = 0,
\ee
and ${T^{\mu\nu}}_{||\nu}=0$ holds by virtue of the matter equations of motion.
Obviously, the spin length $s$ as defined by $2s^2\equiv S_{ab}S^{ab}$ is conserved.

A fully reduced action is obtained by the elimination of all constraints and gauge degrees of freedom.
However, after that the action has still to be transformed into canonical form by certain variable transformations.
To perform this reduction we employ 3+1 splitting of spacetime by spacelike hypersurfaces $t=\textrm{const}$.
The timelike unit covector orthogonal to these hypersurfaces reads $n_{\mu}=(-N,0,0,0)$ or $n^{\mu}=(1,-N^i)/N$.
The three matter constraints can then be solved
in terms of $p_i$, $S_{ij}$, and $\Lambda^{[i](k)}$ as
\begin{align}
np & \equiv n^{\mu} p_{\mu} = -\sqrt{m^2c^2 + \gamma^{ij} p_{i} p_{j}},
\\[1ex]
nS_{i} & \equiv  n^{\mu} S_{\mu i} = \frac{p_{k} \gamma^{kj} S_{ji}}{np} = \gamma_{ij} nS^j,
\\[1ex]
\Lambda^{[j](0)} &= \Lambda^{[j](i)} \frac{p_{(i)}}{p^{(0)}},
\quad
\Lambda^{[0]a} = -\frac{p^a}{mc}.
\end{align}
We take $\mathcal{L}_C=0$ from now on.
A split of the Ricci rotation coefficients results in
\begin{align}
\omega_{kij} &= -\Gamma_{jik} + e_{i,k}^a e_{aj},
\\[1ex]
\label{ric2}
n^{\mu} \omega_{k \mu i} &= K_{ki} - g_{ij} \frac{N^j_{,k}}{N}
+ \frac{e_{ai}}{N} (e^a_{0,k} - e^a_{l,k} N^l),
\\[1ex]
\omega_{0ij} &= N K_{ij} - N_{j;i} + e_{i,0}^a e_{aj},
\\[1ex]
\label{ric4}
n^{\mu} \omega_{0 \mu i} &= K_{ij} N^j - N_{;i} - \gamma_{ij} \frac{N^j_{,0}}{N}
+ \frac{e_{ai}}{N} (e^a_{0,0} - e^a_{l,0} N^l),
\end{align}
where $_{;i}$ denotes the 3-dimensional covariant derivative,
$\Gamma_{jik}$ the 3-dimensional Christoffel symbols,
and the extrinsic curvature $K_{ij}$ is given by $2NK_{ij}=-\gamma_{ij,0}+2N_{(i;j)}$,
where $_{(\cdots)}$ denotes symmetrization.

It is convenient to employ here the time gauge
(see \citealp{schwinger1-63} and also \citealp{dirac-62}, \citealp{kibble-63}, \citealp{nelson-78}),
\be
\label{time-gauge}
e^{\mu}_{(0)} = n^{\mu}.
\ee
Then lapse and shift turn into Lagrange multipliers in the matter action,
like in the ADM formalism for nonspinning matter points.
The condition \eqref{time-gauge} leads to the following relations:
\begin{align}
e_{i}^{(0)} &= 0 = e_{(i)}^0,
\quad
e^{(0)}_0 = N = 1/e_{(0)}^0,
\\[1ex]
N^i &= -N e^{i}_{(0)},
\quad
e^{(i)}_0 = N^j e^{(i)}_j,
\\[1ex]
\gamma_{ij} &= e_i^{(m)} e_{(m)j},
\quad
\gamma^{ij} = e^i_{(m)} e^{(m)j},
\end{align}
which effectively reduce the tetrad $e^{a\mu}$ to a triad $e^{(i)j}$.

The matter part of the Lagrangian density, after making use of the covariant SSC \eqref{covSSC}, turns into
\be
\mathcal{L}_M = \mathcal{L}_{MK} + \mathcal{L}_{MC} + \mathcal{L}_{GK} + (\mbox{td}),
\ee
where $(\mbox{td})$ denotes an irrelevant total divergence.
After fixing the yet arbitrary parameter $\tau$ by choosing $\tau=z^0=ct$, where $t$ is the time coordinate,
the terms attributed to the kinetic matter part are given by
\begin{align}
\mathcal{L}_{MK} &= \bigg[ p_{i} + K_{ij} nS^j + A^{kl} e_{(j)k} e_{l,i}^{(j)}
-\bigg( \frac{1}{2} S_{kj} + \frac{p_{(k} nS_{j)}}{np} \bigg) \Gamma^{kj}_{~~i} \bigg] \dot{z}^{i} \delta
+ \frac{nS^i}{2 np} \dot{p}_i \delta
\nonumber\\[1ex]&\quad
+\bigg[ S_{(i)(j)} + \frac{nS_{(i)} p_{(j)} -  nS_{(j)} p_{(i)}}{np} \bigg]
\frac{\Lambda_{[k]}^{(i)} \dot{\Lambda}^{[k](j)}}{2} \delta,
\end{align}
where $\delta\equiv\delta(x^i - z^i(t))$ and $A^{ij}$ is defined by
\be
\gamma_{ik} \gamma_{jl} A^{kl} = \frac{1}{2} S_{ij} + \frac{nS_i p_j}{2 np}.
\ee
The matter parts of the gravitational constraints result from
\be
\mathcal{L}_{MC} = -N\mathcal{H}^{\rm matter} + N^i \mathcal{H}^{\rm matter}_i,
\ee
where the densities $\mathcal{H}^{\rm matter}$ and $\mathcal{H}^{\rm matter}_i$
are computed from Eqs.\ \eqref{mHden1}--\eqref{mMden1} and \eqref{SET}.
After employing the covariant SSC one gets \citep{steinhoff2-08}
\begin{align}
\label{spinHden1}
\mathcal{H}^{\rm matter} &= \sqrt{\gamma}T_{\mu\nu} n^{\mu}n^{\nu}
= -np \delta - K^{ij} \frac{p_i nS_j}{np} \delta - ( nS^k \delta )_{;k},
\\[1ex]
\label{spinHden2}
\mathcal{H}^{\rm matter}_i &= -\sqrt{\gamma}T_{i \nu}n^{\nu}
= (p_i + K_{ij} nS^j ) \delta + \bigg( \frac{1}{2} \gamma^{mk} S_{ik} \delta
+ \delta_i^{(k} \gamma^{l)m} \frac{p_k nS_l}{np} \delta \bigg)_{;m}.
\end{align}
Further, some terms attributed to the kinetic part of the gravitational field appear as
\be
\mathcal{L}_{GK} = A^{ij} e_{(k)i} \dot{e}_{j}^{(k)} \delta.
\ee

Now we proceed to Newton-Wigner (NW) variables $\hat{z}^i$, $P_i$, $\hat{S}_{(i)(j)}$, and $\hat{\Lambda}^{[i](j)}$,
which turn the kinetic matter part $\mathcal{L}_{MK}$ into canonical form.
The variable transformations read
\begin{align}
z^i &= \hat{z}^i - \frac{nS^i}{mc - np},
\quad
nS_i = -\frac{p_k\gamma^{kj}\hat{S}_{ji}}{mc},
\\[1ex]
S_{ij} &= \hat{S}_{ij} - \frac{p_i nS_{j}}{mc-np} + \frac{p_j nS_i}{mc - np},
\\[1ex]
\Lambda^{[i](j)} &= \hat{\Lambda}^{[i](k)} \bigg( \delta_{kj} + \frac{p_{(k)}p^{(j)}}{mc(mc-np)} \bigg),
\\[1ex]
\label{NWPi}
P_i &= p_i + K_{ij} nS^j + \hat{A}^{kl} e_{(j)k} e_{l,i}^{(j)}
- \bigg( \frac{1}{2} S_{kj} + \frac{p_{(k} nS_{j)}}{np} \bigg) \Gamma^{kj}_{~~i},
\end{align}
where $\hat{A}^{ij}$ is given by
\be
\gamma_{ik} \gamma_{jl} \hat{A}^{kl} = \frac{1}{2} \hat{S}_{ij} + \frac{mc p_{(i} nS_{j)}}{np (mc-np)}.
\ee
The NW variables have the important properties $\hat{S}_{(i)(j)} \hat{S}_{(i)(j)}=2s^2=\mbox{const}$
and $\hat{\Lambda}_{[k]}^{(i)}\hat{\Lambda}^{[k](j)} = \delta_{ij}$,
which implies that $\delta\hat{\theta}^{(i)(j)}\equiv\hat{\Lambda}_{[k]}^{(i)}\md\hat{\Lambda}^{[k](j)}$ is antisymmetric.
The redefinitions of position, spin tensor, and angle-type variables are actually quite natural generalizations
of their Minkowski space versions to curved spacetime, cf.\ \cite{hanson-74} and \cite{fleming-65}.
However, there is no difference between linear momentum $p_i$ and canonical momentum $P_i$ in the Minkowski case.
In these NW variables, one has
\be
\mathcal{L}_{GK} + \mathcal{L}_{MK}
= \hat{\mathcal{L}}_{GK} + \hat{\mathcal{L}}_{MK} + (\mbox{td}),
\ee
with [from now on $\delta = \delta(x^i - \hat{z}^i(t))$]
\begin{align}
\hat{\mathcal{L}}_{MK} &= P_i \dot{\hat{z}}^i \delta + \frac{1}{2} \hat{S}_{(i)(j)} \dot{\hat{\theta}}^{(i)(j)} \delta,
\\[1ex]
\hat{\mathcal{L}}_{GK} &= \hat{A}^{ij} e_{(k)i} e_{j,0}^{(k)} \delta.
\end{align}
Notice that all $\dot{p}_i$ terms in the action have been canceled by the redefinition of the position
and also all $K_{ij}$ terms were eliminated from $\mathcal{L}_{MC}$ and $\mathcal{L}_{MK}$ by the redefinition of the linear momentum.
If the terms explicitly depending on the triad $e^{(i)}_j$ are neglected,
the known source terms of Hamilton and momentum constraints in canonical variables are obtained
[cf. Eqs.\ (4.23) and (4.25) in \cite{steinhoff2-08}].

The final step goes with the ADM action functional of the gravitational field
(\citealp{arnowitt-62}, \citealp{dewitt-67}, \citealp{regge-74}),
but in tetrad form as derived by \cite{deser-76}.
The canonical momentum conjugate to $e_{(k)j}$ is given by
\be
\bar{\pi}^{(k)j} = \frac{8\pi G}{c^3}\frac{\partial \mathcal{L}}{\partial e_{(k)j,0}}
= e_i^{(k)} \pi^{ij} + e_i^{(k)} \frac{8\pi G}{c^3}\hat{A}^{ij} \delta,
\ee
where the momentum $\pi^{ij}$ is given by
\be
\pi^{ij} = \sqrt{\gamma} (\gamma^{ij}\gamma^{kl} - \gamma^{ik}\gamma^{jl})K_{kl}.
\ee
Legendre transformation leads to
\be
\hat{\mathcal{L}}_{GK} + \mathcal{L}_G
= \frac{c^3}{8\pi G} \bar{\pi}^{(k)j} e_{(k)j,0} - \frac{c^4}{16\pi G} \mathcal{E}_{i,i} + \mathcal{L}_{GC} + (\mbox{td}).
\ee
In asymptotically flat spacetimes the quantity $\mathcal{E}_i$ is given by [cf.\ Eq.~\eqref{ADMh}]
\be
\mathcal{E}_i = \gamma_{ij,j} - \gamma_{jj,i}.
\ee
The total energy then reads
\be
E = \frac{c^4}{16\pi G}\oint \md^2s_i\,\mathcal{E}_i.
\ee

The constraint part of the gravitational Lagrangian density takes the form
\be
\mathcal{L}_{GC} = - N \mathcal{H}^{\rm field} +  N^i \mathcal{H}^{\rm field}_i,
\ee
with
\begin{align}
\mathcal{H}^{\rm field} &= - \frac{c^4}{16\pi G\sqrt{\gamma}}
\left[ \gamma R + \frac{1}{2} \left(\gamma_{ij} \pi^{ij} \right)^2 - \gamma_{ij} \gamma_{k l} \pi^{ik} \pi^{jl}\right] ,
\\[1ex]
\mathcal{H}^{\rm field}_i &= \frac{c^3}{8\pi G} \gamma_{ij} \pi^{jk}_{~~ ; k} \,,
\end{align}
where $R$ is the 3-dimensional Ricci scalar.
Due to the symmetry of $\pi^{ij}$, not all components of $\bar{\pi}^{(k)j}$ are independent variables
(i.e., the Legendre map is not invertible), leading to the additional constraint
($[\ldots]$ denotes anti-symmetrization)
\be
\bar{\pi}^{[ij]} = \frac{8\pi G}{c^3} \hat{A}^{[ij]} \delta.
\ee
This constraint will be eliminated by going to the spatial symmetric gauge
(for the frame $e_{(i)j}$)
\be
e_{(i)j} = e_{ij} = e_{ji},
\quad
e^{(i)j} = e^{ij} = e^{ji}.
\ee
Then the triad is fixed as the matrix square-root of the 3-dimensional metric, $e_{ij} e_{jk} = \gamma_{ik}$,
or, in matrix notation,
\be
(e_{ij}) = \sqrt{(\gamma_{ij})}.
\ee
Therefore, we can define a quantity $B^{kl}_{ij}$ as
\be
e_{k[i} e_{j]k,\mu} = B^{kl}_{ij} \gamma_{kl,\mu},
\ee
or, in explicit form,
\be
2 B^{kl}_{ij} = e_{mi} \frac{\partial e_{mj}}{\partial g_{kl}} - e_{mj} \frac{\partial e_{mi}}{\partial g_{kl}}.
\ee
This expression may be evaluated perturbatively, cf.\ \cite{steinhoff2-08}.
One also has $B^{kl}_{ij} \delta_{kl} = 0$. Furthermore,
\be
\label{eliminatee}
e_{(k)i} e_{j,\mu}^{(k)} = B^{kl}_{ij} \gamma_{kl,\mu} + \frac{1}{2}
\gamma_{ij,\mu},
\ee
which yields
\be
\bar{\pi}^{(k)j} e_{(k)j,0} = \frac{1}{2} \pi_{\rm can}^{ij} \gamma_{ij,0},
\ee
with the new canonical field momentum
\be
\pi_{\rm can}^{ij} = \pi^{ij}
+ \frac{8\pi G}{c^3} \hat{A}^{(ij)} \delta
+ \frac{16\pi G}{c^3} B^{ij}_{kl} \hat{A}^{[kl]} \delta.
\ee

The gravitational constraints arising from the variations $\delta N$ and $\delta N^i$ read,
\be
\mathcal{H}^{\rm field} + \mathcal{H}^{\rm matter} = 0,
\quad
\mathcal{H}^{\rm field}_i + \mathcal{H}^{\rm matter}_i = 0.
\ee
They are eliminated by imposing the gauge conditions
\be
3 \gamma_{ij,j} - \gamma_{jj,i} = 0,
\quad
\pi^{ii}_{\rm can} = 0,
\ee
which allow for the decompositions
\be
\label{ADMTT}
\gamma_{ij} = \Psi^4 \delta_{ij} + h^{\rm TT}_{ij},
\quad
\pi^{ij}_{\rm can} = \tilde{\pi}^{ij}_{\rm can} + \pi^{ij\rm TT}_{\rm can},
\ee
where $h^{\rm TT}_{ij}$ and $\pi^{ij\rm TT}_{\rm can}$ are transverse and traceless quantities,
and longitudinal part $\tilde{\pi}^{ij}_{\rm can}$ is related to a vector potential $V^i_{\rm can}$ by
\be
\tilde{\pi}^{ij}_{\rm can} = V^i_{{\rm can},j} + V^j_{{\rm can},i} - \frac{2}{3} \delta_{ij} V^k_{{\rm can},k}.
\ee
Let us note that in the construction of $V^i_{\rm can}$ the operator $\Delta^{-1}$ is employed
[see the text below Eq.~\eqref{defTT3}].

The gravitational constraints can now be solved for $\Psi$ and $\tilde{\pi}^{ij}_{\rm can}$,
leaving $h^{\rm TT}_{ij}$ and $\pi^{ij\rm TT}_{\rm can}$ as the final degrees of freedom of the gravitational field.
Notice that our gauge condition $\pi^{ii}_{\rm can} = 0$ deviates from the original ADM one $\pi^{ii} = 0$
by spin corrections (which enter at 5PN order).
The final fully reduced action reads,
\be
W = \frac{c^4}{16\pi G}\int \md^4 x\, \pi^{ij\rm TT}_{\rm can} h^{\rm TT}_{ij,0}
+ \int \md t \bigg[ P_i \dot{\hat{z}}^i + \frac{1}{2} \hat{S}_{(i)(j)} \dot{\hat{\theta}}^{(i)(j)} - E \bigg].
\ee
The dynamics is completely described by the ADM energy $E$,
which is the total Hamiltonian ($E=H$) once it is expressed in terms of the canonical variables.
This Hamiltonian can be written as the volume integral
\be
\label{spinH}
H[\hat{z}^i, P_i, \hat{S}_{(i)(j)}, h^{\rm TT}_{ij}, \pi^{ij\rm TT}_{\rm can}]
= -\frac{c^4}{2\pi G} \int \md^3 x\, \Delta\Psi[\hat{z}^i, P_i, \hat{S}_{(i)(j)}, h^{\rm TT}_{ij}, \pi^{ij\rm TT}_{\rm can}].
\ee

The equal-time Poisson bracket relations take the standard form,
\begin{align}
\label{spinPB1}
\{ \hat{z}^i, P_j\} = \delta_{ij},
\quad
\{{\hat S}_{(i)}, {\hat S}_{(j)}\} = \epsilon_{ijk} {\hat S}_{(k)},
\\[1ex]
\label{spinPB2}
\{h^{\rm TT}_{ij}({\bf x},t), \pi^{kl\rm TT}_{\rm can}({\bf x}',t)\}
= \frac{16\pi G}{c^3} \delta^{{\rm TT}kl}_{ij}\delta({\bf x} - {\bf x}'),
\end{align}
zero otherwise, where ${\hat S}_{(i)}=\frac{1}{2}\epsilon_{(i)(j)(k)}{\hat S}_{(j)(k)}$,
$\epsilon_{(i)(j)(k)}=\epsilon_{ijk}=(i-j)(j-k)(k-i)/2$,
and $\delta^{{\rm TT}ij}_{mn}$ is the TT-projection operator, see, e.g., \cite{steinhoff2-08}.
Though the commutation relations \eqref{spinPB1} and \eqref{spinPB2}
are sufficient for the variables on which the Hamiltonian \eqref{spinH} depends on,
for completeness we add the non-trivial ones needed when a Hamiltonian, besides $\hat{S}_{(i)(j)}$,
also depends on the 3-dimensional rotation matrix $\hat{\Lambda}_{[i](j)}$ (``angle'' variables).
They read
\be
\{\hat{\Lambda}_{[i](j)}, \hat{S}_{(k)(l)}\} = \hat{\Lambda}_{[i](k)} \delta_{lj} - \hat{\Lambda}_{[i](l)}\delta_{kj}.
\ee
The angular velocity tensor $\hat{\Omega}^{(i)(j)}$, the Legendre dual to $\hat{S}_{(i)(j)}$,
i.e.\ $\hat{\Omega}^{(i)(j)} = 2 \partial H/\partial\hat{S}_{(i)(j)}$,
is defined by $\hat{\Omega}^{(i)(j)} = \delta\hat{\theta}^{(i)(j)}/\md t = \hat{\Lambda}_{[k]}^{~(i)}\dot{\hat{\Lambda}}^{[k](j)}$,
and the time derivative of the spin tensor thus reads
\be
\dot{\hat{S}}_{(i)(j)} = 2\hat{S}_{(k)[(i)}\Omega_{(j)](k)}  + \hat{\Lambda}^{[k](j)}\frac{\partial H}{\partial\hat{\Lambda}^{[k](i)}}
- \hat{\Lambda}^{[k](i)}\frac{\partial H}{\partial\hat{\Lambda}^{[k](j)}}.
\ee

The Hamiltonian $H$ of Eq.~\eqref{spinH} generates the time evolution in the reduced matter+field phase space.
Generalization and application to many-body systems is quite straightforward, see \cite{steinhoff2-08}.
The total linear ($P_i^{\rm tot}$) and angular ($J_{ij}^{\rm tot}$) momenta take the
forms (particle labels are denoted by $a$),
\begin{align}
\label{Ptot}
P_i^{\rm tot} &= \sum_a P_{ai} - \frac{c^3}{16\pi G} \int \md^3x \, \pi_{\rm can}^{kl\rm TT} h^{\rm TT}_{kl,i}, 
\\[1ex]
J_{ij}^{\rm tot} &= \sum_a ( \hat{z}_a^i P_{aj} - \hat{z}_a^j P_{ai}  + \hat{S}_{a(i)(j)})
- \frac{c^3}{8\pi G} \int\md^3x\, ( \pi_{\rm can}^{ik\rm TT} h^{\rm TT}_{kj}- \pi_{\rm can}^{jk\rm TT} h^{\rm TT}_{ki} )
\nonumber\\&\quad
\label{Jtot}
- \frac{c^3}{16\pi G} \int\md^3x\, ( x^i \pi_{\rm can}^{kl\rm TT} h^{\rm TT}_{kl,j} - x^j \pi_{\rm can}^{kl\rm TT} h^{\rm TT}_{kl,i} ),
\end{align}
and are obtained from the reduced action in the standard Noether manner.

\subsection{Spin-squared dynamics}
\label{subsec:spinsquaredADM}

For the construction of the spin-squared terms we resort to the well-known stress-energy tensor  
for pole-dipole particles but augmented by quadrupolar terms.
The stress-energy tensor density then reads \citep{steinhoff3-08}
\be
\label{Tij_ansatz}
\sqrt{-g}\,T^{\mu\nu} = \int \md \tau \bigg[ t^{\mu\nu} \delta_{(4)}
+ ( t^{\mu\nu\alpha} \delta_{(4)} )_{||\alpha}
+ ( t^{\mu\nu\alpha\beta} \delta_{(4)} )_{||\alpha\beta} \bigg].
\ee
The quantities $t^{\mu\nu\dots}=t^{\nu\mu\dots}$ only depend on the four-velocity $u^{\mu}\equiv\md z^{\mu}/{\md \tau}$,
where $z^{\mu}(\tau)$ is the parametrization of the worldline in terms of its proper time $\tau$,
and on the spin and quadrupole tensors.
Notice that, in general, the quadrupole expressions include not only the mass-quadrupole moment,
but also the current-quadrupole moment and the stress-quadrupole moment (see, e.g., \citealp{steinhoff2-10}).
For the pole-dipole particle $t^{\mu\nu\alpha\beta}$ is zero.
In contrast to the stress-energy tensor of pole-dipole particles,
the Riemann tensor shows up at the quadrupolar level.
However, the source terms of the constraints,
\be
{\gamma}^\frac{1}{2} T^{\mu\nu}n_\mu n_\nu = {\mathcal{H}}^{\rm matter},
\quad
-{\gamma}^\frac{1}{2}T^{\;\mu}_i n_\mu = {\mathcal{H}}^{\rm matter}_i,
\ee
at the approximation considered here, do not include the Riemann tensor.

Regarding rotating black holes,
the mass-quadrupole tensor $Q^{ij}_1$ of object 1 is given by \citep{steinhoff3-08}
(also see, e.g., \citealp{thorne-80} and \citealp{damour3-01})
\be
\label{defQ}
m_1 c^2 Q^{ij}_1 \equiv \gamma^{ik}\gamma^{jl}\gamma^{mn}\hat{S}_{1km}\hat{S}_{1nl}
+ \frac{2}{3}{\bf S}^2_1\gamma^{ij}
= e^i_{(k)}e^j_{(l)}\big(S_{1(k)}S_{1(l)} - \frac{1}{3}{\bf S}^2_1\delta_{(k)(l)}\big),
\ee
where $\mathbf{S}_1=(S_{1(i)})$ is the three-dimensional Euclidean spin vector
related to a spin tensor $\hat{S}_{1ij}$ with the help of a dreibein $e_{i(j)}$ by $\hat{S}_{1ij}=e_{i(k)}e_{j(l)}\epsilon_{klm}S_{1(m)}$.
The quantity ${\bf S}_1^2$ is conserved in time,
\be
2{\bf S}_1^2 = \gamma^{ik} \gamma^{jl} \hat{S}_{1ij} \hat{S}_{1kl} = \mbox{const}.
\ee
The source terms of the constraints in the static case
(independent from the linear momenta $P_i$ of the objects,
what means taking $P_i=0$, but $p_i\ne0$) read
\begin{align}
\label{CEss1}
\mathcal{H}^{\rm matter}_{S_1^2,\,{\rm static}} &= c_1 \left( c^2 Q^{ij}_1 \delta_1 \right)_{; ij}
+ \frac{1}{8 m_1} \gamma_{mn} \gamma^{pj} \gamma^{ql} {\gamma^{mi}}_{,p} {\gamma^{nk}}_{,q} \hat{S}_{1 ij} \hat{S}_{1 kl} \delta_1
\nonumber\\[1ex]&\quad
+ \frac{1}{4m_1} \left( \gamma^{ij} \gamma^{mn} \gamma^{kl}_{~~,m} \hat{S}_{1 ln} \hat{S}_{1 jk} \delta_1 \right)_{,i},
\\[1ex]
\label{CEss2}
\mathcal{H}_{i \,\rm static}^{\rm matter} &= \frac{1}{2} \left(\gamma^{mk}{\hat S}_{ik} \delta \right)_{,m} + \mathcal{O}{(\hat{S}^3)}.
\end{align}
The $c_1$ is some constant that must be fixed
by additional considerations, like matching to the Kerr metric.
The noncovariant terms are due to the transition from
three-dimensional covariant linear momentum $p_i$ to canonical linear momentum $P_i$
given by [cf.\ Eq.\ (4.24) in \citealp{steinhoff2-08} or Eq.~\eqref{NWPi} above]
\be
p_i = P_i - \frac{1}{2} \gamma_{ij} \gamma^{lm} \gamma^{jk}_{~~,m} \hat{S}_{kl}
+ \mathcal{O}(P^2) + \mathcal{O}(\hat{S}^2).
\ee
Thus the source terms are indeed covariant when the point-mass and linear-in-spin terms
depending on the (noncovariant) canonical linear momentum are added,
cf.\ Eqs.\ \eqref{spinHden1} and \eqref{spinHden2}.

The simple structure of the $Q_1^{ij}$ term in Eq.~\eqref{CEss1}
is just the structure of minimal coupling of the Minkowski space mass-quadrupole term to gravity.
As shown by \cite{steinhoff3-08}, the most general ansatz for the spin-squared coupling
including the three-dimensional Ricci tensor reduces to the shown term.
Here we may argue that the correct limit to flat space on the one side
and on the other side, an undefined multiplication with a second delta-function,
resulting in that limit from the Ricci tensor of the spinning ``point'' particle, makes the ansatz unique.
A deeper analysis of the structure of nonlinear-in-spin couplings can be found in, e.g., \cite{levi-15}.

\subsection{Approximate Hamiltonians for spinning binaries}
\label{subsec:PNspinH}

All the approximate Hamiltonians presented in this subsection
have been derived or rederived in recent papers by one of the authors and his collaborators
employing canonical formalisms presented in Sects.~\ref{subsec:spinlinearADM} and \ref{subsec:spinsquaredADM}
(\citealp{damour1-08,steinhoff2-08,steinhoff3-08}).
They are two-point-particle Hamiltonians, which can be used to approximately model binaries made of spinning black holes.
For the rest of this section, canonical variables (which are arguments of displayed Hamiltonians) are not hatted any further.
We use $a,b=1,2$ as the bodies labels,
and for $a \ne b$ we define $r_{ab}\mathbf{n}_{ab}\equiv\mathbf{x}_{a}-\mathbf{x}_{b}$
with $\mathbf{n}_{ab}^2=1$.

The Hamiltonian of leading-order (LO) spin-orbit coupling reads
(let us note that in the following $\mathbf{p}_a$ will denote the canonical linear momenta)
\be
H_{\textrm{SO}}^{\textrm{LO}} = \sum_a \sum_{b \neq a} \frac{G}{c^2r_{a b}^2} (\mathbf{S}_a \times \mathbf{n}_{ab})
\cdot \left( \frac{3 m_b}{2 m_a} \mathbf{p}_a - 2 \mathbf{p}_b \right),
\ee
and the one of leading-order spin(1)-spin(2) coupling is given by
\be
H_{{S_1S_2}}^{\textrm{LO}} = \sum_a \sum_{b \neq a} \frac{G}{2 c^2 r_{ab}^3}
\big( 3 (\mathbf{S}_a\cdot\mathbf{n}_{ab})(\mathbf{S}_b\cdot\mathbf{n}_{ab}) - (\mathbf{S}_a\cdot\mathbf{S}_b) \big).
\ee
The more complicated Hamiltonian is the one with spin-squared terms because it relates to the rotational deformation of spinning black holes.
To leading order, say for spin(1), it reads
\be
H_{{S_1^2}}^{\textrm{LO}} = \frac{G m_2}{2 c^2 m_1 r_{12}^3}
\big( 3 (\mathbf{S}_1\cdot\mathbf{n}_{12})^2 - \mathbf{S}_1^2 \big).
\ee
The LO spin-orbit and spin($a$)-spin($b$) centre-of-mass vectors take the form
\be
\mathbf{G}_{\textrm{SO}}^{\textrm{LO}} = \sum_a \frac{1}{2c^2 m_a} (\mathbf{p}_a \times \mathbf{S}_a),
\quad
\mathbf{G}_{{S_1S_2}}^{\textrm{LO}} =0,
\quad
\mathbf{G}_{{S^2_1}}^{\textrm{LO}} = 0.
\ee

The LO spin Hamiltonians have been applied to studies of
binary pulsar and solar system dynamics, including satellites on orbits around the Earth
(see, e.g., \citealp{barker-79} and \citealp{schaefer-04}).
Another application to the coalescence of spinning binary black holes
via the effective-one-body approach is given in \cite{damour3-01}.
The LO spin dynamics was analysed for black holes and other extended objects in external fields
by \cite{death1-75} and \cite{thorne-85},
and for binary black holes in the slow-motion limit by \cite{death2-75}.
In \cite{barausse-09,barausse-09-erratum} the spinning test-particle dynamics in the Kerr metric
has been explored at LO within Hamiltonian formalism based on Dirac brackets.
In the article \cite{kidder-95} the LO spin-orbit and spin1-spin2 dynamics for compact binaries
is treated in full detail, even including their influence on the
gravitational waves and the related gravitational damping,
particularly the quasi-circular inspiraling and the recoil
of the linear momuntum from the LO spin coupling was obtained.

The Hamiltonian of the next-to-leading-order (NLO) spin-orbit coupling reads
\begin{align}
&H_{\textrm{SO}}^{\textrm{NLO}} = -G\frac{\picSin}{c^4r_{12}^2}
\Bigg( \frac{5 m_2 \pipi}{8 m_1^3} + \frac{3 (\pipii+\npi\npii)}{4 m_1^2}
\nonumber\\[1ex]&\quad
- \frac{3(\piipii - 2\npii^2)}{4 m_1 m_2} \Bigg)
+ G \frac{\picSipii}{c^4r_{12}^2} \left( \frac{2 \npii}{m_1 m_2} - \frac{3 \npi}{4 m_1^2} \right)
\nonumber\\[1ex]&\quad
+ G\frac{\piicSin}{c^4r_{12}^2} \frac{\pipii + 3 \npi\npii}{m_1 m_2}
\nonumber\\[1ex]&\quad
- G^2\frac{\picSin}{c^4r_{12}^3} \left(\frac{11 m_2}{2} + \frac{5 m_2^2}{m_1}\right)
\nonumber\\[1ex]&\quad
+ G^2 \frac{\piicSin}{c^4r_{12}^3} \left(6 m_1 + \frac{15 m_2}{2}\right) + (1 \leftrightarrow 2).
\end{align}
This Hamiltonian was derived by \cite{damour1-08}.
The equivalent derivation of the NLO spin-orbit effects in two-body equations of motion
was done in harmonic coordinates by \cite{blanchet2-06,blanchet2-06-e1,blanchet2-06-e2}.

The NLO spin(1)-spin(2) Hamiltonian is given by
\begin{align}
H_{{S_1S_2}}^{\textrm{NLO}} &= \frac{G}{2c^4 m_1 m_2 r_{12}^3} \Big[6 \piicSin \picSiin
\nonumber\\[1ex]&\quad
+ \frac{3}{2} \picSin \piicSiin
\nonumber\\[1ex]&\quad
- 15 \Sin \Siin \npi \npii
\nonumber\\[1ex]&\quad
- 3 \Sin \Siin \pipii
+ 3 \Sipii \Siin \npi
\nonumber\\[1ex]&\quad
+ 3 \Siipi \Sin \npii + 3 \Sipi \Siin \npii
\nonumber\\[1ex]&\quad
+ 3 \Siipii \Sin \npi - 3 \SiSii \npi \npii
\nonumber\\[1ex]&\quad
+ \Sipi \Siipii
- \frac{1}{2} \Sipii \Siipi + \frac{1}{2} \SiSii \pipii\Big]
\nonumber\\[1ex]&\quad
+ \frac{3G}{2c^4 m_1^2 r_{12}^3} \Big[-\picSin \picSiin
\nonumber\\[1ex]&\quad
+ \SiSii \npi^2 - \Sin \Siipi \npi\Big]
\nonumber\\[1ex]&\quad
+ \frac{3G}{2c^4 m_2^2 r_{12}^3} \Big[-\piicSiin \piicSin
\nonumber\\[1ex]&\quad
+ \SiSii \npii^2 - \Siin \Sipii \npii\Big]
\nonumber\\[1ex]&\quad
+ \frac{6G^2(m_1 + m_2)}{c^4 r_{12}^4} [\SiSii - 2\Sin\Siin].
\end{align}

The calculation of the LO and NLO $S_1^2$-Hamiltonians needs employing the source terms \eqref{CEss1}--\eqref{CEss2}.
In the case of polar-dipolar-quadrupolar particles which are to model spinning black holes,
$Q^{ij}_1$ is the quadrupole tensor of the black hole 1 resulting from its rotational deformation
and the value of the constant $c_1$ is fixed by matching to the test-body Hamiltonian in a Kerr background: $c_1=-1/2$.
Additionally one has to use the Poincar\'e algebra for unique fixation of all coefficients in momentum-dependent part of the Hamiltonian.
The NLO $S_1^2$-Hamiltonian was presented for the first time by \cite{steinhoff3-08}.\footnote{
Slightly earlier a fully dynamical calculation of that dynamics was made by \cite{porto2-08}.
This result turned out to be incomplete due to an omitted term in a specific Feynman diagram \citep{porto2-10}.}
It reads
\begin{align}
\label{NLOS1S1Hamiltonian}
H_{S_1^2}^{\textrm{NLO}} &= \frac{G}{c^4r_{12}^3}\bigg\{
\frac{m_{2}}{m_{1}^3} \bigg[
\frac{1}{4}\left({\mathbf p}_{1}\cdot{\mathbf S}_{1}\right)^2
+ \frac{3}{8}\left({\mathbf p}_{1}\cdot{\mathbf n}_{12}\right)^{2}{\mathbf S}_{1}^{2}
- \frac{3}{8} {\mathbf p}_{1}^{2}\left({\mathbf S}_{1}\cdot {\mathbf n}_{12}\right)^2
\nonumber\\[1ex]&\quad
- \frac{3}{4} \left({\mathbf p}_{1}\cdot{\mathbf n}_{12}\right)\left({\mathbf S}_{1}\cdot{\mathbf n}_{12}\right)\left({\mathbf p}_{1}\cdot{\mathbf S}_{1}\right)
\bigg]
+ \frac{3}{4m_{1}m_{2}}\Big[3{\mathbf p}_{2}^{2}\left({\mathbf S}_{1}\cdot{\mathbf n}_{12}\right)^2
\nonumber\\[1ex]&\quad
- {\mathbf p}_{2}^{2}{\mathbf S}_{1}^{2}\Big]
+ \frac{1}{m_1^2} \bigg[ \frac{3}{4}\left({\mathbf p}_{1}\cdot{\mathbf p}_{2}\right){\mathbf S}_{1}^2
-\frac{9}{4}\left({\mathbf p}_{1}\cdot {\mathbf p}_{2}\right)\left({\mathbf S}_{1} \cdot {\mathbf n}_{12}\right)^2
\nonumber\\[1ex]&\quad
- \frac{3}{2}\left({\mathbf p}_{1}\cdot{\mathbf n}_{12}\right)\left({\mathbf p}_{2}\cdot{\mathbf S}_{1}\right)
\left({\mathbf S}_{1}\cdot{\mathbf n}_{12}\right)  
+3\left({\mathbf p}_{2}\cdot{\mathbf n}_{12}\right)\left({\mathbf p}_{1}
\cdot{\mathbf S}_{1}\right)\left({\mathbf S}_{1}\cdot{\mathbf n}_{12}\right)
\nonumber\\[1ex]&\quad
+ \frac{3}{4}\left({\mathbf p}_{1}\cdot{\mathbf n}_{12}\right)\left({\mathbf p}_{2} \cdot{\mathbf n}_{12}\right){\mathbf S}_{1}^2
-\frac{15}{4}\left({\mathbf p}_{1}\cdot{\mathbf n}_{12}\right)\left({\mathbf p}_{2}\cdot{\mathbf n}_{12}\right)\left({\mathbf S}_{1}\cdot{\mathbf n}_{12}\right)^2
\bigg]
\bigg\}
\nonumber\\[1ex]&\quad
- \frac{G^2 m_2}{2 c^4r_{12}^4} \bigg[9({\mathbf S}_1 \cdot {\mathbf n}_{12})^2 - 5 {\mathbf S}_1^2
+ \frac{14 m_2}{m_1} ({\mathbf S}_1 \cdot {\mathbf n}_{12})^2  - \frac{6 m_2}{m_1}{\mathbf S}_1^2 \bigg].
\end{align}

The spin precession equations corresponding to the Hamiltonians $H^{\textrm{NLO}}_{S_1S_2}$ and $H_{S_1^2}^{\textrm{NLO}}$
have been calculated also by \cite{porto1-08}\footnote{The paper \cite{porto1-08} has benefited from \cite{steinhoff1-08}
when forgotten terms from spin-induced velocity corrections in the LO spin-orbit coupling
could be identified (so-called subleading corrections), see Eq.\ (57) in \cite{porto1-08}.}
and \cite{porto2-08}\footnote{
The final spin precession equation of the paper \citep{porto2-08} deviates from the corresponding one in \cite{steinhoff2-08}.
A detailed inspection has shown that the last term in Eq.\ (60) of \cite{porto2-08} has opposite sign (\citealp{steinhoff1-09};
a typo according to \citealp{porto2-10}).
Using the reverse sign, after redefinition of the spin variable,
agreement with the Hamiltonian of \cite{steinhoff2-08} is achieved.}, respectively.

The NLO spin-orbit and spin($a$)-spin($b$) centre-of-mass vectors take the form
\begin{align}
\mathbf{G}_{\textrm{SO}}^{\textrm{NLO}} &= -\sum_a \frac{\mathbf{p}_a^2}{8 c^4m_a^3} (\mathbf{p}_a \times \mathbf{S}_a)
\nonumber\\[1ex]&\quad
+ \sum_a \sum_{b \neq a} \frac{G m_b}{4 c^4m_a r_{ab}} \bigg\{ [(\mathbf{p}_a\times\mathbf{S}_a)\cdot\mathbf{n}_{ab}]
\frac{5\mathbf{x}_a+\mathbf{x}_b}{r_{ab}} - 5 (\mathbf{p}_a \times \mathbf{S}_a) \bigg\}
\nonumber\\[1ex]&\quad
+ \sum_a \sum_{b \neq a} \frac{G}{c^4 r_{ab}} \bigg\{ \frac{3}{2} (\mathbf{p}_b \times \mathbf{S}_a)
- \frac{1}{2} (\mathbf{n}_{ab} \times \mathbf{S}_a) (\mathbf{p}_b \cdot \mathbf{n}_{ab})
\nonumber\\[1ex]&\quad
- [(\mathbf{p}_a \times \mathbf{S}_a) \cdot \mathbf{n}_{ab}] \frac{\mathbf{x}_a+\mathbf{x}_b}{r_{ab}} \bigg\},
\\[2ex]
\mathbf{G}_{{S_1S_2}}^{\textrm{NLO}} &= \frac{G}{2c^4} \sum_a \sum_{b \neq a} \bigg\{
\left[3(\mathbf{S}_{a}\cdot\mathbf{n}_{ab})(\mathbf{S}_{b}\cdot\mathbf{n}_{ab}) -(\mathbf{S}_{a}\cdot\mathbf{S}_{b})\right]
\frac{\mathbf{x}_{a}}{r_{ab}^3} + (\mathbf{S}_{b}\cdot\mathbf{n}_{ab}) \frac{\mathbf{S}_{a}}{r_{ab}^2} \bigg\}\,,
\\[2ex]
\mathbf{G}_{S_1^2}^{\textrm{NLO}} &= \frac{2Gm_2}{c^4m_{1}}
\bigg\{ \frac{3\left(\mathbf{S}_{1}\cdot\mathbf{n}_{12}\right)^2}{8r_{12}^3}\left(\mathbf{x}_{1}+\mathbf{x}_{2}\right)
 +\frac{\mathbf{S}_{1}^2}{8r_{12}^3} \left(3\mathbf{x}_{1}-5\mathbf{x}_{2}\right)
-\frac{\left(\mathbf{S}_{1}\cdot\mathbf{n}_{12}\right)\mathbf{S}_1}{r_{12}^2}\bigg\}.
\end{align}

We can sum up all centre-of-mass vectors displayed in this subsection in the following equation:
\be
\label{spinG}
\mathbf{G} = \mathbf{G}_{\textrm{N}} + \mathbf{G}_{\textrm{1PN}} + \mathbf{G}_{\textrm{2PN}}
+ \mathbf{G}_{\textrm{3PN}} + \mathbf{G}_{\textrm{4PN}}
+ \mathbf{G}_{\textrm{SO}}^{\textrm{LO}} + \mathbf{G}_{\textrm{SO}}^{\textrm{NLO}}
+ \mathbf{G}_{S_{1}S_{2}}^{\textrm{NLO}} + \mathbf{G}_{S_{1}^2}^{\textrm{NLO}} + \mathbf{G}_{S_{2}^2}^{\textrm{NLO}},
\ee
where $\mathbf{G}_{\textrm{N}}$ up to $\mathbf{G}_{\textrm{4PN}}$ represent the pure orbital contributions,
which do not depend on spin variables (the explicit formulae for them one can find in \cite{jaranowski-15}).
The last term in Eq.~\eqref{spinG} can be obtained from the second last one
by means of the exchange $(1 \leftrightarrow 2)$ of the bodies' labels.

The explicitly given above and in Appendices \ref{app:ncom} and \ref{app:spinH} conservative binary Hamiltonians,
modeling binaries made of spinning black holes, can be summarized as follows:
\begin{align}
\label{allH}
H &= H_{\textrm{N}} + H_{\textrm{1PN}} + H_{\textrm{2PN}} + H_{\textrm{3PN}} + H_{\textrm{4PN}}
\nonumber\\[1ex]&\quad
+ H_{\textrm{SO}}^{\textrm{LO}}
+ H^{\textrm{LO}}_{S_1^2} + H^{\textrm{LO}}_{S_{1}S_{2}} + H^{\textrm{LO}}_{S_2^2}
\nonumber\\[1ex]&\quad
+ H_{\textrm{SO}}^{\textrm{NLO}}
+ H^{\textrm{NLO}}_{S_{1}^2} + H^{\textrm{NLO}}_{S_{1}S_{2}} + H^{\textrm{NLO}}_{S_{2}^2}
\nonumber\\[1ex]&\quad
+ H_{\textrm{SO}}^{\textrm{NNLO}}
+ H^{\textrm{NNLO}}_{S_{1}^2} + H^{\textrm{NNLO}}_{S_{1}S_{2}} + H^{\textrm{NNLO}}_{S_{2}^2}
\nonumber\\[1ex]&\quad
+ H^{\textrm{LO}}_{S_1^3} + H^{\textrm{LO}}_{S_1^2 S_2}
+ H^{\textrm{LO}}_{S_1 S_2^2}+ H^{\textrm{LO}}_{S_2^3}
\nonumber\\[1ex]&\quad
+ H^{\textrm{LO}}_{S_{1}^4} + H^{\textrm{LO}}_{S_{1}^3S_{2}}
+ H^{\textrm{LO}}_{S_{1}^2S_{2}^2} + H^{\textrm{LO}}_{S_{1}S_{2}^3} + H^{\textrm{LO}}_{S_{2}^4},
\end{align}
where the first line comprises pure orbital, i.e., spin-independent, Hamiltonians.
The Hamiltonians from the second and the third line are explicitly given above.
The NNLO spin-orbit $H_{\textrm{SO}}^{\textrm{NNLO}}$ and spin1-spin2 $H^{\textrm{NNLO}}_{S_{1}S_{2}}$ Hamiltonians
were obtained by \cite{hartung-13}, their explicit forms can be found in Appendix \ref{app:spinH}.
\cite{levi3-16} derived, applying the EFT method to extended bodies,
the NNLO spin-squared Hamiltonians $H^{\textrm{NNLO}}_{S_{1}^2}$ and $H^{\textrm{NNLO}}_{S_{2}^2}$.
All the Hamiltonians cubic and quartic in the spins were derived by \cite{hergt2-08,hergt1-08}
with the aid of approximate ADMTT coordinates of the Kerr metric and application of the Poincar\'e algebra.\footnote{
The $H^{\textrm{LO}}_{S_1^4}$ and $H^{\textrm{LO}}_{S_2^4}$ terms
were incorrectly claimed to be zero by \cite{hergt2-08}.}
Their generalizations to general extended objects were achieved by \cite{levi-15},
where also for the first time the Hamiltonians $H^{\textrm{LO}}_{S_1^4}$ and $H^{\textrm{LO}}_{S_2^4}$
were obtained (correcting \citealp{hergt2-08}).
All the Hamiltonians cubic and quartic in the spins and displayed in Eq.~\eqref{allH}
are explicitly given in Appendix~\ref{app:spinH}.
Notice that not all Hamiltonians from Eq.~\eqref{allH} are necessarily given in the ADM gauge,
because any use of the equations of motion in their derivation has changed gauge.
E.g., for spinless particles the highest conservative Hamiltonian in ADM gauge is $H_{\textrm{2PN}}$.

For completeness we also give the spin-squared Hamiltonians for neutron stars
through next-to-leading order (\citealp{porto2-08,porto2-10}, \citealp{hergt-10}).
They depend on the quantity $C_Q$, which parametrizes quadrupolar deformation effects induced by spins.
The LO Hamiltonian reads (cf., e.g., \citealp{barker-79})
\be
\label{loHS1S2ns}
H_{S_{1}^2\textrm{(NS)}}^{\textrm{LO}}
= \frac{Gm_{1}m_{2}}{2c^2 r_{12}^3} C_{Q_1} \left(3\frac{\Sin^2}{m_{1}^2}-\frac{\SiSi}{m_{1}^2}\right).
\ee
The NLO Hamiltonian equals
\begin{align}
\label{nloHS1S2ns}
H_{S_{1}^2\textrm{(NS)}}^{\textrm{NLO}}
&= \frac{G}{c^4 r_{12}^3} \Bigg[\frac{m_{2}}{m_{1}^3}\Bigg(
\left(-\frac{21}{8}+\frac{9}{4}C_{Q_1}\right)\pipi\SinNW^2
+\left(\frac{3}{2}C_{Q_1}-\frac{5}{4}\right)\SipiNW^2
\nonumber\\&
+\left(\frac{15}{4}-\frac{9}{2}C_{Q_1}\right)\pinNW\SinNW\SipiNW
\nonumber\\&
+\left(-\frac{9}{8}+\frac{3}{2}C_{Q_1}\right)\pinNW^2\SiSiNW+\left(\frac{5}{4}-\frac{5}{4}C_{Q_1}\right)\pipi\SiSiNW\Bigg)
\nonumber\\&
+\frac{1}{m_{1}^2} \Bigg(-\frac{15}{4}C_{Q_1}\pinNW\piinNW\SinNW^2
\nonumber\\&
+\left(3-\frac{21}{4}C_{Q_1}\right)\pipii\SinNW^2
\nonumber\\&
+\left(-\frac{3}{2}+\frac{9}{2}C_{Q_1}\right)\piinNW\SinNW\SipiNW
\nonumber\\&
+\left(-3+\frac{3}{2}C_{Q_1}\right)\pinNW\SinNW\SipiiNW
\nonumber\\&
+\left(\frac{3}{2}-\frac{3}{2}C_{Q_1}\right)\SipiNW\SipiiNW
\nonumber\\&
+\left(\frac{3}{2}-\frac{3}{4}C_{Q_1}\right)\pinNW\piinNW\SiSiNW
\nonumber\\&
+\left(-\frac{3}{2}+\frac{9}{4}C_{Q_1}\right)\pipii\SiSiNW\Bigg)
\nonumber\\&
+\frac{C_{Q_1}}{m_{1}m_{2}}\Big(\frac{9}{4}\piipii\SinNW^2-\frac{3}{4}\piipii\SiSiNW\Big)\Bigg]
\nonumber\\&
+ \frac{G^2 m_{2}}{c^4 r_{12}^4} \Bigg[\left(2+\frac{1}{2}C_{Q_1}+\frac{m_{2}}{m_{1}}\big(1+2C_{Q_1}\big)\right)\SiSiNW
\nonumber\\&
+\left(-3-\frac{3}{2}C_{Q_1}-\frac{m_{2}}{m_{1}}\big(1+6C_{Q_1}\big)\right)\SinNW^2\Bigg].
\end{align}
This Hamiltonian for $C_{Q_1}=1$ agrees with that given in Eq.~\eqref{NLOS1S1Hamiltonian} describing black-hole binaries
(for neutron stars, $C_{Q_1}=2$--$8$ holds; see, e.g., \citealp{mandal-22b}).
It has been derived fully correctly for the first time by \cite{porto2-10} using the EFT method.
Shortly afterwards, an independent calculation by \cite{hergt-10},
in part based on the Eqs.\ \eqref{CEss1} and \eqref{CEss2} including \eqref{defQ}, has confirmed the result.

The radiation-reaction (or dissipative) Hamiltonians for leading-order spin-orbit and spin1-spin2 couplings
are derived by \cite{steinhoff3-10} and \cite{wang-11}.
All the known dissipative Hamiltonians can thus be summarized as
\be
H^{\rm diss} = H_{\rm 2.5PN} + H_{\rm 3.5PN} +  H^{\text{LO\,diss}}_{\rm SO} + H^{\text{LO\,diss}}_{S_1S_2},
\ee
where $H_{\rm 2.5PN}$ and $H_{\rm 3.5PN}$ are spin-independent (purely orbital) dissipative Hamiltonians.
The leading-order Hamiltonian $H_{\rm 2.5PN}$ is given in Eq.~\eqref{H2.5PN} for two-point-mass
and in Appendix~\ref{app:dissH} for many-point-mass systems,
and the next-to-leading-order Hamiltonian $H_{\rm 3.5PN}$ is explicitly given in the Appendix~\ref{app:dissH}
(also for many-point-mass systems).
The spin-dependent dissipative Hamiltonians $H^{\text{LO\,diss}}_{\rm SO}$ and $H^{\text{LO\,diss}}_{S_1S_2}$
can be read off from the Hamiltonian $H^{\text{spin}}_{\text{3.5PN}}$ given in the Appendix~\ref{app:dissH}
(we keep here the notation of the Hamiltonian used by \citealp{wang-11},
which indicates spin corrections to the spinless 3.5PN dynamics).

\section{Tidal interactions}
\label{sec:tidal}

The work done in this field through higher PN orders
relies on the effective Fokker action with non-minimal matter couplings.
The Hamiltonians are obtained from higher-order Lagrangians in harmonic coordinates
via order reduction and Legendre transforms.
Here we tightly follow \cite{henry-20a,henry-20b};
also see \cite{damour-10,bini-12,steinhoff-16}.

The action for the gravitational field is given in harmonic gauge through
\be
S_g = \frac{c^3}{16\pi G} \int\md^4x \sqrt{-g} \left(R-\frac{1}{2}g_{\mu\nu}\Gamma^\mu\Gamma^\nu\right),
\ee
where $\Gamma^\mu:=g^{\rho\sigma}\Gamma^\mu_{\rho\sigma}$.
The ansatz for the matter action, in sufficient approximation for our intended presentation,
is given by
\begin{align}
S_m = \sum_a \int\md\tau_a \bigg(-m_ac^2 &+ \frac{\mu_a^{(2)}}{4}G^a_{\mu\nu} G_a^{\mu\nu}
+ \frac{\sigma^{(2)}_a}{6c^2}H^a_{\mu\nu}H_a^{\mu\nu}
\nonumber\\
&+ \frac{\mu_a^{(3)}}{12}G^a_{\lambda\mu\nu} G_a^{\lambda\mu\nu}
+ {\cal{O}}\left(\frac{\epsilon_{\rm tidal}}{c^6}\right)\bigg),
\end{align}
with the bodies', labeling $a$, tidal mass quadrupole $G_a^{\mu\nu}$,
tidal current quadrupole $H_a^{\mu\nu}$,
and tidal mass octupole $G_a^{\lambda\mu\nu}$ moments;
$\epsilon_{\rm tidal}\sim1/c^{10}$ denotes order of the dominant tidal effect.
The static (equilibrium) deformability coefficients are denoted, including their orders,
by $\mu_a^{(2)} = {\cal{O}}(\epsilon_{\rm tidal})$, $\sigma_a^{(2)} = {\cal{O}}(\epsilon_{\rm tidal})$,
and $\mu_a^{(3)} = {\cal{O}}(\epsilon_{\rm tidal}/c^4)$.
The first tidal term is leading order plus NLO plus NNLO,
the second is NLO plus NNLO, and the third one is solely NNLO.

The tidal moments are related with the Weyl or Riemann tensor,
centered at the point masses (particles) in the forms
\begin{subequations}
\begin{align}
G^a_{\mu\nu} &= -c^2 [R_{\mu\rho\nu\sigma}]_a u_a^\rho u_a^\sigma,
\\[1ex]
H^a_{\mu\nu} &= 2c^3 [R^*_{(\mu\underline{\rho}\nu)\sigma}]_a u_a^\rho u_a^\sigma,
\\[1ex]
G^a_{\lambda\mu\nu} &= -c^2 [\nabla^\bot_{(\lambda} R_{\mu\underline{\rho}\nu)\sigma}]_a u_a^\rho u_a^\sigma,
\end{align}
\end{subequations}
with the underlined index $\rho$ being excluded from symmetrization
and $\nabla^\bot_\mu:=(\delta^\nu_\mu + u_\mu u^\nu)\nabla_\nu$.

Making use of the gothic metric deviation $h^{\mu\nu} = \sqrt{-g} g^{\mu\nu} - \eta^{\mu\nu}$,
with then defining the vector variable $h\equiv(h^{00ii},h^{0i},h^{ij})$ with $h^{00ii}\equiv h^{00}+\delta_{ij}h^{ij}$,
and decomposing  $h = h_{\rm pp} + h_{\rm tidal}$, where $h_{\rm pp}$ comes from the metric
generated by structureless point particles (pp) and
$h_{\rm tidal}=(\epsilon_{\rm tidal}/c^2,\epsilon_{\rm tidal}/c^3,\epsilon_{\rm tidal}/c^4)$,
the Fokker action $S_{\rm F}[{\rm MV}]$, and as well the Fokker Lagrangian $L_{\rm F}$ with
$\int\md t\,L_{\rm F}({\rm MV}) = S_{\rm F}[{\rm MV}]$,
with MV denoting matter variables, similarly to our Routhian procedure, results in the form
\be
S_{\rm F}[{\rm MV}] = S_{\rm total}[{\rm MV},h_{\rm pp}],
\ee
where we have used
\begin{align}
S_{\rm total}[{\rm MV}, h] &= S_{\rm total}[{\rm MV},h_{\rm pp}]
+ \int\md^4x \frac{\delta S_{\rm total}}{\delta h}[{\rm MV}, h_{\rm pp}]h_{\rm tidal} + {\cal{O}}(h^2_{\rm tidal})
\nonumber\\
&=  S_{\rm total} [{\rm MV}, h_{\rm pp}] + {\cal{O}}(\epsilon^2_{\rm tidal});
\end{align}
also see Appendix C in \cite{damour1-85}.

The explicit form of the NNLO tidal Hamiltonian can be found in \cite{henry-20b}. It reads
\begin{align}
H_\text{tidal} &= \frac{G^2 m_2^2}{r_{12}^6}\Bigg\{ -\frac{3}{2}\mu_1^{(2)}
+ \frac{1}{c^2} w_{\text{NLO}}
+ \frac{1}{c^4} w_{\text{NNLO}}
- \frac{15\mu_1^{(3)}}{2r_{12}^2}
\Bigg\} + (1\leftrightarrow 2)
\nonumber\\&\quad
+ \mathcal{O}\left(\frac{\epsilon_\text{tidal}}{c^{6}}\right),
\end{align}
where
\begin{subequations}
\begin{align}
&w_{\text{NLO}} = -\frac{12 \sigma_1^{(2)} \piipii}{m_{2}^2}
+ \frac{\npii^2}{m_{2}^2} \Bigl(-18 \mu_1^{(2)}
+ 12 \sigma_1^{(2)}\Bigr)
\nonumber\\&\quad
+ \frac{\npi \npii}{m_{1} m_{2}} \Bigl(18 \mu_1^{(2)}
- 24 \sigma_1^{(2)}\Bigr)
+ \frac{\pipii}{m_{1} m_{2}} \Bigl(\frac{9}{2} \mu_1^{(2)}
+ 24 \sigma_1^{(2)}\Bigr)
\nonumber\\&\quad
+ \frac{\npi^2}{m_{1}^2} \Bigl(\frac{9}{2} \mu_1^{(2)}
+ 12 \sigma_1^{(2)}\Bigr)
+ \frac{\pipi}{m_{1}^2} \Bigl(- \frac{15}{4} \mu_1^{(2)}
- 12 \sigma_1^{(2)}\Bigr)
\nonumber\\&\quad
+ \frac{G}{r_{12}} \Bigl(3m_{1} + \frac{21}{2} m_{2}\Bigr) \mu_1^{(2)},
\end{align}
\begin{align}
&w_{\text{NNLO}} = \frac{\npii^4}{m_{2}^4} \Bigl(- \frac{63}{2} \mu_1^{(2)}
- 60 \sigma_1^{(2)}\Bigr)
+ \frac{\piipiip^2}{m_{2}^4} \Bigl(-9 \mu_1^{(2)}
- 12 \sigma_1^{(2)}\Bigr)
\nonumber\\&\quad
+ \frac{\pipii \piipii}{m_{1} m_{2}^3} \Bigl(\frac{99}{4}
\mu_1^{(2)} + 60 \sigma_1^{(2)}\Bigr)
+ \frac{\npii^2}{m_{2}^2} \biggl [\frac{\piipii}{m_{2}^2} \Bigl(54 \mu_1^{(2)}
+ 72 \sigma_1^{(2)}\Bigr)
\nonumber\\&\quad
+ \frac{\pipii}{m_{1} m_{2}} \Bigl(-54 \mu_1^{(2)}
- 144 \sigma_1^{(2)}\Bigr)\biggl]
+ \frac{\pipii^2}{m_{1}^2 m_{2}^2} \Bigl(- \frac{45}{2} \mu_1^{(2)} - 60 \sigma_1^{(2)}\Bigr)
\nonumber\\&\quad
+ \frac{\npi^3 \npii}{m_{1}^3 m_{2}} \Bigl(18 \mu_1^{(2)} + 48 \sigma_1^{(2)}\Bigr)
\nonumber\\&\quad
+ \frac{\pipi}{m_{1}^2} \biggl [\frac{\npii^2}{m_{2}^2} \Bigl(\frac{45}{2} \mu_1^{(2)}
+ 66 \sigma_1^{(2)}\Bigr) + \frac{\piipii}{m_{2}^2} \Bigl(- \frac{45}{4} \mu_1^{(2)} - 30 \sigma_1^{(2)}\Bigr)
\nonumber\\&\quad
+ \frac{\pipii}{m_{1} m_{2}} \Bigl(\frac{81}{4} \mu_1^{(2)} + 48 \sigma_1^{(2)}\Bigr)\biggl]
+ \frac{\npi}{m_{1}} \biggl(\frac{\npii^3}{m_{2}^3} \Bigl(54 \mu_1^{(2)} + 144 \sigma_1^{(2)}\Bigr)
\nonumber\\&\quad
+ \frac{\npii \pipi}{m_{1}^2 m_{2}} \Bigl(- \frac{63}{2}
\mu_1^{(2)} - 48 \sigma_1^{(2)}\Bigr)
\nonumber\\&\quad
+ \frac{\npii}{m_{2}} \biggl [\frac{\piipii}{m_{2}^2} \Bigl(-36 \mu_1^{(2)} - 60 \sigma_1^{(2)}\Bigr)
+ \frac{\pipii}{m_{1} m_{2}} \Bigl(45 \mu_1^{(2)}
+ 120 \sigma_1^{(2)}\Bigr)\biggl]\biggl)
\nonumber\\&\quad
+ \frac{\npi^4}{m_{1}^4} \Bigl(- \frac{9}{2} \mu_1^{(2)}
- 12 \sigma_1^{(2)}\Bigr)
+ \frac{\pipip^2}{m_{1}^4} \Bigl(- \frac{45}{16} \mu_1^{(2)}
- 6 \sigma_1^{(2)}\Bigr)
\nonumber\\&\quad
+ \frac{\npi^2}{m_{1}^2} \biggl [\frac{\npii^2}{m_{2}^2}
\Bigl(-45 \mu_1^{(2)} - 120 \sigma_1^{(2)}\Bigr)
+ \frac{\piipii}{m_{2}^2} \Bigl(9 \mu_1^{(2)}
+ 24 \sigma_1^{(2)}\Bigr)
\nonumber\\&\quad
+ \frac{\pipii}{m_{1} m_{2}} \Bigl(-18 \mu_1^{(2)}
- 48 \sigma_1^{(2)}\Bigr) + \frac{\pipi}{m_{1}^2} \Bigl(\frac{27}{4} \mu_1^{(2)}
+ 18 \sigma_1^{(2)}\Bigr)\biggl]
\nonumber\\&\quad
+ \frac{G}{r_{12}} \biggl(m_{1} \biggl [\frac{\npii^2}{m_{2}^2} \Bigl(207 \mu_1^{(2)}
- 80 \sigma_1^{(2)}\Bigr)
+ \frac{\piipii}{m_{2}^2} \Bigl(- \frac{45}{2} \mu_1^{(2)}
+ 80 \sigma_1^{(2)}\Bigr)
\nonumber\\&\quad
+ \frac{\npi \npii}{m_{1} m_{2}} \Bigl(- \frac{1341}{8}
\mu_1^{(2)} + 172 \sigma_1^{(2)}\Bigr)
+ \frac{\pipii}{m_{1} m_{2}} \Bigl(\frac{3}{8} \mu_1^{(2)}
- 172 \sigma_1^{(2)}\Bigr)
\nonumber\\&\quad
+ \frac{\npi^2}{m_{1}^2} \Bigl(- \frac{183}{2} \mu_1^{(2)}
- 92 \sigma_1^{(2)}\Bigr) + \frac{\pipi}{m_{1}^2} \Bigl(\frac{123}{4} \mu_1^{(2)}
+ 92 \sigma_1^{(2)}\Bigr)\biggl]
\nonumber\\&\quad
+ m_{2} \biggl [\frac{\npii^2}{m_{2}^2} \Bigl(\frac{331}{2} \mu_1^{(2)}
- 120 \sigma_1^{(2)}\Bigr)
+ \frac{\piipii}{m_{2}^2} \Bigl(\frac{61}{4} \mu_1^{(2)}
+ 120 \sigma_1^{(2)}\Bigr)
\nonumber\\&\quad
+ \frac{\npi \npii}{m_{1} m_{2}} \Bigl(- \frac{1189}{8}
\mu_1^{(2)} + 228 \sigma_1^{(2)}\Bigr)
+ \frac{\pipii}{m_{1} m_{2}} \Bigl(- \frac{401}{8} \mu_1^{(2)}
- 228 \sigma_1^{(2)}\Bigr)
\nonumber\\&\quad
+ \frac{\npi^2}{m_{1}^2} \Bigl(- \frac{81}{2} \mu_1^{(2)}
- 108 \sigma_1^{(2)}\Bigr)
+ \frac{\pipi}{m_{1}^2} \Bigl(\frac{135}{4} \mu_1^{(2)} + 108
\sigma_1^{(2)}\Bigr)\biggl]\biggl)
\nonumber\\&\quad
+ \frac{G^2}{r_{12}^2} \Bigl(\frac{303}{28}m_{1}^2 - \frac{455}{8}m_{1}m_{2} - 39m_{2}^2\Bigr)\mu_1^{(2)}.
\end{align}
\end{subequations}
The NNNLO tidal effects were recently computed in \cite{mandal-23}.

\appendix
\normalsize

\section{Hamiltonian dynamics of ideal fluids in Newtonian gravity}
\label{app:Nfluids}

In the Newtonian theory the equations for gravitating ideal fluids
are usually given in the following form:
\begin{enumerate}
\item[(i)] The equation for the conservation of mass,\footnote{
In a Cartesian spatial coordinate system $(x^i)$ and for any vector field $\mathbf{w}$ and any scalar field $\phi$ we define:
$\mathrm{div}\,\mathbf{w}\equiv\partial_iw^i$,
$(\mathrm{curl}\,\mathbf{w})^i\equiv\varepsilon^{ijk}\partial_jw^k$,
$(\mathrm{grad}\,\phi)^i\equiv\partial_i\phi$.}
\be
\partial_t \varrho_* + \mbox{div}( \varrho_* {\bf v}) = 0,
\ee
where $\varrho_*$ is the mass density and ${\bf v}= (v^i)$ is the velocity field of the fluid.
\item[(ii)] The equations of motion,
\be
\varrho_* \partial_t {\bf v} + \frac{\varrho_*}{2}~ \mbox{grad}~{\bf v}^2 -  \varrho_*~ {\bf v} \times  \mbox{curl}~{\bf v} 
= - \grad\, p  + \varrho_*~\mbox{grad}~U,
\ee
where $p$ is the pressure in the fluid and $U$ the gravitational potential.
\item[(iii)] The equation of state,
\be
\epsilon = \epsilon(\varrho_*, s) \quad \mbox{with} \quad
\md\epsilon = h \md\varrho_* + \varrho_*T \md s,  \quad
\mbox{or} \quad \md p = \varrho_* \md h - \varrho_*T \md s,
\ee
with the temperature $T$, the internal energy density $\epsilon$ and the specific enthalpy $h$.
\item[(iv)] The conservation law for the specific entropy $s$ along the flow lines,
\be
\partial_t s + {\bf v}\cdot  \mbox{grad}~s = 0.
\ee
\item[(v)] The Newtonian gravitational field equation,
\be
\Delta U = - 4\pi G \varrho_*,
\ee
where $\Delta$ is the Laplacian.
The gravitational potential hereof reads
\be
U({\bf x},t) = G \int\md^3{\bf x}' \frac{ \varrho_*({\bf x}',t)}{|{\bf x}-{\bf x}'|}.
\ee
\end{enumerate}

Within the Hamilton framework the equations of motion are obtained from the relation
$\partial_t A({\bf x},t)=\{A({\bf x},t),H\}$,
valid for any function $A({\bf x},t)$ living in phase space,
i.e.\ built out of the fundamental variables $\varrho_*$, $\pi_i$, and $s$,
with the Hamiltonian given by $H=H[\varrho_*,\pi_i,s]$,
where $\pi_i$ is the linear momentum density of the fluid \citep{holm-85}.
The brackets $\{\cdot,\cdot\}$ are called Lie-Poisson brackets.
They may be defined by
\be
\left\{ \int\md^3x\, \xi^i \pi_i,\, F[\varrho_*,s,\pi_i] \right\}
= \int\md^3x \left(
\frac{\delta F}{\delta \varrho_*}{\cal{L}}_\xi \varrho_*
+ \frac{\delta F}{\delta s}{\cal{L}}_\xi s
+ \frac{\delta F}{\delta \pi_i}{\cal{L}}_\xi \pi_i \right),
\ee
where $F$ is a functional of $\varrho_*$, $s$, and $\pi_i$,
${\cal{L}}_\xi$ denotes the Lie derivative along the vector field $\xi^i$,
and $\delta F/\delta(\cdots)$ are the Fr\'echet derivatives of the functional $F$
[see, e.g., Appendix C of \cite{blanchet-90} and references therein].

Explicitly, the equations in (i), (ii), and (iv) take the following Hamiltonian form
[the equations in (iii) and (v) remain unchanged]:
\begin{enumerate}
\item[(i)] The mass conservation equation
\be
\frac{\partial \varrho_*}{\partial t} = -\partial_i \left(\frac{\delta H}{\delta \pi_i}\varrho_*\right),
\ee
notice that $\dst v^i = \frac{\delta H}{\delta \pi_i}$.
\item[(ii)] The equations of motion
\be
\frac{\partial \pi_i}{\partial t} = -\partial_j \left(\frac{\delta H}{\delta \pi_j}\pi_i\right)
- \partial_i \left(\frac{\delta H}{\delta \pi_j}\right) \pi_j
- \partial_i \left(\frac{\delta H}{\delta \varrho_*}\right)\varrho_* + \frac{\delta H}{\delta s}  \partial_i s.
\ee
\item[(iv)] The entropy conservation law
\be
\frac{\partial s}{\partial t} = -\frac{\delta H}{\delta \pi_i} \partial_i s.
\ee
\end{enumerate}
The following kinematical Lie-Poisson bracket relations between the fundamental variables are fulfiled:
\begin{align}
\label{eq10}
\{\pi_i({\bf x},t),\varrho_*({\bf x}',t)\} &= \frac{\partial}{\partial x'^i}[\varrho_*({\bf x}',t) \delta ({\bf x}-{\bf x}')],
\\[1ex]
\{\pi_i({\bf x},t),s({\bf x}',t)\} &= \frac{\partial s({\bf x}',t)}{\partial {x}'^i} \delta ({\bf x}-{\bf x}'),
\\[1ex]
\label{eq12}
\{\pi_i({\bf x},t),\pi_j({\bf x}',t)\} &= \pi_i({\bf x}',t)\frac{\partial}{\partial {x}'^j} \delta ({\bf x}-{\bf x}')
- \pi_j({\bf x},t)\frac{\partial}{\partial {x}^i} \delta ({\bf x}-{\bf x}'),
\end{align}
and other brackets are zero.
More explicitly the Hamiltonian of the fluid takes the form,
\be
H = \frac{1}{2}\int\md^3{\bf x} \frac{\pi_i\pi_i}{\varrho_*}
- \frac{G}{2}\int\md^3{\bf x}\,\md^3{\bf x}'\,\frac{\varrho_*({\bf x},t)\varrho_*({\bf x}',t)}{|{\bf x}-{\bf x}'|}
+ \int\md^3{\bf x}\,\epsilon.
\ee

For point masses, the momentum and mass densities are given by
\be
\pi_i = \sum_a p_{ai} \delta({\bf x}-{\bf x}_a),
\qquad
\varrho_* = \sum_a m_a \delta ({\bf x}-{\bf x}_{a}),
\ee
and we have also $h=p=s=0$.
The position and momentum variables fulfill the standard Poisson bracket relations,
\be
\{x^i_a,p_{aj}\} = \delta_{ij}, \quad\quad \mbox{zero otherwise},
\ee
and the Hamiltonian results in
\be
H = \frac{1}{2} \sum_a \frac{\vecp_a^2}{m_a} - \frac{G}{2} \sum_{a \ne b} \frac{m_am_b}{|{\bf x}_a-{\bf x}_b|},
\ee
where the internal and self-energy terms have been dropped
(after performing a proper regularization, see Sect.~\ref{subs:RHreg} in our review).

Let us remark that for fluids a canonical formalism with standard Poisson brackets can be obtained
with the transition to Lagrangian coordinates $b^A(x^i,t)$, such that $\partial_t b^A + {\bf v}\cdot \mbox{grad}~b^A = 0$. 
Then,
\be
p_A = b^i_A \pi_i  \quad  \mbox{with} \quad b^i_A = \frac{\partial x^i}{\partial b^A}.
\ee
The variables $b^A$ and $p_B$ are canonically conjugate to each other, i.e.
\be
\{b^A(x^i,t),\, p_B(y^j,t)\} = \delta^A_B(x^i-y^i).
\ee
The mass density in Lagrangian coordinates, say $\mu(b^A,t)$, is defined by $\varrho_*\,\md^3x=\mu\,\md^3b$
and relates to the usual mass density as $\varrho_*=\mu(b^A,t)\det(b^B_j)$.

\section{Hamiltonian dynamics of ideal fluids in GR}
\label{app:GRfluids}

The general-relativistic equations governing the dynamics of gravitating ideal fluids are as follows
(see, e.g., \citealp{holm-85}, \citealp{blanchet-90}).
\begin{enumerate}
\item[(i)] The equation for the conservation of mass,
\be
\partial_{\mu}(\sqrt{-g}\varrho u^{\mu})=0 \quad \mbox{or} \quad \partial_t \varrho_* + \mbox{div}( \varrho_* {\bf v}) = 0,
\ee
where $\varrho$ denotes the proper rest-mass density
and $u^{\mu}$ the four-velocity field of the fluid ($g_{\mu\nu}u^{\mu}u^{\nu} = -1$),
$\varrho_* = \sqrt{-g} u^0 \varrho$ is the coordinate mass density and ${\bf v}$ the velocity field of the fluid, $v^i = c u^i/u^0$.
\item[(ii)] The equations of motion,
\be
\partial_{\mu}\left(\sqrt{-g}\,T^{\mu}_i\right)
- \frac{1}{2}\, \sqrt{-g}\,T^{\mu\nu}\, \partial_i g_{\mu\nu} = 0,
\ee
where 
\be
T^{\mu\nu} = \varrho(c^2 + h) u^{\mu}u^{\nu} + p g^{\mu\nu}
\ee
is the stress-energy tensor of the fluid with pressure $p$ and specific enthalpy $h$.
\item[(iii)] The equation of state, using the energy density $e= \varrho(c^2 + h) - p$,
\be
e = e(\varrho,s) \quad \mbox{with} \quad \md e=(c^2+h)\md\varrho + \varrho T\md s
\quad \mbox{or}  \quad \md p = \varrho \md h - \varrho T \md s.
\ee
\item[(iv)] The conservation law for the specific entropy $s$ along the flow lines,
\be
u^{\mu}\partial_{\mu}s=0 \quad \mbox{or} \quad \partial_t s + {\bf v}\cdot\grad\,s = 0.
\ee
\item[(v)] The Einsteinian field equations for gravitational potential (or metric) functions $g_{\mu\nu}$,
\be
R^{\mu\nu} = \frac{8\pi G}{c^4} \left(T^{\mu\nu} - \frac{1}{2}g^{\mu\nu}g_{\alpha\beta}T^{\alpha\beta}\right).
\ee
\end{enumerate}

The variables of the canonical formalism get chosen to be
\be
\varrho_* = \sqrt{-g} u^0 \varrho, \quad s, \quad \pi_i = \frac{1}{c} \sqrt{-g} T^0_i.
\ee
They do fulfill the same (universal, free of spacetime metric) kinematical Lie--Poisson bracket relations as in the Newtonian theory
(see \citealp{holm-85} or also \citealp{blanchet-90}),
\begin{align}
\{\pi_i({\bf x},t),\varrho_*({\bf x}',t)\} &= \frac{\partial}{\partial x'^i}[\varrho_*({\bf x}',t) \delta ({\bf x}-{\bf x}')],
\\[1ex]
\{\pi_i({\bf x},t),s({\bf x}',t)\} &= \frac{\partial s({\bf x}',t)}{\partial x'^i} \delta ({\bf x}-{\bf x}'),
\\[1ex]
\{\pi_i({\bf x},t),\pi_j({\bf x}',t)\} &= \pi_i({\bf x}',t)\frac{\partial}{\partial x'^j} \delta ({\bf x}-{\bf x}')
- \pi_j({\bf x},t)\frac{\partial}{\partial x^i} \delta ({\bf x}-{\bf x}').
\end{align}

Written as Hamiltonian equations of motion, i.e.\ $\partial_t A({\bf x},t) = \{A({\bf x},t), H\}$,
the equations in (i), (ii), and (iv) take the following form
[the equations in (iii) and (v) remain unchanged]:
\begin{enumerate}
\item[(i)] The mass conservation equation
\be
\frac{\partial \varrho_*}{\partial t} = -\partial_i \left(\frac{\delta H}{\delta \pi_i}\varrho_*\right),
\ee
notice $\dst v^i = \frac{\delta H}{\delta\pi_i}$.
\item[(ii)] The equations of motion
\be
\frac{\partial \pi_i}{\partial t} = -\partial_j \left(\frac{\delta H}{\delta \pi_j}\pi_i\right)
- \partial_i \left(\frac{\delta H}{\delta \pi_j}\right) \pi_j
- \partial_i \left(\frac{\delta H}{\delta \varrho_*}\right)\varrho_*
+ \frac{\delta H}{\delta s}  \partial_i s.
\ee
\item[(iv)] The entropy conservation law
\be
\frac{\partial s}{\partial t} = -\frac{\delta H}{\delta \pi_i} \partial_i s,
\ee
where the Hamiltonian functional is given by $H=H[\varrho_*,\pi_i,s]$, see \cite{holm-85}.
\end{enumerate}

Point-mass systems fulfill
\be
h = p = s = 0,
\ee
(just as for dust) and the momentum and mass densities read
\be
\pi_i = \sum_a p_{ai} \delta ({\bf x}-{\bf x}_a), \quad\quad \varrho_* = \sum_a m_a \delta ({\bf x}-{\bf x}_{a}),
\quad\quad v^i_a = \frac{\md x^i_a}{\md t}.
\ee
The position and momentum variables again fulfill the standard Poisson bracket relations,
\be
\{x^i_a,p_{aj}\} = \delta_{ij}, \qquad \mbox{zero otherwise}.
\ee
Hereof the standard Hamilton equations are recovered,
\be
\frac{\md p_{ai}}{\md t} = -\frac{\partial H}{\partial x^i_a},
\quad \frac{\md x^i_a}{\md t} = \frac{\partial H}{\partial p_{ai}}.
\ee
Remarkably, the difference to the Newtonian theory solely results from the Hamiltonian,
so the difference between GR and the Newtonian theory is essentially a dynamical and not a kinematical one.
This statement refers to the matter only and not to the gravitational field.
The latter is much more complicated in GR, dynamically and kinematically as well.

\section{4PN-accurate generators of Poincar\'e symmetry for two-point-mass systems}
\label{app:ncom}

Generators of Poincar\'e symmetry for two-point-mass systems are realized
as functions on the two-body phase-space $(\vecx_1,\vecx_2,\vecp_1,\vecp_2)$.
In the $3+1$ splitting the 10 generators are: Hamiltonian $H$, linear momentum $P^i$,
angular momentum $J^i$, and centre-of-energy vector $G^i$ (related to boost vector $K^i$
through $K^i=G^i-tP^i$). They all fulfill the Poincar\'e algebra relations \eqref{PoiRel1}--\eqref{PoiRel6}.
In this Appendix we show 4PN-accurate formulae for these generators derived within the ADM formalism
(see \citealp{bernard-17c} for recent derivation of corresponding and equivalent formulae
for integrals of motion in harmonic coordinates).

The gauge fixing used in the ADM formalism manifestly respects the Euclidean group
(which means that the Hamiltonian $H$ is translationally and rotationally invariant),
therefore the generators $P^i$ and $J^i$ are simply realized as
\be
P^i (\xa,\pa) = \sum_a p_{ai},
\quad
J^i (\xa,\pa) = \sum_a
\varepsilon_{ik\ell} \, x_a^k \, p_{a\ell}.
\ee
These formula are exact (i.e., valid at all PN orders).

The 4PN-accurate conservative Hamiltonian $H_{\le\textrm{4PN}}$
is the sum of local and nonlocal-in-time parts,
\be
H_{\le\textrm{4PN}}[\xa,\pa] = H_\textrm{$\le$4PN}^\textrm{local}(\xa,\pa)
+ H_\textrm{4PN}^\textrm{nonlocal}[\xa,\pa],
\ee
where the nonlocal-in-time piece equals
\be
\label{appH4PNnonloc}
H_{\textrm{4PN}}^{\textrm{nonlocal}}[\xa,\pa]
= -\frac{1}{5}\frac{G^2M}{c^8} \dddot{I}_{\!ij}(t)
\times \Pf_{2r_{12}/c} \int_{-\infty}^{+\infty} \frac{\md \tau}{\vert \tau \vert} \dddot{I}_{\!ij}(t+\tau).
\ee
The third time derivative of $I_{ij}$,
after replacing all time derivatives of $\xa$ by using the Newtonian equations of motion,
can be written as
\begin{align}
\label{dot3I}
\dddot{I}_{\!ij} &= -\frac{2Gm_1m_2}{r_{12}^2}
\left\{4 n^{\langle i}_{12} \left(\frac{p_{1j\rangle}}{m_1}-\frac{p_{2j\rangle}}{m_2}\right)
- 3 \left(\frac{\npi}{m_1}-\frac{\npii}{m_2}\right) n^{\langle i}_{12}n^{j\rangle}_{12} \right\}
\nonumber\\[1ex]
&=  -\frac{2Gm_1m_2}{r_{12}^3}
\left\{4 x^{\langle i}_{12} v_{12}^{j \rangle}
- \frac{3}{r_{12}} (\mathbf{n}_{12}\cdot\mathbf{v}_{12})x^{\langle i}_{12} x_{12}^{j \rangle}\right\},
\end{align}
where the relative velocity $\mathbf{v}_{12}\equiv\vecp_1/m_1-\vecp_2/m_2$
($\langle\cdots\rangle$ denotes a symmetric tracefree projection).
This formula is valid in an arbitrary reference frame and it is obviously Galileo-invariant.
Consequently the nonlocal-in-time Hamiltonian \eqref{appH4PNnonloc} is Galileo-invariant as well.
The local part of the 4PN-accurate Hamiltonian reads
\begin{align}
H_\textrm{$\le$4PN}^\textrm{local}(\xa,\pa)
&= Mc^2 + H_\textrm{N}(\xa,\pa) + H_\textrm{1PN}(\xa,\pa)
+ H_\textrm{2PN}(\xa,\pa)
\nonumber\\[1ex]&\quad
+ H_\textrm{3PN}(\xa,\pa)
+ H_\textrm{4PN}^\textrm{local}(\xa,\pa).
\end{align}
The Hamiltonians $H_\textrm{N}$ to $H_\textrm{3PN}$ in generic, i.e. noncentre-of-mass, reference frame, are equal to
[the operation ``$+\big(1\leftrightarrow 2\big)$'' used below denotes the addition for each term,
including the ones which are symmetric under the exchange of body labels,
of another term obtained by the label permutation $1\leftrightarrow 2$]
\begin{align}
H_{\text{N}}(\xa,\pa) = \frac{\vecp_1^2}{2m_1} - \frac{G m_1 m_2}{2r_{12}}
+ \big(1\leftrightarrow 2\big),
\end{align}
\begin{align}
c^2\,H_{\text{1PN}}(\xa,\pa) &=
- \frac{\pipip^2}{8m_1^3}
+ \frac{Gm_1m_2}{4r_{12}} \Bigg( -\frac{6\pipi}{m_1^2} + \frac{7\pipj}{m_1m_2}
\nonumber\\[1ex]&\quad
+ \frac{\npi\npj}{m_1m_2} \Bigg)
+ \frac{G^2m_1^2m_2}{2r_{12}^2}
+ \big(1\leftrightarrow2\big),
\end{align}
\begin{align}
c^4\,H_{\text{2PN}}&(\xa,\pa) = \frac{\pipip^3}{16m_1^5}
+ \frac{Gm_1m_2}{8r_{12}} \Bigg(
5\,\frac{\pipip^2}{m_1^4}
- \frac{11}{2}\frac{\pipi\,\pjpj}{m_1^2m_2^2}
- \frac{\pipj^2}{m_1^2m_2^2}
\nonumber\\[1ex]&\quad
+ 5\,\frac{\pipi\,\npj^2}{m_1^2m_2^2}
- 6\,\frac{\pipj\,\npi\npj}{m_1^2m_2^2}
\nonumber\\[1ex]&\quad
- \frac{3}{2}\frac{\npi^2\npj^2}{m_1^2m_2^2} 
\Bigg)
+ \frac{G^2m_1m_2}{4r_{12}^2} \Bigg(
m_2\left(10\frac{\pipi}{m_1^2}+19\frac{\pjpj}{m_2^2}\right)
\nonumber\\[1ex]&\quad
- \frac{1}{2}(m_1+m_2)\frac{27\,\pipj+6\,\npi\npj}{m_1m_2} \Bigg)
\nonumber\\[1ex]&\quad
- \frac{G^3m_1m_2 (m_1^2+5m_1m_2+m_2^2)}{8r_{12}^3}
+ \big(1\leftrightarrow2\big),
\end{align}
\begin{align}
c^6\,&H_{\text{3PN}}(\xa,\pa) =
- \frac{5\pipip^4}{128m_1^7}
+ \frac{Gm_1m_2}{32r_{12}} \Bigg(
- \frac{14\pipip^3}{m_1^6}
\nonumber\\[1ex]&
+ 4\frac{\big(\pipj^2+4\,\pipi\,\pjpj\big)\pipi}{m_1^4m_2^2}
+ \frac{6\pipi\npi^2\npj^2}{m_1^4m_2^2}
\nonumber\\[1ex]&
- 10\frac{\big(\pipi\,\npj^2+\pjpj\,\npi^2\big)\pipi}{m_1^4m_2^2}
\nonumber\\[1ex]&
+ 24\frac{\pipi\,\pipj\npi\npj}{m_1^4m_2^2}
+ 2\frac{\pipi\,\pipj\npj^2}{m_1^3m_2^3}
\nonumber\\[1ex]&
+ \frac{\big(7\,\pipi\,\pjpj-10\,\pipj^2\big)\npi\npj}{m_1^3m_2^3}
\nonumber\\[1ex]&
+ \frac{\big(\pipi\,\pjpj-2\,\pipj^2\big)\pipj}{m_1^3m_2^3}
+ 15\frac{\pipj\npi^2\npj^2}{m_1^3m_2^3}
\nonumber\\[1ex]&
- 18\frac{\pipi\,\npi\npj^3}{m_1^3m_2^3}
+ 5\frac{\npi^3\npj^3}{m_1^3m_2^3} \Bigg)
\nonumber\\[1ex]&
+ \frac{G^2m_1m_2}{r_{12}^2} \Bigg(
\frac{1}{16}(m_1-27m_2)\frac{\pipip^2}{m_1^4}
- \frac{115}{16}m_1\frac{\pipi\,\pipj}{m_1^3m_2}
\nonumber\\[1ex]&
+ \frac{1}{48}m_2\frac{25\,\pipj^2+371\,\pipi\,\pjpj}{m_1^2 m_2^2}
+ \frac{17}{16}\frac{\pipi\npi^2}{m_1^3}
+ \frac{5}{12}\frac{\npi^4}{m_1^3}
\nonumber\\[1ex]&
- \frac{1}{8}m_1 \frac{\big(15\,\pipi\,\npj+11\,\pipj\,\npi\big)\npi}{m_1^3 m_2}
\nonumber\\[2ex]&
- \frac{3}{2}m_1\frac{\npi^3\npj}{m_1^3m_2}
+ \frac{125}{12}m_2\frac{\pipj\,\npi\npj}{m_1^2m_2^2}
\nonumber\\[1ex]&
+ \frac{10}{3}m_2\frac{\npi^2\npj^2}{m_1^2m_2^2}
- \frac{1}{48} (220 m_1 + 193 m_2) \frac{\pipi \npj^2}{m_1^2 m_2^2}
\Bigg)
\nonumber\\[1ex]&
+ \frac{G^3m_1m_2}{r_{12}^3} \Bigg(
-\frac{1}{48}
\bigg(425\,m_1^2+\Big(473-\frac{3}{4}\pi^2\Big)m_1m_2+150\,m_2^2\bigg)
\frac{\pipi}{m_1^2}
\nonumber\\[1ex]&
+ \frac{1}{16}
\bigg(77(m_1^2+m_2^2)+\Big(143-\frac{1}{4}\pi^2\Big)m_1m_2\bigg)
\frac{\pipj}{m_1m_2}
\nonumber\\[1ex]&
+ \frac{1}{16}
\bigg(20\,m_1^2-\Big(43+\frac{3}{4}\pi^2\Big)m_1m_2\bigg)
\frac{\npi^2}{m_1^2}
\nonumber\\[1ex]&
+ \frac{1}{16}
\bigg(21(m_1^2+m_2^2)+\Big(119+\frac{3}{4}\pi^2\Big)m_1m_2\bigg)
\frac{\npi\npj}{m_1m_2} \Bigg)
\nonumber\\[1ex]&
+ \frac{G^4m_1m_2^3}{8r_{12}^4} \Bigg( \bigg(\frac{227}{3}-\frac{21}{4}\pi^2\bigg)m_1+m_2 \Bigg)
+ \big(1\leftrightarrow2\big).
\end{align}
The formula for the Hamiltonian $H_\textrm{4PN}^\textrm{local}$ is large,
therefore we display it in smaller pieces:
\begin{align}
c^8\,H_\textrm{4PN}^\textrm{local}(\xa,\pa) &=
\frac{7 \pipip^5}{256 m_1^9}
+ \frac{G m_1 m_2}{r_{12}} H_{48}(\xa,\pa)
+ \frac{G^2 m_1 m_2}{r_{12}^2} m_1\,H_{46}(\xa,\pa)
\nonumber\\[1ex]&\quad
+ \frac{G^3 m_1 m_2}{r_{12}^3} \Big(m_1^2\,H_{441}(\xa,\pa) + m_1 m_2\,H_{442}(\xa,\pa) \Big)
\nonumber\\[1ex]&\quad
+ \frac{G^4 m_1 m_2}{r_{12}^4} \Big(m_1^3\,H_{421}(\xa,\pa) + m_1^2 m_2\,H_{422}(\xa,\pa)\Big)
\nonumber\\[1ex]&\quad
+ \frac{G^5 m_1 m_2}{r_{12}^5} H_{40}(\xa,\pa)
+ \big(1\leftrightarrow 2\big),
\end{align}
where
\begin{align}
&H_{48}(\xa,\pa) = \frac{45 \pipip^4}{128 m_1^8}
-\frac{9 \npi^2 \npii^2 \pipip^2}{64 m_1^6 m_2^2}
+\frac{15 \npii^2 \pipip^3}{64m_1^6 m_2^2}
\nonumber\\[1ex]&
-\frac{9 \npi \npii \pipip^2 \pipii}{16 m_1^6 m_2^2}
-\frac{3 \pipip^2 \pipii^2}{32m_1^6 m_2^2}
\nonumber\\[1ex]&
+\frac{15 \npi^2 \pipip^2 \piipii}{64 m_1^6 m_2^2}
-\frac{21 \pipip^3 \piipii}{64 m_1^6m_2^2}
-\frac{35 \npi^5 \npii^3}{256 m_1^5 m_2^3}
\nonumber\\[1ex]&
+\frac{25 \npi^3 \npii^3 \pipi}{128 m_1^5m_2^3}
+\frac{33 \npi \npii^3 \pipip^2}{256 m_1^5 m_2^3}
\nonumber\\[1ex]&
-\frac{85 \npi^4 \npii^2 \pipii}{256 m_1^5m_2^3}
-\frac{45 \npi^2 \npii^2 \pipi \pipii}{128 m_1^5 m_2^3}
\nonumber\\[1ex]&
-\frac{\npii^2 \pipip^2 \pipii}{256m_1^5 m_2^3}
+\frac{25 \npi^3 \npii \pipii^2}{64 m_1^5 m_2^3}
\nonumber\\[1ex]&
+\frac{7 \npi \npii \pipi\pipii^2}{64 m_1^5 m_2^3}
-\frac{3 \npi^2 \pipii^3}{64 m_1^5 m_2^3}
+\frac{3 \pipi \pipii^3}{64 m_1^5m_2^3}
\nonumber\\[1ex]&
+\frac{55 \npi^5 \npii \piipii}{256 m_1^5 m_2^3}
-\frac{7 \npi^3 \npii \pipi \piipii}{128m_1^5 m_2^3}
\nonumber\\[1ex]&
-\frac{25 \npi \npii \pipip^2 \piipii}{256 m_1^5 m_2^3}
-\frac{23 \npi^4 \pipii\piipii}{256 m_1^5 m_2^3}
\nonumber\\[1ex]&
+\frac{7 \npi^2 \pipi \pipii \piipii}{128 m_1^5 m_2^3}
-\frac{7 \pipip^2\pipii \piipii}{256 m_1^5 m_2^3}
-\frac{5 \npi^2 \npii^4 \pipi}{64 m_1^4 m_2^4}
\nonumber\\[1ex]&
+\frac{7 \npii^4\pipip^2}{64 m_1^4 m_2^4}
-\frac{\npi \npii^3 \pipi \pipii}{4 m_1^4 m_2^4}
\nonumber\\[1ex]&
+\frac{\npii^2 \pipi\pipii^2}{16 m_1^4 m_2^4}
-\frac{5 \npi^4 \npii^2 \piipii}{64 m_1^4 m_2^4}
\nonumber\\[1ex]&
+\frac{21 \npi^2 \npii^2\pipi \piipii}{64 m_1^4 m_2^4}
-\frac{3 \npii^2 \pipip^2 \piipii}{32 m_1^4 m_2^4}
\nonumber\\[1ex]&
-\frac{\npi^3\npii \pipii \piipii}{4 m_1^4 m_2^4}
+\frac{\npi \npii \pipi \pipii \piipii}{16 m_1^4m_2^4}
\nonumber\\[1ex]&
+\frac{\npi^2 \pipii^2 \piipii}{16 m_1^4 m_2^4}
-\frac{\pipi \pipii^2 \piipii}{32 m_1^4m_2^4}
+\frac{7 \npi^4 \piipiip^2}{64 m_1^4 m_2^4}
\nonumber\\[1ex]&
-\frac{3 \npi^2 \pipi \piipiip^2}{32 m_1^4m_2^4}
-\frac{7 \pipip^2 \piipiip^2}{128 m_1^4 m_2^4},
\end{align}
\begin{align}
H_{46}&(\xa,\pa) =
\frac{369 \npi^6}{160 m_1^6}
-\frac{889 \npi^4 \pipi}{192 m_1^6}
+\frac{49 \npi^2 \pipip^2}{16 m_1^6}
\nonumber\\[1ex]&
-\frac{63\pipip^3}{64 m_1^6}
-\frac{549 \npi^5 \npii}{128 m_1^5 m_2}
+\frac{67 \npi^3 \npii \pipi}{16 m_1^5m_2}
\nonumber\\[1ex]&
-\frac{167 \npi \npii \pipip^2}{128 m_1^5 m_2}
+\frac{1547 \npi^4 \pipii}{256 m_1^5m_2}
\nonumber\\[1ex]&
-\frac{851 \npi^2 \pipi \pipii}{128 m_1^5 m_2}
+\frac{1099 \pipip^2 \pipii}{256 m_1^5 m_2}
\nonumber\\[1ex]&
+\frac{3263\npi^4 \npii^2}{1280 m_1^4 m_2^2}
+\frac{1067 \npi^2 \npii^2 \pipi}{480 m_1^4 m_2^2}
\nonumber\\[1ex]&
-\frac{4567\npii^2 \pipip^2}{3840 m_1^4 m_2^2}
-\frac{3571 \npi^3 \npii \pipii}{320 m_1^4 m_2^2}
\nonumber\\[1ex]&
+\frac{3073\npi \npii \pipi \pipii}{480 m_1^4 m_2^2}
+\frac{4349 \npi^2 \pipii^2}{1280 m_1^4 m_2^2}
\nonumber\\[1ex]&
-\frac{3461\pipi \pipii^2}{3840 m_1^4 m_2^2}
+\frac{1673 \npi^4 \piipii}{1920 m_1^4 m_2^2}
-\frac{1999 \npi^2 \pipi\piipii}{3840 m_1^4 m_2^2}
\nonumber\\[1ex]&
+\frac{2081 \pipip^2 \piipii}{3840 m_1^4 m_2^2}
-\frac{13 \npi^3 \npii^3}{8m_1^3 m_2^3}
+\frac{191 \npi \npii^3 \pipi}{192 m_1^3 m_2^3}
\nonumber\\[1ex]&
-\frac{19 \npi^2 \npii^2 \pipii}{384m_1^3 m_2^3}
-\frac{5 \npii^2 \pipi \pipii}{384 m_1^3 m_2^3}
\nonumber\\[1ex]&
+\frac{11 \npi \npii \pipii^2}{192m_1^3 m_2^3}
+\frac{77 \pipii^3}{96 m_1^3 m_2^3}
\nonumber\\[1ex]&
+\frac{233 \npi^3 \npii \piipii}{96 m_1^3m_2^3}
-\frac{47 \npi \npii \pipi \piipii}{32 m_1^3 m_2^3}
\nonumber\\[1ex]&
+\frac{\npi^2 \pipii \piipii}{384 m_1^3m_2^3}
-\frac{185 \pipi \pipii \piipii}{384 m_1^3 m_2^3}
-\frac{7 \npi^2 \npii^4}{4 m_1^2 m_2^4}
\nonumber\\[1ex]&
+\frac{7\npii^4 \pipi}{4 m_1^2 m_2^4}
-\frac{7 \npi \npii^3 \pipii}{2 m_1^2 m_2^4}
\nonumber\\[1ex]&\quad
+\frac{21 \npii^2\pipii^2}{16 m_1^2 m_2^4}
+\frac{7 \npi^2 \npii^2 \piipii}{6 m_1^2 m_2^4}
\nonumber\\[1ex]&
+\frac{49 \npii^2 \pipi\piipii}{48 m_1^2 m_2^4}
-\frac{133 \npi \npii \pipii \piipii}{24 m_1^2 m_2^4}
\nonumber\\[1ex]&
-\frac{77 \pipii^2\piipii}{96 m_1^2 m_2^4}
+\frac{197 \npi^2 \piipiip^2}{96 m_1^2 m_2^4}
-\frac{173 \pipi \piipiip^2}{48 m_1^2m_2^4}
+\frac{13 \piipiip^3}{8 m_2^6},
\end{align}
\begin{align}
H_{441}&(\xa,\pa) =
\frac{5027 \npi^4}{384 m_1^4}
-\frac{22993 \npi^2 \pipi}{960 m_1^4}
-\frac{6695 \pipip^2}{1152 m_1^4}
\nonumber\\[1ex]&\quad
-\frac{3191\npi^3 \npii}{640 m_1^3 m_2}
+\frac{28561 \npi \npii \pipi}{1920 m_1^3 m_2}
\nonumber\\[1ex]&\quad
+\frac{8777 \npi^2\pipii}{384 m_1^3 m_2}
+\frac{752969 \pipi \pipii}{28800 m_1^3 m_2}
\nonumber\\[1ex]&\quad
-\frac{16481 \npi^2 \npii^2}{960m_1^2 m_2^2}
+\frac{94433 \npii^2 \pipi}{4800 m_1^2 m_2^2}
\nonumber\\[1ex]&\quad
-\frac{103957 \npi \npii \pipii}{2400 m_1^2m_2^2}
+\frac{791 \pipii^2}{400 m_1^2 m_2^2}
\nonumber\\[1ex]&\quad
+\frac{26627 \npi^2 \piipii}{1600 m_1^2 m_2^2}
-\frac{118261 \pipi\piipii}{4800 m_1^2 m_2^2}
+\frac{105 \piipiip^2}{32 m_2^4},
\end{align}
\begin{align}
&H_{442}(\xa,\pa) =
\left(\frac{2749\pi^2}{8192}-\frac{211189}{19200}\right) \frac{\pipip^2}{m_1^4}
+\left(\frac{375\pi^2}{8192}-\frac{23533}{1280}\right)\frac{\npi^4}{m_1^4}
\nonumber\\[1ex]&
+\left(\frac{63347}{1600}-\frac{1059\pi^2}{1024}\right)\frac{\npi^2 \pipi}{m_1^4}
+\left(\frac{10631\pi^2}{8192}-\frac{1918349}{57600}\right) \frac{\pipii^2}{m_1^2 m_2^2}
\nonumber\\[1ex]&
+\left(\frac{13723\pi^2}{16384}-\frac{2492417}{57600}\right) \frac{\pipi\piipii}{m_1^2 m_2^2}
+\left(\frac{1411429}{19200}-\frac{1059\pi^2}{512}\right)\frac{\npii^2\pipi}{m_1^2 m_2^2}
\nonumber\\[1ex]&
+\left(\frac{248991}{6400}-\frac{6153\pi^2}{2048}\right) \frac{\npi\npii\pipii}{m_1^2 m_2^2}
\nonumber\\[1ex]&
-\left(\frac{30383}{960}+\frac{36405\pi^2}{16384}\right) \frac{\npi^2 \npii^2}{m_1^2m_2^2}
\nonumber\\[1ex]&
+\left(\frac{2369}{60}+\frac{35655\pi^2}{16384}\right) \frac{\npi^3 \npii}{m_1^3 m_2}
\nonumber\\[1ex]&
+\left(\frac{1243717}{14400}-\frac{40483\pi^2}{16384}\right) \frac{\pipi\pipii}{m_1^3m_2}
\nonumber\\[1ex]&
+\left(\frac{43101\pi^2}{16384}-\frac{391711}{6400}\right) \frac{\npi \npii \pipi}{m_1^3 m_2}
\nonumber\\[1ex]&
+\left(\frac{56955\pi^2}{16384}-\frac{1646983}{19200}\right) \frac{\npi^2 \pipii}{m_1^3 m_2},
\end{align}
\begin{align}
H_{421}(\xa,\pa) &= \frac{64861 \pipi}{4800 m_1^2}
- \frac{91\pipii}{8 m_1 m_2}
+ \frac{105\piipii}{32 m_2^2}
\nonumber\\[1ex]&\quad
- \frac{9841\npi^2}{1600m_1^2}
- \frac{7 \npi \npii}{2 m_1 m_2},
\end{align}
\begin{align}
H_{422}(\xa,\pa) &=
\left(\frac{1937033}{57600}-\frac{199177 \pi ^2}{49152}\right) \frac{\pipi}{m_1^2}
+\left(\frac{282361}{19200}-\frac{21837 \pi ^2}{8192}\right)\frac{\piipii}{m_2^2}
\nonumber\\[1ex]&\quad
+\left(\frac{176033 \pi ^2}{24576}-\frac{2864917}{57600}\right)\frac{\pipii}{m_1 m_2}
\nonumber\\[1ex]&\quad
+\left(\frac{698723}{19200}+\frac{21745 \pi ^2}{16384}\right) \frac{\npi^2}{m_1^2}
\nonumber\\[1ex]&\quad
+\left(\frac{63641 \pi^2}{24576}-\frac{2712013}{19200}\right) \frac{\npi \npii}{m_1 m_2}
\nonumber\\[1ex]&\quad
+\left(\frac{3200179}{57600}-\frac{28691 \pi ^2}{24576}\right)\frac{\npii^2}{m_2^2},
\end{align}
\begin{align}
H_{40}(\xa,\pa) &= -\frac{m_1^4}{16}
+\left(\frac{6237 \pi^2}{1024}-\frac{169799}{2400}\right) m_1^3 m_2
\nonumber\\[1ex]&\quad
+\left(\frac{44825 \pi^2}{6144}-\frac{609427}{7200}\right)m_1^2 m_2^2.
\end{align}

The centre-of-energy vector $G^i(\xa,\pa)$ was constructed with 3PN-accuracy
(using the method of undetermined coefficients) by \cite{damour3-00,damour3-00-e},
and at the 4PN level by \cite{jaranowski-15}. It can be written as\footnote{
Let us note that the centre-of-energy vector $G^i$ does not contain a nonlocal-in-time piece
which would correspond to the nonlocal-in-time tail-related part of the 4PN Hamiltonian.
The very reason for this is that the integrals contributing to $G^i_{\text{4PN}}$
are less singular than those for $H_{\text{4PN}}$,
and the singular structure of terms contributing to $G^i_{\text{4PN}}$
rather relates to the singular structure of terms contributing to $H_{\text{3PN}}$.}
\be
G^i(\xa,\pa) = \sum_a \Big( M_a({\bf x}_b,{\bf p}_b)\,x_a^i
+ N_a({\bf x}_b,{\bf p}_b)\,p_{ai} \Big),
\ee
where the functions $M_a$ and $N_a$ possess the following 4PN-accurate expansions
\begin{align}
M_a(\xa,\pa) &= m_a + \frac{1}{c^2}\,M_a^{\rm 1PN}(\xa,\pa)
+ \frac{1}{c^4} \, M_a^{\rm 2PN}(\xa,\pa)
\nonumber\\[1ex]&\quad
+ \frac{1}{c^6} \, M_a^{\rm 3PN}(\xa,\pa) + \frac{1}{c^8} \, M_a^{\rm 4PN}(\xa,\pa),
\\[1ex]
N_a(\xa,\pa) &= \frac{1}{c^4} \, N_a^{\rm 2PN}(\xa,\pa)
+ \frac{1}{c^6} \, N_a^{\rm 3PN}(\xa,\pa) + \frac{1}{c^8} \, N_a^{\rm 4PN}(\xa,\pa).
\end{align}
The functions $M_1^{\rm 1PN}$ to $M_1^{\rm 3PN}$ read
\be
M_1^{\rm 1PN}(\xa,\pa) = \frac{\pipi}{2m_1} - \frac{Gm_1m_2}{2r_{12}},
\ee
\begin{align}
M^{\text{2PN}}_1(\xa,\pa) &= -\frac{\pipip^2}{8m_1^3}
+ \frac{Gm_1m_2}{4r_{12}} \Bigg( -\frac{5\pipi}{m_1^2} - \frac{\pjpj}{m_2^2} + \frac{7\pipj}{m_1m_2}
\nonumber\\[1ex]&\quad
+ \frac{\npi\npj}{m_1m_2} \Bigg)
+ \frac{G^2m_1m_2(m_1+m_2)}{4r_{12}^2},
\end{align}
\begin{align}
M^{\text{3PN}}_1&(\xa,\pa) = \frac{\pipip^3}{16m_1^5}
+ \frac{Gm_1m_2}{16r_{12}} \Bigg(
9\,\frac{\pipip^2}{m_1^4}
+ \frac{\pjpjp^2}{m_2^4}
- 11\,\frac{\pipi\,\pjpj}{m_1^2m_2^2}
\nonumber\\[1ex]&\quad
- 2\,\frac{\pipj^2}{m_1^2m_2^2}
+ 3\,\frac{\pipi\,\npj^2}{m_1^2m_2^2}
+ 7\,\frac{\pjpj\,\npi^2}{m_1^2m_2^2}
\nonumber\\[1ex]&\quad
- 12\,\frac{\pipj\,\npi\npj}{m_1^2m_2^2}
- 3\,\frac{\npi^2\npj^2}{m_1^2m_2^2} \Bigg)
\nonumber\\[1ex]&\quad
+ \frac{G^2m_1m_2}{24r_{12}^2} \Bigg(
(112m_1+45m_2)\frac{\pipi}{m_1^2}
+ (15m_1+2m_2)\frac{\pjpj}{m_2^2}
\nonumber\\[1ex]&\quad
- \frac{1}{2}(209m_1+115m_2)\frac{\pipj}{m_1m_2}
+ \frac{\npi^2}{m_1}
- \frac{\npj^2}{m_2}
\nonumber\\[1ex]&\quad
- (31m_1+5m_2)\frac{\npi\npj}{m_1m_2}
\Bigg)
\nonumber\\[2ex]&\quad
- \frac{G^3m_1m_2 (m_1^2+5m_1m_2+m_2^2)}{8r_{12}^3}.
\end{align}
The function $M_1^{\textrm{4PN}}$ has the following structure:
\begin{align}
M_1^{\textrm{4PN}}(\xa,\pa) &= -\frac{5\pipip^4}{128 m_1^7}
+ \frac{G m_1 m_2}{r_{12}} M_{46}(\xa,\pa)
\nonumber\\[1ex]&\quad
+ \frac{G^2 m_1 m_2}{r_{12}^2} \Big(m_1\,M_{441}(\xa,\pa) + m_2\,M_{442}(\xa,\pa)\Big)
\nonumber\\[1ex]&\quad
+ \frac{G^3 m_1 m_2}{r_{12}^3} \Big( m_1^2\,M_{421}(\xa,\pa)
+ m_1 m_2\,M_{422}(\xa,\pa)
\nonumber\\[1ex]&\quad
+ m_2^2\,M_{423}(\xa,\pa) \Big)
+ \frac{G^4 m_1 m_2}{r_{12}^4}M_{40}(\xa,\pa),
\end{align}
where
\begin{align}
M_{46}&(\xa,\pa) =
-\frac{13 \pipip^3}{32 m_1^6}
-\frac{15 \npi^4 \npii^2}{256 m_1^4m_2^2}
-\frac{91 \npii^2 \pipip^2}{256 m_1^4 m_2^2}
\nonumber\\[1ex]&\quad
+\frac{45 \npi^2 \npii^2 \pipi}{128 m_1^4m_2^2}
-\frac{5\npi^3 \npii \pipii}{32 m_1^4 m_2^2}
\nonumber\\[1ex]&\quad
+\frac{25 \npi\npii \pipi \pipii}{32 m_1^4 m_2^2}
+\frac{5 \npi^2\pipii^2}{64 m_1^4 m_2^2}
\nonumber\\[1ex]&\quad
+\frac{7 \pipi \pipii^2}{64 m_1^4m_2^2}
+\frac{11 \npi^4 \piipii}{256 m_1^4 m_2^2}
-\frac{47\npi^2 \pipi \piipii}{128 m_1^4 m_2^2}
\nonumber\\[1ex]&\quad
+\frac{91 \pipip^2\piipii}{256 m_1^4 m_2^2}
+\frac{5 \npi^3 \npii^3}{32 m_1^3m_2^3}
-\frac{7 \npi \npii^3 \pipi}{32 m_1^3 m_2^3}
\nonumber\\[1ex]&\quad
+\frac{15 \npi^2 \npii^2 \pipii}{32 m_1^3 m_2^3}
+\frac{7 \npii^2\pipi \pipii}{32 m_1^3 m_2^3}
\nonumber\\[1ex]&\quad
-\frac{5 \npi \npii\pipii^2}{16 m_1^3 m_2^3}
-\frac{11\npi^3 \npii \piipii}{32 m_1^3 m_2^3}
\nonumber\\[1ex]&\quad
-\frac{\pipii^3}{16 m_1^3 m_2^3}
+\frac{7 \npi\npii \pipi \piipii}{32 m_1^3 m_2^3}
-\frac{5 \npi^2 \pipii\piipii}{32 m_1^3 m_2^3}
\nonumber\\[1ex]&\quad
+\frac{\pipi \pipii \piipii}{32 m_1^3m_2^3}
+\frac{15 \npi^2 \npii^4}{256 m_1^2 m_2^4}
-\frac{11\npii^4 \pipi}{256 m_1^2 m_2^4}
\nonumber\\[1ex]&\quad
+\frac{5 \npi \npii^3\pipii}{32 m_1^2 m_2^4}
-\frac{5 \npii^2 \pipii^2}{64 m_1^2m_2^4}
\nonumber\\[1ex]&\quad
-\frac{21 \npi^2 \npii^2 \piipii}{128 m_1^2 m_2^4}
+\frac{7\npii^2 \pipi \piipii}{128 m_1^2 m_2^4}
+\frac{\pipii^2 \piipii}{64 m_1^2m_2^4}
\nonumber\\[1ex]&\quad
-\frac{\npi \npii\pipii \piipii}{32 m_1^2 m_2^4}
+\frac{11 \npi^2 \piipiip^2}{256 m_1^2 m_2^4}
\nonumber\\[1ex]&\quad
+\frac{37 \pipi\piipiip^2}{256 m_1^2 m_2^4}
-\frac{\piipiip^3}{32 m_2^6},
\end{align}
\begin{align}
M_{441}&(\xa,\pa) =
\frac{7711 \npi^4}{3840 m_1^4}
-\frac{2689 \npi^2 \pipi}{3840m_1^4}
+\frac{2683 \pipip^2}{1920 m_1^4}
\nonumber\\[1ex]&
-\frac{67 \npi^3 \npii}{30m_1^3 m_2}
+\frac{1621 \npi \npii \pipi}{1920 m_1^3m_2}
\nonumber\\[1ex]&
-\frac{411 \npi^2 \pipii}{1280 m_1^3 m_2}
-\frac{25021 \pipi\pipii}{3840 m_1^3 m_2}
+\frac{289 \npi^2 \npii^2}{128 m_1^2m_2^2}
\nonumber\\[1ex]&
-\frac{259 \npii^2 \pipi}{128 m_1^2 m_2^2}
+\frac{689\npi \npii \pipii}{192 m_1^2 m_2^2}
+\frac{11 \pipii^2}{48m_1^2 m_2^2}
\nonumber\\[1ex]&
-\frac{147 \npi^2 \piipii}{64 m_1^2 m_2^2}
+\frac{283\pipi \piipii}{64 m_1^2 m_2^2}
+\frac{7 \npi \npii^3}{12m_1 m_2^3}
\nonumber\\[1ex]&
+\frac{49\npii^2\pipii}{48 m_1 m_2^3}
-\frac{7\npi\npii\piipii}{6 m_1 m_2^3}
\nonumber\\[1ex]&
-\frac{7\pipii\piipii}{48m_1 m_2^3}
-\frac{9\piipiip^2}{32 m_2^4},
\end{align}
\begin{align}
M_{442}&(\xa,\pa) =
-\frac{45 \pipip^2}{32 m_1^4}
+\frac{7 \pipi \pipii}{48 m_1^3m_2}
+\frac{7 \npi \npii \pipi}{6 m_1^3 m_2}
\nonumber\\[1ex]&
-\frac{49\npi^2 \pipii}{48 m_1^3 m_2}
-\frac{7 \npi^3 \npii}{12 m_1^3 m_2}
+\frac{7 \pipii^2}{24 m_1^2 m_2^2}
\nonumber\\[1ex]&
+\frac{635 \pipi\piipii}{192 m_1^2 m_2^2}
-\frac{983 \npi^2 \piipii}{384 m_1^2m_2^2}
+\frac{413 \npi^2 \npii^2}{384 m_1^2 m_2^2}
\nonumber\\[1ex]&
-\frac{331\npii^2 \pipi}{192 m_1^2 m_2^2}
+\frac{437 \npi \npii\pipii}{64 m_1^2 m_2^2}
\nonumber\\[1ex]&
+\frac{11 \npi \npii^3}{15 m_1m_2^3}
-\frac{1349 \npii^2 \pipii}{1280 m_1 m_2^3}
\nonumber\\[1ex]&
-\frac{5221\npi \npii \piipii}{1920 m_1 m_2^3}
-\frac{2579 \pipii\piipii}{3840 m_1 m_2^3}
+\frac{6769 \npii^2 \piipii}{3840m_2^4}
\nonumber\\[1ex]&
-\frac{2563 \piipiip^2}{1920 m_2^4}
-\frac{2037 \npii^4}{1280 m_2^4},
\end{align}
\begin{align}
M_{421}(\xa,\pa) &=
-\frac{179843 \pipi}{14400 m_1^2}
+\frac{10223 \pipii}{1200 m_1m_2}
-\frac{15 \piipii}{16 m_2^2}
\nonumber\\[1ex]&\quad
+\frac{8881 \npi \npii}{2400 m_1m_2}
+\frac{17737 \npi^2}{1600 m_1^2},
\end{align}
\begin{align}
& M_{422}(\xa,\pa) =
\left(\frac{8225 \pi ^2}{16384}-\frac{12007}{1152}\right)\frac{\pipi}{m_1^2}
+\left(\frac{143}{16}-\frac{\pi ^2}{64}\right)\frac{\pipii}{m_1m_2}
\qquad\nonumber\\[1ex]&
+\left(\frac{655}{1152}-\frac{7969 \pi ^2}{16384}\right)\frac{\piipii}{m_2^2}
+\left(\frac{6963 \pi^2}{16384}-\frac{40697}{3840}\right)\frac{\npi^2}{m_1^2}
\nonumber\\[1ex]&
+\left(\frac{119}{16}+\frac{3 \pi ^2}{64}\right) \frac{\npi\npii}{m_1 m_2}
+\left(\frac{30377}{3840}-\frac{7731 \pi^2}{16384}\right)\frac{\npii^2}{m_2^2},
\end{align}
\begin{align}
M_{423}(\xa,\pa) &=
- \frac{35\pipi}{16 m_1^2}
+ \frac{1327\pipii}{1200 m_1 m_2}
+ \frac{52343\piipii}{14400 m_2^2}
\nonumber\\[1ex]&\quad
- \frac{2581\npi\npii}{2400 m_1 m_2}
- \frac{15737\npii^2}{1600 m_2^2},
\end{align}
\begin{align}
M_{40}(\xa,\pa) &=
\frac{m_1^3}{16}
+ \left(\frac{3371\pi^2}{6144}-\frac{6701}{1440}\right) m_1^2 m_2
\nonumber\\[1ex]&\quad
+ \left(\frac{20321}{1440}-\frac{7403\pi^2}{6144}\right) m_1 m_2^2
+ \frac{m_2^3}{16}.
\end{align}
The functions $N_1^{\rm 2PN}$ and $N_1^{\rm 3PN}$ equal
\be
N^{\text{2PN}}_1(\xa,\pa) = -\frac{5}{4}\,G\,\npj,
\ee
\begin{align}
N^{\text{3PN}}_1(\xa,\pa) &= \frac{G}{8m_1m_2} \Big( 2\,\pipj\npj - \pjpj\,\npi
\nonumber\\[1ex]&\quad
+ 3\,\npi\npj^2 \Big)
+ \frac{G^2}{48r_{12}}
\Big( 19\,m_2\,\npi
\nonumber\\[1ex]&\quad
+ \left(130\,m_1+137\,m_2\right)\npj \Big).
\end{align}
The more complicated function $N_1^{\textrm{4PN}}$ has the structure:
\begin{align}
N_1^{\rm 4PN}(\xa,\pa) &= G m_2 N_{45}(\xa,\pa)
+ \frac{G^2 m_2}{r_{12}} \Big(m_1\,N_{431}(\xa,\pa)
\nonumber\\&\quad
+ m_2\,N_{432}(\xa,\pa)\Big)
+ \frac{G^3 m_2}{r_{12}^2} \Big(m_1^2\,N_{411}(\xa,\pa)
\nonumber\\&\quad
+ m_1 m_2\,N_{412}(\xa,\pa)
+ m_2^2\,N_{413}(\xa,\pa)\Big),
\end{align}
where
\begin{align}
&N_{45}(\xa,\pa) =
-\frac{5 \npi^3 \npii^2}{64 m_1^3 m_2^2}
+\frac{\npi \npii^2\pipi}{64 m_1^3 m_2^2}
\nonumber\\[1ex]&
+\frac{5 \npi^2 \npii \pipii}{32m_1^3 m_2^2}
-\frac{\npii \pipi \pipii}{32 m_1^3m_2^2}
\nonumber\\[1ex]&
+\frac{3 \npi \pipii^2}{32 m_1^3 m_2^2}
-\frac{\npi^3\piipii}{64 m_1^3 m_2^2}
-\frac{\npi \pipi \piipii}{64 m_1^3m_2^2}
\nonumber\\[1ex]&
+\frac{\npi^2 \npii^3}{32 m_1^2 m_2^3}
-\frac{7 \npii^3\pipi}{32 m_1^2 m_2^3}
+\frac{3 \npi \npii^2 \pipii}{16m_1^2 m_2^3}
\nonumber\\[1ex]&
+\frac{\npii \pipii^2}{16 m_1^2 m_2^3}
-\frac{9\npi^2 \npii \piipii}{32 m_1^2 m_2^3}
+\frac{5 \npii \pipi\piipii}{32 m_1^2 m_2^3}
\nonumber\\[1ex]&
-\frac{3 \npi \pipii \piipii}{16 m_1^2m_2^3}
-\frac{11 \npi \npii^4}{128 m_1 m_2^4}
+\frac{\npii^3\pipii}{32 m_1 m_2^4}
\nonumber\\[1ex]&
+\frac{7 \npi \npii^2 \piipii}{64 m_1m_2^4}
+\frac{\npii \pipii \piipii}{32 m_1 m_2^4}
-\frac{3\npi \piipiip^2}{128 m_1 m_2^4},
\end{align}
\begin{align}
N_{431}&(\xa,\pa) =
-\frac{387 \npi^3}{1280 m_1^3}
+\frac{10429 \npi \pipi}{3840m_1^3}
\nonumber\\[1ex]&
-\frac{751 \npi^2 \npii}{480 m_1^2 m_2}
+\frac{2209\npii \pipi}{640 m_1^2 m_2}
-\frac{6851 \npi \pipii}{1920m_1^2 m_2}
\nonumber\\[1ex]&
+\frac{43 \npi \npii^2}{192 m_1 m_2^2}
-\frac{125\npii \pipii}{192 m_1 m_2^2}
+\frac{25 \npi \piipii}{48 m_1m_2^2}
\nonumber\\[1ex]&
-\frac{7 \npii^3}{8 m_2^3}
+\frac{7 \npii \piipii}{12 m_2^3},
\end{align}
\begin{align}
N_{432}&(\xa,\pa) =
\frac{7 \npii \pipi}{48 m_1^2 m_2}
+\frac{7 \npi \pipii}{24m_1^2 m_2}
-\frac{49 \npi^2 \npii}{48 m_1^2 m_2}
\nonumber\\[1ex]&\quad
+\frac{295\npi \npii^2}{384 m_1 m_2^2}
-\frac{5 \npii \pipii}{24m_1 m_2^2}
-\frac{155 \npi \piipii}{384 m_1 m_2^2}
\nonumber\\[1ex]&\quad
-\frac{5999\npii^3}{3840 m_2^3}
+\frac{11251 \npii \piipii}{3840 m_2^3},
\end{align}
\begin{align}
N_{411}(\xa,\pa) = -\frac{37397\npi}{7200m_1} - \frac{12311\npii}{2400m_2},
\end{align}
\begin{align}
N_{412}(\xa,\pa) &=
\left(\frac{5005\pi^2}{8192}-\frac{81643}{11520}\right)\frac{\npi}{m_1}
\nonumber\\[1ex]&\quad
+ \left(\frac{773\pi^2}{2048}-\frac{61177}{11520}\right)\frac{\npii}{m_2},
\end{align}
\be
N_{413}(\xa,\pa) = -\frac{7073\npii}{1200 m_2}.
\ee

\section{Higher-order spin-dependent conservative Hamiltonians}
\label{app:spinH}

In this appendix we present explicit formulae
for higher-order spin-dependent conservative Hamiltonians
not displayed in the main body of the review.
We start with the next-to-next-to-leading-order spin-orbit Hamiltonian,
which was calculated by \cite{hartung-13} (see also \citealp{hartung1-11}). It reads
\begin{align}
\label{eq:HNNLOSO}
&H^{\text{NNLO}}_{\text{SO}}(\xa,\pa,\mathbf{S}_a) = \frac{G}{c^6\rel^2} \bigg\{
\biggl(
\frac{7 m_2 (\vmom{1}^2)^2}{16 m_1^5}
+ \frac{9 \scpm{\vnun}{\vmom{1}}\scpm{\vnun}{\vmom{2}}\vmom{1}^2}{16 m_1^4}
\nonumber\\&
+ \frac{3 \vmom{1}^2 \scpm{\vnun}{\vmom{2}}^2}{4 m_1^3 m_2}
+ \frac{45 \scpm{\vnun}{\vmom{1}}\scpm{\vnun}{\vmom{2}}^3}{16 m_1^2 m_2^2}
+ \frac{9 \vmom{1}^2 \scpm{\vmom{1}}{\vmom{2}}}{16 m_1^4}
\nonumber\\&
- \frac{3 \scpm{\vnun}{\vmom{2}}^2 \scpm{\vmom{1}}{\vmom{2}}}{16 m_1^2 m_2^2}
- \frac{3 (\vmom{1}^2) (\vmom{2}^2)}{16 m_1^3 m_2}
- \frac{15 \scpm{\vnun}{\vmom{1}}\scpm{\vnun}{\vmom{2}} \vmom{2}^2}{16 m_1^2 m_2^2}
\nonumber\\&
+ \frac{3 \scpm{\vnun}{\vmom{2}}^2 \vmom{2}^2}{4 m_1 m_2^3}
- \frac{3 \scpm{\vmom{1}}{\vmom{2}} \vmom{2}^2}{16 m_1^2 m_2^2}
- \frac{3 (\vmom{2}^2)^2}{16 m_1 m_2^3}
\biggr)((\vnun\times\vmom{1})\cdot\vspin{1})
\nonumber\\&
+ \biggl(
- \frac{3 \scpm{\vnun}{\vmom{1}}\scpm{\vnun}{\vmom{2}}\vmom{1}^2}{2 m_1^3 m_2}
- \frac{15 \scpm{\vnun}{\vmom{1}}^2\scpm{\vnun}{\vmom{2}}^2}{4 m_1^2 m_2^2}
\nonumber\\&
+ \frac{3 \vmom{1}^2 \scpm{\vnun}{\vmom{2}}^2}{4 m_1^2 m_2^2}
- \frac{\vmom{1}^2 \scpm{\vmom{1}}{\vmom{2}}}{2 m_1^3 m_2}
+ \frac{\scpm{\vmom{1}}{\vmom{2}}^2}{2 m_1^2 m_2^2}
\nonumber\\&
+ \frac{3 \scpm{\vnun}{\vmom{1}}^2 \vmom{2}^2}{4 m_1^2 m_2^2}
- \frac{(\vmom{1}^2) (\vmom{2}^2)}{4 m_1^2 m_2^2}
- \frac{3 \scpm{\vnun}{\vmom{1}}\scpm{\vnun}{\vmom{2}}\vmom{2}^2}{2 m_1 m_2^3}
\nonumber\\&
- \frac{\scpm{\vmom{1}}{\vmom{2}} \vmom{2}^2}{2 m_1 m_2^3}
\biggr)((\vnun \times \vmom{2})\cdot\vspin{1})
+ \biggl(
- \frac{9 \scpm{\vnun}{\vmom{1}} \vmom{1}^2}{16 m_1^4}
+ \frac{\vmom{1}^2 \scpm{\vnun}{\vmom{2}}}{m_1^3 m_2}
\nonumber\\&
+ \frac{27 \scpm{\vnun}{\vmom{1}}\scpm{\vnun}{\vmom{2}}^2}{16 m_1^2 m_2^2}
- \frac{\scpm{\vnun}{\vmom{2}}\scpm{\vmom{1}}{\vmom{2}}}{8 m_1^2 m_2^2}
- \frac{5 \scpm{\vnun}{\vmom{1}} \vmom{2}^2}{16 m_1^2 m_2^2}
\nonumber\\&
+ \frac{\scpm{\vnun}{\vmom{2}}\vmom{2}^2}{m_1 m_2^3}
\biggr)((\vmom{1} \times \vmom{2})\cdot\vspin{1})
\bigg\}
\nonumber\\&
+ \frac{G^2}{c^6\rel^3} \Bigg\{
\bigg[
\left(\frac{27m_2^2}{8m_1^3}-\frac{3m_2}{2m_1^2}\right)\vmom{1}^2
-\frac{3m_2\scpm{\vnun}{\vmom{1}}^2}{2 m_1^2}
+\left(
\frac{177}{16 m_1}
+\frac{11}{m_2}
\right) \scpm{\vnun}{\vmom{2}}^2
\nonumber\\&
+\left(
\frac{11}{2 m_1}
+\frac{9 m_2}{2 m_1^2}
\right) \scpm{\vnun}{\vmom{1}} \scpm{\vnun}{\vmom{2}}
+\left(
\frac{23}{4 m_1}
+\frac{9 m_2}{2 m_1^2}
\right) \scpm{\vmom{1}}{\vmom{2}}
\nonumber\\&
-\left(
\frac{159}{16 m_1}
+\frac{37}{8 m_2}
\right) \vmom{2}^2
\bigg]((\vnun\times\vmom{1})\cdot\vspin{1})
+\biggl[
\frac{4 \scpm{\vnun}{\vmom{1}}^2}{m_1}
+\frac{13 \vmom{1}^2}{2 m_1}
\nonumber\\&
+\frac{5 \scpm{\vnun}{\vmom{2}}^2}{m_2}
+\frac{53 \vmom{2}^2}{8 m_2}
- \left(
\frac{211}{8 m_1}
+\frac{22}{m_2}
\right) \scpm{\vnun}{\vmom{1}} \scpm{\vnun}{\vmom{2}}
\nonumber\\&
-\left(
\frac{47}{8 m_1} + \frac{5}{m_2}
\right)\scpm{\vmom{1}}{\vmom{2}}
\biggr]((\vnun \times \vmom{2})\cdot\vspin{1})
+ \bigg[
-\left(
\frac{8}{m_1}
+\frac{9 m_2}{2 m_1^2}
\right)\scpm{\vnun}{\vmom{1}}
\nonumber\\&
+ \left(\frac{59}{4 m_1} + \frac{27}{2 m_2}\right) \scpm{\vnun}{\vmom{2}}
\bigg] ((\vmom{1}\times\vmom{2})\cdot\vspin{1})
\Bigg\}
\nonumber\\&
+ \frac{G^3}{c^6\rel^4} \bigg\{
\left(
\frac{181 m_1 m_2}{16}
+ \frac{95 m_2^2}{4}
+ \frac{75 m_2^3}{8 m_1}
\right) ((\vnun\times\vmom{1})\cdot\vspin{1})
\nonumber\\&
- \left(\frac{21 m_1^2}{2} + \frac{473 m_1 m_2}{16} + \frac{63 m_2^2}{4}
\right)((\vnun\times\vmom{2})\cdot\vspin{1})
\bigg\}
+ (1\leftrightarrow2).
\end{align}

The next-to-next-to-leading-order spin1-spin2 Hamiltonian was calculated
for the first time by \cite{hartung-13}. Its explicit form reads
\begin{align}
\label{eq:HNNLOS1S2}
&H^{\text{NNLO}}_{S_1S_2}(\xa,\pa,\mathbf{S}_a) = \frac{G}{c^6\rel^3} \bigg\{
\frac{((\vmom{1} \times \vmom{2})\cdot\vspin{1})((\vmom{1} \times \vmom{2})\cdot\vspin{2})}{16 m_1^2 m_2^2}
\nonumber\\&\kern-4ex
- \frac{9((\vmom{1} \times \vmom{2})\cdot\vspin{1})((\vnun \times \vmom{2})\cdot\vspin{2})\scpm{\vnun}{\vmom{1}}}{8 m_1^2 m_2^2}
\nonumber\\&\kern-4ex
- \frac{3((\vnun \times \vmom{2})\cdot\vspin{1})((\vmom{1} \times \vmom{2})\cdot\vspin{2})\scpm{\vnun}{\vmom{1}}}{2 m_1^2 m_2^2}
\nonumber\\&\kern-4ex
+ ((\vnun \times \vmom{1})\cdot\vspin{1})((\vnun \times \vmom{1})\cdot\vspin{2})
\biggl(\frac{9 \vmom{1}^2}{8 m_1^4} + \frac{15 \scpm{\vnun}{\vmom{2}}^2}{4 m_1^2 m_2^2} - \frac{3 \vmom{2}^2}{4 m_1^2 m_2^2} \biggr)
\nonumber\\&\kern-4ex
+ ((\vnun \times \vmom{2})\cdot\vspin{1})((\vnun \times \vmom{1})\cdot\vspin{2})
\biggl(
- \frac{3 \vmom{1}^2}{2 m_1^3 m_2}
+ \frac{3 \scpm{\vmom{1}}{\vmom{2}}}{4 m_1^2 m_2^2}
\nonumber\\&\kern-4ex
- \frac{15\scpm{\vnun}{\vmom{1}}\scpm{\vnun}{\vmom{2}}}{4 m_1^2 m_2^2}
\biggr)
+((\vnun \times \vmom{1})\cdot\vspin{1})((\vnun \times \vmom{2})\cdot\vspin{2})
\nonumber\\&\kern-4ex
\times\biggl(
\frac{3\vmom{1}^2}{16 m_1^3 m_2}
-\frac{3 \scpm{\vmom{1}}{\vmom{2}}}{16 m_1^2 m_2^2}
-\frac{15 \scpm{\vnun}{\vmom{1}}\scpm{\vnun}{\vmom{2}}}{16 m_1^2 m_2^2}
\biggr)
\nonumber\\&\kern-4ex
+ \scpm{\vmom{1}}{\vspin{1}} \scpm{\vmom{1}}{\vspin{2}}
\bigg(\frac{3\scpm{\vnun}{\vmom{2}}^2}{4 m_1^2 m_2^2} - \frac{\vmom{2}^2}{4 m_1^2 m_2^2}\bigg)
\nonumber\\&\kern-4ex
+ \scpm{\vmom{1}}{\vspin{1}} \scpm{\vmom{2}}{\vspin{2}}
\bigg(-\frac{\vmom{1}^2}{4 m_1^3 m_2} + \frac{\scpm{\vmom{1}}{\vmom{2}}}{4 m_1^2 m_2^2}\bigg)
\nonumber\\&\kern-4ex
+ \scpm{\vmom{2}}{\vspin{1}}\scpm{\vmom{1}}{\vspin{2}}
\bigg(\frac{5\vmom{1}^2}{16 m_1^3 m_2} - \frac{3\scpm{\vmom{1}}{\vmom{2}}}{16 m_1^2 m_2^2}
- \frac{9\scpm{\vnun}{\vmom{1}}\scpm{\vnun}{\vmom{2}}}{16 m_1^2 m_2^2}\bigg)
\nonumber\\&\kern-4ex
+ \scpm{\vnun}{\vspin{1}}\scpm{\vmom{1}}{\vspin{2}}
\bigg(\frac{9\scpm{\vnun}{\vmom{1}}\vmom{1}^2}{8 m_1^4}
- \frac{3\scpm{\vnun}{\vmom{2}}\vmom{1}^2}{4 m_1^3 m_2}
- \frac{3\scpm{\vnun}{\vmom{2}}\vmom{2}^2}{4 m_1 m_2^3}\bigg)
\nonumber\\&\kern-4ex
+ \scpm{\vmom{1}}{\vspin{1}}\scpm{\vnun}{\vspin{2}}
\bigg(-\frac{3 \scpm{\vnun}{\vmom{2}} \vmom{1}^2}{4 m_1^3 m_2}
- \frac{15 \scpm{\vnun}{\vmom{1}}\scpm{\vnun}{\vmom{2}}^2}{4 m_1^2 m_2^2}
\nonumber\\&\kern-4ex
+ \frac{3\scpm{\vnun}{\vmom{1}}\vmom{2}^2}{4 m_1^2 m_2^2}
- \frac{3\scpm{\vnun}{\vmom{2}}\vmom{2}^2}{4 m_1 m_2^3}\bigg)
+ \scpm{\vnun}{\vspin{1}}\scpm{\vnun}{\vspin{2}}
\bigg(-\frac{3\scpm{\vmom{1}}{\vmom{2}}^2{}}{8 m_1^2 m_2^2}
\nonumber\\&\kern-4ex
+ \frac{105\scpm{\vnun}{\vmom{1}}^2 \scpm{\vnun}{\vmom{2}}^2}{16 m_1^2 m_2^2}
- \frac{15 \scpm{\vnun}{\vmom{2}}^2 \vmom{1}^2}{8 m_1^2 m_2^2}
+ \frac{3 \vmom{1}^2\scpm{\vmom{1}}{\vmom{2}}}{4 m_1^3 m_2}
+ \frac{3 \vmom{1}^2 \vmom{2}^2}{16 m_1^2 m_2^2}
\nonumber\\&\kern-4ex
+ \frac{15 \vmom{1}^2 \scpm{\vnun}{\vmom{1}}\scpm{\vnun}{\vmom{2}}}{4 m_1^3 m_2}\bigg)
+ \scpm{\vspin{1}}{\vspin{2}}
\biggl(\frac{\scpm{\vmom{1}}{\vmom{2}}^2}{16 m_1^2 m_2^2}
- \frac{9 \scpm{\vnun}{\vmom{1}}^2 \vmom{1}^2}{8 m_1^4}
\nonumber\\&\kern-4ex
- \frac{5 \scpm{\vmom{1}}{\vmom{2}} \vmom{1}^2}{16 m_1^3 m_2}
- \frac{3 \scpm{\vnun}{\vmom{2}}^2\vmom{1}^2}{8 m_1^2 m_2^2}
- \frac{15 \scpm{\vnun}{\vmom{1}}^2 \scpm{\vnun}{\vmom{2}}^2}{16 m_1^2 m_2^2}
+ \frac{3 \vmom{1}^2 \vmom{2}^2}{16 m_1^2 m_2^2}
\nonumber\\&\kern-4ex
+ \frac{3\vmom{1}^2\scpm{\vnun}{\vmom{1}}\scpm{\vnun}{\vmom{2}}}{4 m_1^3 m_2}
+ \frac{9\scpm{\vmom{1}}{\vmom{2}}\scpm{\vnun}{\vmom{1}}\scpm{\vnun}{\vmom{2}}}{16 m_1^2 m_2^2}\bigg)
\bigg\}
\nonumber\\&\kern-4ex
+ \frac{G^2}{c^6\rel^4} \bigg\{
((\vnun\times\vmom{1})\cdot\vspin{1})((\vnun\times\vmom{1})\cdot\vspin{2})
\bigg(\frac{12}{m_1} + \frac{9 m_2}{m_1^2}\bigg)
\nonumber\\&\kern-4ex
- \frac{81}{4m_1}((\vnun\times\vmom{2})\cdot\vspin{1})((\vnun\times\vmom{1})\cdot\vspin{2})
\nonumber\\&\kern-4ex
- \frac{27}{4m_1}((\vnun\times\vmom{1})\cdot\vspin{1})((\vnun\times\vmom{2})\cdot\vspin{2})
\nonumber\\&\kern-4ex
- \frac{5}{2 m_1}\scpm{\vmom{1}}{\vspin{1}}\scpm{\vmom{2}}{\vspin{2}}
+ \frac{29}{8 m_1}\scpm{\vmom{2}}{\vspin{1}}\scpm{\vmom{1}}{\vspin{2}}
- \frac{21}{8 m_1}\scpm{\vmom{1}}{\vspin{1}}\scpm{\vmom{1}}{\vspin{2}}
\nonumber\\&\kern-4ex
+ \scpm{\vnun}{\vspin{1}}\scpm{\vmom{1}}{\vspin{2}}\bigg[
\left(\frac{33}{2 m_1} + \frac{9 m_2}{m_1^2}\right)\scpm{\vnun}{\vmom{1}}
- \left(\frac{14}{m_1} + \frac{29}{2 m_2}\right)\scpm{\vnun}{\vmom{2}}\bigg]
\nonumber\\&\kern-4ex
+ \scpm{\vmom{1}}{\vspin{1}} \scpm{\vnun}{\vspin{2}} \bigg[
\frac{4}{m_1}\scpm{\vnun}{\vmom{1}} - \left(\frac{11}{m_1} + \frac{11}{m_2}\right)\scpm{\vnun}{\vmom{2}}\bigg]
\nonumber\\&\kern-4ex
+ \scpm{\vnun}{\vspin{1}} \scpm{\vnun}{\vspin{2}} \bigg[
- \frac{12}{m_1} \scpm{\vnun}{\vmom{1}}^2 - \frac{10}{m_1} \vmom{1}^2 + \frac{37}{4 m_1}\scpm{\vmom{1}}{\vmom{2}}
\nonumber\\&\kern-4ex
+ \frac{255}{4 m_1}\scpm{\vnun}{\vmom{1}}\scpm{\vnun}{\vmom{2}}\bigg]
+ \scpm{\vspin{1}}{\vspin{2}} \bigg[
- \left(\frac{25}{2 m_1} + \frac{9 m_2}{m_1^2}\right) \scpm{\vnun}{\vmom{1}}^2
\nonumber\\&\kern-4ex
+ \frac{49}{8 m_1} \vmom{1}^2
+ \frac{35}{4 m_1}\scpm{\vnun}{\vmom{1}}\scpm{\vnun}{\vmom{2}}
- \frac{43}{8 m_1}\scpm{\vmom{1}}{\vmom{2}}\bigg]
\bigg\}
\nonumber\\&\kern-4ex
+ \frac{G^3}{c^6\rel^5}
\bigg\{
- \scpm{\vspin{1}}{\vspin{2}} \left(\frac{63}{4} m_1^2 + \frac{145}{8} m_1 m_2\right)
\nonumber\\&\kern-4ex
+ \scpm{\vnun}{\vspin{1}} \scpm{\vnun}{\vspin{2}} \left(\frac{105}{4} m_1^2 + \frac{289}{8} m_1 m_2\right)
\bigg\}
+ (1\leftrightarrow2).
\end{align}

Leading-order cubic in spin Hamiltonians (which are also proportional to the linear momenta of the bodies)
were derived by \cite{hergt1-08,hergt2-08} and \cite{levi-15}.
They are collected here into the single Hamiltonian $H_{S^3}^{\text{LO}}$,
which equals
\begin{align}
H_{S^3}^{\text{LO}}&(\xa,\pa,\mathbf{S}_a) \equiv
H^{\textrm{LO}}_{S_1^3} + H^{\textrm{LO}}_{S_1^2 S_2}
+ H^{\textrm{LO}}_{S_1 S_2^2}+ H^{\textrm{LO}}_{S_2^3}
\nonumber\\ &=
\frac{G}{c^4 m_1^2 r_{12}^4} \bigg\{
\frac{3}{2}\bigg[\mathbf{S}_{1}^2\,(\mathbf{S}_{2}\cdot(\mathbf{n}_{12}\times\mathbf{p}_{1}))
+ (\mathbf{S}_{1}\cdot\mathbf{n}_{12})\,(\mathbf{S}_{2}\cdot(\mathbf{S}_{1}\times\mathbf{p}_{1}))
\nonumber\\&\quad
+ (\mathbf{n}_{12}\cdot(\mathbf{S}_{1}\times\mathbf{S}_{2}))\big((\mathbf{S}_{1}\cdot\mathbf{p}_{1})-5(\mathbf{S}_{1}\cdot\mathbf{n}_{12})(\mathbf{p}_{1}\cdot\mathbf{n}_{12})\big)
\nonumber\\&\quad
- 5(\mathbf{S}_{1}\cdot\mathbf{n}_{12})^2\,(\mathbf{S}_{2}\cdot(\mathbf{n}_{12}\times\mathbf{p}_{1}))
-\frac{3m_1}{2m_2} \Big(
\mathbf{S}_{1}^2\,(\mathbf{S}_{2}\cdot(\mathbf{n}_{12}\times\mathbf{p}_2))
\nonumber\\&\quad
+ 2(\mathbf{S}_{1}\cdot\mathbf{n}_{12})(\mathbf{S}_{2}\cdot(\mathbf{S}_{1}\times\mathbf{p}_{2}))
- 5(\mathbf{S}_{1}\cdot\mathbf{n}_{12})^2(\mathbf{S}_{2}\cdot(\mathbf{n}_{12}\times\mathbf{p}_{2}))
\Big)\bigg]
\nonumber\\&\quad
- (\mathbf{S}_{1}\times\mathbf{n}_{12})\cdot\Big(\mathbf{p}_{2} - \frac{m_2}{4m_1}\mathbf{p}_{1}\Big)
\left(\mathbf{S}_{1}^2 - 5\left(\mathbf{S}_{1}\cdot\mathbf{n}_{12}\right)^2 \right)\bigg\}
+ (1\leftrightarrow2).
\end{align}

Leading-order quartic in spin Hamiltonians were derived by \cite{levi-15}.
They are collected here into the single Hamiltonian $H_{S^4}^{\text{LO}}$, which reads
\begin{align}
H_{S^4}^{\text{LO}}&(\xa,\mathbf{S}_a) \equiv
H^{\textrm{LO}}_{S_{1}^2S_{2}^2} + H^{\textrm{LO}}_{S_{1} S_{2}^3}
+ H^{\textrm{LO}}_{S_{2}S_{1}^3} + H^{\textrm{LO}}_{S_{1}^4} + H^{\textrm{LO}}_{S_{2}^4}
\nonumber\\
&= -\frac{3G}{2c^4m_{1}m_{2}r_{12}^5}
\bigg\{ \frac{1}{2}\mathbf{S}_{1}^2\mathbf{S}_{2}^2+\left(\mathbf{S}_{1}\cdot\mathbf{S}_{2}\right)^2
- \frac{5}{2}\left(\mathbf{S}_{1}^2\left(\mathbf{S}_{2}\cdot\mathbf{n}_{12}\right)^2
+ \mathbf{S}_{2}^2\left(\mathbf{S}_{1}\cdot\mathbf{n}_{12}\right)^2\right)
\nonumber\\&\quad
- 10(\mathbf{S}_{1}\cdot\mathbf{n}_{12})\,(\mathbf{S}_{2}\cdot\mathbf{n}_{12})
\left((\mathbf{S}_{1}\cdot\mathbf{S}_{2})
-\frac{7}{4}(\mathbf{S}_{1}\cdot\mathbf{n}_{12})\,(\mathbf{S}_{2}\cdot\mathbf{n}_{12})\right) \bigg\}
\nonumber\\&\quad
- \frac{3G}{2c^4m_{1}^2r_{12}^5} \bigg\{ \mathbf{S}_{1}^2\,(\mathbf{S}_{1}\cdot\mathbf{S}_{2})
- 5(\mathbf{S}_{1}\cdot\mathbf{S}_{2})(\mathbf{S}_{1}\cdot\mathbf{n}_{12})^2
\nonumber\\&\quad
- 5\mathbf{S}_{1}^2\,(\mathbf{S}_{1}\cdot\mathbf{n}_{12})\,(\mathbf{S}_{2}\cdot\mathbf{n}_{12})
+ \frac{35}{3}(\mathbf{S}_{2}\cdot\mathbf{n}_{12})(\mathbf{S}_{1}\cdot\mathbf{n}_{12})^3
\bigg\}
\nonumber\\&\quad
- \frac{3Gm_2}{8c^4m_1^3r_{12}^5} \bigg\{
(\mathbf{S}_{1}^2)^2 - 10\mathbf{S}_{1}^2\left(\mathbf{S}_1\cdot\mathbf{n}_{12}\right)^2
+\frac{35}{3}\left(\mathbf{S}_1\cdot\mathbf{n}_{12}\right)^4 \bigg\}
+ (1\leftrightarrow2).
\end{align}
Let us note that it is possible to compute the leading-order Hamiltonians
to all orders in spin \citep{vines2-16}.

\section{Dissipative many-point-mass Hamiltonians}
\label{app:dissH}

In this appendix we display all known dissipative Hamiltonians
for many-body systems (i.e. for systems comprising any number of components),
made of both spinless or spinning bodies.
We start by displaying the dissipative leading-order 2.5PN
and next-to-leading-order 3.5PN ADM Hamiltonians valid for spinless bodies.
The 2.5PN Hamiltonian is given in Eq.~\eqref{H2.5PN} for two-body systems,
but in this appendix we display formula for it valid for many-body systems.
The 3.5PN Hamiltonian was computed for the first time by \cite{jaranowski-97}.
The Hamiltonians read [in this Appendix we use units in which $c=1$ and $G=1/(16\pi)$]\footnote{
In \cite{jaranowski-97}, Eq.\ (58) for $H_{{\rm 3.5PN}}$ contains misprints,
which were corrected in Eq.\ (2.8) of \cite{koenigsdoerffer-03}.}
\begin{align}
\label{2.5Happ}
H_{{\rm 2.5PN}}(\xa,\pa,t) &= 5\pi\,\dot{\chi}_{(4)ij}(t)\,\chi_{(4)ij}(\xa,\pa),
\\[1ex]
\label{3.5Happ}
H_{{\rm 3.5PN}}(\xa,\pa,t) &=
5\pi\,\chi_{(4)ij}(\xa,\pa) \big(\dot{\Pi}_{1ij}(t) + \dot{\Pi}_{2ij}(t) + \ddot{\Pi}_{3ij}(t)\big)
\nonumber\\&\quad
+ 5\pi\,\dot{\chi}_{(4)ij}(t) \big({\Pi}_{1ij}(\xa,\pa) + \widetilde{\Pi}_{2ij}(\xa,t)\big)
\nonumber\\[1ex]&\quad
- 5\pi\,\ddot{\chi}_{(4)ij}(t) {\Pi}_{3ij}(\xa,\pa)
\nonumber\\[1ex]&\quad
+ \dot{\chi}_{(4)ij}(t) \big(Q'_{ij}(\xa,\pa,t) + Q''_{ij}(\xa,t)\big)
\nonumber\\&\quad
+ \frac{\partial^3}{\partial t^3} \big(R'(\xa,\pa,t) + R''(\xa,t)\big).
\end{align}
To display the building blocks of these Hamiltonians
we adopt the notation that the explicit dependence on time $t$
is through canonical variables with \emph{primed} indices only,
e.g., $\chi_{(4)ij}(t)\equiv\chi_{(4)ij}(\vecx_{a'}(t),\vecp_{a'}(t))$.
We also define $s_{abc}\equiv r_{ab}+r_{bc}+r_{ca}$,
$s_{aa'b'}\equiv r_{aa'} + r_{ab'} + r_{a'b'}$,
and $s_{aba'}\equiv r_{ab}+r_{aa'}+r_{ba'}$.
The building blocks are then defined as follows:\footnote{
In \cite{jaranowski-97}, Eqs.\ (56) and (57) for $Q''_{ij}$ and $R''$, respectively, contain misprints,
which were corrected in Eqs.\ (2.9) of \cite{koenigsdoerffer-03}.}
\begin{align}
\label{chi4def}
\chi_{(4)ij}(\xa,\pa) &\equiv \frac{8}{15}\frac{1}{16\pi}
\sum\limits_a \frac{1}{m_a} \left(\pa^2\delta_{ij}-3 p_{ai} p_{aj}\right)
\nonumber\\&\quad
+ \frac{4}{15}\frac{1}{(16\pi)^2}
\sum\limits_a\sum\limits_{b\ne a} \frac{m_a m_b}{r_{ab}}
\big(3 n_{ab}^i n_{ab}^j - \delta_{ij}\big),
\end{align}
\begin{align}
\Pi_{1ij}&(\xa,\pa) \equiv \frac{4}{15}\frac{1}{16\pi}
\sum\limits_a\frac{\pa^2}{m_a^3}
\left( -\pa^2\delta_{ij} + 3p_{ai}p_{aj} \right)
\nonumber\\&\quad
+ \frac{8}{5} \frac{1}{(16\pi)^2}
\sum\limits_a\sum\limits_{b\ne a}
\frac{m_b}{m_a r_{ab}} \big( - 2 \pa^2\delta_{ij} + 5p_{ai}p_{aj}
+ \pa^2 n_{ab}^i n_{ab}^j \big)
\nonumber\\&\quad
+ \frac{1}{5} \frac{1}{(16\pi)^2}
\sum\limits_a\sum\limits_{b\ne a} \frac{1}{r_{ab}} \Big\{
\big[19 (\pa\cdot\mathbf{p}_b)
- 3 (\mathbf{n}_{ab}\cdot\pa) (\mathbf{n}_{ab}\cdot\mathbf{p}_b)\big]\delta_{ij}
\nonumber\\&\quad
- 42 {p_{ai}p_{bj}}
-3\big[5(\pa\cdot\mathbf{p}_b)
+ (\mathbf{n}_{ab}\cdot\pa) (\mathbf{n}_{ab}\cdot\mathbf{p}_b) \big]
n_{ab}^i n_{ab}^j
\nonumber\\&\quad
+ 6 (\mathbf{n}_{ab}\cdot\mathbf{p}_b)
\big( n_{ab}^i p_{aj} + n_{ab}^j p_{ai} \big) \Big\}
\nonumber\\&\quad
+ \frac{41}{15} \frac{1}{(16\pi)^3}
\sum\limits_a\sum\limits_{b\ne a} \frac{m_a^2 m_b}{r_{ab}^2}
\big( \delta_{ij} - 3 n_{ab}^i n_{ab}^j \big)
\nonumber\\&\quad
+ \frac{1}{45} \frac{1}{(16\pi)^3}
\sum\limits_a\sum\limits_{b\ne a}\sum\limits_{c\ne a,b} m_a m_b m_c
\bigg\{ \frac{18}{r_{ab}r_{ca}}
\big( \delta_{ij} - 3 n_{ab}^i n_{ab}^j \big)
\nonumber\\&\quad
- \frac{180}{s_{abc}} \left[
\left( \frac{1}{r_{ab}} + \frac{1}{s_{abc}} \right) n_{ab}^i n_{ab}^j
+ \frac{1}{s_{abc}} n_{ab}^i n_{bc}^j \right]
\nonumber\\&\quad
+ \frac{10}{s_{abc}} \left[ 4 \left(
\frac{1}{r_{ab}} + \frac{1}{r_{bc}} + \frac{1}{r_{ca}} \right)
- \frac{r_{ab}^2 + r_{bc}^2 + r_{ca}^2}{r_{ab}r_{bc}r_{ca}} \right]\delta_{ij}
\bigg\},
\end{align}
\begin{align}
&\Pi_{2ij}(\xa,\pa) \equiv \frac{1}{5}\frac{1}{(16\pi)^2}
\sum\limits_a\sum\limits_{b\ne a} \frac{m_b}{m_a r_{ab}} \Big\{
\big[5(\mathbf{n}_{ab}\cdot\pa)^2 - \pa^2 \big] \delta_{ij}
- 2p_{ai}p_{aj}
\nonumber\\&
+ \big[5\pa^2 - 3(\mathbf{n}_{ab}\cdot\pa)^2 \big] n_{ab}^i n_{ab}^j
- 6(\mathbf{n}_{ab}\cdot\pa)(n_{ab}^i p_{aj} + n_{ab}^j p_{ai}) \Big\}
\nonumber\\&
+ \frac{6}{5} \frac{1}{(16\pi)^3}
\sum\limits_a\sum\limits_{b\ne a} \frac{m_a^2 m_b}{r_{ab}^2}
\big( 3 n_{ab}^i n_{ab}^j - \delta_{ij} \big)
\nonumber\\&
+ \frac{1}{10}\frac{1}{(16\pi)^3}
\sum\limits_a\sum\limits_{b\ne a}\sum\limits_{c\ne a,b} m_a m_b m_c
\bigg\{
\left[
\frac{5r_{ca}}{r_{ab}^3}\left(1-\frac{r_{ca}}{r_{bc}}\right)
+ \frac{1}{r_{ab}} \left(\frac{13}{r_{ca}}-\frac{40}{s_{abc}}\right)
\right]\delta_{ij}
\nonumber\\[2ex]&
+\left[
3\frac{r_{ab}}{r_{ca}^3}
+\frac{r_{bc}^2}{r_{ab}r_{ca}^3}
-\frac{5}{r_{ab}r_{ca}} 
+\frac{40}{s_{abc}}\left(\frac{1}{r_{ab}}+\frac{1}{s_{abc}}\right)
\right] n_{ab}^i n_{ab}^j
\nonumber\\[2ex]&
+\left[
2 \frac{(r_{ab} + r_{ca})}{r_{bc}^3}
- 16 \left( \frac{1}{r_{ab}^2} + \frac{1}{r_{ca}^2} \right)
+ \frac{88}{s_{abc}^2}
\right] n_{ab}^i n_{ca}^j \bigg\},
\end{align}
\begin{align}
\Pi_{3ij}\left(\xa,\pa\right) &\equiv
\frac{1}{5} \frac{1}{(16\pi)^2}
\sum\limits_a\sum\limits_{b\ne a} m_b \big\{
- 5 (\mathbf{n}_{ab}\cdot\pa) \delta_{ij}
\nonumber\\&\quad
+ (\mathbf{n}_{ab}\cdot\pa) n_{ab}^i n_{ab}^j
+ 7( n_{ab}^i p_{aj} + n_{ab}^j p_{ai}) \big\},
\end{align}
\begin{align}
&\widetilde{\Pi}_{2ij}(\xa,t) \equiv \frac{1}{5}\frac{1}{(16\pi)^2}
\sum\limits_{a}\sum\limits_{a'}\frac{m_a}{m_{a'}r_{aa'}}
\Big\{
\big(5(\mathbf{n}_{a{a'}}\cdot\mathbf{p}_{a'})^2-\mathbf{p}_{a'}^2\big)\delta_{ij} - 2 p_{{a'}i} p_{{a'}j}
\nonumber\\&
+\big(5\mathbf{p}_{a'}^2 - 3 (\mathbf{n}_{{aa'}}\cdot\mathbf{p}_{a'})^2\big)
n_{{aa'}}^i n_{{aa'}}^j
- 6 (\mathbf{n}_{{aa'}}\cdot\mathbf{p}_{a'}) (n_{aa'}^i p_{{a'}j} + n_{aa'}^j p_{{a'}i})
\Big\}
\nonumber\\&
+\frac{1}{10}\frac{1}{(16\pi)^3}
\sum\limits_{a}\sum\limits_{a'}\sum\limits_{{b'}\ne {a'}} m_a m_{a'} m_{b'}
\bigg\{
\frac{32}{s_{a{a'}{b'}}}
\left(\frac{1}{r_{{a'}{b'}}} + \frac{1}{s_{a{a'}{b'}}}\right)
n_{{a'}{b'}}^i n_{{a'}{b'}}^j
\nonumber\\&
+ 16 \bigg( \frac{1}{r_{{a'}{b'}}^2} - \frac{2}{s_{a{a'}{b'}}^2} \bigg)
( n_{aa'}^i n_{{a'}{b'}}^j +  n_{aa'}^j n_{{a'}{b'}}^i )
- 2 \left(\frac{r_{aa'} + r_{ab'}}{r_{{a'}{b'}}^3}
+ \frac{12}{s_{a{a'}{b'}}^2}\right) n_{aa'}^i n_{a{b'}}^j
\nonumber\\&
+\left[
\frac{r_{aa'}}{r_{{a'}{b'}}^3} \left(\frac{r_{aa'}}{r_{ab'}} + 3\right)
-\frac{5}{r_{{a'}{b'}}r_{aa'}}
+\frac{8}{s_{a{a'}{b'}}}
\left(\frac{1}{r_{aa'}} + \frac{1}{s_{a{a'}{b'}}}\right)
\right] n_{aa'}^i n_{{aa'}}^j
\nonumber\\&
+\left[
5\frac{r_{aa'}}{r_{{a'}{b'}}^3}
\left( 1 - \frac{r_{aa'}}{r_{ab'}} \right)
+ \frac{17}{r_{{a'}{b'}}r_{aa'}} - \frac{4}{r_{aa'} r_{ab'}}
- \frac{8}{s_{a{a'}{b'}}}
\left( \frac{1}{r_{aa'}} + \frac{4}{r_{{a'}{b'}}} \right)
\right]\delta_{ij}
\bigg\},
\end{align}
\begin{align}
Q'_{ij}(\xa,\pa,t) &\equiv -\frac{1}{16}\frac{1}{16\pi}
\sum\limits_{a}\sum\limits_{a'}\frac{m_{a'}}{m_{a} r_{aa'}}
\big\{ 2p_{ai}p_{aj}+12(\mathbf{n}_{aa'}\cdot\pa)n_{aa'}^i p_{aj}
\nonumber\\&\quad
- 5\pa^2 n_{aa'}^i n_{aa'}^j
+ 3(\mathbf{n}_{aa'}\cdot\pa)^2 n_{aa'}^i n_{aa'}^j \big\},
\end{align}
\begin{align}
Q''_{ij}&(\xa,t) \equiv
\frac{1}{32} \frac{1}{(16\pi)^2}
\sum\limits_{a}\sum\limits_{b\ne a}\sum\limits_{a'} m_a m_b m_{a'}
\bigg\{ \frac{32}{s_{aba'}} \left( \frac{1}{r_{ab}} + \frac{1}{s_{aba'}}
\right) n_{ab}^i n_{ab}^j
\nonumber\\&
+ \left[ 3 \frac{r_{aa'}}{r_{ab}^3} - \frac{5}{r_{ab}r_{aa'}}
+\frac{r_{ba'}^2}{r_{ab}^3r_{aa'}} + \frac{8}{s_{aba'}}
\left( \frac{1}{r_{aa'}} + \frac{1}{s_{aba'}} \right) \right]
n_{aa'}^in_{aa'}^j
\nonumber\\&
- 2 \left( \frac{r_{aa'}+r_{ba'}}{r_{ab}^3} + \frac{12}{s_{aba'}^2} \right)
n_{aa'}^i n_{ba'}^j
- 32 \left( \frac{1}{r_{ab}^2} - \frac{2}{s_{aba'}^2} \right)
n_{ab}^i n_{aa'}^j \bigg\},
\end{align}
\begin{align}
R'&(\xa,\pa,t) \equiv
\frac{2}{105} \frac{1}{16\pi}
\sum\limits_{a}\sum\limits_{a'} \frac{r_{aa'}^2}{m_{a}m_{a'}}
\big\{
- 5 \pa^2 \mathbf{p}_{a'}^2
+ 11 (\pa\cdot\mathbf{p}_{a'})^2
\nonumber\\&
+ 4(\mathbf{n}_{aa'}\cdot\mathbf{p}_{a'})^2 \pa^2
+ 4 (\mathbf{n}_{aa'}\cdot\pa)^2 \mathbf{p}_{a'}^2
- 12 (\mathbf{n}_{aa'}\cdot\mathbf{p}_{a'}) (\mathbf{n}_{aa'}\cdot\pa)
(\pa\cdot\mathbf{p}_{a'}) \big\}
\nonumber\\&
- \frac{1}{105} \frac{1}{(16\pi)^2}
\sum\limits_{a}\sum\limits_{a'}\sum\limits_{b'\ne a'}
\frac{m_{a'}m_{b'}}{m_a} \bigg\{
\left( 2 \frac{r_{aa'}^4}{r_{a'b'}^3} - 2 \frac{r_{aa'}^2 r_{ab'}^2}{r_{a'b'}^3}
- 5 \frac{r_{aa'}^2}{r_{a'b'}} \right) \pa^2
\nonumber\\&
+ 4 \frac{r_{aa'}^2}{r_{a'b'}} (\mathbf{n}_{aa'}\cdot\pa)^2
+ 17 \left( \frac{r_{aa'}^2}{r_{a'b'}} + r_{a'b'} \right)
(\mathbf{n}_{a'b'}\cdot\pa)^2
\nonumber\\&
+ 2 \left( 6 \frac{r_{aa'}^3}{r_{a'b'}^2} + 17 r_{aa'} \right)
(\mathbf{n}_{aa'}\cdot\pa) (\mathbf{n}_{a'b'}\cdot\pa)
\bigg\},
\end{align}
\begin{align}
R''&(\xa,t) \equiv
\frac{1}{105} \frac{1}{(16\pi)^2}
\sum\limits_{a}\sum\limits_{b\ne a}\sum\limits_{a'} \frac{m_a m_b}{m_{a'}}
\bigg\{ \left( 5 \frac{r_{aa'}^2}{r_{ab}}
+ 2 \frac{r_{aa'}^2r_{ba'}^2}{r_{ab}^3}
- 2 \frac{r_{aa'}^4}{r_{ab}^3} \right) \mathbf{p}_{a'}^2
\nonumber\\&
- 17 \left( \frac{r_{aa'}^2}{r_{ab}} + r_{ab} \right)
(\mathbf{n}_{ab}\cdot\mathbf{p}_{a'})^2
- 4 \frac{r_{aa'}^2}{r_{ab}} (\mathbf{n}_{aa'}\cdot\mathbf{p}_{a'})^2
\nonumber\\&
+ 2 \left( \frac{6r_{aa'}^3}{r_{ab}^2} + 17 r_{aa'} \right)
(\mathbf{n}_{ab}\cdot\mathbf{p}_{a'}) (\mathbf{n}_{aa'}\cdot\mathbf{p}_{a'}) \bigg\}
\nonumber\\&
+ \frac{1}{210} \frac{1}{(16\pi)^3}
\sum\limits_{a}\sum\limits_{b\ne a}\sum\limits_{a'}\sum\limits_{b'\ne a'}
m_a m_b m_{a'} m_{b'} \bigg\{
2 \frac{r_{aa'}^2}{r_{ab}r_{a'b'}^3}\left(r_{aa'}^2 - r_{ab'}^2\right)
\nonumber\\&
+ 2 \frac{r_{aa'}^2}{r_{ab}^3r_{a'b'}}\left(r_{aa'}^2 - r_{ba'}^2 \right)
+ 4 \frac{r_{ab}r_{aa'}^2}{r_{a'b'}^3}
- 5 \frac{r_{aa'}^2}{r_{ab}r_{a'b'}}
- 2 \left( \frac{r_{ab}^3}{r_{a'b'}^3} + \frac{r_{ab}}{r_{a'b'}} \right)
\nonumber\\&
- 4 \frac{r_{ab}r_{aa'}r_{bb'}}{r_{a'b'}^3} (\mathbf{n}_{aa'}\cdot\mathbf{n}_{bb'})
+ 17 \left( \frac{r_{ab}}{r_{a'b'}} + \frac{r_{a'b'}}{r_{ab}}
+ \frac{r_{aa'}^2}{r_{ab}r_{a'b'}} \right) (\mathbf{n}_{ab}\cdot\mathbf{n}_{a'b'})^2
\nonumber\\&
+ 6 \frac{r_{aa'}^4}{r_{ab}^2 r_{a'b'}^2} (\mathbf{n}_{ab}\cdot\mathbf{n}_{a'b'})
+34r_{aa'}^2\left(\frac{1}{r_{ab}^2}+\frac{1}{r_{a'b'}^2}\right)
(\mathbf{n}_{ab}\cdot\mathbf{n}_{a'b'}) \bigg\}.
\end{align}

The leading-order Hamiltonian for systems made of any number of spinning bodies
was derived by \cite{wang-11}. It reads\footnote{
We keep here the total time derivative as given in \cite{wang-11},
though it could be dropped as correspondingly done in the Eq.~\eqref{3.5Happ},
because it can be removed by performing a canonical transformation.}
\begin{align}
\label{LOspinHapp}
H^{\text{spin}}_{\text{3.5PN}}&(\xa,\pa,\mathsf{S}_a,t) = 5\pi \Big(
\chi_{(4)ij}(\xa,\pa) \big(\dot{\Pi}_{1 ij}^{\text{spin}}(t) + \dot{\Pi}_{2 ij}^{\text{spin}}(t) + \ddot{\Pi}_{3 ij}^{\text{spin}}(t)\big)
\nonumber\\&
+ \dot{\chi}_{(4)ij}(t) \big( \Pi_{1ij}^{\text{spin}}(\xa,\pa,\mathsf{S}_a) + \widetilde{\Pi}_{2ij}^{\text{spin}}(\xa,t) \big)
\nonumber\\&
- \ddot{\chi}_{(4)ij}(t) \Pi_{3 ij}^{\text{spin}}(\xa,\mathsf{S}_a)
\Big)
+ \dot{\chi}_{(4)ij}(t) Q'^{\,\text{spin}}_{ij}(\xa,\pa,\mathsf{S}_a,t)
\nonumber\\&
+ \frac{\partial^3}{\partial t^3} \Big( R'^{\,\text{spin}}(\xa,\pa,\mathsf{S}_a,t) + R''^{\,\text{spin}}(\xa,t) \Big)
\nonumber\\&
- \frac{\md}{\md t}\Big(\dot{\chi}_{(4)ij}(t) O_{ij}^{\text{spin}}(\pa,\mathsf{S}_a)\Big),
\end{align}
where $\mathsf{S}_a$ is the spin tensor associated with $a$th body, with components $S_{a(i)(j)}$.
The function $\chi_{(4)ij}$ is defined in Eq.~\eqref{chi4def} above
and the functions ${\Pi}_{1ij}^{\text{spin}}$, ${\Pi}_{2ij}^{\text{spin}}$, ${\Pi}_{3ij}^{\text{spin}}$,
$\widetilde{\Pi}_{2ij}^{\text{spin}}$, $Q'^{\,\text{spin}}_{ij}$, $R'^{\,\text{spin}}$, $R''^{\,\text{spin}}$, and $O_{ij}^{\text{spin}}$
are given by
\begin{align}
&\Pi^{\text{spin}}_{1ij}(\xa,\pa,\mathsf{S}_a)
\equiv \frac{4}{5(16\pi)^2}\sum_a\sum_{b \ne a} \bigg\{
\frac{1}{r_{ab}^2} \Big[
3(\vecn_{ab}\cdot\vecp_b) n_{ab}^k \big(n_{ab}^j S_{a(i)(k)}
\nonumber\\&\quad
+ n_{ab}^iS_{a(j)(k)}\big)
- 3 p_{bk} \big(n_{ab}^j S_{a(i)(k)} + n_{ab}^i S_{a(j)(k)}\big)
- 3 n_{ab}^k \big(p_{bj} S_{a(i)(k)}
\nonumber\\&\quad
+ p_{bi} S_{a(j)(k)}\big)
+ 4 (3n_{ab}^i n_{ab}^j - \delta_{ij}) n_{ab} ^k p_{bl} S_{a(k)(l)}
\Big]
+ \frac{m_b}{m_a} \frac{1}{r_{ab}^2} \Big[p_{ak} (n_{ab}^j S_{a(i)(k)}
\nonumber\\&\quad
+ n_{ab}^i  S_{a(j)(k)})
+ (4 \delta_{ij} - 6 n_{ab}^i  n_{ab}^j ) n_{ab}^k p_{al} S_{a(k)(l)}
+ 4n_{ab}^k \big(p_{aj}  S_{a(i)(k)}
\nonumber\\&\quad
+  p_{ai}  S_{a(j)(k)}\big)
\Big]
- \frac{S_{a(k)(l)}}{r_{ab}^3} \Big[
(3n_{ab}^i n_{ab}^j - \delta_{ij}) S_{b(k)(l)} + 3n_{ab}^k \big(n_{ab}^j S_{b(i)(l)}
\nonumber\\&\quad
+ n_{ab}^i S_{b(j)(l)}\big)
+ 3 (\delta_{ij} - 5 n_{ab}^i  n_{ab}^j ) { n_{ab}^k  n_{ab}^{n}S_{b(n)(l)} }
\Big]
\bigg\},
\end{align}
\begin{align}
&\Pi^{\text{spin}}_{2ij}(\xa,\pa,\mathsf{S}_a) \equiv -\frac{4}{5(16\pi)^2} \sum_a\sum_{b\ne a} \frac{m_b}{m_a}
\frac{1}{r_{ab}^2} \Big\{ -2p_{ak} \big(n_{ab}^i S_{a(j)(k)}
\nonumber\\&
+ n_{ab}^j S_{a(i)(k)}\big)
+ n_{ab}^k (p_{ai} S_{a(j)(k)} + p_{aj} S_{a(i)(k)})
+ 3 (\vecn_{ab}\cdot\vecp_a) n_{ab}^k \big(n_{ab}^i S_{a(j)(k)}
\nonumber\\[1ex]&
+ n_{ab}^j S_{a(i)(k)}\big)
+ (\delta_{ij} + 3 n_{ab}^i n_{ab}^j) n_{ab}^k p_{al} S_{a(k)(l)}
\Big\},
\end{align}
\begin{align}
\Pi^{\text{spin}}_{3ij}(\xa,\pa,\mathsf{S}_a) \equiv \frac{4}{5(16\pi)^2}
\sum_a\sum_{b\ne a} \frac{m_b}{r_{ab}} n_{ab}^k \big(n_{ab}^j S_{a(i)(k)} + n_{ab}^i S_{a(j)(k)}\big),
\end{align}
\begin{align}
{\widetilde \Pi}_{2ij}^{\rm spin}&(\xa,t)
\equiv -\frac{4}{5(16\pi)^2} \sum_a\sum_{a'} \frac{m_a}{\mapr} \frac{1}{\raap^2}
\Big\{
2p_{a'k} (\naap^i S_{a'(j)(k)} + \naap^j S_{a'(i)(k)})
\nonumber\\&
- \naap^k (p_{a'i} S_{a'(j)(k)} + p_{a'j} S_{a'(i)(k)})
- (\delta_{ij} + 3 \naap^i \naap^j) \naap^k  p_{a'l} S_{a'(k)(l)}
\nonumber\\[1ex]&
- 3 (\vecn_{aa'}\cdot\vecp_{a'}) \naap^k (\naap^i S_{a'(j)(k)} +\naap^j S_{a'(i)(k)})
\Big\},
\end{align}
\begin{align}
Q'^{\,\text{spin}}_{ij}&(\xa,\pa,\mathsf{S}_a,t)
\equiv \frac{1}{4(16\pi)} \sum_a\sum_{a'} \frac{\mapr}{m_a} \frac{1}{\raap^2} \Big\{
2 p_{ak} \big(\naap^i S_{a(j)(k)}
\nonumber\\&
+ \naap^j S_{a(i)(k)}\big)
- \naap^k (p_{ai} S_{a(j)(k)} + p_{aj} S_{a(i)(k)})
\nonumber\\[1ex]&
- 3 (\vecn_{aa'}\cdot\vecp_a) \naap^k (\naap^i S_{a(j)(k)} + \naap^j S_{a(i)(k)})
\nonumber\\[1ex]&
- (\delta_{ij} + 3\naap^i \naap^j) \naap^k  p_{al} S_{a(k)(l)}
\Big\},
\end{align}
\begin{align}
&R'^{\,\text{spin}}(\xa,\pa,\mathsf{S}_a,t)
\equiv \frac{1}{15(16\pi)} \sum_a\sum_{a'} S_{a(i)(j)}
\bigg\{
\frac{4 r_{a'a}}{m_{a'} m_a} \big( \vecp_{a'}^2 n_{a'a}^i p_{aj}
\nonumber\\&
- (\vecn_{a'a}\cdot\vecp_{a'}) p_{a'i} p_{aj}
- 2 (\vecp_{a'}\cdot\vecp_a) n_{a'a}^i p_{a'j} \big)
\nonumber\\&
+ \frac{1}{7(16\pi)} \sumbp \frac{m_{a'}m_{b'}}{m_a} \bigg(
17 n_{a'b'}^i p_{aj}
- \frac{2r_{a'a}}{r_{a'b'}}  \big(
17 (\vecn_{a'b'}\cdot\vecp_a) n_{a'b'}^i  n_{a'a}^j
\nonumber\\&
+ 7 n_{a'a}^i p_{aj}
\big)
+ \frac{6 r_{a'a}^2}{r_{a'b'}^2}
\big( n_{a'b'}^i p_{aj} + 2 (\vecn_{a'a}\cdot\vecp_a) n_{a'b'}^i n_{a'a}^j \big)
\nonumber\\&
+ \frac{8 r_{a'a}}{r_{a'b'}^3}
\big( r_{a'a}^2 n_{a'a}^i p_{aj} - r_{b'a}^2 n_{a'a}^i p_{aj} \big)
\bigg) \bigg\}
\nonumber\\&
+ \frac{4}{15(16\pi)} \sum_a\sum_{a'} \frac{r_{aa'}}{m_{a'} m_a} S_{a'(i)(j)}
\Big( \vecp_a^2  n_{a a'}^i p_{a'j} - 2(\vecp_{a'}\cdot\vecp_a) n_{a a'}^i p_{aj}
\nonumber\\&
+ (\vecn_{aa'}\cdot\vecp_a) p_{a'i} p_{aj} \Big)
+ \frac{2}{15(16\pi)} \suma\sum_{a'\neq a} \frac{1}{m_{a'} m_a} S_{a(i)(j)} \Big(
3 p_{a'k} p_{ai} S_{a'(k)(j)}
\nonumber\\&
- 2 (\vecp_{a'}\cdot\vecp_a) S_{a'(i)(j)}
- 2 p_{a'i} p_{ak} S_{a'(k)(j)} \Big),
\end{align}
\begin{align}
R''^{\,\text{spin}}(\xa,t)
&\equiv \frac{2}{15(16\pi)^2} \sum_a\sumb\sum_{a'}
\frac{m_a m_b}{m_{a'}} \frac{r_{a'a}}{r_{ab}} S_{a'(i)(j)}
\Big( n_{a'a}^i p_{a'j}
\nonumber\\&\quad
- 2 (\vecn_{ab}\cdot\vecp_{a'}) n_{a'a}^i n_{ab}^j
- (\vecn_{a'a}\cdot\vecn_{ab}) n_{ab}^i p_{a'j} \Big),
\end{align}
\be
O_{ij}^{\text{spin}}(\pa,\mathsf{S}_a) \equiv \sum_a \frac{1}{8m_a^2} p_{ak}
\big(p_{ai} S_{a(k)(j)} + p_{aj} S_{a(k)(i)}\big).
\ee

\section{Closed-form 1PM Hamiltonian for point-mass systems}
\label{app:1PMH}

The first post-Minkowskian (1PM) closed-form Hamiltonian for point-mass systems
has been derived by \cite{ledvinka-08}.
The starting point is the ADM reduced Hamiltonian describing $N$ gravitationally interacting point masses
with positions $\xa$ and linear momenta $\bp_a$ ($a=1,\ldots,N)$.
The 1PM Hamiltonian is, by definition, accurate through terms linear in $G$ and it reads (setting $c=1$)
\begin{align}
\label{HlinGS}
 H_{\rm lin} &=
\sum_a \bm_a - \frac{1}{2}G\sum_{a,b\ne a} \frac{\bm_a \bm_b}{ r_{ab} }
\left( 1+ \frac{p_a^2}{\bm_a^2}+\frac{p_b^2}{\bm_b^2}\right)
\nonumber\\[1ex]&\quad
+ \frac{1}{4}G\sum_{a,b\ne a} \frac{1}{r_{ab}}\left( 7\, \bp_a \cdot \bp_b + (\bp_a \cdot \bn_{ab})(\bp_b \cdot \bn_{ab}) \right)
\nonumber\\[1ex]&\quad
- \frac{1}{2}\sum_a \frac{p_{ai}p_{aj}}{\bm_a}\,h_{ij}^{\rm TT}(\bx=\bx_a)
+ \frac{1}{16\pi G} \int\md^3x \left( \frac{1}{4} h_{ij,k}^{\rm TT}\, h_{ij,k}^{\rm TT} + \pi^{ij}_{\rm TT} \pi^{ij}_{\rm TT}\right),
\end{align}
where $\bm_a\equiv\left( m_a^2+\bp^2_a \right)^\frac{1}{2}$ and $\bn_{ab} r_{ab}\equiv\bx_a-\bx_b$ (with $|\bn_{ab}|=1$).
The independent degrees of freedom of the gravitational field,  $h_{ij}^{\rm TT}$ and $\pi^{ij}_{\rm TT}$, are treated to linear order in $G$.
Denoting $\bx-\bx_a\equiv\bn_a |\bx-\bx_a|$ and $\cos \theta_a\equiv({\bn_a \cdot \dot \bx_a) /|\dot \bx_a|}$, 
the solution for $h_{ij}^{\rm TT}(\bx)$ was found to be 
\be
\label{LiWi4h}
h_{ij}^{\mathrm{TT}}(\bx) = 
\delta_{ij}^{\mathrm{TT}\,kl} \sum_b 
\frac{4G}{\bm_b} \frac{1}{|\bx-\bx_b|} \frac{p_{bk}p_{bl}}{\sqrt {1-{\dot \bx_b}^2\sin^2 \theta_b}}.
\ee

An autonomous point-mass Hamiltonian needs the field part in the related Routhian,
\be
\label{1PMRfield}
R_f = \frac{1}{16\pi G} \int\md^3x\, \frac{1}{4}\left( h_{ij,k}^{\rm TT}\, h_{ij,k}^{\rm TT} 
- \dot h_{ij}^{\rm TT}\dot h_{ij}^{\rm TT} \right),
\ee
to be transformed into an explicit function of particle variables.
Using the Gauss law in the first term and integrating by parts the term containing 
the time derivatives one arrives at
\begin{align}
\label{HfGauss}
R_f &= -\frac{1}{16\pi G}\int\md^3x\, \frac{1}{4} h_{ij}^{\rm TT}\left( \Delta  h_{ij}^{\rm TT} - \partial_t^2 h_{ij}^{\rm TT} \right) 
+ \frac{1}{64\pi G} \oint\md S_k  (h_{ij}^{\rm TT} h_{ij,k}^{\rm TT})
\nonumber\\[1ex]&\quad
- \frac{1}{64\pi G} \frac{\md}{\md t} \int\md^3 x\,(h_{ij}^{\rm TT} \dot h_{ij}^{\rm TT}).
\end{align}
\sloppy The field equations imply that the first integral directly combines 
with the ``interaction'' term containing 
$\sum\, \bm_a^{-1}\, p_{ai}\, p_{aj} \,h^{\rm TT}_{ij}(\bx_a)$, so only its coefficient gets changed.
The remaining terms in $R_f$, the surface integral and the total time derivative, 
do not modify the dynamics of the system 
since in our approximation of unaccelerated field-generating particles, the surface integral vanishes at large $|\bx|$.
The reduced Routhian thus takes the form,
now referred to as $H$ because it is a Hamiltonian for the particles,
\begin{align}
\label{Hlin}
H_{\rm lin}(\bx_c,\bp_c,\dot \bx_c) &=
\sum_a \bm_a - \frac{1}{2}G\sum_{a,b\ne a} \frac{\bm_a \bm_b }{ r_{ab}}
\left( 1+ 2\frac{p_a^2}{\bm_a^2}\right)
\nonumber\\[1ex]&\quad
+ \frac{1}{4}G\sum_{a,b\ne a} \frac{1}{r_{ab}}\left( 7\, (\bp_a \cdot \bp_b) + (\bp_a \cdot \bn_{ab})(\bp_b \cdot \bn_{ab})   \right)
\nonumber\\[1ex]&\quad
- \frac{1}{4}\sum_a \frac{p_{ai}p_{aj} }{\bm_a}\,h_{ij}^{\rm TT}(\bx=\bx_a;\bx_b,\bp_b,\dot \bx_b).
\end{align}
Though dropping a total time derivative, which implies a canonical transformation, the new canonical coordinates keep their names.
A further change of coordinates has to take place to eliminate the velocities $\dot \bx_a$ in the Hamiltonian.
This can be achieved by simply putting $\dot \bx_a = \bp_a / \bm_a$ (again without changing names of the variables).
Using the shortcut $y_{ba}\equiv\bm_b^{-1} [ m_b^2+ \left (\bn_{ba} \cdot \bp_b\right)^2]^\frac{1}{2}$,
the Hamiltonian comes out in the final form \citep{ledvinka-08}
\begin{align}
\label{H1PM}
H_{\rm lin} &=
\sum_a \bm_a - \frac{1}{2}G\sum_{a,b\ne a} \frac{\bm_a \bm_b }{ r_{ab}}
\left( 1+ \frac{p_a^2}{ \bm_a^2}+\frac{p_b^2}{\bm_b^2}\right)
\nonumber\\[1ex]&\quad
+ \frac{1}{4}G\sum_{a,b\ne a} \frac{1}{r_{ab}}\left( 7\, (\bp_a \cdot \bp_b) + (\bp_a \cdot \bn_{ab})(\bp_b \cdot \bn_{ab}) \right)
\nonumber\\[1ex]&\quad
- \frac{1}{4} G \sum_{a,b\ne a} \frac{1}{r_{ab}} \frac{(\bm_a \bm_b)^{-1}}{ (y_{ba}+1)^2 y_{ba}}
\Bigg\{
2\Big(2 
(\bp_a \cdot  \bp_b)^2 (\bp_b \cdot \bn_{ba})^2
\nonumber\\[1ex]&\quad
- 2 (\bp_a \cdot \bn_{ba}) (\bp_b \cdot \bn_{ba}) (\bp_a \cdot \bp_b) \bp_b^2 
+(\bp_a \cdot \bn_{ba})^2 \bp_b^4
-(\bp_a \cdot  \bp_b)^2 \bp_b^2
\Big ) \frac{1}{\bm_b^2} 
\nonumber\\[1ex]&\quad
+ 2 \Big[
(\bp_a \cdot \bp_b)^2 - \bp_a^2 (\bp_b \cdot \bn_{ba})^2 + (\bp_a \cdot \bn_{ba})^2 (\bp_b \cdot \bn_{ba})^2
\nonumber\\[1ex]&\quad
+ 2 (\bp_a \cdot \bn_{ba}) (\bp_b \cdot \bn_{ba}) (\bp_a \cdot \bp_b)
 - (\bp_a \cdot \bn_{ba})^2 \bp_b^2\Big]
\nonumber\\[1ex]&\quad
+ \Big[
\bp_a^2 \bp_b^2 - 3  \bp_a^2 (\bp_b \cdot \bn_{ba})^2 +(\bp_a \cdot \bn_{ba})^2 (\bp_b \cdot \bn_{ba})^2 
\nonumber\\[1ex]&\quad
+ 8 (\bp_a \cdot \bn_{ba}) (\bp_b \cdot \bn_{ba}) (\bp_a \cdot \bp_b)
- 3 (\bp_a \cdot \bn_{ba})^2 \bp_b^2 \Big]y_{ba}
\Bigg\}.
\end{align}
This is the Hamiltonian for a many-point-mass system through 1PM approximation, i.e., including all terms linear in $G$.
It is given in closed form and entirely in terms of the canonical variables of the particles.

The usefulness of that Hamiltonian has been proved in several applications
(see, e.g., \citealp{foffa-11}, \citealp{jaranowski-12}, \citealp{foffa-13}, \citealp{damour2-16}, \citealp{feng-18}).
Especially in \cite{jaranowski-12} it was checked that the terms linear in $G$
in the 4PN-accurate ADM Hamiltonian derived there, are, up to adding a total time derivative,
compatible with the 4PN-accurate Hamiltonian which can be obtained from the exact 1PM Hamiltonian \eqref{H1PM}.
Let us also note that \cite{damour2-16} has shown that, after a suitable canonical transformation,
the rather complicated Hamiltonian \eqref{H1PM} is equivalent (modulo the EOB energy map)
to the much simpler Hamiltonian of a test particle moving in a (linearized) Schwarzschild metric.
The binary centre-of-mass 2PM Hamiltonian has been derived most recently by \cite{damour2-17} in an EOB-type form
and also the gravitational spin-orbit coupling in binary systems has been achieved at 2PM order by \cite{bini-18}
(for other 2PM results see, e.g., \citealp{bel-81}, \citealp{westpfahl-85}).

\section{Skeleton Hamiltonian for binary black holes}
\label{app:skeleton}

The skeleton approach to GR developed by \cite{faye-04},
is a truncation of GR such that an analytic PN expansion exists to arbitrary orders which, at the same time, is explicitly calculable.
The approach imposes the conformal flat condition for the spatial three-metric for all times 
(not only initially as for the Brill--Lindquist solution),
together with a specific truncation of the field-momentum energy density.
It exactly recovers the general relativity dynamical equations in the limits of test-body and 1PN dynamics.
The usefulness of the skeleton approach in the construction of initial data needed
for numerical solving binary black hole dynamics was studied by \cite{bode-09}.

The conformally flat metric 
\be
\gamma_{ij} = (1+\frac{1}{8}\phi)^4 \delta_{ij}
\ee
straightforwardly results in maximal slicing, using the ADM coordinate conditions, 
\be
\pi^{ij}\gamma_{ij} = 2 \sqrt{\gamma} \gamma^{ij} K_{ij} = 0.
\ee
Our coordinates fit to the both ADM and Dirac coordinate conditions. 
The momentum constraint equations now become
\be
\pi^j_{i,\,j} = -\frac{8 \pi G}{c^3} \sum_a p_{ai}\delta_a.
\ee
The solution of these equations is constructed under the condition that $\pi^j_{i}$ is purely longitudinal, i.e.,
\be
\pi^j_{i} = \partial_i V_j + \partial_j V_i - \frac{2}{3} \delta_{ij} \partial_l V_l.
\ee
This condition is part of the definition of the skeleton model. 

Furthermore, in the Hamiltonian constraint equation, which in our case reads
\be
\Delta\phi = - \frac{ \pi^j_i \pi^i_j }{(1+\frac{1}{8}\phi)^{7} } -\frac{16\pi G}{c^2}\sum_a \frac{ m_a\delta_a}{ (1+\frac{1}{8}\phi)}\,
\biggl ( 1+ \frac{p_a^2}{(1+\frac{1}{8}\phi)^4m_a^2c^2} \biggr )^{1/2}\,,
\ee
a truncation of the numerator of the first term is made in the following form
\be
\pi^j_i \pi^i_j \equiv -2 V_j\partial_i\pi^i_j + \partial_i(2V_j\pi^i_j) \,
 \rightarrow -2 V_j\partial_i\pi^i_j=\frac{16\pi G}{c^3}\sum_a p_{aj}V_j\delta_a\,,
\ee
i.e., dropping from $\pi^j_i \pi^i_j$ the term $\partial_i(2V_j\pi^i_j)$.
This is the second crucial truncation condition additional to the conformal flat one.
Without this truncation neither an explicit analytic solution can be constructed nor a PN expansion is feasible.
From \cite{jaranowski-98,jaranowski-98-e}, it is known that at the 3PN level the $h^{\rm TT}_{ij}$-field is needed
to make the sum $\pi^j_i \pi^i_j$ analytic in $1/c$.

With the aid of the ansatz
\be
\phi = \frac{4G}{c^2}\sum_a \frac{\alpha_a}{r_a}
\ee
and by making use of dimensional regularization,
the energy and momentum constraint equations result in an algebraic equation for $\alpha_a$ of the form \citep{faye-04},
\be
\alpha_a = \frac{m_a}{\dst 1+\frac{G\alpha_b}{2r_{ab}c^2}}
\left[1 + \frac{p_a^2/(m_a^2 c^2)}{\dst\left(1+\frac{G\alpha_b}{2r_{ab}c^2} \right)^{4}} \right]^{1/2}
+ \frac{p_{ai} V_{ai}/c}{\dst\left(1+\frac{G\alpha_b}{2r_{ab}c^2}\right)^{7}},
\quad b \ne a.
\ee
With these inputs the skeleton Hamiltonian for binary black holes results in 
\be
H_{\rm sk} = -\frac{c^4}{16\pi G}\int\md^3x \,\Delta\phi = \sum_a \alpha_ac^2.
\ee
The Hamilton equations of motion read
\be
\dot{\bf x}_a = \frac{\partial H_{\rm sk}}{\partial\pa},
\quad
\dot\vecp_a = -\frac{\partial H_{\rm sk}}{\partial\xa}.
\ee

We will present the more explicit form of the binary skeleton Hamiltonian
in the centre-of-mass reference frame of the binary,
which is defined by the equality $\vecp_1+\vecp_2=\mathbf{0}$.
We define
\be
\vecp \equiv \vecp_1 = -\vecp_2,
\quad
\vecr \equiv \mathbf{x}_1 - \mathbf{x}_2,
\quad
r \equiv |\vecr|.
\ee
It is also convenient to introduce dimensionless quantities\footnote{
Let us note the they differ from the reduced variables introduced in Sect.\ \ref{sec:PNbinaries} in Eq.~\eqref{defRedVar}.}
(here $M\equiv m_1+m_2$ and $\mu\equiv m_1m_2/M$)
\be
\hat\vecr \equiv \frac{\vecr c^2}{GM},
\quad
\hat{\vecp} \equiv \frac{\vecp}{\mu c},
\quad
\mathbf{\hat p}^2 = \hat p_r^2 + \hat j^2/\hat r^2
\quad\textrm{with}\quad
\hat p_r \equiv \frac{p_r}{\mu c}
\quad\textrm{and}\quad
\hat j \equiv \frac{Jc}{GM\mu},
\ee
where $p_r\equiv \mathbf{p} \cdot \mathbf{r}/r$ is the radial linear momentum
and $\mathbf{J}\equiv\mathbf{r} \times \mathbf{p}$ is the orbital angular momentum in the centre-of-mass frame.
The reduced binary skeleton Hamiltonian ${\hat H}_{\rm sk}\equiv H_{\rm sk}/(\mu c^2)$
[it defines equations of motion with respect to dimensionless time $\hat t\equiv t c^3/(GM)$]
can be put into the following form \citep{gopakumar-08}:
\be
{\hat H}_{\rm sk} = 2\,\hat{r}(\psi_1 + \psi_2 - 2),
\ee
where the functions $\psi_1$ and $\psi_2$ are solutions of the following system of coupled equations
\begin{align}
\psi_{1} &= 1+\frac{ \chi_{-} }{4\, \hat r\, \psi_2} \, \sqrt{1+ \frac{ 4\,{\nu}^{2}
\left( {{\hat p_r}}^{2}+{\hat j}^2/{\hat r}^{2} \right) }{ \chi_{-}^2\, \psi_2^4}}
- \frac{ \left( 8\,{{ \hat p_r}}^{2}+7{\hat j}^2/{\hat r}^2 \right)
{\nu}^{2}}{8\, {\hat r}^{2}\psi_2^7 },
\\[1ex]
\psi_{2} &= 1+\frac{ \chi_{+} }{4\, \hat r\, \psi_1} \, \sqrt{1+ \frac{ 4\,{\nu}^{2}
\left( {{\hat p_r}}^{2}+{\hat j}^2/{\hat r}^2 \right) }{ \chi_{+}^2\, \psi_1^4}} - \frac{
\left( 8\,{{ \hat p_r}}^{2}+7{\hat j}^2/{\hat r}^2 \right) {\nu}^{2}}{8\, {\hat r}^{2}\psi_1^7},
\end{align}
where $ \chi_{-}\equiv1-\sqrt {1-4\,\nu}$ and $\chi_+\equiv1+\sqrt {1-4\,\nu}$, with $\nu\equiv\mu/M$.

Beyond the properties mentioned in the beginning,
the conservative skeleton Hamiltonian reproduces the Brill-Lindquist initial-value solution.
It is remarkable that the skeleton Hamiltonian allows a PN expansion in powers of $1/c^2$ to arbitrary orders.
The skeleton Hamiltonian thus describes the evolution of a kind of black holes
under both conformally flat condition and the condition of analyticity in $1/c^2$.
Along circular orbits the two-black-hole skeleton solution is quasistationary
and it satisfies the property of the equality of Komar and ADM masses \citep{komar-59,komar-63}.
Of course, gravitational radiation emission is not included.
It can, however, be added to some reasonable extent, see \cite{gopakumar-08}.

Restricting to circular orbits and defining $x\equiv(GM\omega/c^3)^{2/3}$,
where $\omega$ is the orbital angular frequency, the skeleton Hamiltonian reads explicitly to 3PN order,
\begin{align}
{\hat H}_{\rm sk} &= -\frac{x}{2} + \bigg(\frac{3}{8}+\frac{\nu}{24}\bigg) x^2
+ \bigg(\frac{27}{16}+\frac{29}{16}\nu-\frac{17}{48}\nu^2\bigg) x^3
\nonumber\\[1ex]&\quad
+ \bigg(\frac{675}{128}+\frac{8585}{384}\nu-\frac{7985}{192}\nu^2
+\frac{1115}{10368}\nu^3\bigg) x^4 + \mathcal{O}(x^{5}).
\end{align}
In \cite{faye-04}, the coefficients of this expansion are given to the order $x^{11}$ inclusively.
We recall that the 3PN-accurate result of general relativity reads [cf.\ Eq.~\eqref{COEx}],
\begin{align}
{\hat H}_{\le \rm 3PN} &= -\frac{x}{2} + \bigg( \frac{3}{8} + \frac{\nu}{24} \bigg) x^2
+ \bigg( \frac{27}{16} - \frac{19}{16}\nu + \frac{1}{48}\nu^2 \bigg) x^3
\nonumber\\[1ex]&\quad
+ \Bigg( \frac{675}{128} + \bigg(\frac{205}{192}\pi^2-\frac{34445}{1152}\bigg)\nu 
+ \frac{155}{192} \nu^2 + \frac{35}{10368} \nu^3 \Bigg) x^4.
\end{align}
In the Isenberg--Wilson--Mathews approach to general relativity only the conformal flat condition is employed.
Through 2PN order, the Isenberg--Wilson--Mathews energy of a binary is given by
\be
{\hat H}_{\rm IWM} = -\frac{x}{2} + \bigg( \frac{3}{8} + \frac{\nu}{24} \bigg) x^2
+ \bigg( \frac{27}{16} - \frac{39}{16}\nu - \frac{17}{48} \nu^2 \bigg) x^3.
\ee
The difference between ${\hat H}_{\rm IWM}$ and ${\hat H}_{\rm sk}$ shows
the effect of truncation in the field-momentum part of ${\hat H}_{\rm sk}$ through 2PN order
and the difference between ${\hat H}_{\rm IWM}$ and ${\hat H}_{\le \rm 3PN}$ reveals the effect of conformal flat truncation.
In the test-body limit, $\nu =0$, the three Hamiltonians coincide.

\begin{acknowledgements}

\sloppy We gratefully acknowledge our long-standing collaboration with Thibault Damour.
We thank him for a critical and most constructive reading of the manuscript we had started with.
Thanks also go to Jan Steinhoff for his delivery of LaTeX files with the highest-order conservative Hamiltonians.
Thankfully acknowledged are the critical remarks by anonymous referees which improved the presentations in the article.
The work of P.J.\ was supported in part by the Polish NCN Grants Nos.\ 2014/14/M/ST9/00707 and 2023/49/B/ST9/02777.

\end{acknowledgements}

\phantomsection
\addcontentsline{toc}{section}{References}
\bibliographystyle{spbasic-FS}  
\bibliography{LRRSJarXiv_refs_v5} 

\end{document}